\documentclass[11pt]{article}
\pdfoutput=1 

\usepackage{jheppub} 

\usepackage[T1]{fontenc} 

\usepackage[export]{adjustbox}
\usepackage{color}
\usepackage{xcolor}
\usepackage{autobreak}
\usepackage{multirow}
\usepackage{subfig}
\usepackage{amsmath}
\usepackage[vcentermath]{youngtab}
\usepackage{pifont}
\usepackage{extarrows}
\usepackage{makecell}
\usepackage{diagbox}
\usepackage{lscape}
\usepackage{ulem} 
\usepackage{bbm}
\usepackage{diagbox}
\usepackage{mathtools}

\usepackage{tikz}
\usepackage[compat=1.1.0]{tikz-feynhand}
\usepackage{tikz-cd}
\usetikzlibrary{plotmarks}

\tikzset{thick bos/.style={very thick,decorate,decoration={snake,amplitude=1pt,segment length=4pt}}}
\tikzset{thick sca/.style={very thick,dash=on 4pt off 2pt phase 0pt}}

\newcommand{\sca}[1] {\draw[brown,thick sca] (#1)--(v1)} 
\newcommand{\fer}[3]{\propag[#1,fer] (#2)--(#3);
\draw[#1,style=very thick] (#2)--(#3);} 
\newcommand{\antfer}[3]{\propag[#1,antfer] (#2)--(#3);
\draw[#1,style=very thick] (#2)--(#3);} 
\newcommand{\ferflip}[4]{
\draw[#3,very thick,rotate=#2] (0.7*#1,0)--(0.46*#1,0);
\draw[#4,very thick,decoration={markings,mark=at position 0.56 with {\arrow{Triangle[length=4pt,width=4pt]}}},postaction={decorate},rotate=#2] (0.46*#1,0)--(v1);
\draw[very thick,rotate=#2] plot[mark=x,mark size=2.5] coordinates {(0.46*#1,0)};} 
\newcommand{\antferflip}[4]{
\draw[#3,very thick,rotate=#2] (0.7*#1,0)--(0.46*#1,0);
\draw[#4,very thick,decoration={markings,mark=at position 0.82 with {\arrow{Triangle[length=4pt,width=4pt]}}},postaction={decorate},rotate=#2] (v1)--(0.46*#1,0);
\draw[very thick,rotate=#2] plot[mark=x,mark size=2.5] coordinates {(0.46*#1,0)};
} 
\newcommand{\bos}[2] {\draw[#2,thick bos] (#1)--(v1)} 
\newcommand{\bosflip}[4] {
\begin{scope}[rotate=#2]
\clip (0,-0.1) rectangle (0.44*#1,0.1); 
\draw[#3,thick bos] (0,0)--(#1,0);
\end{scope}
\begin{scope}[rotate=#2]
\clip (0.44*#1,-0.1) rectangle (0.7*#1,0.1); 
\draw[#4,thick bos] (0,0)--(#1,0);
\end{scope}
\draw[very thick,rotate=#2] plot[mark=x,mark size=2.5] coordinates {(0.43*#1,0)};
} 
\newcommand{\bosflipbrown}[2] {
\draw[brown,thick bos,rotate=#2] (0.20*#1,0)--(v1);
\draw[brown,thick bos,rotate=#2] (0.43*#1,0)--(0.20*#1,0);
\draw[brown,thick bos,rotate=#2] (0.7*#1,0)--(0.43*#1,0);
\draw[very thick,rotate=#2] plot[mark=x,mark size=2.5] coordinates {(0.43*#1,0)};
\draw[very thick,rotate=#2] plot[mark=x,mark size=2.5] coordinates {(0.20*#1,0)};
} 
\newcommand{\bosflipflip}[5] {
\begin{scope}[rotate=#2]
\clip (0,-0.1) rectangle (0.35*#1,0.1); 
\draw[#3,thick bos] (0,0)--(#1,0);
\end{scope}
\begin{scope}[rotate=#2]
\clip (0.35*#1,-0.1) rectangle (0.55*#1,0.1); 
\draw[#4,thick bos] (0,0)--(#1,0);
\end{scope}
\begin{scope}[rotate=#2]
\clip (0.55*#1,-0.1) rectangle (0.7*#1,0.1); 
\draw[#5,thick bos] (0,0)--(#1,0);
\end{scope}
\draw[very thick,rotate=#2] plot[mark=x,mark size=2.5] coordinates {(0.55*#1,0)};
\draw[very thick,rotate=#2] plot[mark=x,mark size=2.5] coordinates {(0.35*#1,0)};
} 

\newcommand{\Ampone}[3] {
\begin{tikzpicture}[baseline=-0.1cm] \begin{feynhand}
\setlength{\feynhandarrowsize}{4pt}
\vertex [particle] (i1) at (#1,0) {$#2$}; 
\vertex (v1) at (0,0);
#3;
\vertex[dot] (v1) at (0,0) {};
\end{feynhand} \end{tikzpicture}
}

\newcommand{\Ampthree}[6] {
\begin{tikzpicture}[baseline=-0.1cm] \begin{feynhand}
\setlength{\feynhandarrowsize}{4pt}
\vertex [particle] (i1) at (-1.01,0) {$#1$}; 
\vertex [particle] (i2) at (0.579,0.827) {$#2$}; 
\vertex [particle] (i3) at (0.579,-0.827) {$#3$}; 
\vertex (v1) at (0,0);
#4;
#5;
#6;
\end{feynhand} \end{tikzpicture}
}

\newcommand{\Ampfour}[8] {
\begin{tikzpicture}[baseline=-0.1cm] \begin{feynhand}
\setlength{\feynhandarrowsize}{4pt}
\vertex [particle] (i1) at (-0.9,0.9) {$#1$};
\vertex [particle] (i2) at (-0.9,-0.9) {$#2$};
\vertex [particle] (i3) at (0.9,-0.9) {$#3$};
\vertex [particle] (i4) at (0.9,0.9) {$#4$};
\vertex (v1) at (0,0);
#5;
#6;
#7;
#8;
\end{feynhand} \end{tikzpicture}
}

\usepackage{eucal} 

\usepackage[utf8]{inputenc}
\usepackage{times} 
\usepackage{mathptmx} 

\newcommand{\beq}{\begin {equation}}  
\newcommand{\eeq}{\end   {equation}} 
\newcommand{\bea}{\begin {eqnarray}} 
\newcommand{\eea}{\end   {eqnarray}}  
\newcommand{\baa}{\begin {array}   } 
\newcommand{\eaa}{\end   {array}   }     
\newcommand{\bit}{\begin {itemize} }
\newcommand{\eit}{\end   {itemize} }
\newcommand{\be }{\begin {equation}} 
\newcommand{\ee }{\end   {equation}}

\allowdisplaybreaks

\title{\boldmath Massless-Massive Amplitude Correspondence II: Constructive Massive Amplitudes in Standard Model}

\author[a,b,e]{Yu-Han Ni, }
\author[a]{Chao Wu, }
\author[a, b, c, d]{Jiang-Hao Yu, }

\affiliation[a]{Institute of Theoretical Physics, Chinese Academy of Sciences, Beijing 100190, China}
\affiliation[b]{School of Physical Sciences, University of Chinese Academy of Sciences, Beijing 100049, China}
\affiliation[c]{School of Fundamental Physics and Mathematical Sciences, Hangzhou Institute for Advanced
Study, UCAS, Hangzhou 310024, China}
\affiliation[d]{International Center for Theoretical Physics Asia-Pacific, Beijing/Hangzhou, China}
\affiliation[e]{School of Science and Engineering, Chinese University of Hong Kong, Shenzhen, Shenzhen 518172, China}

\emailAdd{niyuhan@cuhk.edu.cn}
\emailAdd{wuch7@itp.ac.cn}
\emailAdd{jhyu@itp.ac.cn}

\abstract{

In the minimal helicity-chirality formalism, we systematically construct higher-point massive amplitudes from the fundamental building blocks: the contact three-point and four-point massive amplitudes. The inclusion of four-point contact amplitudes is essential to maintain gauge invariance in the spontaneously broken Standard Model. We construct all the standard model massive contact amplitudes and identify the physical light-cone gauge nature of massive amplitudes. Then only using the contact minimal helicity-chirality amplitudes {\it at the leading order}, we show both bootstrap techniques and on-shell recursion relations can be utilized to compute higher-point massive amplitudes. This provides a systematic framework for constructing various higher-point electroweak amplitudes, analogous to established on-shell methods for massless theories. Finally by deforming the gauge-invariant $n$-point amplitudes, we extend the massless-massive correspondence from three-and-four point contact amplitudes to general $n$-point factorized amplitudes.

}

\begin{document} 
\maketitle
\flushbottom


\section{Introduction}
\label{sec:intro}


Over recent decades, the on-shell method has become a central and powerful technique in scattering amplitude calculations. It uses the spinor-helicity formalism to represent scattering amplitudes purely in terms of on-shell states, which automatically enforces equations of motion and makes gauge invariance explicit for massless particles~\cite{Parke:1986gb,Xu:1986xb,Bern:1996je,Dixon:1996wi,Elvang:2013cua,Cheung:2017pzi,Travaglini:2022uwo,Badger:2023eqz}. This physical transparency stems from a key distinction: while a massless gauge boson possesses only two physical helicity states, its corresponding gauge field in a local Lagrangian carries redundant degrees of freedom. The on-shell method, by construction, avoids these off-shell redundancies. Consequently, in contrast to the traditional Feynman diagrammatic approach built from Lagrangians and off-shell fields, the on-shell methodology offers a more efficient and physically transparent route to computing scattering amplitudes in gauge theories and gravity.

In the real world, scattering processes involve a variety of massive particles, creating a demand for on-shell methods to compute massive amplitudes. Unlike massless particles, which have one or two helicity states, massive spin-$s$ particles possess $(2s+1)$ physical states. Consequently, for massive particles there is no distinction between on-shell and off-shell degrees of freedom, nor an underlying gauge symmetry to exploit. This would seem to negate the primary advantages of the on-shell approach, potentially rendering it no more efficient than traditional Feynman diagram methods. However, an important exception exists when the massive theory descends from a massless ultraviolet theory via spontaneous symmetry breaking—as in the Standard Model (SM) and its extensions. In these theories, the underlying gauge symmetry, though hidden, provides a powerful organizing principle for simplifying scattering amplitudes.

In the on-shell method, the fundamental three-point massless amplitudes are fully determined by their scaling behavior under little group transformations~\cite{Benincasa:2007xk,Witten:2003nn}. Guided by first principles, symmetry, locality, unitarity, and analyticity, tree-level amplitudes can be constructed recursively from these three-point building blocks to arbitrary multiplicity and extended to loop level via generalized unitarity, which builds loop integrands directly from tree-level amplitudes~\cite{Bern:1994zx,Bern:1994cg,Britto:2004nc,Britto:2004ap,Britto:2005fq}. This foundational logic has been successfully extended to massive scattering amplitudes. Arkani-Hamed, Huang, and Huang (AHH)~\cite{Arkani-Hamed:2017jhn} introduced $SU(2)$ little-group covariant states within a spinor-helicity formalism for massive particles, enabling a systematic enumeration of three-point amplitudes through little-group. {The AHH formalism has since been applied to a wide range of areas, including electroweak theory~\cite{Franken:2019wqr, Bachu:2019ehv, Ballav:2020ese, Wu:2021nmq, Ballav:2021ahg, Liu:2022alx, Bachu:2023fjn, Ema:2024rss}, effective field theories~\cite{Shadmi:2018xan, Aoude:2019tzn, Durieux:2019eor, Ma:2019gtx, Durieux:2019siw, Li:2020gnx, Li:2020xlh, Li:2020tsi, Li:2020zfq, Durieux:2020gip, Li:2021tsq, AccettulliHuber:2021uoa, Dong:2021vxo, Li:2022tec, Balkin:2021dko, DeAngelis:2022qco, Dong:2022mcv, Ren:2022tvi, Liu:2023jbq, Goldberg:2024eot}, and gravitational wave physics~\cite{Guevara:2018wpp, Chung:2018kqs, Guevara:2019fsj, Bern:2019nnu, Maybee:2019jus, Arkani-Hamed:2019ymq}.} Bootstrap techniques, such as gluing these three-point amplitudes, were then employed to construct several four-point massive amplitudes, involving in massless particles in external leg~\cite{Bachu:2019ehv,AccettulliHuber:2021uoa,Liu:2022alx}. Subsequent work has further demonstrated that on-shell recursive relations, most notably the BCFW recursion, can be effectively generalized to compute scattering amplitudes involving both external massless and massive particles~\cite{Ochirov:2018uyq,Franken:2019wqr,Aoude:2019tzn,Falkowski:2020aso,Ballav:2020ese,Ballav:2021ahg,Lazopoulos:2021mna,Wu:2021nmq,Campbell:2023fjg,Ema:2024vww,Ema:2024rss}.

The on-shell constructive approach encounters significant challenges in processes where all external particles are massive, particularly when at least one particle carries spin of one or higher. In such cases, standard techniques cannot be directly applied and require specific modifications. For instance, within bootstrap methods, high-energy unitarity constraints~\cite{Liu:2022alx} often necessitate introducing additional four-point contact amplitudes alongside the standard gluing procedure. Similarly, when applying BCFW recursion to amplitudes involving massive gauge bosons, one must carefully select distinct momentum shifts for longitudinal and transverse polarizations, while also imposing Ward identities to remove spurious boundary terms~\cite{Ema:2024vww,Ema:2024rss}. In previous work, we build the massless-massive correspondence for 3-point amplitude, and would extend this correspondence for general n-point amplitudes, allowing the full suite of techniques developed for massless amplitudes to be applied directly to the calculation of massive amplitudes.

In this work, we employ the Minimal Helicity-Chirality (MHC) formalism~\cite{Ni:2026zaa} to systematically construct higher-point massive amplitudes from their fundamental building blocks, the contact 3-point and 4-point massive amplitudes. The massive AHH amplitude can be decomposed into primary and descendant MHC amplitudes and the leading MHC amplitudes carry all the information of the massive AHH amplitudes. At the same time, the leading MHC amplitudes possess a definite helicity and are equivalent to massless amplitudes in the light-front gauge, thereby inheriting the calculational advantages of the massless theory. In the massless gauge theory, the on-shell calculation in the light-cone gauge needs the contact four-point amplitudes to ensure the gauge invariance. From massless-massive correspondence, the 3-point leading MHC amplitudes in the SM are the on-shell amplitudes within the light-cone gauge, and thus the AHH massive amplitudes in the SM should be viewed as the massive amplitudes in the light-cone gauge. Therefore, the contact 4-point massive amplitudes, should be naturally included in the massive higher-point construction.

Given the fact that both the contact 3-point and 4-point massive amplitudes are in the light-cone gauge, all of them should be treated as the fundamental building blocks for higher point amplitude construction~\footnote{No five-point or higher contact massive amplitudes are needed for renormalizable theories like the SM.}.
Once the correct building blocks are identified, both the bootstrap techniques and on-shell recursion relations can be adapted to compute higher-point massive amplitudes, analogous to massless constructive methods. This procedure applies to both the primary MHC amplitudes, and the AHH amplitudes. Compared to the massless construction, there are several differences: 
\begin{itemize}

    \item In the massless construction, the fundamental 3-point amplitudes are gauge invariant, and thus there is no need to introduce 4-point contact massless amplitudes to ensure gauge invariance. This is the advantage of the massless construction. On the other hand, the massless construction usually involves the $\phi^4$ interaction in the Higgs sector, which would induce non-vanishing boundary terms in the recursive construction.

    \item In the massive construction, the 3-point amplitudes are in the light-cone gauge, and thus 4-point contact massive amplitudes are introduced, but no higher contact one is needed for renormalizable SM. At the same time, the massive construction can avoid the $\phi^4$ interaction due to the little group covariance. Unlike independent massless helicity amplitudes, a single MHC amplitude implicitly encodes information about all helicity states of the massive particle. Therefore selecting the MHC helicity without the $\phi^4$ interaction would avoid such non-vanishing boundary term.   
    
\end{itemize}
Let us further specify the common feature and differences between the primary MHC amplitudes, and the AHH amplitudes:
\begin{itemize}

    \item Both the MHC and AHH amplitudes need the 4-point contact amplitudes as building blocks to obtain higher-point amplitudes in the bootstrap method. For the MHC amplitude construction, choosing one helicity to perform bootstrap is enough to determine the whole massive amplitudes. Since the full set of MHC amplitudes can be combined to form a single, little-group covariant massive amplitude,  obtaining the leading MHC amplitude in principle determines all others and allows for the recovery of the complete AHH amplitude. So the leading contact 3-and 4-pt MHC amplitudes are enough to serve as the building blocks for constructive SM. 

    \item   In the recursive methods for massive amplitudes, distinct momentum shifts are typically required for different vector boson modes. The MHC framework naturally accommodates different shifts for transverse and longitudinal modes. Furthermore, since little-group covariance ensures that calculating the amplitude for one specific MHC polarization is sufficient, the full amplitude can then be reconstructed without the separate shifts for all helicity states.

\end{itemize}
To demonstrate the advantage of these bootstrap and recursive techniques within the MHC formalism, we apply them to three exemplary massive scattering processes:  $ff \to Wh$, $WW \to hh$ and $WW \to ZZ$, and other processes and higher point calculation should follow the same procedure. Specifically, we calculate the $WW \to hh$ amplitudes in three different ways: the MHC bootstrap, the AHH bootstrap, and recursive relation.

Finally we have built the massless-massive matching for 3-point amplitudes through the MHC amplitude deformation. It should be extended to higher point amplitudes: first construct the n-point massless amplitudes, and then deform to the n-point leading MHC amplitudes via amplitude deformation. At the same time, this matching offers an alternative route for constructing massive amplitudes, but with some differences:
\begin{itemize}
    \item The building blocks for gluing amplitudes are different. In the massive constructive method, the building blocks are 3-point and 4-point contact massive amplitudes in the light-cone gauge. However, in the massless-massive matching, the building blocks are 3-point contact massless amplitudes, without the gauge dependence. 

    \item In the massive constructive method, there is no further treatments after construction. But in the massless-massive matching, constructing the higher point massless amplitudes is just the starting point, and further amplitude deformation to the MHC amplitudes in the light-cone gauge, with the pole separation, is needed. 
    
\end{itemize}
Furthermore, the massless-massive matching is for the leading MHC amplitudes. Beyond the leading matching, sub-leading contributions can be incorporated through established Higgsing rules, {which extend the on-shell Higgsing method introduced in Ref.~\cite{Balkin:2021dko}}. However, this extension requires further refinement: for processes involving internal massive propagators, a new set of internal particle Higgsing rules should be formulated and applied to correctly perform the matching throughout the entire amplitude.

The paper is organized as follows. Section 2 reviews the Minimal Helicity-Chirality (MHC) formalism, introducing MHC one-particle states, two-particle currents, and three-particle amplitudes. This establishes the massless-massive correspondence at the single-particle, two-particle, and three-particle levels. In Section 3, we turn to the SM, summarizing all three-point MHC amplitudes through the identification of conserved currents and the method of amplitude deformation. Section 4 then demonstrates how three-point MHC amplitudes can be "glued" together to construct four-point amplitudes. We further derive the complete set of four-point contact amplitudes in the light-cone gauge. Building on these results, Section 5 illustrates three complementary approaches to obtaining four-point amplitudes, by direct assembly of the basic three-point and contact four-point MHC and AHH amplitudes, and via BCFW recursion. In Section 6, we detail the procedure for obtaining the massless-massive matching through amplitude deformation and pole separation, and extend the matching to sub-leading contributions using Higgsing rules for both external and internal particles. Finally, Section 7 provides a summary and outlook.

\section{Minimal Helicity-Chirality Formalism}
\label{sec:spinor}

\subsection{Spinor-helicity and spin-transversality}

We employ the spinor-helicity formalism to construct scattering amplitudes, utilizing the isomorphism between the Lorentz group $SO(3,1)$ and $SL(2,\mathbb{C}) \sim SU(2)_l \times SU(2)_r$. The mapping from the fundamental representation of the $SO(3,1)$, such as the four-momentum $p^\mu$, to the fundamental representation of the $SU(2)_l \times SU(2)_r$ is implemented via the Pauli matrices $\sigma^\mu_{\alpha \dot{\alpha}}$. For a massless particle, the matrix $p_{\alpha \dot{\alpha}}$ is of rank one and can be factorized as the direct product of two Weyl spinors:
\begin{eqnarray}
p_{\alpha \dot{\alpha}} = p_\mu \sigma^{\mu} _{\alpha \dot{\alpha}} \equiv \lambda_{\alpha} \tilde{\lambda}_{\dot{\alpha}}, 
\end{eqnarray}
where $\lambda_\alpha$ transforms under $SU(2)_l$, and $\tilde{\lambda}_{\dot{\alpha}} $ under $SU(2)_r$.  
For a massive particle, the momentum matrix $\mathbf{p}_{\alpha \dot{\alpha}}$ is of rank two. This necessitates introducing an additional $SU(2)$ index $I = 1, 2$ to form a pair of massive spinors:
\begin{eqnarray}
\mathbf{p}_{\alpha \dot{\alpha}} \equiv \lambda_{\alpha}^I \tilde{\lambda}_{I\dot{\alpha}},
\label{eq:massivep}
\end{eqnarray}
where summation over the index $I$ is implied. The index $I$ labels the two spin states of a spin-$\frac12$ particle.

A single-particle state is specified by its momentum and spin, forming an induced representation of the Poincar\'e group. By definition, a little-group transformation leaves the momentum $p$ invariant while rotating the spinors. For a massless particle, the little group is $U(1)$ and acts as a phase rotation:
\begin{eqnarray}
\lambda_\alpha \rightarrow e^{-i\phi} \lambda_{\alpha},\ \tilde{\lambda}_{\dot{\alpha}} \rightarrow e^{i\phi} \tilde{\lambda}_{\dot{\alpha}},
\end{eqnarray}
where $\phi$ is the little-group parameter. Consequently, a massless state of helicity $h$ can be represented as:
\begin{equation}
|p,h\rangle=
\begin{cases}
\lambda_{\alpha_1}...\lambda_{\alpha_{2h}},\;\;\;\quad h\ge 0, \\
\tilde{\lambda}_{\dot{\alpha}_1}... \tilde{\lambda}_{\dot{\alpha}_{(-2h)}},\quad h<0,
\end{cases}
\end{equation}
establishing a one-to-one correspondence between the particle state with different helicity, and the spinor with different Lorentz representation.

For a massive particle, the little group is $SU(2)$. Its action on the massive spinors is given by an $SU(2)$ Wigner rotation $W$:
\begin{eqnarray} \label{eq:wmatrix}
\lambda_\alpha^I \rightarrow (W^{-1})^I_J \lambda_\alpha^J,\ \tilde{\lambda}_{\dot{\alpha},I} \rightarrow W_{I}^J \tilde{\lambda}_{\dot{\alpha},J},
\end{eqnarray}
where $\lambda_\alpha^I$ and $\tilde{\lambda}_{\dot{\alpha},I}$ belong to the left- and right-handed Lorentz representations, $(\frac12, 0)$ and $(0, \frac12)$, respectively. 
Analogous to the massless case, the massive spinors can be used to build Poincar\'e representations for a particle of spin $s$
\begin{equation}
|\mathbf{p},s\rangle=
\begin{cases}
\lambda^{(I_1}_{\alpha_1}\lambda^{I_2}_{\alpha_2}\dots\lambda^{I_{2s})}_{\alpha_{2s}},\\
\tilde\lambda^{(I_1}_{\dot\alpha_1}\lambda^{I_2}_{\alpha_2}\dots\lambda^{I_{2s})}_{\alpha_{2s}},\\
\cdots,\\
\tilde\lambda^{(I_1}_{\dot\alpha_1}\tilde\lambda^{I_2}_{\alpha_2}\dots\tilde\lambda^{I_{2s})}_{\alpha_{2s}},
\end{cases}
\end{equation}
where the indices $(I_1 I_2 \dots I_{2s})$ are fully symmetrized. This formulation makes the $SU(2)$ little-group structure manifest, ensuring that the massive state carries the correct $(2s+1)$ spin degrees of freedom. However, in the massive case, the spinors $\lambda^I$ and $\tilde{\lambda}^I$ are related by the equation of motion $\mathbf{p}_{\alpha \dot{\alpha}} \tilde{\lambda}^{\dot{\alpha} I} = \mathbf{m} \lambda_{\alpha}^{I}$. As a result, all representations in the set are equivalent under the little group. This implies that, unlike for massless particles, there is no one-to-one correspondence between a Lorentz representation, labeled by $(j_1, j_2)$, and a Poincar\'e representation, labeled by $|\mathbf{p}, s\rangle$. For example, for a massive spin-$\frac12$ particle, one Poincar\'e representation would have two corresponding Lorentz representations
\begin{equation}
\begin{tabular}{c|c|c|c }
\hline
Poincaré: $|\mathbf{p},s\rangle$ & Lorentz: $|j_1,j_2\rangle$ & massive spinor & chirality \\
\hline
\multirow{2}{*}{$|\mathbf{p},\frac{1}{2}\rangle$} & $|\frac{1}{2},0\rangle$ & $\lambda_{\alpha}^I$ & $c = -
\frac12$ \\
& $|0,\frac{1}{2}\rangle$ & $\tilde\lambda_{\dot\alpha}^I$ & $c = +\frac12$ \\
\hline
\end{tabular}
\end{equation}
where $c = \pm \frac12$ denotes the left and right chirality, determined by their Lorentz representations. We classify these chirality representations as follows,
\begin{equation} 
c\equiv j_2-j_1\quad\to\quad
\left\{\begin{aligned}
&c<0:\quad \text{chiral representation},\\
&c=0:\quad \text{symmetric representation},\\
&c>0:\quad \text{anti-chiral representation}.\\
\end{aligned}\right.
\end{equation}
For spin-$\frac{1}{2}$ particles, both chiral and anti-chiral representations exist.

To establish the massless-massive correspondence, it is crucial to maintain a one-to-one mapping between the Lorentz and Poincaré representations for massive particles. This can be achieved by introducing an additional $U(1)$ transformation~\cite{Ni:2024yrr, Ni:2025xkg}, and under which the massive spinors transform as
\begin{equation}
\lambda_{\alpha}^{I}\rightarrow e^{-\frac{i}{2}\phi}\lambda_{\alpha}^{J}\;,\quad\tilde{\lambda}_{\dot{\alpha}I}\rightarrow e^{\frac{i}{2}\phi}\tilde{\lambda}_{\dot{\alpha}J}. 
\end{equation}
Correspondingly the mass parameter would become complex and transform under this additional $U(1)$ symmetry
\begin{equation}
m \equiv \frac{1}{2}\lambda_{\alpha I}\lambda^{\alpha I} \rightarrow e^{-i\phi}m,\quad \tilde{m} \equiv \frac{1}{2} \tilde{\lambda}_{\dot{\alpha}I}\tilde{\lambda}^{\dot{\alpha}I}\rightarrow e^{i\phi}\tilde{m}.
\end{equation}
This $U(1)$ symmetry identify the quantum number {\it transversality} $t$. In this case, the Lorentz group extends to $U(1) \times SO(3,1)$, with the associated representation now characterized by $[t,j_1,j_2]$, where $|t| \leq s$.

Similar to the Lorentz representation, the Poincare symmetry is also extended. The $U(1)$ generator $D_-$, together with $m$ and $\tilde{m}$, composes the Lie algebra of a $ISO(2)$ group:
\begin{equation}
    [D_-,m]=-m\;,\quad [D_-,\tilde{m}]=+\tilde{m}\;,\quad [m,\tilde{m}]=0.
\end{equation}
This extends the Poincare symmetry to be $ISO(3,1) \times ISO(2)$. Note that the $ISO(2)$ symmetry is restricted due to the on-shell condition $\mathbf{p}^2 = \mathbf{m}^2 = m \tilde{m}$. The massive little group becomes $U(2)=SU(2)\times U(1)$ because $\mathbf{p}$ is also invariant under the $U(1)$. Correspondingly, the particle states become $|\mathbf{p},s,t\rangle$.
 When $t=j_2-j_1$, this establishes a one-to-one correspondence between the extended Poincaré and Lorentz representations $[t,j_1,j_2]\sim |\mathbf{p},s,t\rangle$ with
\begin{equation}
\begin{tabular}{c|c|c|c}
\hline
Poincaré: $|\mathbf{p},s, t\rangle$ & Lorentz: $[t,j_1,j_2]$ & massive spinor & chirality \\
\hline
$|\mathbf{p},\frac{1}{2},\frac{1}{2}\rangle$ & $[\frac{1}{2},\frac{1}{2},0]$ & $\lambda_{\alpha}^I$ & $c= -\frac12$ \\
$|\mathbf{p},\frac{1}{2},-\frac{1}{2}\rangle$ & $[-\frac{1}{2},0,\frac{1}{2}]$ & $\tilde\lambda_{\dot\alpha}^I$ & $c = +\frac12$ \\
\hline
\end{tabular}
\end{equation}
where the transversality $t$ plays the role of the chirality in the representation of a particle states. 
A general spin-$s$ particle state $|\mathbf{p},s, t\rangle$ can be presented by the massive spinors, transforming under the $[t, j_1, j_2]$ irrep of the $SO(3,1) \times U(1)$
\begin{equation}
|\mathbf{p},s,t\rangle=
\begin{cases}
\lambda^{(I_1}_{\alpha_1}\lambda^{I_2}_{\alpha_2}\dots\lambda^{I_{2s})}_{\alpha_{2s}}, & t = -s,\\
\tilde\lambda^{(I_1}_{\dot\alpha_1}\lambda^{I_2}_{\alpha_2}\dots\lambda^{I_{2s})}_{\alpha_{2s}},& t = -(s-1),\\
\cdots,\\
\tilde\lambda^{(I_1}_{\dot\alpha_1}\tilde\lambda^{I_2}_{\alpha_2}\dots\tilde\lambda^{I_{2s})}_{\alpha_{2s}}, & t =s.
\end{cases}
\end{equation}

The $ISO(2)$ symmetry suggests the particle states $|\mathbf{p},s,t\rangle$ have both the primary and descendant representations. For a spin-$\frac12$ particle, the primary representations are the massive spinors $\lambda_\alpha^I$ and $\tilde{\lambda}_{\dot{\alpha}}^I$, belonging to different chirality. The descendant representations are expressed as
\begin{equation}
\begin{tabular}{c|c|c|c}
\hline
Poincaré: $|\mathbf{p},s, t\rangle$ & Lorentz: $[t,j_1,j_2]$ & massive spinor & chirality \\
\hline
$|\mathbf{p},\frac{1}{2},-\frac{1}{2}\rangle$ & $[-\frac{1}{2}\frac{1}{2},0]$ & $\tilde{m}\lambda_{\alpha}^I$ & $c= -\frac12$ \\
$|\mathbf{p},\frac{1}{2},\frac{1}{2}\rangle$ & $[\frac{1}{2},0,\frac{1}{2}]$ & $m\tilde\lambda_{\dot\alpha}^I$ & $c = +\frac12$ \\
\hline
\end{tabular}
\end{equation}
where the transversality is flipped but the chirality does not change. Therefore, the particle states contains both the primary and descendant spinors
\begin{equation}
\begin{tabular}{c|c|c}
\hline
& $c = -\frac12 $ & $c = +\frac12 $ \\
\hline
primary & $\lambda^I_\alpha$ ($t = -\frac12$)& $\tilde\lambda^I_{\dot\alpha}$ ($t = \frac12$) \\
descendant & $\tilde{m}\lambda^I_\alpha$ ($t = +\frac12$) &  $m \tilde\lambda^I_{\dot\alpha}$ ($t = -\frac12$) \\
\hline
\end{tabular}
\end{equation}
which belongs to different transversality and chirality. Given $t \le s$, there is no 2nd descendant representation for the spin-half particle. 
For spin-one particle, the particle states are
\begin{eqnarray}
\begin{tabular}{c|ccc}
\hline
\diagbox[width=2.7cm]{$\Delta$}{$c$} & $-1$ & $0$ & $1$ \\
\hline
primary & $\lambda^{(I}_{\alpha} \lambda^{J)}_{\beta}$ ($t= -1$) & $\lambda^{(I}_{\alpha} \tilde{\lambda}^{J)}_{\dot{\beta}}$ ($t= 0$) & $\tilde{\lambda}^{(I}_{\dot{\alpha}} \tilde{\lambda}^{J)}_{\dot{\beta}}$ ($t= 1$) \\
1st descendant & $\tilde{m} \lambda^{(I}_{\alpha} \lambda^{J)}_{\beta}$ ($t= 0$)  & $m \lambda^{(I}_{\alpha} \tilde{\lambda}^{J)}_{\dot{\beta}}$ ($-1$),$\tilde{m} \lambda^{(I}_{\alpha} \tilde{\lambda}^{J)}_{\dot{\beta}}$ ($+1$) & $ \ m \tilde{\lambda}^{(I}_{\dot{\alpha}} \tilde{\lambda}^{J)}_{\dot{\beta}}$  ($t= 0$) \\
2nd descendant & $\tilde{m}^2 \lambda^{(I}_{\alpha} \lambda^{J)}_{\beta}$ ($t= 1$) & ${\color{gray} \mathbf{m}^2 \lambda^{(I}_{\alpha} \tilde{\lambda}^{J)}_{\dot{\beta}}}$ & $m^2 \tilde{\lambda}^{(I}_{\dot{\alpha}} \tilde{\lambda}^{J)}_{\dot{\beta}}$  ($t= -1$)\\
\hline
\end{tabular}
\end{eqnarray}
The one colored in gray, $\mathbf{m}^2 \lambda^{(I}_{\alpha} \tilde{\lambda}^{J)}_{\dot{\beta}}$, is considered equivalent to $\lambda^{(I}_{\alpha} \tilde{\lambda}^{J)}_{\dot{\beta}}$, because we mode out the mass shell $\mathbf{m}^2$ for massive representations. For spin-one particle, since $t \le s$, there is no 3rd descendant representation.

For the primary representations in a particle states, they can be related by introducing new generators $T^{+}_{\alpha \dot{\alpha}}\equiv \tilde{\lambda}_{\dot{\alpha}}^{I}\frac{\partial}{\partial \lambda^{\alpha I}}$ and $
T^{-}_{\alpha \dot{\alpha}}\equiv \lambda_{\alpha}^{I}\frac{\partial}{\partial \tilde{\lambda}^{\dot{\alpha}I}}$:
\begin{equation}
T^{+}\circ [t,j_1,j_2]=[t+1,j_1,j_2], \ \  
T^{-}\circ [t,j_1,j_2]=[t-1,j_1,j_2]. \nonumber
\end{equation}
Similarly for each descendant representation. 
Note that these generators play different roles than the equation of motion (EOM), in which the transversality remains unchanged but the chirality is flipped
\begin{equation} \begin{aligned} \label{eq:EOM}
t&=+\frac{1}{2}:& \mathbf{p}_{\alpha \dot{\alpha}} \tilde{\lambda}^{\dot{\alpha} I} &= \tilde{m} \lambda_{\alpha}^{I}, & [\frac12, 0,\frac12] &\to [\frac12, \frac12,0],\\
t&=-\frac{1}{2}:& \mathbf{p}_{\alpha \dot{\alpha}} \lambda^{\alpha I} &= -m \tilde{\lambda}_{\dot{\alpha}}^{I}, & [-\frac12, \frac12,0] &\to [-\frac12, 0, \frac12].    
\end{aligned} \end{equation}
These EOM would relate different chirality representations in the same $t$ states. Similar to the generators $T^\pm$, we can treat $m$ and $\tilde{m}$ as new generators, which trivialize the equation of motion.

Consider an $n$-particle scattering process involving massless particles, with the corresponding amplitude denoted by $\mathcal{A}(1^{h_1},\dots,n^{h_n})$, where $h_i$ is the helicity of particle $i$. The amplitude must respect covariance under little-group scaling transformations, which act as
\begin{eqnarray}
\mathcal{A}(1^{h_1},\dots,n^{h_n}) \;\longrightarrow\; e^{2i h_1 \phi_1} \cdots e^{2i h_n \phi_n} \,\mathcal{A}(1^{h_1},\dots,n^{h_n}),
\end{eqnarray}
where $\phi_i$ is the scaling parameter for particle $i$. In a general theory, the mass dimension of an amplitude is not fixed by helicity alone, since the number of momentum factors is unconstrained. However, within the renormalizable framework of the Standard Model (SM), amplitudes are required to satisfy the specific dimensional constraint
\begin{eqnarray}\label{eq:dimension}
\dim\big[\mathcal{A}(1,\dots,n)\big] = 4 - n,
\end{eqnarray}
for both massless and massive cases. If an amplitude is expressed in its simplest kinematic form and still violates Eq.~\eqref{eq:dimension}, it must be excluded on physical grounds: $\dim[\mathcal{A}] < 4-n$ indicates a non-local interaction, while $\dim[\mathcal{A}] > 4-n$ signals a violation of unitarity or the presence of higher-dimensional operators beyond the renormalizable SM.

Three-point massless amplitudes are fully constrained by spacetime symmetry, specifically, by their transformation properties under the $U(1)_{\text{LG}}$ little group, along with the requirement of locality. These amplitudes serve as the fundamental building blocks for recursively constructing higher-multiplicity scattering amplitudes. For massless three-particle kinematics, momentum conservation forces the holomorphic spinors to be pairwise proportional, $\lambda_1 \propto \lambda_2 \propto \lambda_3$, or, equivalently, the anti‑holomorphic spinors to be pairwise proportional, $\tilde{\lambda}_1 \propto \tilde{\lambda}_2 \propto \tilde{\lambda}_3$. Consequently, a three‑point amplitude can depend only on angle brackets $\langle i j \rangle$ or only on square brackets $[i j]$. Combining this observation with little‑group covariance and dimensional analysis uniquely determines the functional form of the amplitude, up to an overall coupling constant:
\begin{eqnarray} \label{eq:scaling3pt}
\mathcal{A}(1^{h_1},2^{h_2},3^{h_3}) =
\begin{cases}
\langle12\rangle^{h_1+h_2-h_3}\,
\langle23\rangle^{h_2+h_3-h_1}\,
\langle31\rangle^{h_3+h_1-h_2}, & h_1+h_2+h_3 < 0, \\[6pt]
[12]^{-h_1-h_2+h_3}\,
[23]^{-h_2-h_3+h_1}\,
[31]^{-h_3-h_1+h_2}, & h_1+h_2+h_3 > 0,
\end{cases}
\end{eqnarray}
where we employ the standard spinor‑helicity notation $|i\rangle_{\alpha} \equiv {\lambda_i}_{\alpha}$ and $|i]^{\dot{\alpha}} \equiv {\tilde{\lambda}_i}^{\dot{\alpha}}$. The case $h_1+h_2+h_3 = 0$ corresponds to a constant amplitude, which is admissible only when all three particles are scalars.

Three-point massive amplitudes are similarly constrained by extended Poincar\'e symmetry~\cite{Ni:2024yrr, Ni:2025xkg}. For a given set of spins $(s_1,s_2,s_3)$, the theory admits a set of amplitudes distinguished by their transversality quantum numbers $[t_1,t_2,t_3]$. The total transversality $t_1+t_2+t_3$ acts as a weight under the action of the raising and lowering operators $T^\pm$. This structure provides a systematic procedure for generating all primary and descendant three-point amplitudes:
\begin{itemize}
    \item First, determine the highest-weight amplitude, which corresponds to maximal total transversality $t_1+t_2+t_3 = s_1+s_2+s_3$. This amplitude exhibits simple scaling properties analogous to those of a massless three-point amplitude. Within the SM, its kinematic structure takes the compact form
    \begin{eqnarray} \label{eq:HW_3pt}
    [\mathbf{12}]^{s_1+s_2-s_3}\,
    [\mathbf{23}]^{s_2+s_3-s_1}\,
    [\mathbf{31}]^{s_3+s_1-s_2}.
    \end{eqnarray}
    \item Next, apply successive lowering operations $T_i^- T_j^-$ (with $i \ne j$) to systematically generate amplitudes with lower total weight. This process is repeated until the lowest-weight representation, with total transversality $t_1+t_2+t_3 = -s_1-s_2-s_3$, is reached.
    \item The steps above yield all primary massive amplitudes. To obtain the corresponding descendant amplitudes, one then dresses each primary structure with appropriate powers of the mass spurion, thereby constructing the full tower of descendants up to order $2(s_1+s_2+s_3)$.
\end{itemize}

For the massive $FFS$ amplitude, the highest-weight amplitude is $[\mathbf{12}]$, which has total transversality $+1$. Only $T_1^-\cdot T_2^-$ gives a non-vanishing result
\begin{equation} \begin{aligned}
T_1^-\cdot T_2^- \circ[\mathbf{12}]=\langle\mathbf{12}\rangle,
\end{aligned} \end{equation}
producing the lowest-weight amplitude with total transversality $-1$. Therefore, the $FFS$ amplitude has two Lorentz structures $[\mathbf{12}]$ and $\langle\mathbf{12}\rangle$, related as
\begin{equation} \begin{aligned}
[\mathbf{12}] \xleftrightarrow[]{T_1^-\cdot T_2^-}\langle\mathbf{12}\rangle.
\end{aligned} \end{equation}
As a more involved example, consider the $FFV$ amplitude. In the $FFV$ case, the highest-weight amplitude is $[\mathbf{23}][\mathbf{31}]$. All three products $T^-_1\cdot T^-_2$, $T^-_2\cdot T^-_3$ and $T^-_3\cdot T^-_1$ will generate massive structures with lower weight. Its Lorentz structures are organized as follows,
\begin{equation}
\begin{tikzpicture}[baseline=-0.1cm]
\path
(0,0) node(C1) [rectangle] {$[\mathbf{23}][\mathbf{31}]$}
(3,1) node(C2) [rectangle] {$[\mathbf{23}]\langle\mathbf{31}\rangle$}
(3,-1) node(C3) [rectangle] {$\langle\mathbf{23}\rangle[\mathbf{31}]$}
(6,0) node(C4) [rectangle] {$\langle\mathbf{23}\rangle\langle\mathbf{31}\rangle$};
\draw [thick,->] (C1)--node[above]{\small $T_1^-\cdot T_3^-$} (C2);
\draw [thick,->] (C1)--node[above]{\small $T_2^-\cdot T_3^-$} (C3);
\draw [thick,->] (C2)--node[above]{\small $T_2^-\cdot T_3^-$} (C4);
\draw [thick,->] (C3)--node[above]{\small $T_1^-\cdot T_3^-$} (C4);
\end{tikzpicture}
\end{equation}
Note that the highest-weight amplitude is $[\mathbf{23}][\mathbf{31}]$ and the lowest-weight amplitude is $\langle\mathbf{23}\rangle\langle\mathbf{31}\rangle$. The two intermediate structures, $[\mathbf{23}]\langle\mathbf{31}\rangle$ and $\langle\mathbf{23}\rangle[\mathbf{31}]$, both have total transversality $0$.

In the SM, the most complicated amplitude is $VVV$, whose highest-weight amplitude is $[\mathbf{12}][\mathbf{23}][\mathbf{31}]$. In this case, all three products $T^-_1\cdot T^-_2$, $T^-_2\cdot T^-_3$ and $T^-_3\cdot T^-_1$ will generate massive structures with lower weight:
\begin{equation}
\begin{tikzpicture}[baseline=-0.1cm]
\path
(0,0) node(C1) [rectangle] {$[\mathbf{12}][\mathbf{23}][\mathbf{31}]$}
(3.5,2) node(C2) [rectangle] {$[\mathbf{12}][\mathbf{23}]\langle\mathbf{31}\rangle$}
(3.5,0) node(C3) [rectangle] {$[\mathbf{12}]\langle\mathbf{23}\rangle[\mathbf{31}]$}
(3.5,-2) node(C4) [rectangle] {$\langle\mathbf{12}\rangle[\mathbf{23}][\mathbf{31}]$}
(7,2) node(C5) [rectangle] {$[\mathbf{12}]\langle\mathbf{23}\rangle\langle\mathbf{31}\rangle$}
(7,0) node(C6) [rectangle] {$\langle\mathbf{12}\rangle[\mathbf{23}]\langle\mathbf{31}\rangle$}
(7,-2) node(C7) [rectangle] {$\langle\mathbf{12}\rangle\langle\mathbf{23}\rangle[\mathbf{31}]$}
(10.5,0) node(C8) [rectangle] {$\langle\mathbf{12}\rangle\langle\mathbf{23}\rangle\langle\mathbf{31}\rangle$};
\draw [thick,->] (C1)--node[left, xshift=-10pt, yshift=-4pt]{\small $T_1^-\cdot T_3^-$} (C2);
\draw [thick,->] (C1)--node[above]{\small $T_2^-\cdot T_3^-$} (C3);
\draw [thick,->] (C1)--node[above, xshift=-6pt, yshift=8pt]{\small $T_1^-\cdot T_2^-$} (C4);
\draw [thick,->] (C2)--node[above]{\small $T_2^-\cdot T_3^-$} (C5);
\draw [thick,->] (C2)--node[above, xshift=-6pt, yshift=8pt]{\small $T_1^-\cdot T_2^-$} (C6);
\draw [thick,->] (C3)--node[left, xshift=-10pt, yshift=-4pt]{\small $T_1^-\cdot T_3^-$} (C5);
\draw [thick,->] (C3)--node[above, xshift=-6pt, yshift=8pt]{\small $T_1^-\cdot T_2^-$} (C7);
\draw [thick,->] (C4)--node[left, xshift=-10pt, yshift=-4pt]{\small $T_1^-\cdot T_3^-$} (C6);
\draw [thick,->] (C4)--node[above]{\small $T_2^-\cdot T_3^-$} (C7);
\draw [thick,->] (C5)--node[above, xshift=-6pt, yshift=8pt]{\small $T_1^-\cdot T_2^-$} (C8);
\draw [thick,->] (C6)--node[above]{\small $T_2^-\cdot T_3^-$} (C8);
\draw [thick,->] (C7)--node[below right]{\small $T_1^-\cdot T_3^-$} (C8);
\end{tikzpicture}
\end{equation}
This method allows us to systematically enumerate all independent three-point massive amplitudes for particles of spin $0$, $\frac{1}{2}$, and $1$:
\begin{eqnarray}
\begin{array}{c|c|c}
\mbox{external particles} & \mbox{SM structures} & \mbox{EFT structures} \\
\hline
(\mathbf{1}^{1/2}, \mathbf{2}^{1/2}, \mathbf{3}^{1}) &  \langle\mathbf{23}\rangle [\mathbf{31}], [\mathbf{23}] \langle\mathbf{31}\rangle & [\mathbf{23}][\mathbf{31}],\langle\mathbf{23}\rangle \langle\mathbf{31}\rangle \\
\hline
(\mathbf{1}^{1/2}, \mathbf{2}^{1/2}, \mathbf{3}^{0}) & [\mathbf{12}], \langle\mathbf{12}\rangle &  \\
\hline
(\mathbf{1}^{1}, \mathbf{2}^{1}, \mathbf{3}^{1}) & \makecell{ [\mathbf{12}] [\mathbf{23}] \langle\mathbf{31}\rangle, [\mathbf{12}] \langle\mathbf{23}\rangle [\mathbf{31}], \langle\mathbf{12}\rangle [\mathbf{23}] [\mathbf{31}],  \\  \langle\mathbf{12}\rangle [\mathbf{23}] \langle\mathbf{31}\rangle,[\mathbf{12}]\langle\mathbf{23}\rangle \langle\mathbf{31}\rangle,\langle\mathbf{12}\rangle \langle\mathbf{23}\rangle [\mathbf{31}] }
& \makecell{[\mathbf{12}] [\mathbf{23}] [\mathbf{31}],\\\langle\mathbf{12}\rangle \langle\mathbf{23}\rangle \langle\mathbf{31}\rangle} \\
\hline
(\mathbf{1}^{1}, \mathbf{2}^{1}, \mathbf{3}^{0}) &  \langle\mathbf{12}\rangle [\mathbf{12}] & \langle\mathbf{12}\rangle^2,[\mathbf{12}]^2 \\
\hline
(\mathbf{1}^{0}, \mathbf{2}^{0}, \mathbf{3}^{0}) & 1 &
\end{array}\label{eq:3-ptmassive}
\end{eqnarray}
Here, we distinguish between structures that appear in the SM and those that only arise in effective field theory.

After obtaining the primary massive three-point amplitudes, we are ready to obtain the descendant massive amplitudes. We use the massive amplitudes $FFV$ to illustrate the expansion. Take one of the primary amplitudes is $\langle\mathbf{13}\rangle [\mathbf{32}]$, we can obtain its descendant amplitudes
\begin{eqnarray}
\mbox{primary: }& \langle\mathbf{13}\rangle [\mathbf{32}]; \nonumber \\
\mbox{1st descendant: }& \tilde{m}_1 \langle\mathbf{13}\rangle [\mathbf{32}], m_2 \langle\mathbf{13}\rangle [\mathbf{32}], m_3 \langle\mathbf{13}\rangle [\mathbf{32}], \tilde{m}_3 \langle\mathbf{13}\rangle [\mathbf{32}]; \nonumber \\
\mbox{2nd descendant: }& \tilde{m}_1 m_2 \langle\mathbf{13}\rangle [\mathbf{32}], \tilde{m}_1 m_3 \langle\mathbf{13}\rangle [\mathbf{32}], \tilde{m}_1 \tilde{m}_3 \langle\mathbf{13}\rangle [\mathbf{32}], m_2 m_3 \langle\mathbf{13}\rangle [\mathbf{32}], m_2 \tilde{m}_3 \langle\mathbf{13}\rangle [\mathbf{32}]; \nonumber \\
\mbox{3rd descendant: }& \tilde{m}_1 m_2 m_3 \langle\mathbf{13}\rangle [\mathbf{32}], \tilde{m}_1 m_2 \tilde{m}_3 \langle\mathbf{13}\rangle [\mathbf{32}].
\end{eqnarray}
Similarly other descendant amplitudes can be obtained from other primary amplitudes in eq.~\eqref{eq:3-ptmassive}.

\subsection{Minimal Helicity-Chirality one-particle states}

The massive momentum $\mathbf{p}$ can always be decomposed to two massless momenta
\begin{equation} \begin{aligned}
\mathbf{p}_{\alpha\dot\alpha}=p_{\alpha\dot\alpha}+\eta_{\alpha\dot\alpha}=\lambda_{\alpha}\tilde\lambda_{\dot\alpha}+\eta_{\alpha}\tilde\eta_{\dot\alpha},
\end{aligned} \end{equation}
and the massive spinor can be decomposed into two massless helicity spinors:
\begin{equation}
\begin{cases}
\lambda_{\alpha}^{I} = -\lambda_{\alpha} \zeta^{-I} +\eta_{\alpha} \zeta^{+I}, \\
\tilde{\lambda}_{\dot{\alpha}}^I = \tilde{\lambda}_{\dot{\alpha}} \zeta^{+I} +\tilde{\eta}_{\dot{\alpha}} \zeta^{-I},
\end{cases}
\label{eq:massless-decompose}    
\end{equation}
where $\zeta^{\pm I}$ are the basis vectors of the $SU(2)$ little group space, satisfying $\epsilon_{IJ} \zeta^{+I} \zeta^{-J} = 1$. 
The EOMs are then reduced to the following form,
\begin{align}
p_{\alpha \dot{\alpha}} \tilde{\eta}^{\dot{\alpha}} &= \tilde{m} \lambda_{\alpha},\qquad
p_{\alpha \dot{\alpha}} \eta^{\alpha} = -m \tilde{\lambda}_{\dot{\alpha}}.\label{eq:EOM_LO}\\
\eta_{\alpha \dot{\alpha}} \tilde{\lambda}^{\dot{\alpha}} &= \tilde{m} \eta_{\alpha},\qquad
\eta_{\alpha \dot{\alpha}} \lambda^{\alpha} = -m \tilde{\eta}_{\dot{\alpha}}.\label{eq:EOM_NLO}
\end{align}

The high energy decomposition induces a reduction of the extended little group $SU(2)\times ISO(2)$ to its subgroup $U(1)_w\times U(1)_z$. Here, $U(1)_w$ is associated with the generator $J^3$ and the helicity quantum number $h$ while $U(1)_z$ corresponds to the generator $D_-$ and the transversality $t$. 
The particle state is described by three quantum numbers: spin $s$, helicity $h$ and transversality $t$, corresponding to representations in the extended Little group $SU(2)\times ISO(2)$. The single-particle state with the quantum numbers is $| s, h, t \rangle $, corresponding to the generators $[(J)^2, J^3, D_-]$. The generators of $SU(2)$ are
\begin{eqnarray}
J^{3} &=& \frac{1}{2} \left[ \Big( \eta_{\alpha} \frac{\partial}{\partial\eta_{\alpha}} +\tilde{\lambda}_{\dot{\alpha}} \frac{\partial}{\partial \tilde{\lambda}_{\dot{\alpha}}} \Big) -\Big( \lambda_{\alpha} \frac{\partial}{\partial\lambda_{\alpha}} +\tilde{\eta}_{\dot{\alpha}} \frac{\partial}{\partial \tilde{\eta}_{\dot{\alpha}}} \Big) \right], \\
J^{+} &=&  -\eta_{\alpha} \frac{\partial}{\partial \lambda_{\alpha}} +\tilde{\lambda}_{\dot{\alpha}} \frac{\partial}{\partial \tilde{\eta}_{\dot{\alpha}}}, \label{eq:J+}\\
J^{-} &=&  -\lambda_{\alpha} \frac{\partial}{\partial\eta_{\alpha}} +\tilde{\eta}_{\dot{\alpha}} \frac{\partial}{\partial\tilde{\lambda}_{\dot{\alpha}}}, \label{eq:J-}
\end{eqnarray}
and their action on a particle state yields
\begin{equation} \begin{aligned} \label{eq:J_act}
J^{3}|s,h,t\rangle &=h|s,h,t\rangle,\\
J^+|s,h,t\rangle &=\big(s(s+1)-h(h+1)\big)|s,h+1,t\rangle,\\
J^-|s,h,t\rangle &=\big(s(s+1)-h(h-1)\big)|s,h-1,t\rangle.\\
\end{aligned} \end{equation}
Therefore, $J^{\pm}$ act as ladder operators for the $SU(2)$ group.

Meanwhile, the additional little group $ISO(2)$ corresponds to the transversality $t$. Its generators are
\begin{eqnarray}
D_- &=& \frac{1}{2} \left[ \Big( \tilde{\lambda}_{\dot{\alpha}} \frac{\partial}{\partial \tilde{\lambda}_{\dot{\alpha}}} + \tilde{\eta}_{\dot{\alpha}} \frac{\partial}{\partial \tilde{\eta}_{\dot{\alpha}}} \Big) - \Big( \lambda_{\alpha} \frac{\partial}{\partial \lambda_{\alpha}} + \eta_{\alpha} \frac{\partial}{\partial \eta_{\alpha}} \Big) \right], \\
m &=& \lambda^{\alpha}\eta_{\alpha}, \\
\tilde m &=& \tilde\lambda^{\dot\alpha}\tilde\eta_{\dot\alpha},
\end{eqnarray}
and they act as
\begin{equation} \begin{aligned}
D_-|s,h,t\rangle &=t|s,h,t\rangle,\\
m|s,h,t\rangle &=\mathbf m|s,h,t-1\rangle,\\
\tilde m|s,h,t\rangle &=\mathbf m |s,h,t+1\rangle.\\
\end{aligned} \end{equation}
Here, $\mathbf m$ is introduced on the right-hand side to ensure that the states $|s,h,t\pm1\rangle$ have the same mass dimension as $|s,h,t\rangle$. Therefore, $m$ and $\tilde m$ serve as ladder operators for the $ISO(2)$ group.
With the ladder operators $J^\pm, m, \tilde{m}$, the subgroup can be recovered with $U(1)_w \times U(1)_z \rightarrow SU(2)\times ISO(2)$.

This construction establishes the \textit{helicity-chirality formalism}, which enables simultaneous tracking of both helicity and chirality information in the high-energy expansion of a massive particles. In this work, we focus on the massive representations that correspond directly to massless states. This would select the \textit{minimal helicity-chirality} (MHC) representations, satisfying the {\it helicity-transversality unification}. In the high energy limit, $U(1)_w\times U(1)_z$ are unified into one $U(1)_w$, indicating that the massive spinors with both the transversality and the helicity quantum numbers should be unified into the massless one with only the helicity quantum number.

The energy scaling behaviors of the large-component spinors ($\lambda,\tilde\lambda$) and small-component spinors ($\eta,\tilde\eta$), gives
\begin{equation} \begin{aligned} \label{eq:spinor_scaling}
\lambda_{\alpha} &\sim \sqrt{E},& \tilde{\lambda}_{\dot{\alpha}} &\sim \sqrt{E}, \\
\tilde{m}\eta_{\alpha} &\sim \frac{\mathbf{m}^2}{\sqrt{E}},& m\tilde{\eta}_{\dot{\alpha}} &\sim \frac{\mathbf{m}^2}{\sqrt{E}},
\end{aligned} \end{equation}
where $\mathbf{m}$ is the absolute value of mass spurions $m$ and $\tilde m$. 
For the massive spinors, the primary and descendant representations gives the following decomposed states 
\begin{eqnarray}
\begin{array}{c|c|c}
\hline
 & h=-\frac12 & h=+\frac12 \\
\hline
t=-\frac12 & \lambda_{\alpha}, m \tilde{\eta}_{\dot{\alpha}}  &  \eta_{\alpha}, m \tilde{\lambda}_{\dot{\alpha}}   \\
t=+\frac12 & \tilde{\eta}_{\dot{\alpha}}, \tilde{m}\lambda_{\alpha}  & \tilde{\lambda}_{\dot{\alpha}}, \tilde{m} \eta_{\alpha}  \\
\hline
\end{array}
\label{eq:spin-half-states}
\end{eqnarray} 
In the MHC formalism, the condition $h=t$ is imposed. Thus among all the above massless states, the MHC spinors are selected to be 
\begin{equation}
\begin{tabular}{c|c|c}
\hline
& $c=-\frac12$ & $c=+\frac12$  \\
\hline
primary  & $\lambda_\alpha$ ($h=t=-\frac12$) &   $\tilde\lambda_{\dot\alpha}$ ($h=t=+\frac12$) \\
descendant &  $\tilde m\eta_\alpha$ ($h=t=+\frac12$)  &   $m\tilde\eta_{\dot\alpha}$ ($h=t=-\frac12$) \\
\hline
\end{tabular}
\end{equation}
They can be related by applying $J^{-}$ and $m$, or $J^+$ and $\tilde m$.
Therefore,  the composite ladder operator is utilized
\begin{equation}
m J^-,\quad \tilde m J^+.
\end{equation}
to change the helicity $h$ and transversality $t$ simultaneously. 
These operators acting on the spinors gives the chirality flips
\begin{eqnarray}
\begin{array}{c|cc}
 &  c=-\frac12  &  c=+\frac12  \\
\hline
h=-\frac{1}{2} & \multirow{3}{*}{\begin{tikzpicture}
\node (A) at (0,1) {$\lambda_{\alpha}$};
\node (B) at (0,0) {$\tilde{m}\eta_{\alpha}$};
\draw [->] (-0.1,0.7) -- (-0.1,0.3);
\node at (-0.6,0.5) {$\tilde mJ^+$};
\draw [->] (0.1,0.3) -- (0.1,0.7);
\node at (0.6,0.5) {$m J^-$};
\end{tikzpicture}} & \multirow{3}{*}{\begin{tikzpicture}
\node (A) at (0,1) {$m\tilde{\eta}_{\dot{\alpha}}$};
\node (B) at (0,0) {$\tilde{\lambda}_{\dot{\alpha}}$};
\draw [->] (-0.1,0.7) -- (-0.1,0.3);
\node at (-0.6,0.5) {$\tilde m J^+$};
\draw [->] (0.1,0.3) -- (0.1,0.7);
\node at (0.6,0.5) {$m J^-$};
\end{tikzpicture}} \\
&& \\
h = +\frac{1}{2} &&
\end{array} \label{eq:helicityflip} 
\end{eqnarray}

Let us use diagram to illustrate the chirality flip. For the primary state with $c=h=+\tfrac12$, acting with ${m} J^-$ yields the corresponding descendant state with $h=-\tfrac12$,
\begin{equation} \begin{aligned}
\Ampone{1.5}{+}{\fer{cyan}{i1}{v1}}=\tilde\lambda_{\dot\alpha}
\quad&\xrightarrow{ m J^-}\quad
\Ampone{1.5}{-}{\ferflip{1.5}{0}{red}{cyan}}= m \tilde\eta_{\dot\alpha} .
\end{aligned} \end{equation}
Further action of $ m J^-$ annihilates this state, consistent with the condition $|h|\le s$. For the primary state with $c=h=-\tfrac12$, we can act with $ {\tilde m}  J^+$ and obtain a similar descendant state
\footnote{
In Ref.~\cite{Ni:2026zaa}, we have introduced a slash and a cross to denote respectively the action of spurion masses $m,\tilde m$ and ladder operators $J^\pm$: 
\begin{equation}
\left\{\begin{aligned}
\begin{tikzpicture}[baseline=-0.1cm] \begin{feynhand}
\setlength{\feynhandarrowsize}{4pt}
\vertex [particle] (i1) at (1.5,0) {$+$}; 
\vertex (v1) at (0,0);
\draw[cyan,very thick] (0.7*1.5,0)--(0.6*1.5,0);
\draw[cyan,very thick,decoration={markings,mark=at position 0.56 with {\arrow{Triangle[length=4pt,width=4pt]}}},postaction={decorate}] (0.6*1.5,0)--(v1);
\draw (0.2*1.5,-0.08) -- (0.26*1.5,0.08);
\vertex[dot] (v1) at (0,0) {};
\end{feynhand} \end{tikzpicture}&=J^-\circ\tilde\lambda_{\dot\alpha}\\
\begin{tikzpicture}[baseline=-0.1cm] \begin{feynhand}
\setlength{\feynhandarrowsize}{4pt}
\vertex [particle] (i1) at (1.5,0) {$-$}; 
\vertex (v1) at (0,0);
\draw[red,very thick] (0.7*1.5,0)--(0.46*1.5,0);
\draw[cyan,very thick,decoration={markings,mark=at position 0.56 with {\arrow{Triangle[length=4pt,width=4pt]}}},postaction={decorate}] (0.46*1.5,0)--(v1);
\draw[very thick] plot[mark=x,mark size=2.5] coordinates {(0.46*1.5,0)};
\vertex[dot] (v1) at (0,0) {};
\end{feynhand} \end{tikzpicture}&=m\circ\tilde\lambda_{\dot\alpha}
\end{aligned}\right.
\quad\to\quad
\begin{tikzpicture}[baseline=-0.1cm] \begin{feynhand}
\setlength{\feynhandarrowsize}{4pt}
\vertex [particle] (i1) at (1.5,0) {$-$}; 
\vertex (v1) at (0,0);
\draw[red,very thick] (0.7*1.5,0)--(0.46*1.5,0);
\draw[cyan,very thick,decoration={markings,mark=at position 0.45 with {\arrow{Triangle[length=4pt,width=4pt]}}},postaction={decorate}] (0.46*1.5,0)--(v1);
\draw[very thick] plot[mark=x,mark size=2.5] coordinates {(0.46*1.5,0)};
\draw (0.1*1.5,-0.08) -- (0.16*1.5,+0.08);
\vertex[dot] (v1) at (0,0) {};
\end{feynhand} \end{tikzpicture}=mJ^-\circ\tilde\lambda_{\dot\alpha}.
\end{equation}
In the present work, only the combinations $mJ^-$ and $\tilde m J^+$ are used to convert MHC particle states. Therefore, we will omit the slash and simply employ the cross to represent descendant MHC states in this paper.
} 
with $h=+\tfrac12$
\begin{equation} \begin{aligned}
\Ampone{1.5}{+}{\fer{red}{i1}{v1}}= \lambda_{ \alpha}
\quad&\xrightarrow{ {\tilde m}  J^+}\quad
\Ampone{1.5}{-}{\ferflip{1.5}{0}{cyan}{red}}= {\tilde m}  \eta_{ \alpha} 
\end{aligned} \end{equation}

Starting from a particle state $|s,h=t\rangle$, applying these operators ensures the state continues to satisfy $h=t$. By acting with them on external or internal particle lines, we can generate all relevant structures from a single MHC state.
Similar to the fermion case, we could also obtain the MHC particle states for the vector boson
\begin{equation}
\begin{tabular}{c|c|c|c}
\hline
&  $c=-1$ &$c=0$  & $c=+1$ \\
\hline
$h=t=-1$ & $\lambda_{\alpha_1}\lambda_{\alpha_2}$ &   $m\lambda_{\alpha} \tilde \eta_{\dot\alpha}$  & $m^2\tilde\eta_{\dot\alpha_1} \tilde\eta_{\dot\alpha_2}$\\
\hline
 $h=t=0$ & $\tilde m\lambda_{\alpha_1}\eta_{\alpha_2}$  & \makecell{  $\tilde\lambda_{\dot\alpha}\lambda_{\alpha}$ \\ $m\tilde m\eta_{\alpha}\tilde\eta_{\dot\alpha}$ } & $m\tilde\lambda_{\dot\alpha_1}\tilde\eta_{\dot\alpha_2}$ \\
\hline
$h=t=+1$ &  $\tilde m^2\eta_{\alpha_1}\eta_{\alpha_2}$  &  $\tilde m\tilde\lambda_{\dot\alpha}\eta_{\alpha}$ & $\tilde\lambda_{\dot\alpha_1}\tilde\lambda_{\dot\alpha_2}$  \\
\hline
\end{tabular}
\end{equation}
This can be reorganized by the primary and descendant states
\begin{equation}
\begin{tabular}{c|c|c|c}
\hline
&  $c=-1$ &$c=0$  & $c=+1$ \\
\hline
primary & $\lambda_{\alpha_1}\lambda_{\alpha_2}$ ($h=t=-1$) &  $\tilde\lambda_{\dot\alpha}\lambda_{\alpha}$($h=t=0$)   & $\tilde\lambda_{\dot\alpha_1}\tilde\lambda_{\dot\alpha_2}$ ($h=t=+1$) \\
\hline
1st descendant & $\tilde m\lambda_{\alpha_1}\eta_{\alpha_2}$ ($h=t=0$ ) & \makecell{ $m\lambda_{\alpha} \tilde \eta_{\dot\alpha}$  ($h=t=-1$)\\ $\tilde m\tilde\lambda_{\dot\alpha}\eta_{\alpha}$ ($h=t=+1$)} & $m\tilde\lambda_{\dot\alpha_1}\tilde\eta_{\dot\alpha_2}$ ($h=t=0$) \\
\hline
second descendant  &  $\tilde m^2\eta_{\alpha_1}\eta_{\alpha_2}$ ($h=t=+1$) & $m\tilde m\eta_{\alpha}\tilde\eta_{\dot\alpha}$($h=t=0$)  & $m^2\tilde\eta_{\dot\alpha_1} \tilde\eta_{\dot\alpha_2}$ ($h=t=-1$) \\
\hline
\end{tabular}
\end{equation}

In particular, Let us look at the vector boson with the chirality $c=0$. If the vector boson is the massive gauge boson, it should recover the transverse gauge boson for $h=t = \pm 1$ and the Goldstone boson for $h=t=0$, in the massless limit:
\begin{itemize}
    \item For the massless spin-1 particle, the only zero helicity state is the derivative of massless scalar $\partial_\mu \phi$, which can be recognize to be the Goldstone boson. Thus the $\tilde\lambda_{\dot\alpha}\lambda_{\alpha}$ should be matched to the zero helicity massless spin-1 state
\begin{eqnarray} 
\partial_{\alpha, \dot{\alpha}} \phi  \equiv \lambda_{\alpha}\tilde{\lambda}_{\dot{\alpha}},
\end{eqnarray}
which denotes a massless Goldstone boson. It couples to the conserved current.

\item For the massive gauge boson with $h =\pm 1$, its massless limit should recover the massless gauge boson. For positive polarization, the transverse massive gauge boson can be deformed as follows
\begin{eqnarray}
 \tilde m\eta_{\alpha}\tilde{\lambda}_{\dot{\alpha}} 
= \mathbf m^2 \frac{\eta_{\alpha}\tilde{\lambda}_{\dot{\alpha}}}{\langle \eta \lambda \rangle }.
\end{eqnarray}
On the other hand, the massless gauge boson has two helicity states, with the polarization vectors 
\begin{eqnarray} \label{eq:massless_polarization}
A_{\alpha\dot{\alpha}}^+: \frac{\xi_{\alpha}\tilde{\lambda}_{\dot{\alpha}}}{\langle \xi \lambda \rangle }, \quad 
A_{\alpha\dot{\alpha}}^-: -\frac{\lambda_{\alpha}\tilde{\xi}_{\dot{\alpha}}}{[ \tilde{\lambda} \tilde{\xi} ] }.
\end{eqnarray}
where superscript $\pm$ denotes the helicity of the gauge boson. The reference spinor $\xi$ can be interpreted as fixing the light-cone gauge condition $A^\mu \xi_\mu=0$. By taking $\xi$ to be $\eta$, these two massless helicity states are identified as 
\begin{eqnarray}
A_{\alpha\dot{\alpha}}^+: \frac{\eta_{\alpha}\tilde{\lambda}_{\dot{\alpha}}}{\langle \eta \lambda \rangle }, 
\quad 
A_{\alpha\dot{\alpha}}^-: -\frac{\lambda_{\alpha}\tilde{\eta}_{\dot{\alpha}}}{[ \tilde{\lambda} \tilde{\eta} ] }.
\end{eqnarray}
Therefore, the transverse massive gauge boson state can be identified as the massless gauge boson in the light cone gauge. 

\end{itemize}

\begin{figure}[htbp]
\centering
\subfloat[\hspace*{3em}]{ \label{fig:state_flip1}
\includegraphics[scale=1.1,valign=c]{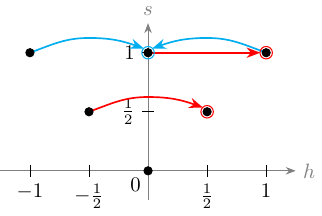}}
\subfloat[\hspace*{3em}]{ \label{fig:state_flip2}
\includegraphics[scale=1.1,valign=c]{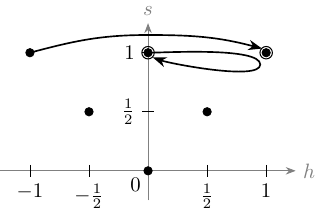}}
\caption{Illustration of the ladder operators $mJ^-$ and $\tilde{m}J^+$ acting on the particle state. Diagram (a) shows a single application of the ladder operators for spin-0, spin-1/2, and spin-1 particles. Diagram (b) shows the result of two successive applications.}
\label{fig:state_flip}
\end{figure}

Let us organize all the MHC states using the ladder operators. We begin with a primary MHC state composed solely of $\lambda$ and $\tilde\lambda$, and then apply $m J^-$ or $\tilde m J^+$ to generate descendant MHC states
\begin{equation}
\text{primary}\xrightarrow{mJ^-,\tilde mJ^+}
\text{1st descendant}\xrightarrow{mJ^-,\tilde mJ^+}
\cdots
\end{equation}
After applying these operations $2s$ times, we obtain all MHC states, as shown in Fig.~\ref{fig:state_flip}. In such representations, both $h$-flip and $m$-flip occur when we  apply $J^{-}$ and $m$, or $J^+$ and $\tilde m$, to the primary states. These two types of operations are interchangeable because the commutators satisfy $[m,J^{\pm}]=[\tilde{m},J^{\pm}]=0$. For example, applying either $mJ^-$ or $J^- m$ to the state $(s,h,t)=(\frac12,+\frac12,+\frac12)$ yields the same result. Although both operations lead to the same physical state, we adopt the convention of using the first type of operation (i.e. $mJ^-$ or $\tilde m J^+$)  to represent states with the chirality flips.

Let us organize the ladder operations of the primary and descendant particle states. To maintain the extended Little group (ELG) covariance, we will utilize the normalized ladder operators $\frac{m}{\mathbf m} J^\pm$, which would maintain the energy power of different states.   
\begin{enumerate}
\item \textbf{primary $\to$ descendant:} We begin with a primary state whose chirality equals its helicity, $c=h$. Acting with ladder operators generates descendant states with $c\neq h$. Scalar particles have no descendant states, so we first consider fermions. For the primary state with $c=h=+\tfrac12$, acting with $\frac{m}{\mathbf m} J^-$ yields the corresponding descendant state with $h=-\tfrac12$,
\begin{equation} \begin{aligned}
\Ampone{1.5}{+}{\fer{cyan}{i1}{v1}}=\tilde\lambda_{\dot\alpha}
\quad&\xrightarrow{\frac{m}{\mathbf m} J^-}\quad
\Ampone{1.5}{-}{\ferflip{1.5}{0}{red}{cyan}}=\frac{m}{\mathbf m}\tilde\eta_{\dot\alpha} .
\end{aligned} \end{equation}
Further action of $\frac{m}{\mathbf m} J^-$ annihilates this state, consistent with the condition $|h|\le s$. For the primary state with $c=h=-\tfrac12$, we can act with $\frac{\tilde m}{\mathbf m} J^+$ and obtain a similar descendant state with $h=+\tfrac12$ (and opposite color).

For vector states, we first consider the primary state with $c=h=+1$. Acting once and twice with the ladder operator gives the 1st and 2nd descendant states with $|c-h|=1$ and $2$, respectively:
\begin{align}
\Ampone{1.5}{+}{\bos{i1}{cyan}}=\tilde{\lambda}_{\dot{\alpha}_1}\tilde{\lambda}_{\dot{\alpha}_2}
\quad&\xrightarrow{\frac{m}{\mathbf m} J^-}\quad
\Ampone{1.5}{0}{\bosflip{1.5}{0}{cyan}{brown}}=2\frac{m}{\mathbf m}\tilde{\eta}_{(\dot{\alpha}_1}\tilde{\lambda}_{\dot{\alpha}_2)},\\
\Ampone{1.5}{+}{\bos{i1}{cyan}}=\tilde{\lambda}_{\dot{\alpha}_1}\tilde{\lambda}_{\dot{\alpha}_2}
\quad&\xrightarrow{(\frac{m}{\mathbf m} J^-)^2}\quad
\Ampone{1.5}{-}{\bosflipflip{1.5}{0}{cyan}{brown}{red}}=2\frac{m^2}{\mathbf m^2}\tilde{\eta}_{\dot{\alpha}_1}\tilde{\eta}_{\dot{\alpha}_2}.
\end{align}
Here the factor 2 arises from Clebsch–Gordan coefficients as shown in Eq.~\eqref{eq:J_act}. It should be dropped when transforming the scattering amplitude into different helicity amplitudes. The primary state with $c=h=-1$ gives analogous results with opposite helicity and color. For the primary state with $c=h=0$, the situation is slightly different. Acting with ladder operators yields two distinct 1st descendant states with $|c-h|=1$:
\begin{equation} \begin{aligned}
\Ampone{1.5}{0}{\bos{i1}{brown}}=\lambda_{\alpha}\tilde{\lambda}_{\dot{\alpha}}
\quad&\xrightarrow{\frac{\tilde m}{\mathbf m} J^+}\quad
\Ampone{1.5}{+}{\bosflip{1.5}{0}{brown}{cyan}}=\frac{\tilde m}{\mathbf m}\eta_{\alpha}\tilde{\lambda}_{\dot{\alpha}},\\ 
\Ampone{1.5}{0}{\bos{i1}{brown}}=\lambda_{\alpha}\tilde{\lambda}_{\dot{\alpha}}
\quad&\xrightarrow{\frac{m}{\mathbf m} J^-}\quad
\Ampone{1.5}{-}{\bosflip{1.5}{0}{brown}{red}}=\frac{m}{\mathbf m} \lambda_{\alpha}\tilde{\eta}_{\dot{\alpha}}.
\end{aligned} \end{equation}
In this case, the 2nd descendant state has the same helicity as the primary state. TTherefore, we act with both ladder operators $\frac{\tilde m}{\mathbf m} J^+$ and $\frac{m}{\mathbf m} J^-$ to obtain
\begin{equation} \begin{aligned}
\Ampone{1.5}{0}{\bos{i1}{brown}}=\lambda_{\alpha}\tilde{\lambda}_{\dot{\alpha}}
\quad&\xrightarrow{\frac{\tilde m}{\mathbf m} J^+ \frac{m}{\mathbf m} J^-}\quad
\Ampone{1.5}{0}{\bos{i1}{brown}}
+\Ampone{1.5}{0}{\bosflipflip{1.5}{0}{brown}{brown}{brown}}=
\lambda_{\alpha}\tilde{\lambda}_{\dot{\alpha}}+\frac{m\tilde m}{\mathbf m^2}\eta_{\alpha}\tilde{\eta}_{\dot{\alpha}}.
\end{aligned} \end{equation}
The result contains both the primary and the 2nd descendant state. Acting the two ladder operators in the opposite order, $\frac{m}{\mathbf m} J^-\frac{\tilde m}{\mathbf m} J^+$, gives the same result.

\item \textbf{descendant $\to$ primary:} Conversely, we can apply ladder operators on descendant states to recover primary states. For a 1st descendant state, one action suffices:
\begin{equation} \begin{aligned}
\Ampone{1.5}{-}{\ferflip{1.5}{0}{red}{cyan}}&=\frac{m}{\mathbf m}\tilde\eta_{\dot\alpha}&
&\xrightarrow{\frac{\tilde m}{\mathbf m} J^+}&
\Ampone{1.5}{+}{\fer{cyan}{i1}{v1}}&=\tilde\lambda_{\dot\alpha},& \\
\Ampone{1.5}{0}{\bosflip{1.5}{0}{cyan}{brown}}&=\frac{m}{\mathbf m}\tilde{\eta}_{(\dot{\alpha}_1}\tilde{\lambda}_{\dot{\alpha}_2)}&
&\xrightarrow{\frac{\tilde m}{\mathbf m} J^+}&
\Ampone{1.5}{+}{\bos{i1}{cyan}}&=\tilde{\lambda}_{\dot{\alpha}_1}\tilde{\lambda}_{\dot{\alpha}_2},& \\
\Ampone{1.5}{+}{\bosflip{1.5}{0}{brown}{cyan}}&=\frac{\tilde m}{\mathbf m}\eta_{\alpha}\tilde{\lambda}_{\dot{\alpha}}&
&\xrightarrow{\frac{m}{\mathbf m} J^-}&
\Ampone{1.5}{0}{\bos{i1}{brown}}&=\lambda_{\alpha}\tilde{\lambda}_{\dot{\alpha}},& \\
\Ampone{1.5}{-}{\bosflip{1.5}{0}{brown}{red}}&=\frac{m}{\mathbf m} \lambda_{\alpha}\tilde{\eta}_{\dot{\alpha}}&
&\xrightarrow{\frac{\tilde m}{\mathbf m} J^+}&
\Ampone{1.5}{0}{\bos{i1}{brown}}&=\lambda_{\alpha}\tilde{\lambda}_{\dot{\alpha}}.&
\end{aligned} \end{equation}
For a 2nd descendant state, two actions are required,
\begin{equation} \begin{aligned}
\Ampone{1.5}{-}{\bosflipflip{1.5}{0}{cyan}{brown}{red}}&=\frac{m^2}{\mathbf m^2}\tilde{\eta}_{\dot{\alpha}_1}\tilde{\eta}_{\dot{\alpha}_2}&
&\xrightarrow{(\frac{\tilde m}{\mathbf m} J^+)^2}&
&\Ampone{1.5}{+}{\bos{i1}{cyan}}=2\tilde{\lambda}_{\dot{\alpha}_1}\tilde{\lambda}_{\dot{\alpha}_2},&\\
\Ampone{1.5}{0}{\bosflipflip{1.5}{0}{brown}{brown}{brown}}&=
\frac{m\tilde m}{\mathbf m^2}\eta_{\alpha}\tilde{\eta}_{\dot{\alpha}}&
&\xrightarrow{\frac{\tilde m}{\mathbf m} J^+ \frac{m}{\mathbf m} J^-}&
&\left\{\begin{aligned}
\Ampone{1.5}{0}{\bos{i1}{brown}}&=\lambda_{\alpha}\tilde{\lambda}_{\dot{\alpha}},\\
\Ampone{1.5}{0}{\bosflipflip{1.5}{0}{brown}{brown}{brown}}&=\frac{m\tilde m}{\mathbf m^2}\eta_{\alpha}\tilde{\eta}_{\dot{\alpha}}.
\end{aligned}\right.&
\end{aligned} \end{equation}
For states with $c=0$, the descendant and primary appear together.

\item \textbf{descendant $\to$ descendant:} A fermion of given chirality $c$ has only one kind of descendant state, so conversion between descendant states is not needed. For vector states, let $h^{\text{1st}}$ and $h^{\text{2nd}}$ denote the helicity of the 1st and 2nd descendant states.  When $h^{\text{1st}}>h^{\text{2nd}}$, we can act with $\frac{m}{\mathbf m} J^-$ on the 1st descendant state to obtain the 2nd one:
\begin{equation} \begin{aligned}
\Ampone{1.5}{+}{\bosflip{1.5}{0}{brown}{cyan}}=\frac{\tilde m}{\mathbf m}\eta_{\alpha}\tilde{\lambda}_{\dot{\alpha}}
\quad&\overset{\frac{m}{\mathbf m} J^-}{\to}\quad
\left\{\begin{aligned}
\Ampone{1.5}{0}{\bos{i1}{brown}}&=\lambda_{\alpha}\tilde{\lambda}_{\dot{\alpha}},\\
\Ampone{1.5}{0}{\bosflipflip{1.5}{0}{brown}{brown}{brown}}&=\frac{m\tilde m}{\mathbf m^2}\eta_{\alpha}\tilde{\eta}_{\dot{\alpha}},
\end{aligned}\right.\\
\Ampone{1.5}{0}{\bosflip{1.5}{0}{cyan}{brown}}=\frac{m}{\mathbf m}\tilde{\eta}_{(\dot{\alpha}_1}\tilde{\lambda}_{\dot{\alpha}_2)}
\quad&\overset{\frac{\tilde m}{\mathbf m} J^+}{\to}\quad
\Ampone{1.5}{-}{\bosflipflip{1.5}{0}{cyan}{brown}{red}}=\frac{m^2}{\mathbf m^2}\tilde{\eta}_{\dot{\alpha}_1}\tilde{\eta}_{\dot{\alpha}_2}.
\end{aligned} \end{equation}
In the first line, the primary and 2nd descendant states for $c=0$ still appear together. Conversely, acting with $\frac{\tilde m}{\mathbf m} J^+$ on the 2nd descendant state yields the 1st one:
\begin{equation} \begin{aligned}
\Ampone{1.5}{0}{\bosflipflip{1.5}{0}{brown}{brown}{brown}}=\frac{m\tilde m}{\mathbf m^2}\eta_{\alpha}\tilde{\eta}_{\dot{\alpha}}
\quad&\overset{\frac{\tilde m}{\mathbf m} J^+}{\to}\quad
\Ampone{1.5}{+}{\bosflip{1.5}{0}{brown}{cyan}}=\frac{\tilde m}{\mathbf m}\eta_{\alpha}\tilde{\lambda}_{\dot{\alpha}},\\
\Ampone{1.5}{-}{\bosflipflip{1.5}{0}{cyan}{brown}{red}}=\frac{m^2}{\mathbf m^2}\tilde{\eta}_{\dot{\alpha}_1}\tilde{\eta}_{\dot{\alpha}_2}
\quad&\overset{\frac{\tilde m}{\mathbf m} J^+}{\to}\quad
\Ampone{1.5}{0}{\bosflip{1.5}{0}{cyan}{brown}}=2\frac{m}{\mathbf m}\tilde{\eta}_{(\dot{\alpha}_1}\tilde{\lambda}_{\dot{\alpha}_2)}.
\end{aligned} \end{equation}
For the case $h^{\text{1st}}<h^{\text{2nd}}$, analogous results follow by exchanging the roles of $\frac{\tilde m}{\mathbf m} J^+$ and $\frac{m}{\mathbf m} J^-$.

\end{enumerate}

Finally, let us summarize the MHC states for particle with spin up to one. For spins $s=0,\frac{1}{2},1$, the primary states are the same as the massless ones, while all the descendant states are obtained from the chirality flip, marked as cross sign. Diagrammatically, the resulting states can be listed as follow
\begin{equation}  \label{eq:MHC_particle}
\includegraphics[width=\linewidth]{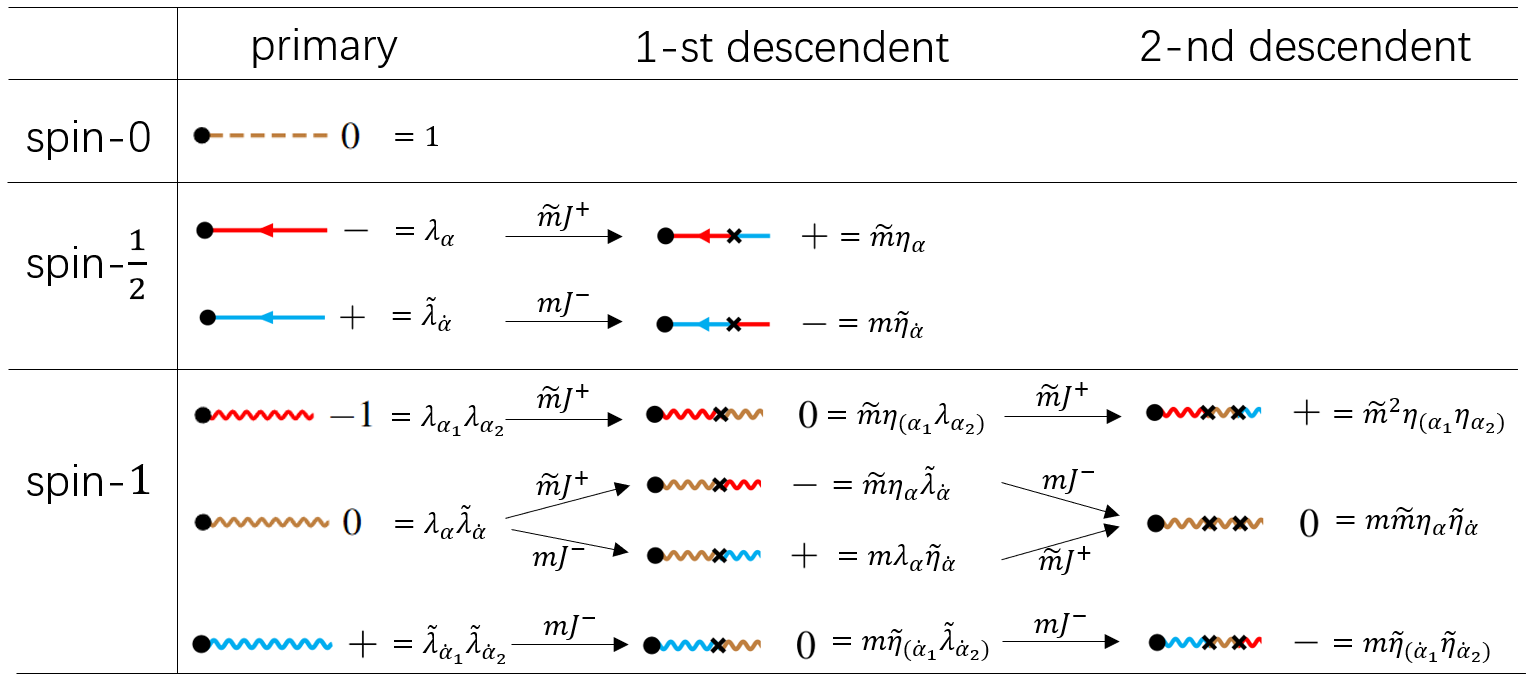}
\end{equation}
where the color labels the chirality, and the number on the right labels the helicity. The difference between chirality and transversality is that chirality is also related to particle and anti-particle while transversality is not.

\subsection{MHC 2-point currents}

Having established the MHC one-particle states for particles of different spins, we now proceed to construct two-particle states. Given that a single particle transforms under the Lorentz representation $[t, j_1, j_2]$, the two-particle states are found in the representations obtained from the tensor product:
\begin{eqnarray}
[t, j_1, j_2]_1 \otimes [t, j_1, j_2]_2 = \sum [t, j_1, j_2]_{12}.
\end{eqnarray}
In this work, we focus specifically on the $\left(\frac{1}{2}, \frac{1}{2}\right)$ Lorentz representation, in which the two-particle structure---the current---plays a central role in analyzing amplitude deformations. We define the MHC two-point current as~\footnote{
The massive current is generally defined as $\mathbf{J}_{\alpha\dot{\alpha}}(1,2,\dots,N) = \langle\Omega|\mathcal{O}_{\alpha\dot{\alpha}}|\{1,2,\dots,N\}\rangle$, which is an $N$-point form factor in the $(\frac{1}{2}, \frac{1}{2})$ Lorentz representation, characterized by the spin and transversality of each particle.}
\begin{eqnarray}
\mathbf{J}^{\dot{\alpha}\alpha}(1,2).
\end{eqnarray}
This object is a two-particle structure transforming in the $(\frac{1}{2}, \frac{1}{2})$ representation.

Following the logic of the MHC state expansion, the full MHC current can be expanded into a series of components of increasing descendant order:
\begin{eqnarray}
\mathbf{J}=[\mathbf{J}]_0+[\mathbf{J}]_1+[\mathbf{J}]_2+\cdots
\end{eqnarray}
These components are classified as follows:
\begin{eqnarray}
\begin{aligned}
\text{primary current} &: \quad[\mathbf{J}]_0,\\
\text{1st descendant current} &: \quad[\mathbf{J}]_1,\\
\text{2nd descendant current} &: \quad[\mathbf{J}]_2,\\
&\; \vdots
\end{aligned}
\end{eqnarray}
The primary current $[\mathbf{J}]_0$ contains only primary particle states, whereas the descendant currents incorporate descendant particle states.

Analogous to the construction of MHC states, descendant currents are derived by repeatedly applying the ladder operators $mJ^-$ or $\tilde{m}J^+$ to the primary states. Applying one such operator to a specific particle flips its helicity and yields the corresponding first descendant current; successive applications generate higher descendants:
\begin{eqnarray}
\begin{aligned}
[\mathbf{J}]_1 &= mJ^- \circ [\mathbf{J}]_0 \;+\; \tilde{m}J^+ \circ [\mathbf{J}]_0,\\
[\mathbf{J}]_2 &= (mJ^-)^2 \circ [\mathbf{J}]_0 \;+\; (\tilde{m}J^+ mJ^-) \circ [\mathbf{J}]_0 \;+\; (\tilde{m}J^+)^2 \circ [\mathbf{J}]_0,\\
&\; \vdots
\end{aligned}
\end{eqnarray}
We note that, depending on the spins $s_1$ and $s_2$ of the particles in the current, the MHC expansion terminates at the $2(s_1 + s_2)$-th descendant current.

\paragraph{\text{FF} current}

We first consider the $FF$ current, constructed from the tensor product of two fermion MHC states. Because the ladder operator can be applied at most twice, the MHC expansion for this current terminates at the second descendant level. We begin with the primary current $[\mathbf{J}]_0$. For clarity, we decompose it into distinct helicity components:
\begin{eqnarray}
[\mathbf{J}]_0 = \sum_{h_1,h_2} \left[\mathbf{J}\left(1^{h_1}, 2^{h_2}\right)\right]_0,
\end{eqnarray}
where $h_1$ and $h_2$ denote the helicities of particles 1 and 2. For the $FF$ primary current, there are two helicity components: $(-1/2, +1/2)$ and $(+1/2, -1/2)$, while other helicity components do not correspond to the current $\mathbf{J}^{\dot{\alpha}\alpha}$, in the $(\frac12, \frac12)$ representation. Explicitly, they are given by
\begin{eqnarray} \label{eq:current_FF_primary}
(- \tfrac{1}{2}, + \tfrac{1}{2}) &: \quad \left[\mathbf{J}\left(1^{-}, 2^{+}\right)\right]_0 = -c_1 |2]^{\dot{\alpha}} \langle 1|^{\alpha}, \\[4pt]
(+ \tfrac{1}{2}, - \tfrac{1}{2}) &: \quad \left[\mathbf{J}\left(1^{+}, 2^{-}\right)\right]_0 = -c_2 |1]^{\dot{\alpha}} \langle 2|^{\alpha},
\end{eqnarray}
where the superscripts $\pm$ indicate helicity $\pm 1/2$, and $c_1, c_2$ are dependent coefficients.

\begin{figure}[htbp]
\centering
\subfloat[\hspace*{3em}]{ \label{fig:FF_flip1}
\includegraphics[scale=1.2,valign=c]{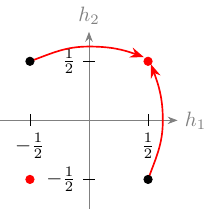}}
\hspace{2em}
\subfloat[\hspace*{3em}]{ \label{fig:FF_flip2}
\includegraphics[scale=1.2,valign=c]{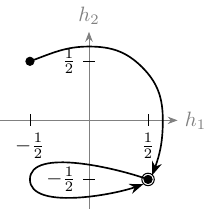}}
\caption{Illustration of the ladder operators $mJ^-$ and $\tilde{m}J^+$ acting on the $FF$ current, denoted by the helicity of the two particles in the current. On the left (a), the two black dots denote the two primary currents $\left[\mathbf{J}\left(1^{-}, 2^{+}\right)\right]_0$ and $\left[\mathbf{J}\left(1^{+}, 2^{-}\right)\right]_0$, while two red dots denote the two descendant currents $[\mathbf J(1^{+},2^{+})]_1$ and $[\mathbf J(1^{-},2^{-})]_1$.  One the right (b), the second descendant current is obtained from the two primary currents. }
\end{figure}

As shown in Fig.~\ref{fig:FF_flip1}, 
each primary current with helicity $(\mp\frac12,\pm\frac12)$ can be transformed via ladder operators into a first descendant current of uniform helicity, $(+\frac12,+\frac12)$ or $(-\frac12,-\frac12)$:
\begin{equation} \begin{aligned}
\relax
[\mathbf J(1^{+},2^{+})]_1&=
\tilde m_2J_2^+\circ [\mathbf J(1^{+},2^{-})]_0+
\tilde m_1J_1^+\circ [\mathbf J(1^{-},2^{+})]_0,\\
[\mathbf J(1^{-},2^{-})]_1&=
m_1J^-_1\circ [\mathbf J(1^{+},2^{-})]_0+
m_2J^-_2\circ [\mathbf J(1^{-},2^{+})]_0.
\end{aligned}\end{equation}
Substituting the explicit forms from Eq.~\eqref{eq:current_FF_primary} yields the explicit first descendant currents:
\begin{align}
(+\tfrac{1}{2},+\tfrac{1}{2}):\quad [\mathbf J(1^{+},2^{+})]_1&=c_1\tilde m_1|2]^{\dot\alpha}\langle\eta_1|^{\alpha}+c_2\tilde m_2|1]^{\dot\alpha}\langle\eta_2|^{\alpha},\\
(-\tfrac{1}{2},-\tfrac{1}{2}):\quad [\mathbf J(1^{-},2^{-})]_1&=-c_1 m_2|\eta_2]^{\dot\alpha}\langle1|^{\alpha}-c_2 m_1|\eta_1]^{\dot\alpha}\langle2|^{\alpha}.
\end{align}
The coefficients $c_1$ and $c_2$ allow each term to be traced back to its specific primary current origin.

As illustrated in Fig.~\ref{fig:FF_flip2} . Applying two ladder operators converts the primary currents with helicity $(\pm\frac12, \mp\frac12)$ into second descendant currents with the opposite helicity configuration, $(\mp\frac12, \pm\frac12)$:
\begin{equation} \begin{aligned}
\relax
[\mathbf J(1^{+},2^{-})]_2&=
m_2J^-_2 \tilde m_1 J_1^+\circ [\mathbf J(1^{-},2^{+})]_0,\\
[\mathbf J(1^{-},2^{+})]_2&=
m_1J^-_1 \tilde m_2 J_2^+\circ [\mathbf J(1^{+},2^{-})]_0.
\end{aligned}\end{equation}
In this case, each resulting component receives contributions from only one primary current. Evaluating these expressions gives the second descendant currents:
\begin{align}
(+\frac{1}{2},-\frac{1}{2}):\quad [\mathbf J(1^{+},2^{-})]_2&=c_1 \tilde m_1 m_2|\eta_2]^{\dot\alpha}\langle\eta_1|^{\alpha},\\
(-\frac{1}{2},+\frac{1}{2}):\quad [\mathbf J(1^{-},2^{+})]_2&=c_2 m_1\tilde m_2|\eta_1]^{\dot\alpha}\langle\eta_2|^{\alpha}.
\end{align}
Thus, all MHC current components for the $FF$ case have been systematically derived.

\paragraph{\text{VS} current}

We next consider the $VS$ current, where particle 1 is a vector boson and particle 2 is a scalar boson. In this case, there is a single primary current, corresponding to the helicity configuration $(0,0)$:
\begin{align}
[\mathbf J(1^{0},2^{0})]_0=-|1]^{\dot\alpha}\langle1|^{\alpha},
\end{align}
Given that the primary current contains only one helicity component, we omit an explicit coefficient here for simplicity.

Applying the ladder operators to this primary current generates two first descendant currents with helicity configurations $(\pm1,0)$:
\begin{align}
[\mathbf J(1^{0},2^{0})]_0
&\xrightarrow{m_1 J_1^-}
[\mathbf J(1^{-},2^{0})]_1=-m_1|\eta_1]^{\dot\alpha}\langle1|^{\alpha},\\
[\mathbf J(1^{0},2^{0})]_0
&\xrightarrow{\tilde m_1 J_1^+}
[\mathbf J(1^{+},2^{0})]_1=\tilde m_1|1]^{\dot\alpha}\langle\eta_1|^{\alpha}. 
\end{align}
Applying a second ladder flip returns the system to the $(0,0)$ helicity configuration, producing the second descendant current:
\begin{align}
[\mathbf J(1^{0},2^{0})]_2=m_1\tilde m_1|\eta_1]^{\dot\alpha}\langle\eta_1|^{\alpha}.
\end{align}
Since no contributions exist beyond $[\mathbf{J}]_2$ for the $VS$ current, we have now systematically derived all MHC current components in this case.

\paragraph{\text{VV} current}

Finally, we analyze the $VV$ current. Its primary component consists of four independent helicity configurations: $(\pm1,0)$ and $(0,\pm1)$. These are expressed as
\begin{align}
[\mathbf J(1^+,2^0)]_0&=-c_1[12]|1]^{\dot\alpha}\langle2|^{\alpha},\\
[\mathbf J(1^-,2^0)]_0&=-c_2\langle12\rangle |2]^{\dot\alpha}\langle1|^{\alpha},\\
[\mathbf J(1^0,2^+)]_0&=-c_3[12]|2]^{\dot\alpha}\langle1|^{\alpha},\\
[\mathbf J(1^0,2^-)]_0&=-c_4\langle12\rangle |1]^{\dot\alpha}\langle2|^{\alpha},
\end{align}
where $c_i$ ($i=1,\dots,4$) are independent the coefficients. All the primary currents are conserved currents.

\begin{figure}[htbp]
\centering
\subfloat[\hspace*{3em}]{ \label{fig:VV_flip1}
\includegraphics[scale=1.5,valign=c]{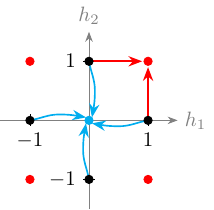}}
\hspace{2em}
\subfloat[\hspace*{3em}]{ \label{fig:VV_flip2}
\includegraphics[scale=1.5,valign=c]{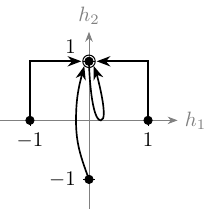}}
\caption{Illustration of the ladder operators $mJ^-$ and $\tilde{m}J^+$ acting on the $VV$ current. Black dots represent the primary current $[\mathbf J]_0$. Diagram (a) shows the flip from $[\mathbf J]_0$ to the 1st descendant current $[\mathbf J]_1$, indicated by red and cyan dots. Diagram (b) shows the flip from $[\mathbf J]_0$ to the 2nd descendant current $[\mathbf J]_2$, which occupies the same position as one of the $[\mathbf J]_0$ currents, represented by a black dot with a circle.}
\end{figure}

Applying ladder operators to these primary configurations yields five helicity components of the first descendant current: $(\pm1,\pm1)$, $(\pm1,\mp1)$, and $(0,0)$. As illustrated in Fig.~\ref{fig:VV_flip1}, the $(0,0)$ component (represented by a cyan dot) receives contributions from all four primary currents. Each of the other helicity components (marked by red dots) receives contributions from two distinct primary currents. 
The resulting 1st descendant currents are 
\begin{align}
[\mathbf J(1^+,2^+)]_1=&
c_1\tilde m_2 [12]|1]^{\dot\alpha}\langle\eta_2|^{\alpha}+
c_3\tilde m_1 [12] |2]^{\dot\alpha}\langle\eta_1|^{\alpha} ,\\
[\mathbf J(1^+,2^-)]_1=&
-c_1 m_2[1\eta_2] |1]^{\dot\alpha}\langle2|^{\alpha}+c_4\tilde m_1\langle\eta_12\rangle |1]^{\dot\alpha}\langle2|^{\alpha},\\
[\mathbf J(1^-,2^+)]_1=& 
c_2 \tilde m_2\langle1\eta_2\rangle |2]^{\dot\alpha}\langle1|^{\alpha}
-c_3 m_1[\eta_1 2] |2]^{\dot\alpha}\langle1|^{\alpha},\\
[\mathbf J(1^-,2^-)]_1=& 
-c_2 m_2\langle12\rangle |\eta_2]^{\dot\alpha}\langle1|^{\alpha}+
-c_4 m_1\langle12\rangle |\eta_1]^{\dot\alpha}\langle2|^{\alpha},\\
[\mathbf J(1^0,2^0)]_1=&
-c_1 m_1([\eta_12]|1]^{\dot\alpha}\langle2|^{\alpha}+[12]|\eta_1]^{\dot\alpha}\langle2|^{\alpha}) \nonumber\\
&+c_2 \tilde m_1(\langle\eta_12\rangle |2]^{\dot\alpha}\langle1|^{\alpha}+\langle12\rangle |2]^{\dot\alpha}\langle\eta_1|^{\alpha}) \nonumber\\
&-c_3 m_2([1\eta_2]|2]^{\dot\alpha}\langle1|^{\alpha}+[12]|\eta_2]^{\dot\alpha}\langle1|^{\alpha}) \nonumber\\
&+c_4 \tilde m_2(\langle1\eta_2\rangle |1]^{\dot\alpha}\langle2|^{\alpha}+\langle12\rangle |1]^{\dot\alpha}\langle\eta_2|^{\alpha}).
\end{align}
Here all the currents with $(\pm1,\mp1)$, and $(0,0)$ are conserved currents. For the helicity $(\pm1,\pm1)$, only one linear combinations of each helicity currents are conserved
\begin{eqnarray}
    [\mathbf{J}(1^-,2^-)]_1 &=& -\frac{1}{\tilde{m}_2} \langle12\rangle |\eta_2]^{\dot{\alpha}} \langle1|^{\alpha} -\frac{1}{\tilde{m}_1} \langle12\rangle |\eta_1]^{\dot{\alpha}} \langle2|^{\alpha}, \nonumber \\
    {[\mathbf{J}(1^+,2^+)]}_1 &=& \frac{1}{m_2} [12] |1]^{\dot{\alpha}} \langle\eta_2|^{\alpha} +\frac{1}{m_1} [12] |2]^{\dot{\alpha}} \langle\eta_1|^{\alpha},
\end{eqnarray}
while another combinations are not conserved.

Applying two ladder flips generates the second descendant currents, which return to the helicity configurations $(\pm1,0)$ and $(0,\pm1)$. As shown in Fig.~\ref{fig:VV_flip2}, each of these components receives contributions from all four primary currents:
\begin{align}
[\mathbf J(1^+,2^0)]_2=&
c_1\tilde m_2 m_2[1\eta_2]|1]^{\dot\alpha}\langle\eta_2|^{\alpha}
-c_2\tilde m_1^2\langle\eta_12\rangle |2]^{\dot\alpha}\langle\eta_1|^{\alpha} \nonumber\\
&+c_3\tilde m_1 m_2([1\eta_2]|2]^{\dot\alpha}\langle\eta_1|^{\alpha}+[12]|\eta_2]^{\dot\alpha}\langle\eta_1|^{\alpha})\nonumber\\
&-c_4\tilde m_1 \tilde m_2(\langle\eta_1\eta_2\rangle |1]^{\dot\alpha}\langle2|^{\alpha}+\langle\eta_12\rangle |1]^{\dot\alpha}\langle\eta_2|^{\alpha}),\\
[\mathbf J(1^-,2^0)]_2=&
-c_1 m_1^2[\eta_12]|\eta_1]^{\dot\alpha}\langle2|^{\alpha}
+c_2\tilde m_2 m_2\langle1\eta_2\rangle |\eta_2]^{\dot\alpha}\langle1|^{\alpha} \nonumber\\
&-c_3 m_1 m_2([\eta_1\eta_2]|2]^{\dot\alpha}\langle1|^{\alpha}+[\eta_12]|\eta_2]^{\dot\alpha}\langle1|^{\alpha})\nonumber \\
&+c_4 m_1\tilde m_2(\langle1\eta_2\rangle |\eta_1]^{\dot\alpha}\langle2|^{\alpha}+\langle1\eta_2\rangle |\eta_1]^{\dot\alpha}\langle2|^{\alpha}),\\
[\mathbf J(1^0,2^+)]_2=&
c_1 m_1\tilde m_2([\eta_12]|1]^{\dot\alpha}\langle\eta_2|^{\alpha}+[12]|\eta_1]^{\dot\alpha}\langle\eta_2|^{\alpha}) \nonumber\\
&-c_2 m_1m_2(\langle\eta_1\eta_2\rangle |2]^{\dot\alpha}\langle1|^{\alpha}+\langle1\eta_2\rangle |2]^{\dot\alpha}\langle\eta_1|^{\alpha}) \nonumber\\
&+c_3 \tilde m_1 m_1[\eta_12]|2]^{\dot\alpha}\langle\eta_1|^{\alpha}
-c_4 \tilde m_2^2\langle1\eta_2\rangle |1]^{\dot\alpha}\langle\eta_2|^{\alpha},\\
[\mathbf J(1^0,2^-)]_2=&
-c_1 m_1 m_2([\eta_1\eta_2]|1]^{\dot\alpha}\langle2|^{\alpha}+ [1\eta_2]|\eta_1]^{\dot\alpha}\langle2|^{\alpha}) \nonumber\\
&+c_2 \tilde m_1 m_2(\langle\eta_12\rangle |\eta_2]^{\dot\alpha}\langle1|^{\alpha}+\langle12\rangle |\eta_2]^{\dot\alpha}\langle\eta_1|^{\alpha}) \nonumber\\
&-c_3 m_2^2[1\eta_2]|\eta_2]^{\dot\alpha}\langle1|^{\alpha}
+c_4 \tilde m_1^2\langle\eta_12\rangle |\eta_1]^{\dot\alpha}\langle2|^{\alpha},
\end{align}
where all the 2nd descendant currents are not conserved currents.

The $VV$ current expansion also contains higher-order contributions $[\mathbf{J}]_3$ and $[\mathbf{J}]_4$, obtained by applying the ladder operators three and four times, respectively. However, these higher-order terms do not contribute to the leading-order massless-massive matching and are therefore not discussed further in this analysis.

\subsection{MHC 3-point amplitudes}

Beginning with three-particle states, we focus on the MHC amplitudes, which belong to the $(0,0)$ Lorentz representation. This can be obtained by the high-energy expansion of massive three-point amplitude $\mathbf{M}(\mathbf{1}^{s_1}, \mathbf{2}^{s_2}, \mathbf{3}^{s_3})$. In this expansion, the full amplitude, characterized by spins $(s_1, s_2, s_3)$, decomposes into a sum of helicity-chirality components $\mathcal{M}^{\mathcal{H},\mathcal{T}}$ according to Eq.~\eqref{eq:massless-decompose}, and then select the terms that satisfy the chirality-helicity unification
\begin{eqnarray}
\mathbf{M}(\mathbf{1}^{s_1}, \mathbf{2}^{s_2}, \mathbf{3}^{s_3})
&\xrightarrow[]{MHC} &
\sum_{\mathcal{T} = \mathcal{H}} 
 \mathcal{M}^{\mathcal{H},\mathcal{T}}
= \sum_{l=0}^{2s} [\mathcal{M}^{\mathcal{T} = \mathcal{H}}]_l
\;\sim\; \sum_{l=0}^{2s}
\underbrace{\frac{1}{\mathbf{m}^{\,s-1+l}}}_{\text{coefficient}} \times
\underbrace{E^{\,s}\left(\frac{\mathbf{m}^2}{E}\right)^{l}}_{\text{spinor structure}},
\end{eqnarray}
where $\mathcal{H} = (h_1, h_2, h_3)$ denotes the helicity configuration and $\mathcal{T} = (t_1, t_2, t_3)$ the corresponding transversality. Here $[\mathcal{M}^{\mathcal{T} = \mathcal{H}}]_l$ denotes the component of order $l$ in the high-energy expansion, and $s = s_1 + s_2 + s_3$ is the total spin. The number of terms is determined by $s$. By analogy with the classification of particle states, we refer to these structures as primary and descendant MHC amplitudes. 
These components are classified as follows:
\begin{eqnarray}
\begin{aligned}
\text{primary amplitude} &: \quad \mathcal{M}^{\rm pri} =  [\mathcal{M}]_0,\\
\text{1st descendant amplitude} &: \quad \mathcal{M}^{\rm 1st} =  [\mathcal{M}]_1,\\
\text{2nd descendant amplitudes} &: \quad \mathcal{M}^{\rm 2nd} =  [\mathcal{M}]_2,\\
&\; \vdots
\end{aligned}
\end{eqnarray}
The primary amplitude $[\mathcal{M}]_0$ contains only primary particle states, whereas the descendant currents incorporate descendant particle states.

To illustrate this expansion, we examine the $FFV$ massive amplitude $\langle\mathbf{13}\rangle [\mathbf{32}]$, which can be decomposed into $2\times 2 \times 3 = 12$ helicity categories. The $FFV$ amplitude consists of five terms
\begin{equation}
\mathcal{M}(FFV)=[\mathcal{M}]_{0}+[\mathcal{M}]_{1}+[\mathcal{M}]_{2} + [\mathcal{M}]_{3}+[\mathcal{M}]_{4},
\end{equation}
where all the primary and descendant amplitudes are 
\begin{equation}
\begin{tabular}{c|c|c|c|c}
\hline
order & \multicolumn{4}{c}{amplitude} \\
\hline
primary $[\mathcal{M}]_0$ & \multicolumn{4}{c}{$(-\frac12,+\frac12,0)$:\quad $\langle13\rangle [32]$} \\
\hline
1st des. $[\mathcal{M}]_1$ &  \makecell{$(+\frac12,+\frac12,0)$\\$-\tilde{m}_1 \langle\eta_13\rangle [32]$} & 
\makecell{$(-\frac12,-\frac12,0)$\\$m_2 \langle13\rangle [3\eta_2]$} & \makecell{$(-\frac12,+\frac12,+1)$\\$-m_3 \langle1\eta_3\rangle [32] $} 
& \makecell{$(-\frac12,+\frac12,-1)$\\$\tilde{m}_3 \langle13\rangle [\eta_32]$} \\
\hline
2nd des. $[\mathcal{M}]_2$ &  \makecell{$(-\frac12,+\frac12,0)$\\$-\tilde{m}_1 \langle\eta_13\rangle [32]$} & \makecell{$(+\frac12,-\frac12,0)$\\$m_2 \langle13\rangle [3\eta_2]$} & \makecell{$(\pm\frac12,\pm\frac12,\pm 1)$\\ $-m_3 \langle1\eta_3\rangle [32] $} 
& \makecell{$(\pm \frac12,\pm \frac12,\mp 1)$\\$\tilde{m}_3 \langle13\rangle [\eta_32]$} \\
\hline
3rd des. $[\mathcal{M}]_3$ & \makecell{$(+\frac12,+\frac12,0)$\\$-\tilde{m}_1 \langle\eta_13\rangle [32]$} & \makecell{$(-\frac12,-\frac12,0)$\\$m_2 \langle13\rangle [3\eta_2]$} & \makecell{$(+\frac12,-\frac12,+ 1)$\\$-m_3 \langle1\eta_3\rangle [32]$}
& \makecell{$(+ \frac12,- \frac12,- 1)$\\$\tilde{m}_3 \langle13\rangle [\eta_32]$} \\
\hline
4th des. $[\mathcal{M}]_4$ & \multicolumn{4}{c}{$(+\frac12,-\frac12,0)$:\quad${\color{gray} \tilde{m}_1 m_2 m_3 \tilde{m}_3 \langle\eta_1\eta_3\rangle [\eta_3\eta_2]}$} \\
\hline
\end{tabular}
\end{equation}
The amplitudes would only expand up to the 4th descendant order.

\paragraph{Extended Little Group Covariance}

Analogous to the construction of MHC states, first start with the primary amplitude, and then descendant ones are derived by repeatedly applying the ladder operators $mJ^-$ or $\tilde{m}J^+$ to the primary one. Applying one such operator to a specific particle flips its helicity and yields the corresponding first descendant ones; successive applications generate higher descendants:
\begin{eqnarray}
\begin{aligned}
[\mathcal{M}]_1 &= mJ^- \circ [\mathcal{M}]_0 \;+\; \tilde{m}J^+ \circ [\mathcal{M}]_0,\\
[\mathcal{M}]_2 &= (mJ^-)^2 \circ [\mathcal{M}]_0 \;+\; (\tilde{m}J^+ mJ^-) \circ [\mathcal{M}]_0 \;+\; (\tilde{m}J^+)^2 \circ [\mathcal{M}]_0,\\
&\; \vdots
\end{aligned}
\end{eqnarray}
We note that, depending on the spins $s_1$ and $s_2$ of the particles in the amplitude, the MHC expansion terminates at the $2(s_1 + s_2+s_3)$-th descendant amplitudes.

All the MHC amplitudes can be obtained by applying the ladder operator to the primary one. We note that all the MHC amplitudes should have the extended Little group covariance. We utilize the $\frac{m}{\mathbf m} J^-$ and $\frac{\tilde m}{\mathbf m} J^+$ to precisely obtain the ELG covariant form. Starting from a primary amplitude with total helicity $h^{\text{pri}}$, this action can be illustrated as follows
\begin{eqnarray}
\includegraphics[scale=1,valign=c]{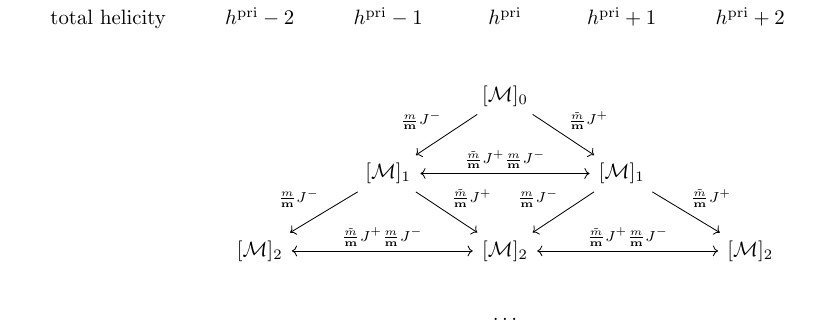}
\end{eqnarray}

If the primary amplitude involve in the vector boson, the MHC amplitudes can also be identified to combination of the two-particle currents and vector boson states, derived in previous subsection
\begin{eqnarray}
    [\mathcal{M}]_{\text{Leading}} = \mathbb{J} \cdot \mathbb{A}.
\end{eqnarray}
In the following, we will start from the primary MHC amplitudes, obtained by simply unbolding the massive amplitudes, and then use the ladder operators to determine the descendant ones, each decomposed into two-particle MHC current and MHC vector boson order by order.

\paragraph{FFV amplitudes}

The first example is the $FFV$ amplitudes. The the primary and descendant amplitudes can be decomposed into two-particle MHC current and MHC vector boson order by order, 
\begin{eqnarray}
\begin{array}{c|c|c|c}
\hline
\mbox{order} & \mbox{amplitude} & \mathbb{J} & \mathbb{A} \\
\hline
\mbox{primary} & \langle 13 \rangle [23] & |{2}]^{\dot{\alpha}} \langle{1}|^{\alpha} & |{3}\rangle_{\alpha} [{3}|_{\dot{\alpha}} \\
\hline
\multirow{4}{*}{1st descendant} & \tilde{m}_1 \langle\eta_1 3\rangle [23] & \tilde{m}_1 |{2}]^{\dot{\alpha}} \langle \eta_1|^{\alpha} & \multirow{2}{*}{ $|3\rangle_{\alpha} [3|_{\dot{\alpha}}$ } \\
& m_2 \langle13\rangle [\eta_23] & m_2 
|\eta_2]^{\dot{\alpha}} \langle 1|^{\alpha} & \\
\cline{2-4}
& m_3 \langle13\rangle [\eta_32] & 
\multirow{2}{*}{$|2]^{\dot{\alpha}} \langle1|^{\alpha}$} & m_3 |\eta_3\rangle_{\alpha} [3|_{\dot{\alpha}} \\
& \tilde m_3 \langle1\eta_3\rangle [32] & & \tilde m_3 |3\rangle_{\alpha} [\eta_3|_{\dot{\alpha}} \\
\hline
\multirow{3}{*}{2nd descendant} & \tilde{m}_1 m_3 \langle\eta_13\rangle [\eta_32] & \tilde{m}_1 |2]^{\dot{\alpha}} \langle\eta_1|^{\alpha} & m_3 |3\rangle_{\alpha} [\eta_3|_{\dot{\alpha}} \\
& \tilde{m}_1 m_2 \langle \eta_13\rangle [3\eta_2] & \tilde{m}_1 m_2 |\eta_2]^{\dot{\alpha}} \langle\eta_1|^{\alpha} & |3\rangle_{\alpha} [3|_{\dot{\alpha}} \\
& \vdots & \vdots & \vdots \\
\hline
\end{array}
\end{eqnarray}
In the following only the primary and first descendant MHC amplitudes are needed, and all other descendant amplitudes can be neglected.

For the $FFV$ amplitude, the primary amplitudes are obtained by simply unbolding the $FFV$ massive amplitudes. The primary amplitudes fall into two helicity categories with total helicity $h=0$,
\begin{equation} \begin{aligned}
h=0:\quad &(-\tfrac12,+\tfrac12,0),& &(+\tfrac12,-\tfrac12,0).& \\
\end{aligned} \end{equation}
The corresponding amplitude are
\begin{equation} \begin{aligned}
\langle13\rangle[32]=\Ampthree{1^-}{2^+}{3^0}{\fer{red}{i1}{v1}}{\antfer{red}{i2}{v1}}{\bos{i3}{brown}},\quad
[13]\langle32\rangle=\Ampthree{1^+}{2^-}{3^0}{\fer{cyan}{i1}{v1}}{\antfer{cyan}{i2}{v1}}{\bos{i3}{brown}}.
\end{aligned} \end{equation}
These are vanishing structures because they contain both $\langle ij\rangle$ and $[ij]$. Due to momentum conservation $\sum p_i = 0$, we have $\lambda_1 \propto \lambda_2 \propto \lambda_3$ or $\tilde\lambda_1 \propto \tilde\lambda_2 \propto \tilde\lambda_3$, which implies that any MHC amplitude involving both $\langle ij\rangle$ and $[ij]$ must vanish.

These vanishing primary amplitudes correspond to the combination between a primary $FF$ current $[\mathbf J]_0$ and a priamry vector state $[\mathbf A]_0$. 
\begin{equation} \begin{aligned}
\langle13\rangle[32]&=[\mathbf J(1^{-\frac12},2^{+\frac12})]_0\cdot[\mathbf A(3^0)]_0,\\
[13]\langle32\rangle&=[\mathbf J(1^{+\frac12},2^{-\frac12})]_0\cdot[\mathbf A(3^0)]_0,
\end{aligned} \end{equation}
where the primary currents are
\begin{align}
[\mathbf J(1^{-\frac12},2^{+\frac12})]_0=|2]^{\dot\alpha}\langle1|^{\alpha},\quad
[\mathbf J(1^{+\frac12},2^{-\frac12})]_0=|1]^{\dot\alpha}\langle2|^{\alpha}.
\end{align}
These currents correspond to the Noether current of massless fermion, $\bar\psi\gamma^\mu\psi$. They satisfy the conserved current condition
\begin{equation} \begin{aligned}
\partial\cdot[\mathbf J(1^{-\frac12},2^{+\frac12})]_0=\langle1|1+2|2]=0,\quad
\partial\cdot[\mathbf J(1^{+\frac12},2^{-\frac12})]_0=\langle2|1+2|1]=0.
\end{aligned} \end{equation}
Thus the conserved current condition is closely associated with the vanishing MHC amplitudes.

\begin{figure}[htbp]
\centering
\subfloat[\hspace*{3em}]{ \label{fig:FFV_flip1}
\includegraphics[scale=1.2,valign=c]{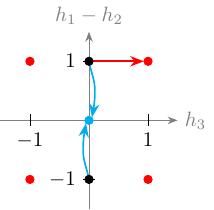}}
\hspace{2em}
\subfloat[\hspace*{3em}]{ \label{fig:FFV_flip2}
\includegraphics[scale=1.2,valign=c]{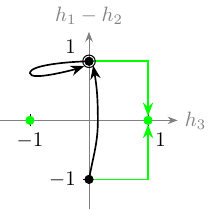}}
\caption{Illustration of the ladder operators $mJ^-$ and $\tilde{m}J^+$ acting on the $FFV$ amplitude. The black dots
represent the primary amplitudes. Diagram (a) shows the flip from the primary amplitude to its 1st descendant, indicated by red and cyan dots. Diagram (b) shows the flip from primary amplitude to its 2nd, represented by a green dot and a circled black dot.}
\end{figure}

Then we can apply the ladder operator to the primary amplitude, as shown in Fig.~\ref{fig:FFV_flip1} and \ref{fig:FFV_flip2}, and obtain the 1st descendant amplitudes, which have six helicity categories,
\begin{equation} \begin{aligned}
h=+1:\quad&(+\tfrac12,+\tfrac12,0),(+\tfrac12,-\tfrac12,+1),(-\tfrac12,+\tfrac12,+1),& \\
h=-1:\quad&(-\tfrac12,-\tfrac12,0),(+\tfrac12,-\tfrac12,-1),(-\tfrac12,+\tfrac12,-1).&
\end{aligned} \end{equation}
They satisfy the condition $h=\pm1$, so the 1st descendant amplitude can provide the leading contribution. As examples, we consider two typical helicity categories: $(-\tfrac12,+\tfrac12,+1)$ with one transverse vector, and $(+\tfrac12,+\tfrac12,0)$ with zero transverse vectors. Their corresponding amplitudes are
\begin{equation} \begin{aligned}
(-\tfrac12,+\tfrac12,+1):&\quad\tilde m_3[23]\langle\eta_31\rangle= \Ampthree{1^-}{2^+}{3^+}{\fer{red}{i1}{v1}}{\antfer{red}{i2}{v1}}{\bosflip{1}{-55}{brown}{cyan}},\\
(+\tfrac12,+\tfrac12,0):&\quad\tilde m_1[23]\langle3\eta_1\rangle=
\Ampthree{1^+}{2^+}{3^0}{\ferflip{1}{180}{cyan}{red}}{\antfer{red}{i2}{v1}}{\bos{i3}{brown}},\quad
\tilde m_2\langle\eta_23\rangle[31]=
\Ampthree{1^+}{2^+}{3^0}{\fer{cyan}{i1}{v1}}{\antferflip{1}{55}{red}{cyan}}{\bos{i3}{brown}}.
\end{aligned} \end{equation}
Others can be similarly obtained.

Note that the above leading MHC amplitudes does not have extended little group covariance. 
To recover the extended little group covariance, let us apply the normalized ladder operator, which would relate the four amplitudes. 
\begin{equation}
\includegraphics[scale=1,valign=c]{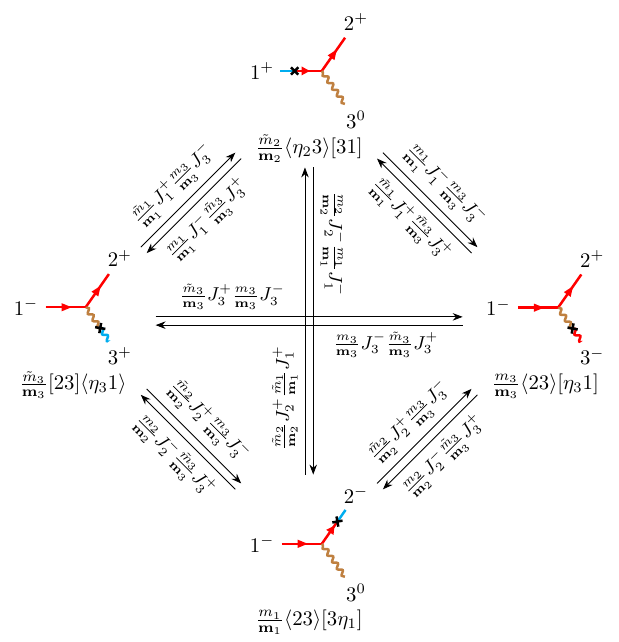}
\end{equation}
Similarly, this method can be used to obtain higher-order descendant amplitudes.

\paragraph{FFS \& SSS Amplitudes}

For the $FFS$ amplitude, the primary MHC amplitudes has two helicity categories
\begin{equation} \begin{aligned}
h=-1:\quad &(-\tfrac12,-\tfrac12,0),& \\
h=+1:\quad &(+\tfrac12,+\tfrac12,0),& 
\end{aligned} \end{equation}
They satisfy the condition $h=\pm1$. The corresponding primary amplitude are
\begin{equation} \begin{aligned}
(-\tfrac12,-\tfrac12,0):&\quad\langle12\rangle=\Ampthree{1^-}{2^-}{3^0}{\fer{red}{i1}{v1}}{\antfer{cyan}{i2}{v1}}{\sca{i3}},\\
(+\tfrac12,+\tfrac12,0):&\quad [12]=\Ampthree{1^+}{2^+}{3^0}{\fer{cyan}{i1}{v1}}{\antfer{red}{i2}{v1}}{\sca{i3}}.
\end{aligned}\end{equation}
They do not vanish, they consist of $\lambda$ or $\tilde\lambda$ only. Thus, the priamry $FFS$ amplitude gives the leading contributions.

For the $SSS$ amplitude, it only has one helicity category $(0,0,0)$. The primary MHC amplitude corresponds to the trivial Lorentz structure
\begin{equation} \begin{aligned}
(0,0,0):\quad 1=\Ampthree{1^0}{2^0}{3^0}{\sca{i1}}{\sca{i2}}{\sca{i3}}.
\end{aligned}\end{equation}
It gives the leading contribution obviously.

\paragraph{VVS amplitudes} 
For the $VVS$ amplitude, the primary amplitude has only one  helicity category,
\begin{equation} \begin{aligned}
h=0:\quad (0,0,0).
\end{aligned}\end{equation}
The corresponding amplitude is given by
\begin{equation} \begin{aligned}
(0,0,0):\quad \langle12\rangle[21]=\Ampthree{1^0}{2^0}{3^0}{\bos{i1}{brown}}{\bos{i2}{brown}}{\sca{i3}}
\end{aligned}\end{equation}
This amplitude vanishes due to the simultaneous presence of $\langle12\rangle$ and $[21]$. 

Applying the ladder operator to the primary $VVS$ amplitude yields the first descendant amplitudes, which fall into four helicity categories,
\begin{equation} \begin{aligned}
h=+1:\quad&(-1,0,0),(0,-1,0),& \\
h=-1:\quad&(+1,0,0),(0,+1,0).&
\end{aligned} \end{equation}
Since these satisfy $h = \pm 1$, the corresponding amplitudes are leading contributions:
\begin{equation} \begin{aligned}
m_1\langle12\rangle[2\eta_1]&=\Ampthree{1^-}{2^0}{3^0}{\bosflip{1}{180}{brown}{red}}{\bos{i2}{brown}}{\sca{i3}},&
-\tilde m_1\langle\eta_12\rangle[21]&=\Ampthree{1^+}{2^0}{3^0}{\bosflip{1}{180}{brown}{cyan}}{\bos{i2}{brown}}{\sca{i3}}, \\
m_2\langle12\rangle[\eta_21]&=\Ampthree{1^0}{2^-}{3^0}{\bos{i1}{brown}}{\bosflip{1}{55}{brown}{red}}{\sca{i3}},&
-\tilde m_2\langle1\eta_2\rangle[21]&=\Ampthree{1^0}{2^+}{3^0}{\bos{i1}{brown}}{\bosflip{1}{55}{brown}{cyan}}{\sca{i3}}.\\
\end{aligned}\end{equation}

For the $VVS$ amplitude, the primary amplitude has only one  helicity category,
\begin{equation} \begin{aligned}
h=0:\quad (0,0,0).
\end{aligned}\end{equation}
The corresponding amplitude is given by
\begin{equation} \begin{aligned}
(0,0,0):\quad \langle12\rangle[21]=\Ampthree{1^0}{2^0}{3^0}{\bos{i1}{brown}}{\bos{i2}{brown}}{\sca{i3}}.
\end{aligned}\end{equation}
This amplitude vanishes due to the simultaneous presence of $\langle12\rangle$ and $[21]$. The vanishing primary amplitude can also be expressed as a coupling of the form $[\mathbf{J}]_0 \cdot [\mathbf{A}]_0$. There are two equivalent ways to perform the decomposition: 
\begin{equation} \begin{aligned}
\langle12\rangle[21]=[\mathbf J(2^{0},3^{0})]_0\cdot[\mathbf A(1^{0})]_0=[\mathbf J(1^{0},3^{0})]_0\cdot[\mathbf A(2^{0})]_0,
\end{aligned}\end{equation}
where the primary $VS$ current $[\mathbf{J}]_0$ is defined as
\begin{equation} \begin{aligned}
\relax
[\mathbf{J}(2^{0},3^{0})]_0&=|2]^{\dot\alpha}\langle2|^{\alpha}, \\
[\mathbf{J}(1^{0},3^{0})]_0&=|1]^{\dot\alpha}\langle1|^{\alpha}.
\end{aligned}\end{equation}
This current is conserved and can be interpreted as the Noether current for a massless scalar boson, $\phi^* (D^\mu \phi) - (D^\mu \phi) \phi$~\footnote{Strictly speaking, the Noether current $\phi^* (D^\mu \phi) - (D^\mu \phi^*) \phi$ corresponds to the spinor structure $\tfrac12(|2]^{\dot\alpha}\langle2|^{\alpha} - |3]^{\dot\alpha}\langle3|^{\alpha})$, which is not identical to the primary MHC current $[\mathbf{J}]_0$. However, when coupled to a vector $[\mathbf{A}]_0$, the two become equivalent: $[\mathbf{J}]_0 \cdot [\mathbf{A}]_0 = \langle12\rangle[21] = \tfrac12(\langle12\rangle[21] - \langle13\rangle[31])$.}.

Applying the ladder operator to the primary $VVS$ amplitude yields the first descendant amplitudes, which fall into four helicity categories,
\begin{equation} \begin{aligned}
h=+1:\quad&(-1,0,0),(0,-1,0),& \\
h=-1:\quad&(+1,0,0),(0,+1,0).&
\end{aligned} \end{equation}
Since these satisfy $h = \pm 1$, the corresponding amplitudes are leading contributions:
\begin{equation} \begin{aligned}
m_1\langle12\rangle[2\eta_1]&=\Ampthree{1^-}{2^0}{3^0}{\bosflip{1}{180}{brown}{red}}{\bos{i2}{brown}}{\sca{i3}},&
-\tilde m_1\langle\eta_12\rangle[21]&=\Ampthree{1^+}{2^0}{3^0}{\bosflip{1}{180}{brown}{cyan}}{\bos{i2}{brown}}{\sca{i3}}, \\
m_2\langle12\rangle[\eta_21]&=\Ampthree{1^0}{2^-}{3^0}{\bos{i1}{brown}}{\bosflip{1}{55}{brown}{red}}{\sca{i3}},&
-\tilde m_2\langle1\eta_2\rangle[21]&=\Ampthree{1^0}{2^+}{3^0}{\bos{i1}{brown}}{\bosflip{1}{55}{brown}{cyan}}{\sca{i3}}.\\
\end{aligned}\end{equation}

For the $VVS$ case, we can also apply the ladder operators to relate the descendants of the same order, recovering the extended little group covariance. For the four 1st descendant amplitudes, we have
\begin{equation}
\includegraphics[scale=1,valign=c]{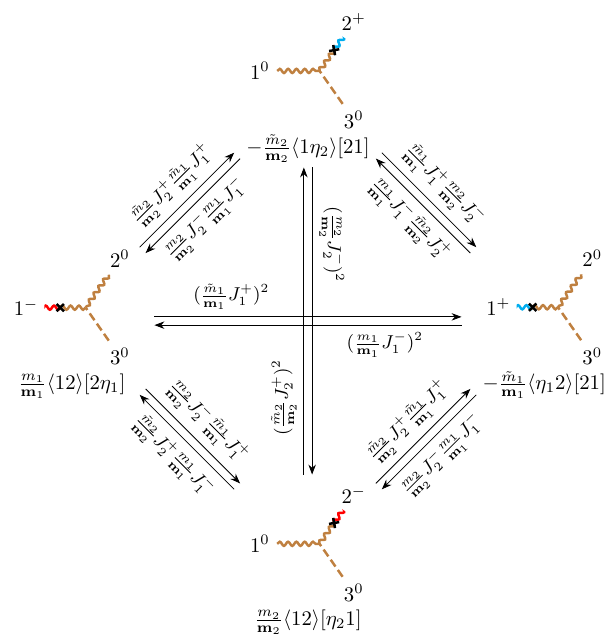}
\end{equation}
The higher-order descendant amplitudes can be obtained in a similar way.

\paragraph{VVV amplitudes}

For the $VVV$ amplitude, given the six massive amplitudes, the primary amplitudes fall into six helicity categories~\footnote{There are another two helicity categories $(-1,-1,-1)$ and $(+1, +1, +1)$ for primary amplitudes, belonging to the EFT operators. }, classified by the total helicity $h$,
\begin{equation} \begin{aligned}
h=-1:\quad &(-1,0,0),(0,-1,0),(0,0,-1),& \\
h=+1:\quad &(+1,0,0),(0,+1,0),(0,0,+1).& \\
\end{aligned} \end{equation}
The corresponding amplitudes are
\begin{equation} \begin{aligned} \label{eq:VVV_primary}
&\langle12\rangle[23]\langle31\rangle,&
&\langle12\rangle\langle23\rangle[31],& 
&[12]\langle23\rangle\langle31\rangle, \\  
&[12]\langle23\rangle[31],&
&[12][23]\langle31\rangle,&
&\langle12\rangle[23][31].
\end{aligned} \end{equation}
Although these primary amplitudes carry total helicity $h = \pm 1$, they vanish because each contains both angle and square brackets $\langle ij\rangle$ and $[ij]$. Diagrammatically, these vanishing primary amplitudes can be represented as
\begin{equation} \begin{aligned}
&\Ampthree{1^-}{2^0}{3^0}{\bos{i1}{red}}{\bos{i2}{brown}}{\bos{i3}{brown}},\quad
\Ampthree{1^0}{2^-}{3^0}{\bos{i1}{brown}}{\bos{i2}{red}}{\bos{i3}{brown}}, \quad
\Ampthree{1^0}{2^0}{3^-}{\bos{i1}{brown}}{\bos{i2}{brown}}{\bos{i3}{red}},\quad \\  
&\Ampthree{1^+}{2^0}{3^0}{\bos{i1}{cyan}}{\bos{i2}{brown}}{\bos{i3}{brown}},\quad
\Ampthree{1^0}{2^+}{3^0}{\bos{i1}{brown}}{\bos{i2}{cyan}}{\bos{i3}{brown}},\quad
\Ampthree{1^0}{2^0}{3^+}{\bos{i1}{brown}}{\bos{i2}{brown}}{\bos{i3}{cyan}}.
\end{aligned} \end{equation}
Each primary $VVV$ amplitude can be expressed in two equivalent ways as the current form $[\mathbf{J}]_0 \cdot [\mathbf{A}]_0$. For the helicity category $(-1,0,0)$: 
\begin{equation} \begin{aligned}
\langle12\rangle[23]\langle31\rangle=[\mathbf J(1^{-1},3^{0})]_0\cdot[\mathbf A(2^{0})]_0=[\mathbf J(1^{-1},2^{0})]_0\cdot[\mathbf A(3^{0})]_0,
\end{aligned}\end{equation}
where the primary $VV$ current $[\mathbf{J}]_0$ is conserved current and defined as
\begin{equation} \begin{aligned}
\relax
{[\mathbf{J}(1^-,2^0)]}_0 &= -\langle12\rangle |2]^{\dot\alpha}\langle1|^{\alpha}, \\
{[\mathbf{J}(1^0,2^-)]}_0 &= -\langle12\rangle |1]^{\dot\alpha}\langle2|^{\alpha},\\
{[\mathbf{J}(1^+,2^0)]}_0 &= -[12] |1]^{\dot\alpha} \langle2|^{\alpha}, \\
{[\mathbf{J}(1^0,2^+)]}_0 &= -[12] |2]^{\dot\alpha} \langle1|^{\alpha}.
\end{aligned}\end{equation}

For the 1st descendant MHC amplitudes, there are thirteen helicity categories,
\begin{equation} \begin{aligned}
h&=+2:& &(+1, +1,0),\quad(0,+1,+1),\quad(+1,0,+1),\\
h&=0:& &(\pm1, \mp1,0),\quad(0,\pm1,\mp1),\quad(\pm1,0,\mp1),\quad (0,0,0),\\
h&=-2:& &(-1, -1,0),\quad(0,-1,-1),\quad(-1,0,-1).
\end{aligned} \end{equation}
These do not satisfy the condition $h=\pm1$, so the 1st descendant $VVV$ amplitude do not correspond to the massless amplitude in the SM, and they should either vanish or belong to the EFT category. 
Among the 1st descendant currents, the helicity-$(\pm\mp)$ currents are all conserved
\begin{align}
    [\mathbf J(1^+,2^-)]_1 =&
-c_1 m_2[1\eta_2] |1]^{\dot\alpha}\langle2|^{\alpha}+c_4\tilde m_1\langle\eta_12\rangle |1]^{\dot\alpha}\langle2|^{\alpha},\\
[\mathbf J(1^-,2^+)]_1 =& 
c_2 \tilde m_2\langle1\eta_2\rangle |2]^{\dot\alpha}\langle1|^{\alpha}
-c_3 m_1[\eta_1 2] |2]^{\dot\alpha}\langle1|^{\alpha},
\end{align}
corresponding to the $F_{\mu\nu} A^{\nu}$ (and $\tilde{F}_{\mu\nu} A^{\nu}$) Noether current. 
While the helicity-$(\pm\pm)$ currents, in contrast, 
\begin{align}
[\mathbf J(1^+,2^+)]_1=&
c_1\tilde m_2 [12]|1]^{\dot\alpha}\langle\eta_2|^{\alpha}+
c_3\tilde m_1 [12] |2]^{\dot\alpha}\langle\eta_1|^{\alpha} ,\\
[\mathbf J(1^-,2^-)]_1=& 
-c_2 m_2\langle12\rangle |\eta_2]^{\dot\alpha}\langle1|^{\alpha}+
-c_4 m_1\langle12\rangle |\eta_1]^{\dot\alpha}\langle2|^{\alpha},
\end{align}
lead to either the $F_{\mu\nu} A^{\nu}$ Noether current or non-conserved ones.

We classify them into three cases, in which only in the first case some combination of amplitudes are non-vanishing EFT ones, and all other are vanishing: 
\begin{itemize}
\item First consider the helicity categories with $h=\pm2$, which have two transverse vectors of the same helicity. In this case, certain combinations of amplitudes vanish, but another non-vanishing combination belongs to the EFT amplitudes. For example, the helicity $(-1,-1,0)$ includes two terms,
\begin{equation} \begin{aligned}
(-1,-1,0):&\quad 
m_2\langle12\rangle[\eta_2 3]\langle31\rangle
=\Ampthree{1^-}{2^-}{3^0}{\bos{i1}{red}}{\bosflip{1}{55}{brown}{red}}{\bos{i3}{brown}},
m_1\langle12\rangle\langle23\rangle[3\eta_1]
=\Ampthree{1^-}{2^-}{3^0}{\bosflip{1}{180}{brown}{red}}{\bos{i2}{red}}{\bos{i3}{brown}}.
\end{aligned} \end{equation}
Each of the two terms can be identified as $[\mathbf{J}]_1 \cdot [\mathbf{A}]_0$, while together they can be identified as $[\mathbf{J}]_0 \cdot [\mathbf{A}]_1$:
\begin{equation} \begin{aligned}
[\mathbf J(1^{-1},2^{-1})]_1\cdot[\mathbf A(3^0)]_0=
c_1 m_2\langle12\rangle[\eta_2 3]\langle31\rangle
+c_2 m_1\langle12\rangle\langle23\rangle[3\eta_1],
\end{aligned} \end{equation}
where $c_i$ are coefficients. Although the two terms are individually non-zero, their sum can vanish
\begin{equation} \begin{aligned}
\frac{1}{\mathbf m^2_2} m_2\langle12\rangle[\eta_2 3]\langle31\rangle
+\frac{1}{\mathbf m^2_1}m_1\langle12\rangle\langle23\rangle[3\eta_1]=0.
\end{aligned} \end{equation}
Hence, the 1st descendant current $[\mathbf{J}(1^{-1},2^{-1})]_1$ is conserved only if takes the form 
\begin{equation} \label{eq:conserved_descendant}
[\mathbf J(1^{-1},2^{-1})]_1= -\frac{1}{\mathbf m_2^2} m_2\langle12\rangle |\eta_2]^{\dot\alpha}\langle1|^{\alpha}
-\frac{1}{\mathbf m_1^2} m_1\langle12\rangle |\eta_1]^{\dot\alpha}\langle2|^{\alpha}.
\end{equation}
For the other helicity category with $h=\pm2$, each category corresponds to one such conserved current.

\item For helicity categories with $h=0$, there are two cases, and in both cases, the amplitudes vanish. We consider the helicity categories with two transverse vector of opposite helicity. For example, the helicity $(-1,+1,0)$ includes two terms,
\begin{equation} \begin{aligned}
(-1,+1,0):&\quad \tilde m_2\langle1\eta_2\rangle[23]\langle31\rangle 
=\Ampthree{1^-}{2^+}{3^0}{\bos{i1}{red}}{\bosflip{1}{55}{brown}{cyan}}{\bos{i3}{brown}},
m_1[\eta_1 2][23]\langle31\rangle
=\Ampthree{1^-}{2^+}{3^0}{\bosflip{1}{180}{brown}{red}}{\bos{i2}{cyan}}{\bos{i3}{brown}},\\
\end{aligned} \end{equation}
which are derived by acting ladder operators on the primary amplitudes with helicities $(-1,+1,0)$ and $(+1,-1,0)$. We identify these two terms as the coupling $[\mathbf J]_1\cdot[\mathbf A]_0$:
\begin{equation} \begin{aligned}
[\mathbf J(1^{+1},2^{-1})]_1\cdot[\mathbf A(3^0)]_0=
c_1\tilde m_2\langle1\eta_2\rangle[23]\langle31\rangle
+c_2 m_1[\eta_1 2][23]\langle31\rangle,
\end{aligned} \end{equation}
Both terms contain $[23]$ and $\langle 31 \rangle$, and thus vanish individually.
\begin{equation} \begin{aligned}
\tilde m_2\langle1\eta_2\rangle[23]\langle31\rangle
=m_1[\eta_1 2][23]\langle31\rangle=0.
\end{aligned} \end{equation}
Therefore, the current $[\mathbf{J}(1^{+1},2^{-1})]_1$ is conserved, with no constraints on the coefficients $c_1$ and $c_2$. The same holds for currents in other helicity categories with $h = \pm 2$.

\item Finally we consider the helicity $(0,0,0)$. It has six terms,
\begin{align}
(0,0,0):\quad&\tilde m_1(\langle \eta_12\rangle[23]\langle31\rangle+\langle 12\rangle[23]\langle3\eta_1\rangle)
=\Ampthree{1^0}{2^0}{3^0}{\bosflip{1}{180}{red}{brown}}{\bos{i2}{brown}}{\bos{i3}{brown}},& \nonumber \\
&\tilde m_2(\langle1\eta_2\rangle\langle23\rangle[31]+\langle12\rangle\langle\eta_23\rangle[31])
=\Ampthree{1^0}{2^0}{3^0}{\bos{i1}{brown}}{\bosflip{1}{55}{red}{brown}}{\bos{i3}{brown}},& \nonumber \\
&\tilde m_3([12]\langle2\eta_3\rangle\langle31\rangle+[12]\langle23\rangle\langle\eta_31\rangle)
=\Ampthree{1^0}{2^0}{3^0}{\bos{i1}{brown}}{\bos{i2}{brown}}{\bosflip{1}{-55}{red}{brown}}, \nonumber \\  
&m_1([\eta_12]\langle23\rangle[31]+[12]\langle23\rangle[3\eta_1])
=\Ampthree{1^0}{2^0}{3^0}{\bosflip{1}{180}{cyan}{brown}}{\bos{i2}{brown}}{\bos{i3}{brown}},& \nonumber \\
&m_2([1\eta_2][23]\langle31\rangle+[12][\eta_23]\langle31\rangle)
=\Ampthree{1^0}{2^0}{3^0}{\bos{i1}{brown}}{\bosflip{1}{55}{cyan}{brown}}{\bos{i3}{brown}},& \nonumber \\
&m_3(\langle12\rangle[2\eta_3][31]+\langle12\rangle[23][\eta_31])
=\Ampthree{1^0}{2^0}{3^0}{\bos{i1}{brown}}{\bos{i2}{brown}}{\bosflip{1}{-55}{cyan}{brown}}. 
\end{align}
If we identify particle 3 as the vector $[\mathbf A]_0$, there are four terms can compose coupling $[\mathbf J]_1\cdot [\mathbf A]_0$, 
\begin{equation} \begin{aligned}
[\mathbf J(1^{0},2^{0})]_1\cdot[\mathbf A(3^0)]_0=
&c_1\tilde m_1(\langle \eta_12\rangle[23]\langle31\rangle+\langle 12\rangle[23]\langle3\eta_1\rangle)\\
&+c_2\tilde m_2(\langle1\eta_2\rangle\langle23\rangle[31]+\langle12\rangle\langle\eta_23\rangle[31])\\
&+c_3 m_1([\eta_12]\langle23\rangle[31]+[12]\langle23\rangle[3\eta_1])\\
&+c_4 m_2([1\eta_2][23]\langle31\rangle+[12][\eta_23]\langle31\rangle),
\end{aligned} \end{equation}
These terms contain $[ij]$ and $\langle ij \rangle$ simultaneously, and thus vanish individually. Therefore, the current $[\mathbf{J}(1^{0},2^{0})]_1$ is also conserved, with no constraints on the four coefficients.

\end{itemize}

For 2nd descendant MHC amplitudes, it has twelve helicity categories,
\begin{equation} \begin{aligned}
h&=+1:& &(-1,+1,+1),& &(+1,-1,+1),& &(+1,+1,-1),&\\
&& &(0,0,+1),& &(0,+1,0),& &(+1,0,0),&\\
h&=-1:& &(-1, -1,+1),& &(+1,-1,-1),& &(-1,+1,-1),&\\
&& &(0,0,-1),& &(0,-1,0),& &(-1,0,0),&
\end{aligned} \end{equation}
They satisfy the condition $h=\pm1$, and one can verify that the MHC amplitudes do not vanish. Therefore, the leading $VVV$ contribution arises from 2nd descendant MHC amplitudes. Due to the large number of such amplitudes, we restrict our consideration to two typical helicity categories $(+1,+1,-1)$ and $(-1,0,0)$, which has three and one transverse vector bosons. The other helicity categories can be obtained in a similar way, whose diagram can be derived by change color and location of mass insertion. The two typical helicity categories have
\begin{itemize}
    \item For the $(+1,+1,-1)$ helicity category, the only amplitudes are the $[\mathbf{J}]_1\cdot [\mathbf{A}]_1$ case, since the currents belong to $(\pm1, \pm 1)$ or $(\pm1, \mp 1)$. The amplitudes are constructed by $[\mathbf{J}(1^+,2^+)]_1\cdot[\mathbf{A}(3^-)]_1$ or $[\mathbf{J}(1^+,2^-)]_1\cdot[\mathbf{A}(3^+)]_1$. Note that $[\mathbf{J}(1^+,2^-)]_1$ are all conserved current, while $[\mathbf{J}(1^+,2^+)]_1$ are partially conserved. The non-conserved combination would give rise to the descendant EFT amplitudes. While all other amplitudes with the conserved currents give rise to the leading SM amplitudes. First the three conserved currents give  
\begin{equation} \begin{aligned} \label{eq:VVV_leading_1}
\tilde m_1 m_3 [12][2 \eta_3]\langle3\eta_1\rangle
&=\Ampthree{1^+}{2^+}{3^-}{\bosflip{1}{180}{brown}{cyan}}{\bos{i2}{cyan}}{\bosflip{1}{-55}{brown}{red}},&
\tilde m_2 m_3 [12]\langle\eta_23\rangle[\eta_31]
&=\Ampthree{1^+}{2^+}{3^-}{\bos{i1}{cyan}}{\bosflip{1}{55}{brown}{cyan}}{\bosflip{1}{-55}{brown}{red}},& \\
\tilde m_1 \tilde m_2 [12]\langle\eta_23\rangle\langle 3 \eta_1\rangle
&=\Ampthree{1^+}{2^+}{3^-}{\bosflip{1}{180}{brown}{cyan}}{\bosflip{1}{55}{brown}{cyan}}{\bos{i3}{red}}.&
\end{aligned}\end{equation}
Then the conserved current combination gives
\begin{eqnarray}
    [\mathbf{J}(1^-,2^-)]_1\cdot[\mathbf{A}(3^+)]_1 =& \frac{\tilde{m}_3}{\tilde{m}_2} \langle12\rangle [3\eta_2] \langle1\eta_3\rangle +\frac{\tilde{m}_3}{\tilde{m}_1} \langle12\rangle [3\eta_1] \langle2\eta_3\rangle \nonumber \\
=& \Ampthree{1^-}{2^-}{3^+}{\bos{i1}{red}}{\bosflip{1}{55}{brown}{red}}{\bosflip{1}{-55}{brown}{cyan}} +\Ampthree{1^-}{2^-}{3^+}{\bosflip{1}{180}{brown}{red}}{\bos{i2}{red}}{\bosflip{1}{-55}{brown}{cyan}} .
\end{eqnarray}

\item
For the $(-1,0,0)$ helicity category, there are three kinds of current structures: $[\mathbf{J}(1^-,2^0)]_2\cdot [\mathbf{A}(3^0)]_0$, $[\mathbf{J}(1^-,2^0)]_0\cdot [\mathbf{A}(3^0)]_2$, $[\mathbf{J}(2^0,3^0)]_1\cdot [\mathbf{A}(1^-)]_1$. For the $[\mathbf{J}]_1\cdot [\mathbf{A}]_1$ case, $[\mathbf{J}(2^0,3^0)]_1$ is the conserved current, which gives
\begin{eqnarray}
    [\mathbf{J}(2^0,3^0)]_1 \cdot [\mathbf{A}(1^-)]_1
&=& -c_{31} m_1 \tilde{m}_3 (\langle\eta_32\rangle [\eta_12] \langle31\rangle +\langle32\rangle [\eta_12] \langle\eta_31\rangle) \nonumber \\
&&-c_{32} m_1 \tilde{m}_2 (\langle3\eta_2\rangle [\eta_13] \langle21\rangle +\langle32\rangle [\eta_13] \langle\eta_21\rangle) \nonumber \\ 
&&+c_{33} m_1 m_3 ([\eta_32] [\eta_13] \langle21\rangle +[32] [\eta_1\eta_3] \langle21\rangle) \nonumber \\
&&+c_{34} m_1 m_2 ([3\eta_2] [\eta_12] \langle31\rangle +[32] [\eta_1\eta_2] \langle31\rangle).
\end{eqnarray}
Note that the MHC amplitudes contains both the leading SM amplitudes and 1st descendant EFT amplitudes.   
For the $[\mathbf{J}(1^-,2^0)]_2\cdot [\mathbf{A}(3^0)]_0$, there is no conserved current. Thus the amplitudes should belong to either the leading SM one or 1st descendant EFT amplitudes. 
\begin{eqnarray}
   && [\mathbf{J}(1^-,2^0)]_2\cdot [\mathbf{A}(3^0)]_0
+[\mathbf{J}(1^-,3^0)]_2\cdot [\mathbf{A}(2^0)]_0 \nonumber \\
&& = -c_1 \langle1\eta_2\rangle [\eta_23] \langle31\rangle
-c_2 m_1 \tilde{m}_2 (\langle1\eta_2\rangle \langle23\rangle [3\eta_1] +\langle12\rangle \langle\eta_23\rangle [3\eta_1]) \nonumber \\
&&+c_3 m_1^2 [\eta_12] \langle23\rangle [3\eta_1] 
+c_4 m_1 m_2 ([\eta_1\eta_2] [23] \langle31\rangle +[\eta_12] [\eta_23] \langle31\rangle) \nonumber \\
&&-c_5 \langle1\eta_3\rangle [\eta_32] \langle21\rangle 
-c_6 m_1 \tilde{m}_3 (\langle1\eta_3\rangle \langle32\rangle [2\eta_1] +\langle13\rangle \langle\eta_32\rangle [2\eta_1]) \nonumber \\
&&+c_7 m_1^2 [\eta_13] \langle32\rangle [2\eta_1] 
+c_8 m_1 m_3 ([\eta_1\eta_3] [32] \langle21\rangle +[\eta_13] [\eta_32] \langle21\rangle).
\end{eqnarray}
For $[\mathbf{J}(1^-,2^0)]_0\cdot [\mathbf{A}(3^0)]_2$, $[\mathbf{J}(1^-,2^0)]_0$ is the conserved current, which would contribute to the leading SM amplitudes and descendant EFT amplitudes.  
Finally let us select the seven MHC amplitudes which contribute to the SM
\begin{align}
&-m_1 \tilde m_3 [\eta_1 2]\langle2\eta_3\rangle\langle31\rangle
=\Ampthree{1^-}{2^0}{3^0}{\bosflip{1}{180}{brown}{red}}{\bos{i2}{brown}}{\bosflip{1}{-55}{red}{brown}},\qquad 
-m_1 \tilde m_2 \langle12\rangle\langle\eta_23\rangle[3\eta_1]
=\Ampthree{1^-}{2^0}{3^0}{\bosflip{1}{180}{brown}{red}}{\bosflip{1}{55}{red}{brown}}{\bos{i3}{brown}}, \\
&-m_3\tilde m_3\langle12\rangle[2\eta_3]\langle\eta_31\rangle
=\Ampthree{1^-}{2^0}{3^0}{\bos{i1}{red}}{\bos{i2}{brown}}{\bosflipflip{1}{-55}{brown}{red}{brown}},\qquad
-m_2\tilde m_2\langle1\eta_2\rangle[\eta_23]\langle31\rangle
=\Ampthree{1^-}{2^0}{3^0}{\bos{i1}{red}}{\bosflipflip{1}{55}{brown}{red}{brown}}{\bos{i3}{brown}},  \\
&m_1 m_3 (\langle12\rangle[2\eta_3][3\eta_1]+\langle12\rangle[23][\eta_3\eta_1])
=\Ampthree{1^-}{2^0}{3^0}{\bosflip{1}{180}{brown}{red}}{\bos{i2}{brown}}{\bosflip{1}{-55}{cyan}{brown}},\\
&m_1 m_2 ([\eta_1 2][\eta_2 3]\langle31\rangle+[\eta_1 \eta_2][2 3]\langle31\rangle)
=\Ampthree{1^-}{2^0}{3^0}{\bosflip{1}{180}{brown}{red}}{\bosflip{1}{55}{cyan}{brown}}{\bos{i3}{brown}}, \\
&m_1^2[\eta_1 2]\langle23\rangle[3 \eta_1]
=\Ampthree{1^-}{2^0}{3^0}{\bosflipflip{1}{180}{cyan}{brown}{red}}{\bos{i2}{brown}}{\bos{i3}{brown}}.
\end{align}

\end{itemize}

The $VVV$ descendant amplitudes at the same order can also be related through the ladder operators. Given the large number of terms in the $VVV$ amplitude, we only list for conciseness all leading (2nd descendant) diagrams originating from a primary amplitude, without showing their corresponding analytic expressions:
\begin{equation}
\includegraphics[width=0.9\linewidth,valign=c]{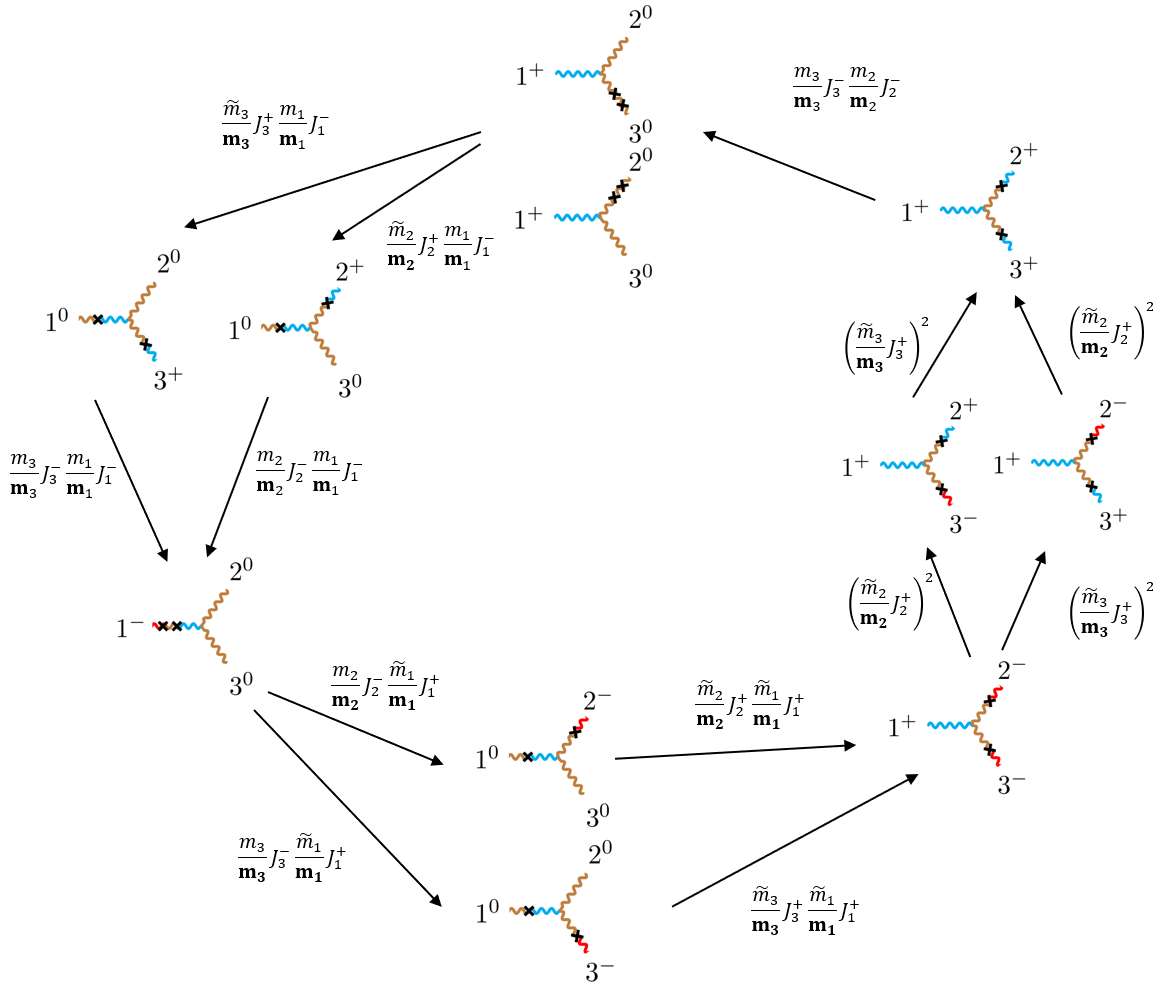}
\end{equation}

\section{The 3-point Amplitude Matching For SM Amplitudes}
\label{sec:SMcontact}

In previous section, both the 3-point massless and massive amplitudes are determined by the little group scaling and extended little group covariance. The primary MHC amplitudes are obtained from the massive amplitudes, and all the descendant MHC ones are followed by action of the ladder operators. In this section, we will discuss the gauge structures of the massless and massive amplitudes, and then match the SM massless and massive amplitudes via the MHC amplitude deformation.

\subsection{Massless and massive 3-point SM amplitudes}

In the unbroken phase of the renormalizable SM, massless amplitudes are required to satisfy the dimensional constraint $\dim\big[\mathcal{A}(1,\dots,n)\big] = 4 - n$. This imposes a constraint on the coupling constant $\mathcal{G}$,
\begin{equation}
\text{dim }\mathcal{G}=1-|h_1+h_2+h_3|.
\end{equation}
In the massless SM, the 3-point coupling constant should be $0$, indicating the total helicity of the 3-pt massless amplitude is constrained to
\begin{equation}
h_1+h_2+h_3=\pm 1.
\end{equation}
This determines the total helicity for the 3-pt amplitude in the SM. Therefore, the amplitudes such as $(h_1,h_2,h_3)=(+\frac12,-\frac12,0)$, $(+1,-1,0)$ or $(0,0,0)$ are forbidden. Therefore, the massless amplitudes in the SM belong to the $SSSS$, $VVV$, $FFS$, $FFV$ and $VSS$ categories, while the $VVS$ one is excluded.

The 3-point massless amplitudes contains the group structure in the unbroken phase. Here the notation of the group indices are 
\begin{eqnarray}
\begin{array}{c|c|c|c}
\hline
\text{unbroken phase} & \text{gauge boson} & \text{fermion} & \text{scalar boson} \\
\hline
\text{index} & I & \hat{i} & i \\
\hline
\text{upper index value} & W^I=\{W^1,W^2,W^3\};B & \{u_L,d_L\},\{\nu_L,e_L\};u_R,d_R,e_R & \{\phi^+,\phi^0\} \\
\hline
\text{lower index value} & \mbox{-} & \{\bar{u}_L,\bar{d}_L\},\{\bar{\nu}_L,\bar{e}_L\};\bar{u}_R,\bar{d}_R,\bar{e}_R & \{\phi^-,\phi^{0*}\} \\
\hline
\end{array} 
\end{eqnarray}
where $i$ represents the $SU(2)_W$ fundamental indices, denoting the first or second component of $H_i$; $\hat{i}$ stands for the reducible representation for fermion, including both $SU(2)_W$-doublets $\{u_L,d_L\}$, $\{\nu_L, e_L\}$ and $SU(2)_W$-singlets $u_R,d_R,e_R$. The upper index and lower index represent particle and anti-particle separately. For simplicity, here we only consider one generation of fermions.

For 3-pt amplitude, the group structure should have three indices, labeling a covariant tensor under the $SU(2)_W\times U(1)$ group.  There are four types of 3-pt massless amplitude, and we can define their coefficient to be
\begin{align} 
FFS&:\quad 7\times7\times2\text{ tensors} \quad \left(Y^{i_3}\right)_{\hat{i}_2}^{\hat{i}_1 }, \left(\tilde{Y}^{i_3}\right)_{\hat{i}_2}^{\hat{i}_1 }, \left(Y_{i_3}\right)_{\hat{i}_2}^{\hat{i}_1 }, \left(\tilde{Y}_{i_3}\right)_{\hat{i}_2}^{\hat{i}_1 }, \\
FFV&:\quad 7\times7\times4\text{ tensor} \quad (T_f^{I_3})^{\hat{i}_1}_{\hat{i}_2}, (\tilde{T}_f^{I_3})^{\hat{i}_1}_{\hat{i}_2},  \\
VSS&:\quad 4\times2\times2\text{ tensor} \quad (T_s^{I_1})^{i_2}_{i_3}, \\
VVV&:\quad 4\times4\times4\text{ tensor} \quad f^{I_1 I_2 I_3}.
\end{align}

\paragraph{Massless VVV} 
The electroweak $SU(2)$ Yang-Mills theory has the following 3-pt amplitude
\begin{eqnarray}
    \includegraphics[scale=1.1,valign=c]{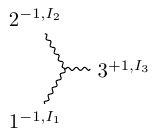} 
&=& \mathcal{A}(1^{-1,I_1}, 2^{-1,I_2}, 3^{+1,I_3}) = f^{I_1 I_2 I_3} \frac{\langle12\rangle^3}{\langle23\rangle \langle31\rangle}, 
\end{eqnarray}
where all three indices are fixed and the coupling takes the form
\begin{align}
f^{W^I W^J W^K} = -i\sqrt{2} g \epsilon^{IJK}.
\end{align}
where $\epsilon^{IJK}$ is the Levi-Civita tensor. The $U(1)$ structure with the $B$ boson does not participate in the gauge boson self interaction. In the off-shell Lagrangian description, the
3-vertex is an oﬀ-shell non-gauge invariant object, and thus the contact quartic $VVVV$ is fully determined from the off-shell non-gauge invariant $VVV$ to ensure the off-shell gauge invariance of the Lagrangian. On the other hand, in the on-shell method, since the 3-point on-shell amplitude is gauge
invariant, there is no need to add contact 4-pt amplitudes.

\paragraph{Massless VSS}
The unbroken SM contains the scalar QED theory, in which the 3-point $VSS$ amplitudes take the form
\begin{eqnarray}
    \includegraphics[scale=1.1,valign=c]{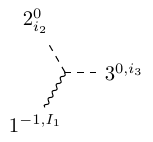}
&=& \mathcal{A}(1^{-1,I_1}, 2^{0}_{i_2}, 3^{0,i_3}) = ({T_s}^{I_1})^{i_2}_{i_3} \frac{\langle12\rangle \langle31\rangle}{\langle23\rangle}, \label{eq:vss}
\end{eqnarray}
where $I_1$ denote $(B, W^I)$, and $i_2, i_3$ denotes $(h^+, h^0)$.
The coupling structures are
\begin{align}
(T_s^{B})^{i_2}_{i_3} = -\sqrt{2} g'  
\begin{pmatrix}
Y_{\phi^+} & 0 \\
0 & Y_{\phi^0}
\end{pmatrix},\quad (T_s^{W^I})^{i_2}_{i_3} = -\frac{g}{\sqrt{2}} 
\begin{pmatrix}
\sigma^I     
\end{pmatrix},
\end{align}
where the hypercharge for scalar boson is $Y_{\phi^+}=Y_{\phi^0}=\frac12$. Note that again the contact quartic $VVSS$ is not needed because the gauge invariance of the 3-pt amplitudes in the on-shell method. 

\paragraph{Massless FFV}
Of course, the SM contains the gauge fermion couplings. We consider two massless amplitudes with opposite helicity for the fermion and antifermion with helicity $(\mp\frac12,\pm\frac12,+1)$
\begin{equation} \begin{aligned}
\begin{tikzpicture}[baseline=0.7cm] \begin{feynhand}
\setlength{\feynhandarrowsize}{3.5pt}
\vertex [particle] (i1) at (-0.3,0.8) {$1^{-1/2,\hat i_1}$}; 
\vertex [particle] (i2) at (1.6,1.6) {$2^{+1/2}_{\hat i_2}$}; 
\vertex [particle] (i3) at (1.6,0) {$3^{+1, I_3}$};  
\vertex (v1) at (0.9,0.8); 
\graph{(i1)--[fer](v1)--[fer](i2)};
\graph{(i3)--[bos] (v1)};  
\end{feynhand} \end{tikzpicture} & = \mathcal{A}(1^{-\frac12,\hat i_1},2^{+\frac12}_{\hat i_2},3^{+1, I_3})&=(\tilde{T}_f^{I_3})^{\hat{i}_1}_{\hat{i}_2}\frac{[23]^2}{[12]},\\
\begin{tikzpicture}[baseline=0.7cm] \begin{feynhand}
\setlength{\feynhandarrowsize}{3.5pt}
\vertex [particle] (i1) at (-0.3,0.8) {$1^{+1/2,\hat i_1}$}; 
\vertex [particle] (i2) at (1.6,1.6) {$2^{-1/2}_{\hat i_2}$}; 
\vertex [particle] (i3) at (1.6,0) {$3^{+1, I_3}$};  
\vertex (v1) at (0.9,0.8); 
\graph{(i1)--[fer](v1)--[fer](i2)};
\graph{(i3)--[bos] (v1)};  
\end{feynhand} \end{tikzpicture} & = \mathcal{A}(1^{+\frac12,\hat i_1},2^{-\frac12}_{\hat i_2},3^{+1, I_3})
&=(T_f^{I_3})^{\hat{i}_1}_{\hat{i}_2}\frac{[13]^2}{[12]},
\end{aligned} \end{equation}
where $\tilde T_f$ denotes the coupling for the right-handed current, and $T_f$ for left-handed current. $T_f$ represents the coefficients of $\frac{\langle23\rangle^2}{\langle12\rangle},\frac{[13]^2}{[12]}$, and $\tilde{T}_f$ represents the ones of $\frac{\langle13\rangle^2}{\langle12\rangle},\frac{[23]^2}{[12]}$.

For $W^I$ boson, the two diagonal blocks are 
\begin{equation} \begin{aligned}
(T_f^{W^I})^{\hat{i}_1}_{\hat{i}_2} &= -\frac{g}{\sqrt{2}} 
\begin{pmatrix}
\sigma^I & \\
 & \sigma^I \\
\end{pmatrix},& 
\hat{i}_1&=u_L,d_L,\nu_L,e_L,\quad \hat{i}_2=\bar u_L,\bar d_L,\bar \nu_L,\bar e_L, \\
(\tilde{T}_f^{W^I})^{\hat{i}_1}_{\hat{i}_2} &=
\begin{pmatrix}
 0 & & \\
 & 0 & \\
 & & 0 \\
\end{pmatrix},&
\hat{i}_1&=u_R,d_R,e_R,\quad \hat{i}_2=\bar u_R,\bar d_R,\bar e_R,  \\
\end{aligned} \end{equation}
where $g$ is the $SU(2)_W$ coupling constant, and $\sigma^I$ are the Pauli matrices acting as $SU(2)W$ generators. For $B$ boson, we have
\begin{equation} \begin{aligned}
(T_f^{B})^{\hat{i}_1}_{\hat{i}_2} =-{\sqrt{2}} {g'}
\begin{pmatrix}
Y_{u_L} & & & \\
 & Y_{d_L} & & \\
 & & Y_{\nu_L} & \\
 &  &  & Y_{e_L} \\ 
\end{pmatrix},\quad
(\tilde{T}_f^{B})^{\hat{i}_1}_{\hat{i}_2} &=-{\sqrt{2}} {g'}
\begin{pmatrix}
 Y_{u_R} & & \\
 & Y_{d_R} & \\
 & & Y_{e_R} \\
\end{pmatrix},
\end{aligned} \end{equation}
where $g^{\prime}$ is $U(1)_Y$ coupling constant. In the SM, the hypercharge $Y$ of $U(1)_Y$ group is set by
\begin{equation} \begin{aligned}
Y_{u_L}&=Y_{d_L}=\frac16,& Y_{u_R}&=\frac23,& Y_{d_R}&=-\frac13, \\
Y_{e_L}&=Y_{\nu_L}=-\frac12,& Y_{e_R}&=-1.& 
\end{aligned}
\end{equation}

\paragraph{Massless FFS}
The SM contains the scalar Yukawa interactions, and their amplitudes take the form
\begin{equation} \begin{aligned} \label{eq:FFS_leading}
\begin{tikzpicture}[baseline=0.7cm] \begin{feynhand}
\setlength{\feynhandarrowsize}{3.5pt}
\vertex [particle] (i1) at (-0.3,0.8) {$1^{-1/2,\hat i_1}$}; 
\vertex [particle] (i2) at (1.6,1.6) {$2^{-1/2}_{\hat i_2}$}; 
\vertex [particle] (i3) at (1.6,0) {$3^{0,i_3}$};  
\vertex (v1) at (0.9,0.8); 
\graph{(i1)--[fer](v1)--[fer](i2)};
\graph{(i3)--[sca] (v1)};  
\end{feynhand} \end{tikzpicture}=Y_{\hat{i}_2}^{\hat{i}_1 i_3} \langle12\rangle, \quad 
\begin{tikzpicture}[baseline=0.7cm] \begin{feynhand}
\setlength{\feynhandarrowsize}{3.5pt}
\vertex [particle] (i1) at (0,0.8) {$1^{+1/2,\hat i_1}$}; 
\vertex [particle] (i2) at (1.6,1.6) {$2^{+1/2}_{\hat i_2}$}; 
\vertex [particle] (i3) at (1.6,0) {$3^{0,i_3}$};  
\vertex (v1) at (0.9,0.8); 
\graph{(i1)--[fer](v1)--[fer](i2)};
\graph{(i3)--[sca] (v1)};  
\end{feynhand} \end{tikzpicture}=\tilde{Y}_{\hat{i}_2}^{\hat{i}_1 i_3}[12].
\end{aligned} \end{equation}
Here $Y$ and $\tilde{Y}$ represents the coefficients of $\langle12\rangle$ and $[12]$, respectively. The two Yukawa matrices  are
\begin{align}
Y_{\hat{i}_2}^{\hat{i}_1 i_3} &=-
\begin{pmatrix}
\mathcal{Y}^{(u)} \delta_{\phi^0}^{i_3} & 0 & 0 \\
\mathcal{Y}^{(u)} \delta_{\phi^+}^{i_3} & 0 & 0 \\
0 & 0 & 0 \\
0 & 0 & 0 \\
\end{pmatrix},& &\hat{i}_1=u_L,d_L,\nu_L,e_L\quad \hat{i}_2=\bar u_R,\bar d_R,\bar e_R, \\
\tilde{Y}_{\hat{i}_2}^{\hat{i}_1 i_3} &=-
\begin{pmatrix}
0 & 0 & 0 & 0 \\
\mathcal{Y}^{(d)} \delta^{i_3}_{\phi^+} & \mathcal{Y}^{(d)} \delta^{i_3}_{\phi^0} & 0 & 0 \\
0 & 0 & \mathcal{Y}^{(e)} \delta^{i_3}_{\phi^+} & \mathcal{Y}^{(e)} \delta^{i_3}_{\phi^0} \\
\end{pmatrix},& &\hat{i}_1=u_R,d_R,e_R,\quad \hat{i}_2=\bar u_L,\bar d_L,\bar \nu_L,\bar e_L, 
\end{align}
where $i_3=\phi^+,\phi^0$ and $\mathcal{Y}^{(f)}$ denotes the Yukawa coupling for fermion $f$. There is no right-handed neutrino in the SM, so we do not have a coupling like $\mathcal{Y}^{(\nu)}$ and the above blocks are not square matrices. Similarly we have 
\begin{align}
Y_{\hat{i}_2 i_3}^{\hat{i}_1} &=-
\begin{pmatrix}
 0 & \mathcal{Y}^{(d)} \delta_{i_3}^{\phi^-} & 0 \\
 0 & \mathcal{Y}^{(d)} \delta_{i_3}^{\phi^0} & 0 \\
 0 & 0 & \mathcal{Y}^{(e)} \delta_{i_3}^{\phi^-} \\
 0 & 0 & \mathcal{Y}^{(e)} \delta_{i_3}^{\phi^0} \\ 
\end{pmatrix},&
\tilde{Y}_{\hat{i}_2 i_3}^{\hat{i}_1} &=-
\begin{pmatrix}
\mathcal{Y}^{(u)} \delta^{\phi^0}_{i_3} & \mathcal{Y}^{(u)} \delta^{\phi^-}_{i_3} & 0 & 0  \\
0 & 0 & 0 & 0  \\
0 & 0 & 0 & 0  \\
\end{pmatrix}.&
\end{align}

\paragraph{Massless SSSS}
Finally the SM contains the single 4-point amplitudes, the $\lambda \phi^4$ amplitude 
\begin{eqnarray}
\includegraphics[scale=1.1,valign=c]{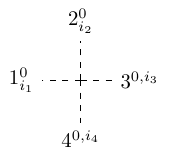}
&=& \mathcal{A}(1^{0}_{i_1}, 2^{0}_{i_2}, 3^{0,i_3}, 4^{0,i_4}) = -4\lambda \delta^{(i_1}_{i_3} \delta^{i_2)}_{i_4}. \label{eq:4H}
\end{eqnarray}

\paragraph{Massive 3-point SM Amplitudes}

After spontaneously symmetry breaking, we use the indices $\mathbf{I}$, $\hat{\mathbf{i}}$ and $h$ to label vector boson, fermion and scalar boson. Their possible values are
\begin{eqnarray}
\begin{array}{c|c|c|c}
\hline
\text{broken phase} & \text{vector boson} & \text{fermion} & \text{scalar boson} \\
\hline
\text{index} & \mathbf{I} & \hat{\mathbf{i}} & {h} \\
\hline
\text{upper index value} & \{W^+,W^-,Z,A\} & \{u,d;\nu,e\} & \{h\} \\
\hline
\text{lower index value} & \mbox{-} & \{\bar{u},\bar{d};\bar{\nu},\bar{e}\} & \mbox{-} \\
\hline
\end{array}
\end{eqnarray}

Similarly the massive 3-point AHH amplitudes are summarized. In the broken phase, identify the gauge structure and we should obtain the following results
\begin{eqnarray}
\includegraphics[scale=1.1,valign=c]{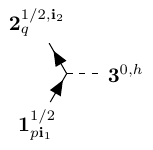}
&=& \mathcal{M}({\mathbf{1}}^{1/2}_{\mathbf{i}_1}, \mathbf{2}^{1/2,\mathbf{i}_2}, \mathbf{3}^{0,h}) = {y}^{\mathbf{i}_1}_{\mathbf{i}_2} \langle\mathbf{12}\rangle +{y'}^{\mathbf{i}_1}_{\mathbf{i}_2} [\mathbf{12}], \\
\includegraphics[scale=1.1,valign=c]{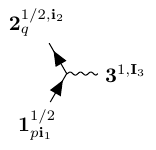}
&=& \mathcal{M}(\mathbf{1}^{1/2}_{\mathbf{i}_1}, \mathbf{2}^{1/2,\mathbf{i}_2}, \mathbf{3}^{1,\mathbf{I}_3})
= (X_1^{\mathbf{I}_3} )^{\hat{\mathbf{i}}_1}_{\hat{\mathbf{i}}_2} \frac{\langle\mathbf{13}\rangle [\mathbf{23}]}{\mathbf{m}_3} +(X_2^{\mathbf{I}_3} )^{\hat{\mathbf{i}}_1}_{\hat{\mathbf{i}}_2} \frac{\langle\mathbf{23}\rangle [\mathbf{13}]}{\mathbf{m}_3}, \label{eq:ffV} \\
\includegraphics[scale=1.1,valign=c]{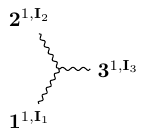}
&=& \mathcal{M}(\mathbf{1}^{1,\mathbf{I}_1}, \mathbf{2}^{1,\mathbf{I}_2}, \mathbf{3}^{1,\mathbf{I}_3}) \nonumber \\ 
&=& 
\mathbf{f}^{\mathbf{I}_1 \mathbf{I}_2 \mathbf{I}_3} \left(\frac{[\mathbf{12}]\langle\mathbf{23}\rangle\langle\mathbf{31}\rangle}{\mathbf{m}_1 \mathbf{m}_2}+\frac{[\mathbf{12}][\mathbf{23}]\langle\mathbf{31}\rangle }{\mathbf{m}_1 \mathbf{m}_3}+\frac{[\mathbf{12}]\langle\mathbf{23}\rangle[\mathbf{31}]}{\mathbf{m}_2 \mathbf{m}_3} \right) \nonumber \\
&&+ \mathbf{f}^{\mathbf{I}_1 \mathbf{I}_2 \mathbf{I}_3} \left(\frac{\langle\mathbf{12}\rangle[\mathbf{23}][\mathbf{31}]}{\mathbf{m}_1 \mathbf{m}_2}+\frac{\langle\mathbf{12}\rangle\langle\mathbf{23}\rangle[\mathbf{31}]}{\mathbf{m}_1 \mathbf{m}_3}+\frac{\langle\mathbf{12}\rangle[\mathbf{23}]\langle\mathbf{31}\rangle}{\mathbf{m}_2 \mathbf{m}_3} \right),
\label{eq:3V}\\
\includegraphics[scale=1.1,valign=c]{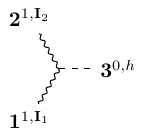}
&=& \mathcal{M}(\mathbf{1}^{1,\mathbf{I}_1}, \mathbf{2}^{1,\mathbf{I}_2}, \mathbf{3}^{0,h}) = 
{\mathbf{g}}^{\mathbf{I}_1 \mathbf{I}_2} \frac{\langle\mathbf{12}\rangle [\mathbf{12}]}{\mathbf m_1 \mathbf m_2} , 
\label{eq:VVh} \\
\includegraphics[scale=1.1,valign=c]{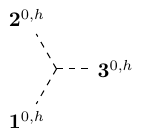}
&=& \mathcal{M}(\mathbf{1}^{0,h}, \mathbf{2}^{0,h}, \mathbf{3}^{0,h}) = \lambda_3, \label{eq:3h} \\
\includegraphics[scale=1.1,valign=c]{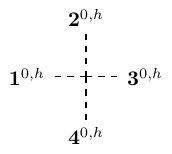}
&=& \mathcal{M}(\mathbf{1}^{0,h}, \mathbf{2}^{0,h}, \mathbf{3}^{0,h}, \mathbf{4}^{0,h}) = \lambda_4, \label{eq:4h}
\end{eqnarray}
Unlike the massless amplitudes, here all the couplings are to be determined.

\subsection{3-point matching from MHC amplitude deformation}

Given the massless and massive amplitudes, let us perform their matching by identifying the helcity-transversality unification at the amplitude level. Note that all the helicity categories of the massless amplitudes are identified by the condition $h_1+h_2+h_3=\pm1$:
\begin{eqnarray}
\begin{array}{c|cc|cc}
\multirow{2}{*}{\text{Massless}} & \multicolumn{2}{c|}{h_1+h_2+h_3=1} & \multicolumn{2}{c}{h_1+h_2+h_3=-1} \\
\cline{2-5}
& (h_1,h_2,h_3) & \text{Lorentz structures} & (h_1,h_2,h_3) & \text{Lorentz structures} \\
\hline
FFV & (-\frac12,+\frac12,+1) & \frac{[23]^2}{[12]} & (-\frac12,+\frac12,-1) & \frac{\langle13\rangle^2}{\langle12\rangle} \\
\hline
FFS & (+\frac12,+\frac12,0) & [12] & (-\frac12,-\frac12,0) & \langle12\rangle \\
\hline
VVV & (+1,+1,-1) & \frac{[12]^3}{[23][31]} & (-1,-1,+1) & \frac{\langle12\rangle^3}{\langle23\rangle\langle31\rangle} \\
\hline
VSS & (+1,0,0) & \frac{[12][31]}{[23]} & (-1,0,0) & \frac{\langle12\rangle\langle31\rangle}{\langle23\rangle} \\
\end{array}
\end{eqnarray}
The only exception in the SM is $SSS$, which corresponds to the 4-point massless amplitude. 
On the other hand, simply unbolding the massive amplitudes gives the following helicity categories 
\begin{eqnarray}
\begin{array}{c|cc|cc}
\multirow{2}{*}{\text{Primary MHC}} & \multicolumn{2}{c|}{h_1+h_2+h_3} & \multicolumn{2}{c}{h_1+h_2+h_3} \\
\cline{2-5}
& (h_1,h_2,h_3) & \text{Lorentz structures} & (h_1,h_2,h_3) & \text{Lorentz structures} \\
\hline
FFV & (-\frac12,+\frac12,0) & \langle13\rangle[32] & (+\frac12,-\frac12,0) & [13]\langle32\rangle \\
\hline
FFS & (+\frac12,+\frac12,0) & [12] & (-\frac12,-\frac12,0) & \langle12\rangle \\
\hline
VVV & (-1,0,0) & \langle12\rangle[23]\langle31\rangle & (+1,0,0) & [12]\langle23\rangle[31] \\
\hline
 &  (0,-1,0) & \langle12\rangle\langle23\rangle[31] & (0,+1,0) & [12][23]\langle31\rangle \\
\hline
 & (0,0,-1) & [12]\langle23\rangle\langle31\rangle & (0,0,+1) & \langle12\rangle[23][31] \\
\hline
VSS & (0,0,0) & \langle12\rangle[21] &   &  \\
\end{array}
\end{eqnarray}
Comparing the helicity structures, we note that the helicity does not match except the $FFS$ structure. This agrees with the results from the previous section, that all the primary MHC amplitudes except the $FFS$ vanish due to conserved currents. Therefore, we expect for those vanishing primary case the matching should be the first non-vanishing descendant amplitudes, named as the leading MHC amplitudes.

For a massive amplitude with spins $(s_1,s_2,s_3)$, all massless amplitudes with $|h_i|\le s_i$ can contribute to the massive structure. The leading MHC amplitude receive the contribution from 3-pt massless amplitude, so the corresponding helicity category must satisfy $h_1+h_2+h_3=\pm1$.
To obtain the leading MHC amplitudes, let us start from the primary helicity categories in the SM, and check the total helicity 
\begin{equation}
h=h_1+h_2+h_3.
\end{equation}
If they satisfy $h=\pm 1$ and they do not vanishes, we identify these primary amplitude is leading. Otherwise, we will use the ladder operator to obtain the descendant MHC amplitudes with total helicity $h+1$ or $h-1$, until we find the amplitudes with total helicity $\pm1$, as the leading contribution. 
For convenience, we define the \textit{helicity class} as the set of helicity categories sharing the same number of transverse vector bosons, denoted as $n_T$. In the SM, the helicity classes for all 3-pt massive amplitudes are
\begin{equation}
\begin{tabular}{c|c|c}
\hline
Leading MHC & helicity class & helicity category \\
\hline
SSS & $n_T=0$ & $(0,0,0)$ \\
\hline
FFS & $n_T=0$ & $(\pm\frac12,\pm\frac12,0)$ \\
\hline
FFV & \makecell{$n_T=0$\\$n_T=1$} & \makecell{$(\pm\frac12,\pm\frac12,0)$\\$(\pm\frac12,\mp\frac12,+1),(\pm\frac12,\mp\frac12,-1)$} \\
\hline
VVS & $n_T=1$ & $(\pm 1,0,0),(0,\pm 1,0)$ \\
\hline
VVV & \makecell{$n_T=1$\\$n_T=3$} & \makecell{$(\pm 1,0,0),(0,\pm 1,0),(0,0,\pm 1)$\\$(\pm 1,\mp 1,\mp 1),(\mp 1,\pm 1,\mp 1),(\mp 1,\mp 1,\pm 1)$} \\
\hline
\end{tabular}
\end{equation}
Notice that the leading MHC $FFV$ would contain massless $FFS$ structures, and the leading $VVV$ also contains $VSS$ structures.  
In the following we will match the massless amplitudes with the leading MHC ones.

\paragraph{1. FFS \& SSS}

For the $FFS$ amplitude, the primary MHC amplitudes has two helicity categories
\begin{equation} \begin{aligned}
h=-1:\quad &(-\tfrac12,-\tfrac12,0),& \\
h=+1:\quad &(+\tfrac12,+\tfrac12,0),& 
\end{aligned} \end{equation}
They satisfy the condition $h=\pm1$. Since the primary amplitude is non-vanishing and thus there is no conserved current, the matching is quite simple
\begin{equation} \begin{aligned}
(-\tfrac12,-\tfrac12,0):&\quad\langle12\rangle=\Ampthree{1^-}{2^-}{3^0}{\fer{red}{i1}{v1}}{\antfer{cyan}{i2}{v1}}{\sca{i3}},\\
(+\tfrac12,+\tfrac12,0):&\quad [12]=\Ampthree{1^+}{2^+}{3^0}{\fer{cyan}{i1}{v1}}{\antfer{red}{i2}{v1}}{\sca{i3}}.
\end{aligned}\end{equation}

For the $SSS$ amplitude, it only has one helicity category $(0,0,0)$. The primary MHC amplitude is non-vanishing and thus the matching is also simple
\begin{equation} \begin{aligned}
(0,0,0):\quad 1=\Ampthree{1^0}{2^0}{3^0}{\sca{i1}}{\sca{i2}}{\sca{i3}}.
\end{aligned}\end{equation}

\paragraph{2. FFV}

From the MHC leading amplitudes 
\begin{equation} \begin{aligned}
&(+\tfrac12,+\tfrac12,0):\quad \tilde m_1 [23]\langle3 \eta_1\rangle,\quad \tilde m_2 \langle\eta_23\rangle[31], \\ 
&(-\tfrac12,-\tfrac12,0):\quad  m_2 [\eta_23]\langle31\rangle, \quad m_1 \langle23\rangle[3\eta_1] , \\
&(-\tfrac12,+\tfrac12,+1):\quad  \tilde m_3 [23]\langle\eta_31\rangle, \qquad 
(-\tfrac12,+\tfrac12,-1):\quad  m_3 [2\eta_3]\langle31\rangle \\ 
&(+\tfrac12,-\tfrac12,+1):\quad  \tilde m_3 \langle2\eta_3\rangle[31],\qquad 
(+\tfrac12,-\tfrac12,-1):\quad  m_3 \langle23\rangle[\eta_31],
\end{aligned}
\end{equation}
Therefore, the 1st descendant amplitudes can be expressed in the following two forms
\begin{equation}
\begin{tabular}{c|c|c}
\hline
helicity & current form & diagram \\
\hline
$(-\tfrac12,+\tfrac12,+1)$ & $[\mathbf{J}(1^{-\frac12},2^{+\frac12})]_0\cdot [\mathbf{A}(3^+)]_1$ & $\Ampthree{1^-}{2^+}{3^+}{\fer{red}{i1}{v1}}{\antfer{red}{i2}{v1}}{\bosflip{1}{-55}{brown}{cyan}}$\\
\hline
$(+\tfrac12,+\tfrac12,0)$ & $[\mathbf{J}(1^{+\frac12},2^{+\frac12})]_1\cdot [\mathbf{A}(3^0)]_0$ & 
$\Ampthree{1^+}{2^+}{3^0}{\ferflip{1}{180}{cyan}{red}}{\antfer{red}{i2}{v1}}{\bos{i3}{brown}}+\Ampthree{1^+}{2^+}{3^0}{\fer{cyan}{i1}{v1}}{\antferflip{1}{55}{red}{cyan}}{\bos{i3}{brown}}$\\
\hline 
\end{tabular}
\end{equation}
Note that the last line contains two MHC amplitudes, since the descendant current $[\mathbf{J}(1^{+\frac12}, 2^{+\frac12})]_1$ consists of two terms. Similarly, MHC amplitudes in other helicity categories can also be rewritten as $[\mathbf{J}]_1 \cdot [\mathbf{A}]_0$ or $[\mathbf{J}]_0 \cdot [\mathbf{A}]_1$.

\textit{Gauge Deformation}: For the massless $FFV$ amplitude with helicity $(-\frac12,+\frac12,+1)$, we should flip particle 3. Diagrammatically, we have
\begin{equation} \begin{aligned} \label{eq:FFV_to_FFV}
\begin{tikzpicture}[baseline=0.7cm] \begin{feynhand}
\setlength{\feynhandarrowsize}{3.5pt}
\vertex [particle] (i1) at (0,0.8) {$1^-$}; 
\vertex [particle] (i2) at (1.6,1.6) {$2^+$}; 
\vertex [particle] (i3) at (1.6,0) {$3^+$};  
\vertex (v1) at (0.9,0.8); 
\graph{(i1)--[fer](v1)--[fer](i2)};
\graph{(i3)--[bos] (v1)};  
\end{feynhand} \end{tikzpicture}
\quad \Longrightarrow \quad [\mathbf{J}^{-+}]_0\cdot [\mathbf{A}^+]_{1}
=
\Ampthree{1^-}{2^+}{3^+}{\fer{red}{i1}{v1}}{\antfer{red}{i2}{v1}}{\bosflip{1}{-55}{brown}{cyan}}
.
\end{aligned} \end{equation}
Applying the gauge deformation, the massless amplitude match to
\begin{equation}
\frac{[23]^2}{[12]}\xrightarrow{\text{Gauge}} \frac{\tilde m_3\langle1\eta_3\rangle[23]}{\mathbf m_3^2}.
\end{equation}

\textit{Goldstone Deformation}: When we match the Goldstone boson to a symmetric vector boson, we should flip the chirality of particles in the current. For massless $FFS$ amplitude with helicity $(-\frac12,-\frac12,0)$, particle 3 is a Goldstone boson, so we can only flip particle 1 or 2 in the current $\mathbb J$.  
In this case, one massless diagram corresponds to two MHC diagrams
\begin{equation} \begin{aligned}
\begin{tikzpicture}[baseline=0.7cm] \begin{feynhand}
\setlength{\feynhandarrowsize}{3.5pt}
\vertex [particle] (i1) at (0,0.8) {$1^-$}; 
\vertex [particle] (i2) at (1.6,1.6) {$2^-$}; 
\vertex [particle] (i3) at (1.6,0) {$3^0$};  
\vertex (v1) at (0.9,0.8); 
\graph{(i1)--[fer](v1)--[fer](i2)};
\graph{(i3)--[sca] (v1)};  
\end{feynhand} \end{tikzpicture} =  \langle12\rangle \rightarrow [\mathbf{J}^{--}]_{1}\cdot [\mathbf{A}^0]_0 =
\Ampthree{1^-}{2^-}{3^0}{\fer{red}{i1}{v1}}{\antferflip{1}{55}{cyan}{red}}{\bos{i3}{brown}}
=m_2\langle13\rangle[3\eta_2],
\end{aligned} \end{equation}
Thus, applying Goldstone deformation can gives two MHC terms
\begin{equation}
[12]\xrightarrow{\text{Goldstone}} -b_1\frac{\tilde m_1\langle\eta_13\rangle[32]}{\mathbf m_1^2}-b_2\frac{\tilde m_2 [13]\langle3\eta_2\rangle}{\mathbf m_2^2},
\end{equation}
where $b_1+b_2=1$.

Finally taking the little group covariance, we obtain the AHH formalism.

\paragraph{3. VVS}

The primary amplitude is also vanishing and thus the leading MHC amplitudes are 
\begin{equation} \begin{aligned}
(-1, 0, 0): \quad m_1\langle12\rangle[2\eta_1], \qquad & (+1, 0, 0): \quad 
-\tilde m_1\langle\eta_12\rangle[21], \\
(0, -1, 0): \quad m_2\langle12\rangle[\eta_21], \qquad &
(0, +1, 0): \quad -\tilde m_2\langle1\eta_2\rangle[21].
\end{aligned}\end{equation}
Therefore the massless amplitudes belong to the same helicity, and the diagrammatic matching is straightforward.

These can also be rewritten in terms of MHC currents and vector states. For each helicity, two decomposition forms remain available. As an example, consider the helicity category $(-1,0,0)$:
\begin{equation}
\begin{tabular}{c|c|c}
\hline
helicity & current form  & diagram \\
\hline
$(-1,0,0)$ & \makecell{$[\mathbf{J}(2^{0},3^{0})]_0\cdot [\mathbf{A}(1^-)]_1$\\$[\mathbf{J}(1^{-1},3^{0})]_1\cdot [\mathbf{A}(1^0)]_0$} & $\Ampthree{1^-}{2^0}{3^0}{\bosflip{1}{180}{brown}{red}}{\bos{i2}{brown}}{\sca{i3}}$\\
\hline
\end{tabular}
\end{equation}
Amplitudes for other helicities can similarly be rewritten as $[\mathbf{J}]_1 \cdot [\mathbf{A}]_0$ or $[\mathbf{J}]_0 \cdot [\mathbf{A}]_1$. In analogy with the $FFV$ case, the 1st descendant current $[\mathbf{J}]_1$ is not conserved.

From the first MHC amplitude, the massless matching has
\begin{equation} \begin{aligned}
\begin{tikzpicture}[baseline=0.7cm] \begin{feynhand}
\setlength{\feynhandarrowsize}{3.5pt}
\vertex [particle] (i1) at (0,0.8) {$1^-$}; 
\vertex [particle] (i2) at (1.6,1.6) {$2^0$}; 
\vertex [particle] (i3) at (1.6,0) {$3^0$};  
\vertex (v1) at (0.9,0.8); 
\graph{(i1)--[bos](v1)--[sca](i2)};
\graph{(i3)--[sca](v1)};  
\end{feynhand} \end{tikzpicture}=\frac{\langle12\rangle\langle31\rangle}{\langle23\rangle}
\quad\rightarrow\quad
\Ampthree{1^-}{2^0}{3^0}{\bosflip{1}{180}{brown}{red}}{\bos{i2}{brown}}{\sca{i3}}=m_1\langle12\rangle[2\eta_1],
\end{aligned} \end{equation}

\paragraph{4. VVV}

To apply the matching procedure, the first step is analyzing the MHC amplitudes. The primary amplitudes are vanishing, the 1st descendant amplitudes are either vanishing or belong to the EFT amplitude, and thus the leading amplitudes should be the 2nd descendant amplitues with the following helicity categories: 
\begin{equation} \begin{aligned}
&(\pm, \pm, \mp), 
(\pm, \mp, \pm),
(\mp, \pm, \pm), \\
&(\pm, 0, 0), 
(0, \pm, 0),
(0, 0, \pm).
\end{aligned}\end{equation}

Let us separate the leading matching into two different categories. The first category is the matching from massless $VVV$ to massive $VVV$ amplitudes. As shown in eq.~\eqref{eq:VVV_leading_1}, a massless $VVV$ diagram corresponds to three MHC diagrams. Because the massless amplitude has exchange symmetry, it is more convenient to treat the combination of MHC diagrams as the matching object. For helicity $(+1,+1,-1)$, we can consider the exchange symmetry between particles 1 and 2, so the matching yields two cases:  
\begin{equation}
\begin{tikzpicture}[baseline=0.7cm] \begin{feynhand}
\setlength{\feynhandarrowsize}{3.5pt}
\vertex [particle] (i1) at (0,0.8) {$1^+$}; 
\vertex [particle] (i2) at (1.6,1.6) {$2^+$}; 
\vertex [particle] (i3) at (1.6,0) {$3^-$};  
\vertex (v1) at (0.9,0.8); 
\graph{(i1)--[bos](v1)--[bos](i2)};
\graph{(i3)--[bos](v1)};  
\end{feynhand} \end{tikzpicture}
=\frac{[12]^3}{[23][31]}\to
\left\{\begin{aligned}
\Ampthree{1^+}{2^+}{3^-}{\bosflip{1}{180}{brown}{cyan}}{\bosflip{1}{55}{brown}{cyan}}{\bos{i3}{red}}, \\
\Ampthree{1^+}{2^+}{3^-}{\bosflip{1}{180}{brown}{cyan}}{\bos{i2}{cyan}}{\bosflip{1}{-55}{brown}{red}}+
\Ampthree{1^+}{2^+}{3^-}{\bos{i1}{cyan}}{\bosflip{1}{55}{brown}{cyan}}{\bosflip{1}{-55}{brown}{red}}.
\end{aligned}\right.
\end{equation}

In this case, any two of the above 2nd descendant amplitudes can be identified as a massive current $[\mathbf{J}]_1$ that couples to a vector $[\mathbf{A}]_1$:
\begin{equation}
\begin{tabular}{c|c}
\hline
current form  & diagram \\
\hline
$[\mathbf{J}(1^+,2^+)]_1\cdot [\mathbf{A}(3^-)]_1$ & 
$\Ampthree{1^+}{2^+}{3^-}{\bosflip{1}{180}{brown}{cyan}}{\bos{i2}{cyan}}{\bosflip{1}{-55}{brown}{red}}+\Ampthree{1^+}{2^+}{3^-}{\bos{i1}{cyan}}{\bosflip{1}{55}{brown}{cyan}}{\bosflip{1}{-55}{brown}{red}}$\\
\hline
$[\mathbf{J}(1^+,3^-)]_1\cdot [\mathbf{A}(2^+)]_1$ & $\Ampthree{1^+}{2^+}{3^-}{\bos{i1}{cyan}}{\bosflip{1}{55}{brown}{cyan}}{\bosflip{1}{-55}{brown}{red}}+\Ampthree{1^+}{2^+}{3^-}{\bosflip{1}{180}{brown}{cyan}}{\bosflip{1}{55}{brown}{cyan}}{\bos{i3}{red}}$\\
\hline
$[\mathbf{J}(2^+,3^-)]_1\cdot [\mathbf{A}(1^+)]_1$ & $\Ampthree{1^+}{2^+}{3^-}{\bosflip{1}{180}{brown}{cyan}}{\bos{i2}{cyan}}{\bosflip{1}{-55}{brown}{red}}+\Ampthree{1^+}{2^+}{3^-}{\bosflip{1}{180}{brown}{cyan}}{\bosflip{1}{55}{brown}{cyan}}{\bos{i3}{red}}$\\
\hline 
\end{tabular}
\end{equation}
where $[\mathbf{J}]_1$ is the conserved current we previously derived from the vanishing 1st descendant MHC amplitudes.~\footnote{There are also non-conserved $[\mathbf{J}]_1$ currents which contribute to the EFT amplitudes. Here we identify that $[\mathbf{J}(1^\pm,2^\pm)]_1$ contribute to the SM amplitudes while $[\mathbf{J}(1^\pm,2^\mp)]_1$ contribute to the EFT amplitudes.}

After amplitude deformation, we obtain the matching results as
\begin{equation}
\includegraphics[width=0.9\linewidth]{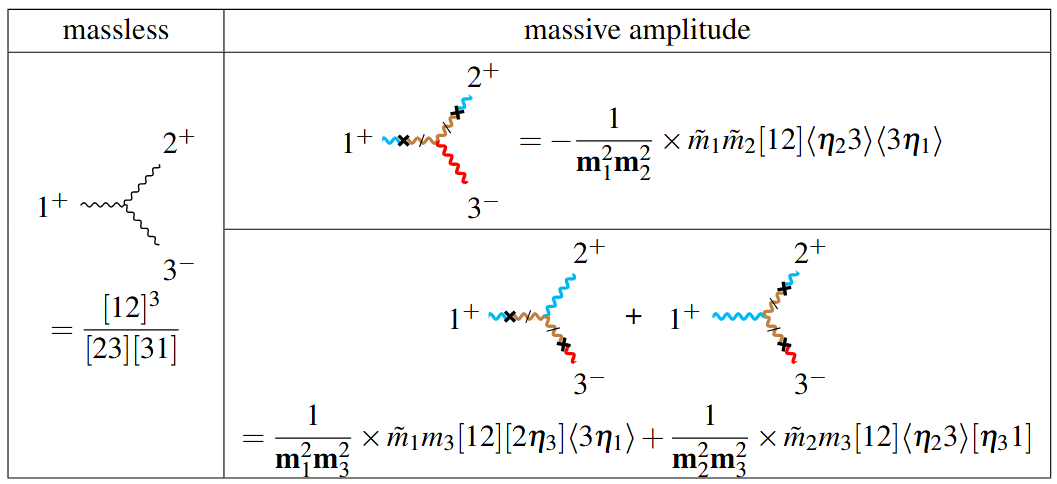}
\end{equation}

The second category is the matching from the
massless $VSS$ amplitude to massive $VVV$ amplitudes. According to exchange symmetry of two scalar boson, there are four cases
\begin{equation}
\begin{tikzpicture}[baseline=0.7cm] \begin{feynhand}
\setlength{\feynhandarrowsize}{3.5pt}
\vertex [particle] (i1) at (0,0.8) {$1^-$}; 
\vertex [particle] (i2) at (1.6,1.6) {$2^0$}; 
\vertex [particle] (i3) at (1.6,0) {$3^0$};  
\vertex (v1) at (0.9,0.8); 
\graph{(i1)--[bos](v1)--[sca](i2)};
\graph{(i3)--[sca](v1)};  
\end{feynhand} \end{tikzpicture}
=\frac{\langle12\rangle\langle31\rangle}{\langle23\rangle}
\to \left\{\begin{aligned}
\Ampthree{1^-}{2^0}{3^0}{\bosflip{1}{180}{brown}{red}}{\bos{i2}{brown}}{\bosflip{1}{-55}{cyan}{brown}}
+\Ampthree{1^-}{2^0}{3^0}{\bosflip{1}{180}{brown}{red}}{\bosflip{1}{55}{cyan}{brown}}{\bos{i3}{brown}}, \\
\Ampthree{1^-}{2^0}{3^0}{\bosflip{1}{180}{brown}{red}}{\bos{i2}{brown}}{\bosflip{1}{-55}{red}{brown}}
+\Ampthree{1^-}{2^0}{3^0}{\bosflip{1}{180}{brown}{red}}{\bosflip{1}{55}{red}{brown}}{\bos{i3}{brown}}, \\
\Ampthree{1^-}{2^0}{3^0}{\bos{i1}{red}}{\bos{i2}{brown}}{\bosflipflip{1}{-55}{brown}{red}{brown}}+
\Ampthree{1^-}{2^0}{3^0}{\bos{i1}{red}}{\bosflipflip{1}{55}{brown}{red}{brown}}{\bos{i3}{brown}}, \\
\Ampthree{1^-}{2^0}{3^0}{\bosflipflip{1}{180}{cyan}{brown}{red}}{\bos{i2}{brown}}{\bos{i3}{brown}}.
\end{aligned}\right.
\end{equation}

In this case, we can identify several 2nd descendant amplitudes as the coupling $[\mathbf{J}]_0\cdot[\mathbf{A}]_2$, $[\mathbf{J}]_1\cdot[\mathbf{A}]_1$ or $[\mathbf{J}]_2\cdot[\mathbf{A}]_0$:
\begin{equation}
\begin{tabular}{c|c}
\hline
current form  & diagram \\
\hline
$[\mathbf{J}(1^-,2^0)]_2\cdot [\mathbf{A}(3^0)]_0$ & 
$\Ampthree{1^-}{2^0}{3^0}{\bosflip{1}{180}{brown}{red}}{\bosflip{1}{55}{red}{brown}}{\bos{i3}{brown}}
+\Ampthree{1^-}{2^0}{3^0}{\bos{i1}{red}}{\bosflipflip{1}{55}{brown}{red}{brown}}{\bos{i3}{brown}}
+\Ampthree{1^-}{2^0}{3^0}{\bosflip{1}{180}{brown}{red}}{\bosflip{1}{55}{cyan}{brown}}{\bos{i3}{brown}}
+\Ampthree{1^-}{2^0}{3^0}{\bosflipflip{1}{180}{cyan}{brown}{red}}{\bos{i2}{brown}}{\bos{i3}{brown}}$\\
$[\mathbf{J}(1^-,3^0)]_2\cdot [\mathbf{A}(2^0)]_0$ & $\Ampthree{1^-}{2^0}{3^0}{\bosflip{1}{180}{brown}{red}}{\bos{i2}{brown}}{\bosflip{1}{-55}{red}{brown}}+\Ampthree{1^-}{2^0}{3^0}{\bos{i1}{red}}{\bos{i2}{brown}}{\bosflipflip{1}{-55}{brown}{red}{brown}}+\Ampthree{1^-}{2^0}{3^0}{\bosflip{1}{180}{brown}{red}}{\bos{i2}{brown}}{\bosflip{1}{-55}{cyan}{brown}}
+\Ampthree{1^-}{2^0}{3^0}{\bosflipflip{1}{180}{cyan}{brown}{red}}{\bos{i2}{brown}}{\bos{i3}{brown}}$ \\
\hline
$[\mathbf{J}(1^-,2^0)]_0\cdot [\mathbf{A}(3^0)]_2$ & 
$\Ampthree{1^-}{2^0}{3^0}{\bos{i1}{red}}{\bos{i2}{brown}}{\bosflipflip{1}{-55}{brown}{red}{brown}}$\\
\makecell{$[\mathbf{J}(1^-,3^0)]_0\cdot [\mathbf{A}(2^0)]_2$} & \makecell{
\Ampthree{1^-}{2^0}{3^0}{\bos{i1}{red}}{\bosflipflip{1}{55}{brown}{red}{brown}}{\bos{i3}{brown}}} \\
\hline
$[\mathbf{J}(2^0,3^0)]_1\cdot [\mathbf{A}(1^-)]_1$ & $\Ampthree{1^-}{2^0}{3^0}{\bosflip{1}{180}{brown}{red}}{\bos{i2}{brown}}{\bosflip{1}{-55}{red}{brown}}
+\Ampthree{1^-}{2^0}{3^0}{\bosflip{1}{180}{brown}{red}}{\bosflip{1}{55}{red}{brown}}{\bos{i3}{brown}}
+\Ampthree{1^-}{2^0}{3^0}{\bosflip{1}{180}{brown}{red}}{\bos{i2}{brown}}{\bosflip{1}{-55}{cyan}{brown}}
+\Ampthree{1^-}{2^0}{3^0}{\bosflip{1}{180}{brown}{red}}{\bosflip{1}{55}{cyan}{brown}}{\bos{i3}{brown}}$\\
\hline 
\end{tabular}
\end{equation}
Here $[\mathbf J]_2$ is not conserved and is absent from the primary and first-descendant $VVV$ amplitudes. Determining the current form of each diagram will help us analyze the leading matching in a later subsection.

The above matching result is summarized in the following table,
\begin{equation}
\includegraphics[width=0.9\linewidth]{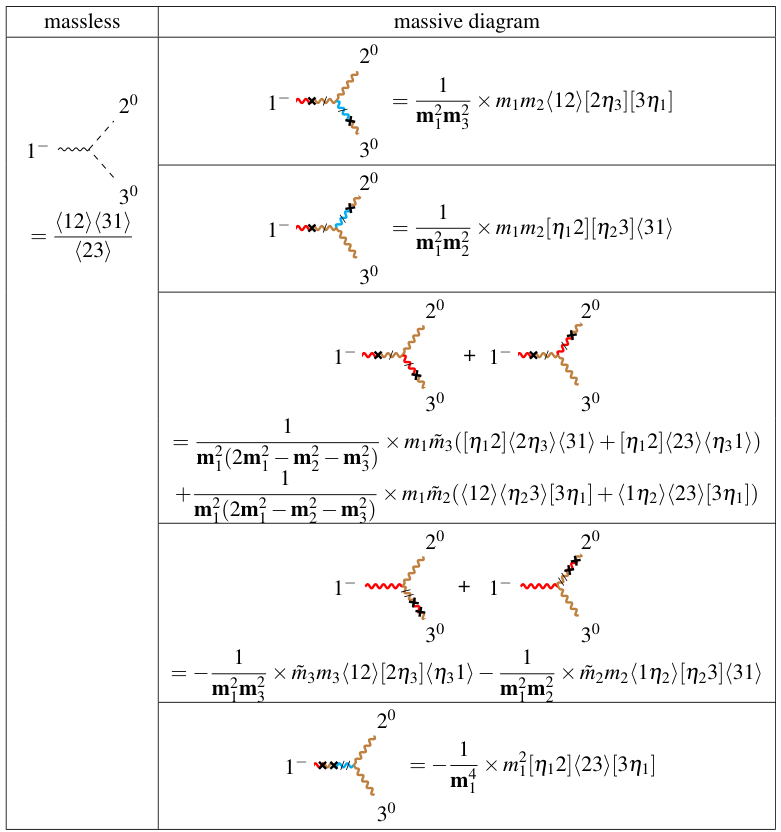}
\end{equation}
Finally taking the little group covariance, we obtain the AHH formalism
\begin{equation} \begin{aligned} \label{eq:VVV_result}
\mathbf{M}(\mathbf{1}^{1}, \mathbf{2}^{1}, \mathbf{3}^{1}) 
=& c_1 \left(\frac{[\mathbf{12}]\langle\mathbf{23}\rangle\langle\mathbf{31}\rangle}{\mathbf{m}_1 \mathbf{m}_2}+\frac{[\mathbf{12}][\mathbf{23}]\langle\mathbf{31}\rangle }{\mathbf{m}_1 \mathbf{m}_3}+\frac{[\mathbf{12}]\langle\mathbf{23}\rangle[\mathbf{31}]}{\mathbf{m}_2 \mathbf{m}_3} \right)\\
&+ c_2 \left(\frac{\langle\mathbf{12}\rangle[\mathbf{23}][\mathbf{31}]}{\mathbf{m}_1 \mathbf{m}_2}+\frac{\langle\mathbf{12}\rangle\langle\mathbf{23}\rangle[\mathbf{31}]}{\mathbf{m}_1 \mathbf{m}_3}+\frac{\langle\mathbf{12}\rangle[\mathbf{23}]\langle\mathbf{31}\rangle}{\mathbf{m}_2 \mathbf{m}_3} \right),
\end{aligned} \end{equation}
where $c_i$ should be determined by the UV physics from the massless amplitudes.

\subsection{Matching results for SM amplitudes}

The amplitude deformation from massless to leading MHC amplitudes provides the matching for the Lorentz structure. In this subsection, we would perform the matching for group structure. The group structure for the massless and massive amplitude are given in the first subsection. Now let us address how the indices changes from the unbroken to broken phase through the transformation matrices. There are four types of transformation matrices, summarized as follows
\begin{eqnarray}
\begin{tikzpicture}[baseline=-1.2cm]
 \node (unb) at (0,0) {unbroken};
 \node (b) at (0,-2.5) {broken};
 \node (us) at (2.5,0) {scalar $i$};
 \node (uv) at (6,0) {gauge boson $I$};
 \node (uf) at (9,0) {fermion $\hat{i}$};
 \node (bs) at (2.5,-2.5) {Higgs $h$};
 \node (bv) at (6,-2.5) {massive vector $\mathbf{I}$};
 \node (bf) at (9,-2.5) {fermion $\hat{\mathbf{i}}$};
 \draw [->] (unb) -- (b)
               node[midway] {};
 \draw [->] (us) -- (bs)
               node[midway] {$\mathcal{U}_{i}^{h}, \mathcal{U}^{i h}$};
 \draw [->] (us) -- (bv)
               node[midway] {$\mathcal{U}_{i}^{\mathbf{I}}, \mathcal{U}^{i \mathbf{I}}$};
 \draw [->] (uv) -- (bv)
               node[midway] {$O^{I \mathbf{I}}$};
 \draw [->] (uf) -- (bf)
               node[midway] {$\Omega_{\hat{i}}^{\mathbf{i}}, \Omega^{\hat{i}}_{\mathbf{i}}$};
\end{tikzpicture}
\end{eqnarray}
Note that for $S\to V$ and $S\to S$ cases, there are two kinds of transformation matrices, from the Higgs mechanism.

We classify four types of transformation matrices between massless and massive states in the SM:
\begin{itemize}

\item $S\to S$ and $S \to V$ transitions:

We use the matrices $\mathcal{U}$ to match massless scalar to massive Higgs boson and the vector boson. 
The projector $\mathcal{U}$ applies to the Higgs field. On one hand, the Higgs field matches the massive scalar field $h$ via the projector $\mathcal{U}^h$, 
\begin{equation}
\label{eq:projectors2}
    \mathcal{U}^{ih} = \mathcal{U}_i^h = \left(\begin{array}{c}
0\\\frac{1}{\sqrt{2}}\,
    \end{array}\right).
\end{equation}
Here, the complex scalars $\phi^0$ or $\phi^{0*}$ possesses 2 DOF, while the real Higgs boson $h$ has 1 DOF. This leads to a factor of $\frac{1}{\sqrt{2}}$.

On the other hand, the Higgs field matches the massive vector fields because of the Higgs mechanism. Since of the charge conservation, the projections on the Higgs field and its hermitian conjugate are different,
\begin{equation}
\label{eq:projectors3}
\mathcal{U}^{I\,\mathbf{I}} = \left(\begin{array}{cc}
\mathcal{U}^{1W^-} & \mathcal{U}^{1Z} \\
\mathcal{U}^{2W^-} & \mathcal{U}^{2Z} 
\end{array}\right) = \left(\begin{array}{cc}
    \sqrt{2} & \\
     & 1
\end{array}\right)\,,\quad \mathcal{U}_I^{\mathbf{I}} = \left(\begin{array}{cc}
\mathcal{U}^{W^+}_1 & \mathcal{U}^{Z}_1 \\
\mathcal{U}^{W^+}_2 & \mathcal{U}^{Z}_2 
\end{array}\right) = -\left(\begin{array}{cc}
    \sqrt{2} & \\
     & 1
\end{array}\right)\,.
\end{equation}
The factor $\sqrt 2$ reflects the DOF difference between the complex scalar $\phi^{\pm}$ and the longitudinal modes of the massive $W^{\pm}$ boson. For the neutral sector, matching the complex $\phi^0$ to the real $Z$ boson modifies the factor to $\sqrt{2} \times 1/\sqrt{2} = 1$.

\item $F\rightarrow F$ and $V\rightarrow V$ transitions: 

The conversion between massless and massive fermion is described by matrices $\Omega^{\hat{i}\hat{\mathbf{i}}}$ and $\Omega_{\hat{i}\hat{\mathbf{i}}}$. Since there is only one generation, we do not consider the mixing of fermion. Here $\Omega$ is diagonal and thus trivial
\begin{equation}
\label{eq:projectors1}
    \Omega_i^{\hat{\mathbf{i}}} = \Omega^i_{\hat{\mathbf{i}}} = \left(\begin{array}{cccc}
1 & & & \\
 & 1 & & \\
& & 1 & \\
& & & 1 \\
    \end{array}\right),
\end{equation}

The $4\times 4$ square matrix $O^{I \mathbf{I}}$ maps massless gauge bosons $(W^1, W^2, W^3, B)$ to massive vectors $W^+, W^-, Z, \gamma$, incorporating the gauge boson mixing. Under the symmetry breaking $SU(2)_W \times U(1)Y \to U(1){\text{em}}$, the transformation matrices are
\begin{equation}
\label{eq:projectors2}
    O^{I\,\mathbf{I}} = O_I^{\mathbf{I}} = \left(\begin{array}{cccc}
\frac{1}{\sqrt{2}} & \frac{1}{\sqrt{2}} & & \\
\frac{i}{\sqrt{2}} & -\frac{i}{\sqrt{2}} & & \\
& & \cos\theta_W & \sin\theta_W \\
& & -\sin\theta_W & \cos\theta_W \\
    \end{array}\right)\,,
\end{equation}
where $\theta_W$ is the Weinberg angle, and 
\begin{equation}
    \cos\theta_W = \frac{g}{\sqrt{g^2+g'^2}}\,,\quad\sin\theta_W=\frac{g'}{\sqrt{g^2+g'^2}}\,.
\end{equation}

\end{itemize}

Using this strategy, we determine all massive coefficients for 3-pt amplitudes and establish the relation between massless and massive coefficient.

\textbf{1. SSS and SSSS:} This is the simplest case. the $\lambda\phi^4$ amplitude takes the form
\begin{equation} \begin{aligned}
\mathcal{A}^{\text{ct}}(1^{0},2^{0},3^{0},4^{0})&=-4\lambda \delta^{(i_1}_{i_3} \delta^{i_2)}_{i_4}.
\end{aligned} \end{equation}
where the superscript $0$ indicates the helicity of massless scalars.
Then match to $SSSS$ is trivial and take VEV $v$ to match to the massive $SSS$ amplitude 
\begin{equation} \begin{aligned}
\mathbf{M}({\mathbf{1}}^{0}, \mathbf{2}^{0}, \mathbf{3}^{0}) =-6\lambda v,
\end{aligned} \end{equation}
where the superscript $0$ now denotes the spin of massive scalars.

\textbf{2. FFS:} The massless Yukawa amplitudes take the form,
\begin{equation} \begin{aligned}
\mathcal{A}(1^{-\frac12},2^{-\frac12},3^{0})&={Y}^{\hat{i}_1 i_3}_{\hat{i}_2}\langle12\rangle,\\
\mathcal{A}(1^{+\frac12},2^{+\frac12},3^{0})&=\tilde{Y}^{\hat{i}_1}_{\hat{i}_2 i_3}[12],
\end{aligned} \end{equation}
where $Y$ and $\tilde Y$ represent the massless matrices. To match the massless $FFS$ amplitude to the massive one, we utilize the transformation matrices $\Omega$ and $\mathcal{U}$ acting on the massless couplings $Y$ and $\tilde{Y}$, and they become the massive couplings $y$ and $y'$ as follows,
\begin{equation} \begin{aligned}
{Y}^{\hat{i}_1 i_3}_{\hat{i}_2}\times \Omega^{\hat{\mathbf{i}}_1}_{\hat{i}_1} \Omega_{\hat{\mathbf{i}}_2}^{\hat{i}_2} \mathcal{U}_{i_3}^h + {Y}^{\hat{i}_1}_{\hat{i}_2 i_3} \times \Omega^{\hat{\mathbf{i}}_1}_{\hat{i}_1} \Omega_{\hat{\mathbf{i}}_2}^{\hat{i}_2} \mathcal{U}^{i_3 h}
\quad&\Rightarrow\quad {y}^{\hat{\mathbf{i}}_1}_{\hat{\mathbf{i}}_2},\\
\tilde{Y}^{\hat{i}_1 i_3}_{\hat{i}_2}\times \Omega^{\hat{\mathbf{i}}_1}_{\hat{i}_1} \Omega_{\hat{\mathbf{i}}_2}^{\hat{i}_2} \mathcal{U}_{i_3}^h + \tilde{Y}^{\hat{i}_1}_{\hat{i}_2 i_3} \times \Omega^{\hat{\mathbf{i}}_1}_{\hat{i}_1} \Omega_{\hat{\mathbf{i}}_2}^{\hat{i}_2} \mathcal{U}^{i_3 h}
\quad&\Rightarrow\quad {y'}^{\hat{\mathbf{i}}_1}_{\hat{\mathbf{i}}_2} .
\end{aligned} \end{equation}
Here the index for massive Higgs boson is suppressed for simplicity. By bolding the massless spinors, we obtain the massive $FFS$ amplitude
\begin{equation} \begin{aligned}
\mathbf{M}({\mathbf{1}}^{1/2}, \mathbf{2}^{1/2}, \mathbf{3}^{0}) = 
{y}^{\hat{\mathbf{i}}_1}_{\hat{\mathbf{i}}_2} \langle\mathbf{12}\rangle +{y'}^{\hat{\mathbf{i}}_1}_{\hat{\mathbf{i}}_2} [\mathbf{12}].
\end{aligned} \end{equation}
The non-zero values of the massive coefficients $y$ and $y'$ are given by
\begin{equation} \begin{aligned}
{y}^{u}_{\bar{u}} = {y'}^{u}_{\bar{u}} &= \frac{1}{\sqrt{2}}\mathcal{Y}^{(u)},\\
{y}^{d}_{\bar{d}} = {y'}^{d}_{\bar{d}} &= \frac{1}{\sqrt{2}}\mathcal{Y}^{(d)},\\
{y}^{e}_{\bar{e}} = {y'}^{e}_{\bar{e}} &= \frac{1}{\sqrt{2}}\mathcal{Y}^{(e)}.\\
\end{aligned} \end{equation}
The result agrees with the Yukawa matrix for massive fermions in the single-flavor case.

\textbf{3. FFV:} In this case, both massless $FFS$ and $FFV$ match to the massive $FFV$. Because of the little group covariance, we can take only one massless $FFV$ to perform the matching, and the massless $FFS$ matching is implicitly taken. Here we consider two massless amplitudes with opposite helicity for the fermion and anti-fermion $(\mp\frac12,\pm\frac12,+1)$. The deformation tells
\begin{equation} \begin{aligned}
\mathcal{A}(1^{-\frac12},2^{+\frac12},3^{+1})&=(\tilde{T}_f^{I_3})^{\hat{i}_1}_{\hat{i}_2}\frac{[23]^2}{[12]}
\quad \to \quad (\tilde{T}_f^{I_3})^{\hat{i}_1}_{\hat{i}_2}\frac{[23]\langle1\eta_3\rangle}{m_3},\\
\mathcal{A}(1^{+\frac12},2^{-\frac12},3^{+1})&=(T_f^{I_3})^{\hat{i}_1}_{\hat{i}_2}\frac{[13]^2}{[12]}
\quad \to \quad -(T_f^{I_3})^{\hat{i}_1}_{\hat{i}_2}\frac{[13]\langle2\eta_3\rangle}{m_3},
\end{aligned} \end{equation}
where $\tilde T_f$ denotes the coupling for the right-handed current, and $T_f$ for the left-handed current. Applying the transformation matrices $O$ and $\Omega$ to convert massless particles to massive ones, we obtain the massive coefficients $X_1$ and $X_2$:
\begin{equation} \begin{aligned}
(T_f^{I_3})^{\hat{i}_1}_{\hat{i}_2}\times O^{I_3 \mathbf{I}_3} \Omega^{\hat{\mathbf{i}}_1}_{\hat{i}_1} \Omega_{\hat{\mathbf{i}}_2}^{\hat{i}_2}
\quad&\Rightarrow\quad (X_1^{\mathbf{I}_3} )^{\hat{\mathbf{i}}_1}_{\hat{\mathbf{i}}_2},\\
(\tilde{T}_f^{I_3})^{\hat{i}_1}_{\hat{i}_2}\times O^{I_3 \mathbf{I}_3} \Omega^{\hat{\mathbf{i}}_1}_{\hat{i}_1} \Omega_{\hat{\mathbf{i}}_2}^{\hat{i}_2}
\quad&\Rightarrow\quad (X_2^{\mathbf{I}_3} )^{\hat{\mathbf{i}}_1}_{\hat{\mathbf{i}}_2}.
\end{aligned} \end{equation}
By bolding the massless spinors, we obtain the massive amplitudes
\begin{equation} \label{eq:FFV_result}
\mathbf{M}(\mathbf{1}^{1/2}, \mathbf{2}^{1/2}, \mathbf{3}^{1}) = \sqrt2(X_1^{\mathbf{I}_3} )^{\hat{\mathbf{i}}_1}_{\hat{\mathbf{i}}_2} \frac{\langle\mathbf{13}\rangle [\mathbf{23}]}{\mathbf{m}_3} + \sqrt2(X_2^{\mathbf{I}_3} )^{\hat{\mathbf{i}}_1}_{\hat{\mathbf{i}}_2} \frac{\langle\mathbf{23}\rangle [\mathbf{13}]}{\mathbf{m}_3}.
\end{equation}
Here for the charged vector boson $\mathbf I_3=W^\pm$, only $X_1$ exists,
\begin{equation} \begin{aligned}
(X_1^{W^{+}} )^{d}_{\bar{u}}&=(X_1^{W^{+}} )^{\nu}_{\bar{e}}=\frac{e}{\sqrt2\sin\theta_W},\\
(X_1^{W^{-}} )^{u}_{\bar{d}}&=(X_1^{W^{-}} )^{e}_{\bar{\nu}}=\frac{e}{\sqrt2\sin\theta_W},\\
(X_2^{W^{\pm}} )^{\hat{\mathbf{i}}_1}_{\hat{\mathbf{i}}_2} &=0,
\end{aligned} \end{equation}
where $e=g\sin\theta_W$ is the coupling of $U(1)_{\text{em}}$. This describes the coupling between the $W$ boson and the charged current. For the neutral vector boson $\mathbf{I}_3 = Z$, both coefficients contribute, with $X_1$ and $X_2$ taking different values: 
\begin{equation} \begin{aligned}
(X_1^{Z} )^{u}_{\bar{u}}&=\frac{e}{2}\left(\frac13 \tan\theta_W-\cot\theta_W\right),&
(X_1^{Z} )^{d}_{\bar{d}}&=\frac{e}{2}\left(\frac13 \tan\theta_W+\cot\theta_W\right),&\\
(X_1^{Z} )^{\nu}_{\bar{\nu}}&=\frac{e}{2}(-\tan\theta_W-\cot\theta_W),&
(X_1^{Z} )^{e}_{\bar{e}}&=\frac{e}{2}\left(-\tan\theta_W+\cot\theta_W\right),&\\
(X_2^{Z} )^{u}_{\bar{u}}&=\frac{2e}{3}\tan\theta_W,&
(X_2^{Z} )^{d}_{\bar{d}}&=-\frac{e}{2}\tan\theta_W,& \\
(X_2^{Z} )^{e}_{\bar{e}}&=-e\tan\theta_W.&
\end{aligned} \end{equation}
For the photon $\mathbf{I}_3 = A$, the two coefficients $X_1$ and $X_2$ are identical: 
\begin{equation}
(X_1^{A} )^{u}_{\bar{u}}=(X_2^{A} )^{u}_{\bar{u}}=-\frac{2}{3}e,\quad
(X_1^{A} )^{d}_{\bar{d}}=(X_2^{A} )^{d}_{\bar{d}}=\frac{1}{3}e,\quad
(X_1^{A} )^{e}_{\bar{e}}=(X_2^{A} )^{e}_{\bar{e}}=e.
\end{equation}

\textbf{4. VVS:} In this case, a single massless $VSS$ amplitude with helicity category $(-1,0,0)$ is sufficient, 
\begin{equation} \begin{aligned} \label{eq:massless_VSS}
\mathcal{A}(1^{-1},2^{0},3^{0})&=(T_s^{I_1})^{i_2}_{i_3}\frac{\langle12\rangle\langle31\rangle}{\langle23\rangle}
\quad\to\quad 
\frac{[\eta_1 2]\langle12\rangle}{\tilde m_1},
\end{aligned} \end{equation}
The massless coefficient $T_s$ maps to the massive coupling $\mathbf{g}$ via
\begin{equation}
{({T_s}^{I_1})}_{i_3}^{i_2} O^{I_1 \mathbf{I}_1} \mathcal{U}_{i_2}^{\mathbf{I}_2} \mathcal{U}^{i_3 h} -{({T_s}^{I_1})}_{i_2}^{i_3} O^{I_1 \mathbf{I}_1} \mathcal{U}^{i_2 \mathbf{I}_2} \mathcal{U}_{i_3}^{h}
\quad\Rightarrow\quad \mathbf{g}^{\mathbf{I}_1 \mathbf{I}_2}.
\end{equation}
Bolding the massless spinors, we obtain the massive amplitude
\begin{equation}
\mathbf{M}(\mathbf{1}^{1}, \mathbf{2}^{1}, \mathbf{3}^{0}) = 
\mathbf{g}^{\mathbf{I}_1 \mathbf{I}_2} \frac{\langle\mathbf{12}\rangle [\mathbf{12}]}{\mathbf{m}_1},
\end{equation}
where we enforce $\mathbf{m}_1 = \mathbf{m}_2$ for consistency. 
The non-vanishing components of the massive coupling $\mathbf{g}$ are
\begin{equation} \begin{aligned}
\mathbf{g}^{W^+ W^-} = \mathbf{g}^{W^- W^+} = g,
\quad \mathbf{g}^{Z Z} = \frac{g}{\cos\theta_W}.
\end{aligned} \end{equation}

\textbf{5. VVV:} Similar to the $FFV$ case, the matching can be taken from only the massless $VVV$ amplitudes to their massive counterparts, and the matching from massless $VSS$ to massive $VVV$ can be recovered by the little group covariance. Note that, one massless $VVV$ amplitude cannot account for all massive terms. We select two massless amplitudes with opposite helicities in the matching procedure. Specifically, we choose:
\begin{equation} \begin{aligned}
\mathcal{A}(1^{+1},2^{+1},3^{-1})&=f^{I_1 I_2 I_3}\frac{[12]^3}{[23][31]},\\
\mathcal{A}(1^{-1},2^{-1},3^{+1})&=f^{I_1 I_2 I_3} \frac{\langle12\rangle^3}{\langle23\rangle\langle31\rangle}.
\end{aligned} \end{equation}
Each massless amplitude admits two distinct deformations to the MHC amplitude, which gives equal contributions with the factor $\frac12$. The massless deformation thus yields
\begin{equation} \begin{aligned}
f^{I_1 I_2 I_3}\frac{[12]^3}{[23][31]} &\rightarrow 
\frac{1}{2}f^{I_1 I_2 I_3}\frac{[12]\langle\eta_23\rangle\langle 3 \eta_1\rangle}{m_1 m_2}+\frac{1}{2}f^{I_1 I_2 I_3}\left(\frac{[12][2 \eta_3]\langle3\eta_1\rangle}{m_1 \tilde{m}_3}+\frac{[12]\langle\eta_23\rangle[\eta_31]}{m_2 \tilde{m}_3}\right), \\
f^{I_1 I_2 I_3}\frac{\langle12\rangle^3}{\langle23\rangle\langle31\rangle} 
&\rightarrow \frac{1}{2}f^{I_1 I_2 I_3}\frac{\langle 12\rangle[\eta_23][3 \eta_1]}{\tilde m_1 \tilde m_2}+\frac{1}{2}f^{I_1 I_2 I_3}\left(\frac{\langle12\rangle\langle2 \eta_3\rangle[3\eta_1]}{\tilde m_1 m_3}+\frac{\langle12\rangle[\eta_23]\langle\eta_31\rangle}{\tilde m_2 m_3}\right).
\end{aligned} \end{equation}
The massless coefficient $f$ maps to the massive coupling $\mathbf{f}$ via
\begin{equation}
\frac{1}{2}f^{I_1 I_2 I_3} O^{I_1 \mathbf{I}_1} O^{I_2 \mathbf{I}_2} O^{I_3 \mathbf{I}_3}
\quad\Rightarrow\quad \mathbf{f}^{\mathbf{I}_1 \mathbf{I}_2 \mathbf{I}_3} .
\end{equation}
Summing both contributions and restoring $SU(2)_{\text{LG}}$ covariance yields the complete massive $VVV$ amplitude
\begin{equation} \begin{aligned} \label{eq:VVV_result}
\mathbf{M}(\mathbf{1}^{1}, \mathbf{2}^{1}, \mathbf{3}^{1}) 
=& \mathbf{f}^{\mathbf{I}_1 \mathbf{I}_2 \mathbf{I}_3} \left(\frac{[\mathbf{12}]\langle\mathbf{23}\rangle\langle\mathbf{31}\rangle}{\mathbf{m}_1 \mathbf{m}_2}+\frac{[\mathbf{12}][\mathbf{23}]\langle\mathbf{31}\rangle }{\mathbf{m}_1 \mathbf{m}_3}+\frac{[\mathbf{12}]\langle\mathbf{23}\rangle[\mathbf{31}]}{\mathbf{m}_2 \mathbf{m}_3} \right)\\
&+ \mathbf{f}^{\mathbf{I}_1 \mathbf{I}_2 \mathbf{I}_3} \left(\frac{\langle\mathbf{12}\rangle[\mathbf{23}][\mathbf{31}]}{\mathbf{m}_1 \mathbf{m}_2}+\frac{\langle\mathbf{12}\rangle\langle\mathbf{23}\rangle[\mathbf{31}]}{\mathbf{m}_1 \mathbf{m}_3}+\frac{\langle\mathbf{12}\rangle[\mathbf{23}]\langle\mathbf{31}\rangle}{\mathbf{m}_2 \mathbf{m}_3} \right),
\end{aligned} \end{equation}
where all terms share the same massive coefficient.
In analogy to the massless case, $\mathbf{f}^{\mathbf{I}_1 \mathbf{I}_2 \mathbf{I}_3}$ is totally antisymmetric. Substituting the relevant particle types, we obtain the non-zero values,
\begin{equation}
\begin{tabular}{c|c}
\hline
$\mathbf{I}_1\mathbf{I}_2\mathbf{I}_3$ & $\mathbf{f}^{\mathbf{I}_1 \mathbf{I}_2 \mathbf{I}_3}$ \\
\hline
\makecell{$W^+ W^- Z$\\ $W^- Z W^+$\\ $Z W^+ W^-$} & $\frac12 g \cos\theta_W$ \\
\hline
\makecell{$W^- W^+ Z$\\ $W^+ Z W^-$\\ $Z W^- W^+$} & $-\frac12 g \cos\theta_W$  \\
\hline
\end{tabular}
\end{equation}

\section{MHC Amplitude Bootstrapping and 4-point Contact MHC amplitudes}
\label{sec:bootstrap}

In the previous section, the 3-pt MHC amplitudes are the necessary ingredients to build the massless-massive 3-point amplitude correspondence in the SM. In order to describe the higher-point amplitudes, one can suitably glue the 3-pt amplitudes together, which is based on the following properties of the singularity structure of scattering amplitudes with the total $\delta^{(4)}$-momentum conservation stripped out:
\begin{itemize}
\item Analyticity: the singularities are restricted to only having poles and branch points;
\item Locality: the singularities are further restricted to having only simple poles and branch cuts;
\item Unitarity: when approaching the singularities, one or more particles go on-shell and the amplitude factorizes into lower-point amplitudes.
\end{itemize}
In the factorization limit, the amplitudes are decomposed into a product of sub-amplitudes corresponding to two distinct processes. 
This factorization property places stringent conditions on the simple pole structure of the massless amplitude
\begin{equation} \label{eq:FactorizeMassless}
\lim_{P^2\rightarrow 0} P^2 \mathcal{M} = M^L\times M^R,
\end{equation}
where $P^2$ is the on-shell propagator of the immediate massless particle, and the $M^L$ and $M^R$ are sub-amplitudes.

For the massive amplitudes, the pole structure is different from the massless ones, which has only one simple pole structure $1/p^2$. 
The massive pole structure with propagator momentum $\mathbf P$ takes the form $\mathbf{P}^2-\mathbf m^2$. Since the MHC amplitudes decompose massive structure into large and small components, the propagator momentum should also be expanded as $\mathbf{P} = P + \eta$. Therefore, the massive pole structure decomposes as
\begin{equation} \label{eq:expand_pole}
\mathbf{P}^2-\mathbf m^2=P^2+(2P\cdot\eta-\mathbf m^2)+\eta^2.
\end{equation}
This gives three contributions at different orders. Thus, a 4-point MHC amplitude $\mathcal{M}_4$ with such a pole structure should has the form
\begin{equation} \label{eq:MHC_higher_point}
\mathcal{M}_4\sim\frac{N}{P^2}+\frac{N}{P^2}\frac{2P\cdot\eta-\mathbf m^2}{P^2}+\frac{N}{P^2}(\frac{(2P\cdot\eta-\mathbf m^2)^2}{P^4}+\frac{\eta^2}{P^2})+\cdots,
\end{equation}
where $N = M^L\times M^R$ is the numerator, with the $M^L$ and $M^R$ are massive sub-amplitudes. This expansion is infinite due to the propagator. For higher-point amplitudes, each massive pole will have a corresponding expansion.

The term $P^2$ corresponds to a massless pole. To isolate this contribution in the MHC amplitude, we take the limit
\begin{equation}
\lim_{\substack{2P\cdot\eta \rightarrow \mathbf{m}^2\\ \eta^2\rightarrow 0}} \mathcal{M}_4\sim\frac{M^L\times M^R}{P^2},
\end{equation}
which extracts the single-pole term, and $M^L$ and $M^R$ are 3-pt MHC sub-amplitudes. Using the gluing technique, we can obtain this term by further taking the limit $P^2\to 0$, leading to
\begin{equation} \label{eq:FactorizeMassless}
\lim_{P^2\rightarrow 0} \lim_{\substack{2P\cdot\eta \rightarrow \mathbf{m}^2\\ \eta^2\rightarrow 0}} P^2 \mathcal{M}_4
=\sum_{h_P} M^L\times M^R,
\end{equation}
where $h_P$ is the helicity of the on-shell propagator. Once the single-pole term is obtained, the higher-order terms in eq.~\eqref{eq:MHC_higher_point} can be systematically derived by expanding the massive pole $\mathbf{P}^2 - \mathbf{m}^2$.

\subsection{Gluing 3-point MHC amplitudes: $u\bar{d}\to Wh$ as example}
\label{sec:glue_MHC}

In this subsection, let us consider the massive $u\bar{d}Wh$ amplitude to illustrate the gluing technique. This amplitude has twelve helicity categories:
\begin{equation} \begin{aligned}
&(\pm\frac{1}{2}, \pm\frac{1}{2}, +1, 0),\quad
(\pm\frac{1}{2}, \pm\frac{1}{2}, 0, 0),\quad
(\pm\frac{1}{2}, \pm\frac{1}{2}, -1, 0),\\
&(\pm\frac{1}{2}, \mp\frac{1}{2}, +1, 0),\quad
(\pm\frac{1}{2}, \mp\frac{1}{2}, 0, 0),\quad
(\pm\frac{1}{2}, \mp\frac{1}{2}, -1, 0).
\end{aligned} \end{equation}
In what follows, we choose $(+\frac{1}{2}, -\frac{1}{2}, +1, 0)$ to show the gluing procedure, and the other helicity categories can be obtained in a similar way. The amplitude has three channel
\begin{equation} \begin{aligned}
\mathcal{M}(1^{+\frac{1}{2}},2^{-\frac{1}{2}},3^{+1},4^0)
=\mathcal{M}_{(12)}+\mathcal{M}_{(13)}+\mathcal{M}_{(14)}.
\end{aligned} \end{equation}
The term $\mathcal{M}_{(ij)}$ corresponds to the term with single pole $P_{ij}=p_i+p_j$. Let us consider them one by one.

\paragraph{$(14)$-channel}

We first consider the $(14)$ channel, where  the internal line corresponds to a $u$ quark propagator.
The relevant 3-pt sub-amplitudes are
\begin{align}
\begin{tikzpicture}[baseline=-0.1cm] \begin{feynhand}
\setlength{\feynhandarrowsize}{4pt}
\vertex [particle] (i1) at (-0.827,0.579) {$1^-$}; 
\vertex [particle] (i2) at (0.827,0.579) {$4^0$}; 
\vertex [particle] (i3) at (0,-1.01) {$P^-$}; 
\vertex (v1) at (0,0);
\fer{red}{i1}{v1};
\sca{i2};
\antfer{cyan}{i3}{v1};
\end{feynhand} \end{tikzpicture} &={y}^{u}_{\bar u}\langle 1P_{14}\rangle, &
\begin{tikzpicture}[baseline=-0.1cm] \begin{feynhand}
\setlength{\feynhandarrowsize}{4pt}
\vertex [particle] (i1) at (-0.827,0.579) {$1^-$}; 
\vertex [particle] (i2) at (0.827,0.579) {$4^0$}; 
\vertex [particle] (i3) at (0,-1.01) {$P^+$}; 
\vertex (v1) at (0,0);
\fer{red}{i1}{v1};
\sca{i2};
\antferflip{1}{-90}{red}{cyan};
\end{feynhand} \end{tikzpicture} &={y}^{u}_{\bar u}\frac{\tilde m_{14}}{\mathbf m_{14}}\langle1\eta_{14}\rangle,& \label{eq:FFS_sub1}\\
\begin{tikzpicture}[baseline=-0.1cm] \begin{feynhand}
\setlength{\feynhandarrowsize}{4pt}
\vertex [particle] (i1) at (-0.827,0.579) {$1^-$}; 
\vertex [particle] (i2) at (0.827,0.579) {$4^0$}; 
\vertex [particle] (i3) at (0,-1.01) {$P^-$}; 
\vertex (v1) at (0,0);
\ferflip{1}{145}{red}{cyan};
\sca{i2};
\antferflip{1}{-90}{cyan}{red};
\end{feynhand} \end{tikzpicture} &={y'}^{u}_{\bar u}\frac{m_1}{\mathbf m_1} \frac{m_{14}}{\mathbf m_{14}}[\eta_1 \eta_{14}], &
\begin{tikzpicture}[baseline=-0.1cm] \begin{feynhand}
\setlength{\feynhandarrowsize}{4pt}
\vertex [particle] (i1) at (-0.827,0.579) {$1^-$}; 
\vertex [particle] (i2) at (0.827,0.579) {$4^0$}; 
\vertex [particle] (i3) at (0,-1.01) {$P^+$}; 
\vertex (v1) at (0,0);
\ferflip{1}{145}{red}{cyan};
\sca{i2};
\antfer{red}{i3}{v1};
\end{feynhand} \end{tikzpicture} &={y'}^{u}_{\bar u}\frac{m_1}{\mathbf m_1}[\eta_1 P_{14}],& \label{eq:FFS_sub2}\\
\begin{tikzpicture}[baseline=-0.1cm] \begin{feynhand}
\setlength{\feynhandarrowsize}{4pt}
\vertex [particle] (i1) at (-0.827,-0.579) {$2^+$}; 
\vertex [particle] (i2) at (0.827,-0.579) {$3^-$}; 
\vertex [particle] (i3) at (0,1.01) {$P^+$}; 
\vertex (v1) at (0,0);
\ferflip{1}{90}{cyan}{red};
\antfer{red}{i1}{v1};
\bosflip{1}{-35}{brown}{red};
\end{feynhand} \end{tikzpicture}
&=-\sqrt2(X_1^{W^-} )^{u}_{\bar d}\frac{\tilde m_3}{\mathbf m_3}\frac{\tilde m_{14}}{\mathbf m_{14}}\frac{[2 \eta_3] \langle 3\eta_{14}\rangle}{\mathbf m_3},& 
\begin{tikzpicture}[baseline=-0.1cm] \begin{feynhand}
\setlength{\feynhandarrowsize}{4pt}
\vertex [particle] (i1) at (-0.827,-0.579) {$2^+$}; 
\vertex [particle] (i2) at (0.827,-0.579) {$3^-$}; 
\vertex [particle] (i3) at (0,1.01) {$P^-$}; 
\vertex (v1) at (0,0);
\fer{red}{i3}{v1};
\antfer{red}{i1}{v1};
\bosflip{1}{-35}{brown}{red};
\end{feynhand} \end{tikzpicture}
&=-\sqrt2(X_1^{W^-} )^{u}_{\bar d}\frac{\tilde m_3}{\mathbf m_3}\frac{[2 \eta_3]\langle 3P_{14}\rangle}{\mathbf m_3},&  \label{eq:FFV_sub}
\end{align}
where $P_{14}\equiv p_1+p_4$, $\eta_{14}\equiv \eta_1+\eta_4$, and particle $P$ is taken to be the internal fermion. The first line and second lines correspond to the $u\bar{u}h$ amplitude, while the third line corresponds to the $u\bar{d}W^-$ amplitude.

In the factorized limit, we sum over helicities of internal particle. When a particle line with a chirality flip is glued to one without chirality flip, two helicity combinations contribute to one propagator
\begin{equation} \label{eq:glue_1}
\begin{cases}
\begin{tikzpicture}[baseline=0.7cm] \begin{feynhand}
\setlength{\feynhandarrowsize}{4pt}
\vertex [dot] (v1) at (0,0.8) {};
\vertex (i1) at (1.4,0.8) {$P^-$};
\antfer{cyan}{i1}{v1}
\end{feynhand} \end{tikzpicture}
\begin{tikzpicture}[baseline=0.7cm] \begin{feynhand}
\vertex [dot] (v1) at (1.4,0.8) {};
\vertex (v2) at (0.9,0.8);
\vertex (i1) at (0,0.8) {$P^+$};
\draw[cyan,very thick] (i1)--(v2);
\draw[red,very thick,decoration={markings,mark=at position 0.56 with {\arrow{Triangle[length=4pt,width=4pt]}}},postaction={decorate}] (v2)--(v1);
\draw[very thick] plot[mark=x,mark size=2.5] coordinates {(v2)};
\end{feynhand} \end{tikzpicture}=\tilde m \eta_{\alpha}\times \lambda^{\beta}\\
\begin{tikzpicture}[baseline=0.7cm] \begin{feynhand}
\vertex [dot] (v1) at (0,0.8) {};
\vertex (v2) at (0.6,0.8);
\vertex (i1) at (1.4,0.8) {$P^+$};
\draw[red,very thick] (v2)--(i1);
\draw[cyan,very thick,decoration={markings,mark=at position 0.82 with {\arrow{Triangle[length=4pt,width=4pt]}}},postaction={decorate}] (v1)--(v2);
\draw[very thick] plot[mark=x,mark size=2.5] coordinates {(v2)};
\end{feynhand} \end{tikzpicture}
\begin{tikzpicture}[baseline=0.7cm] \begin{feynhand}
\setlength{\feynhandarrowsize}{4pt}
\vertex [dot] (v1) at (1.4,0.8) {};
\vertex (i1) at (0,0.8) {$P^-$};
\fer{red}{i1}{v1};
\end{feynhand} \end{tikzpicture}=-\lambda_{\alpha}\times \tilde m \eta_{\beta}
\end{cases}
\quad \Rightarrow \quad
\begin{tikzpicture}[baseline=0.7cm] \begin{feynhand}
\setlength{\feynhandarrowsize}{4pt}
\vertex [dot] (i1) at (0,0.8) {};
\vertex (v2) at (0.8,0.8);
\vertex [dot] (i2) at (1.6,0.8) {};
\fer{cyan}{i1}{v2};
\fer{red}{v2}{i2};
\draw[very thick] plot[mark=x,mark size=2.5] coordinates {(v2)};
\end{feynhand} \end{tikzpicture}=\mathbf m^2\epsilon_{\alpha\beta}.
\end{equation}
The resulting structure is for the propagator, which carries zero helicity and transversality, and its chirality-flip nature is encoded in the Levi-Civita tensor $\epsilon_{\alpha\beta}$. Using this method, we glue the 3-pt sub-amplitudes from eq.~\eqref{eq:FFS_sub1} and \eqref{eq:FFV_sub} to obtain
\begin{equation}
\begin{tikzpicture}[baseline=-0.1cm] \begin{feynhand}
\setlength{\feynhandarrowsize}{4pt}
\vertex [particle] (i1) at (-0.9,0.9) {$1^-$};
\vertex [particle] (i2) at (-0.9,-0.9) {$2^+$};
\vertex [particle] (i3) at (0.9,-0.9) {$3^-$};
\vertex [particle] (i4) at (0.9,0.9) {$4^0$};
\vertex (v1) at (0,-0.4);
\vertex (v2) at (0,0.4);
\vertex (v3) at (0,0);
\vertex (v4) at (0.9*0.3,-0.4-0.5*0.3);
\fer{red}{i1}{v2};
\fer{cyan}{v2}{v3};
\fer{red}{v3}{v1};
\fer{red}{v1}{i2};
\draw[brown,thick sca] (v2)--(i4);
\draw[brown,thick bos] (v1)--(v4);
\draw[red,thick bos] (v4)--(i3);
\draw[very thick,rotate=-25] plot[mark=x,mark size=2.5] coordinates {(v4)};
\draw[very thick] plot[mark=x,mark size=2.7] coordinates {(v3)};
\end{feynhand} \end{tikzpicture}
\quad\Rightarrow\quad
-\sqrt2{y}^{u}_{\bar u}(X_1^{W^-} )^{u}_{\bar d}\frac{\tilde m_3}{\mathbf m_3^2}\mathbf m_{14}\langle13\rangle[\eta_3 2],
\end{equation}
where $\mathbf m_{14}$ is the u quark mass. It exhibit a different power of $\mathbf m_{14}$ from eq.~\eqref{eq:glue_1}, due to a $1/\mathbf{m}_{14}$ factor in the 3-point sub-amplitude.

If we glue two particle lines both without chirality flips or both with chirality flips, we obtain the internal fermion in other two orders
\begin{equation} \begin{aligned}
\begin{tikzpicture}[baseline=0.7cm] \begin{feynhand}
\setlength{\feynhandarrowsize}{4pt}
\vertex [dot] (v1) at (0,0.8) {};
\vertex (i1) at (1.4,0.8) {$P^+$};
\antfer{red}{i1}{v1}
\end{feynhand} \end{tikzpicture}
\begin{tikzpicture}[baseline=0.7cm] \begin{feynhand}
\setlength{\feynhandarrowsize}{4pt}
\vertex [dot] (v1) at (1.4,0.8) {};
\vertex (i1) at (0,0.8) {$P^-$};
\fer{red}{i1}{v1};
\end{feynhand} \end{tikzpicture}&=\tilde\lambda_{\dot\alpha}\times \lambda_\beta&
&\Rightarrow&
\begin{tikzpicture}[baseline=0.7cm] \begin{feynhand}
\setlength{\feynhandarrowsize}{4pt}
\vertex [dot] (i1) at (0,0.8) {};
\vertex [dot] (i2) at (1.6,0.8) {};
\fer{red}{i1}{i2};
\end{feynhand} \end{tikzpicture}&=P_{\beta\dot\alpha},& \\
\begin{tikzpicture}[baseline=0.7cm] \begin{feynhand}
\vertex [dot] (v1) at (0,0.8) {};
\vertex (v2) at (0.6,0.8);
\vertex (i1) at (1.4,0.8) {$P^-$};
\draw[cyan,very thick] (v2)--(i1);
\draw[red,very thick,decoration={markings,mark=at position 0.82 with {\arrow{Triangle[length=4pt,width=4pt]}}},postaction={decorate}] (v1)--(v2);
\draw[very thick] plot[mark=x,mark size=2.5] coordinates {(v2)};
\end{feynhand} \end{tikzpicture}
\begin{tikzpicture}[baseline=0.7cm] \begin{feynhand}
\vertex [dot] (v1) at (1.4,0.8) {};
\vertex (v2) at (0.9,0.8);
\vertex (i1) at (0,0.8) {$P^+$};
\draw[cyan,very thick] (i1)--(v2);
\draw[red,very thick,decoration={markings,mark=at position 0.56 with {\arrow{Triangle[length=4pt,width=4pt]}}},postaction={decorate}] (v2)--(v1);
\draw[very thick] plot[mark=x,mark size=2.5] coordinates {(v2)};
\end{feynhand} \end{tikzpicture}&=m\tilde\eta_{\dot\alpha}\times \tilde m\eta_\beta&
&\Rightarrow&
\begin{tikzpicture}[baseline=0.7cm] \begin{feynhand}
\setlength{\feynhandarrowsize}{4pt}
\vertex [dot] (i1) at (0,0.8) {};
\vertex [dot] (i2) at (1.6,0.8) {};
\fer{red}{i1}{i2};
\draw[very thick] plot[mark=x,mark size=2.5] coordinates {(0.5,0.8)};
\draw[very thick] plot[mark=x,mark size=2.5] coordinates {(1.1,0.8)};
\end{feynhand} \end{tikzpicture}&=\mathbf m^2\eta_{\beta\dot\alpha}.&\\
\end{aligned} \end{equation} 
In the right-hand side of the second line, the glued internal line is shown without color change for convenience, since two chirality flips on a fermion do not produce a Levi-Civita tensor, making it equivalent to the non-flip representation. These correspond to the 3-pt sub-amplitudes given in eqs.~\eqref{eq:FFS_sub2} and \eqref{eq:FFV_sub}. Gluing them, we obtain
\begin{equation} \begin{aligned}
\begin{tikzpicture}[baseline=-0.1cm] \begin{feynhand}
\setlength{\feynhandarrowsize}{4pt}
\vertex [particle] (i1) at (-0.9,0.9) {$1^-$};
\vertex [particle] (i2) at (-0.9,-0.9) {$2^+$};
\vertex [particle] (i3) at (0.9,-0.9) {$3^-$};
\vertex [particle] (i4) at (0.9,0.9) {$4^0$};
\vertex (v1) at (0,-0.4);
\vertex (v2) at (0,0.4);
\vertex (v3) at (-0.9*0.3,0.4+0.5*0.3);
\vertex (v4) at (0.9*0.3,-0.4-0.5*0.3);
\fer{red}{i1}{v3};
\fer{cyan}{v3}{v2};
\fer{red}{v2}{v1};
\fer{red}{v1}{i2};
\draw[brown,thick sca] (v2)--(i4);
\draw[brown,thick bos] (v1)--(v4);
\draw[red,thick bos] (v4)--(i3);
\draw[very thick,rotate=-25] plot[mark=x,mark size=2.5] coordinates {(v4)};
\draw[very thick,rotate=-25] plot[mark=x,mark size=2.7] coordinates {(v3)};
\end{feynhand} \end{tikzpicture}\quad
&\Rightarrow\quad
\sqrt2{y'}^{u}_{\bar u}(X_1^{W^-} )^{u}_{\bar d}\frac{\tilde m_3}{\mathbf m_3^2}\frac{m_1}{\mathbf m_1}[\eta_1|P_{14}|3\rangle[\eta_32],\\
\begin{tikzpicture}[baseline=-0.1cm] \begin{feynhand}
\setlength{\feynhandarrowsize}{4pt}
\vertex [particle] (i1) at (-0.9,0.9) {$1^-$};
\vertex [particle] (i2) at (-0.9,-0.9) {$2^+$};
\vertex [particle] (i3) at (0.9,-0.9) {$3^-$};
\vertex [particle] (i4) at (0.9,0.9) {$4^0$};
\vertex (v1) at (0,-0.4);
\vertex (v2) at (0,0.4);
\vertex (v3) at (-0.9*0.3,0.4+0.5*0.3);
\vertex (v4) at (0.9*0.3,-0.4-0.5*0.3);
\fer{red}{i1}{v3};
\fer{cyan}{v3}{v2};
\fer{red}{v2}{v1};
\fer{red}{v1}{i2};
\draw[brown,thick sca] (v2)--(i4);
\draw[brown,thick bos] (v1)--(v4);
\draw[red,thick bos] (v4)--(i3);
\draw[very thick,rotate=-25] plot[mark=x,mark size=2.5] coordinates {(v4)};
\draw[very thick,rotate=-25] plot[mark=x,mark size=2.7] coordinates {(v3)};
\draw[very thick,rotate=0] plot[mark=x,mark size=2.7] coordinates {(0,0.2)};
\draw[very thick,rotate=0] plot[mark=x,mark size=2.7] coordinates {(0,-0.2)};
\end{feynhand} \end{tikzpicture}\quad
&\Rightarrow\quad
\sqrt2{y'}^{u}_{\bar u}(X_1^{W^-} )^{u}_{\bar d}\frac{\tilde m_3}{\mathbf m_3^2}\frac{m_1}{\mathbf m_1}[\eta_1|\eta_{14}|3\rangle[\eta_32].
\end{aligned} \end{equation}
After accounting for the $1/\mathbf{m}$ factors in the 3-pt sub-amplitudes, neither of these results contain a factor of $\mathbf{m}_{14}$.
Summing over contributions from all three diagrams and restoring the denominator, the final result takes the form
\begin{equation} 
\begin{aligned} \label{eq:FFVS-s13-s14}
\mathcal{M}_t=\sqrt2{y}^{u}_{\bar u}(X_1^{W^-} )^{u}_{\bar d}\frac{\tilde m_3}{\mathbf m_3^2}\frac{\mathbf m_{14}\langle13\rangle[\eta_3 2]}{s_{14}}
+\sqrt2{y'}^{u}_{\bar u}(X_1^{W^-} )^{u}_{\bar d}\frac{\tilde m_3}{\mathbf m_3^2}\frac{m_1}{\mathbf m_1}\frac{[\eta_1|P_{14}+\eta_{14}|3\rangle[\eta_32]}{s_{14}},
\end{aligned} 
\end{equation}
where the massless pole is given by $s_{14}=P_{14}^2$.

\paragraph{$(13)$-channel}

We now turn to the $(13)$ channel, where the internal line also corresponds to a $d$ quark propagator. In this case, we  glue the $d\bar dh$ and $u\bar dW^-$ amplitudes. The gluing results are 
\begin{align}
\begin{tikzpicture}[baseline=-0.1cm] \begin{feynhand}
\setlength{\feynhandarrowsize}{4pt}
\vertex [particle] (i1) at (-0.9,0.9) {$1^-$};
\vertex [particle] (i2) at (-0.9,-0.9) {$2^+$};
\vertex [particle] (i3) at (0.9,-0.9) {$4^0$};
\vertex [particle] (i4) at (0.9,0.9) {$3^-$};
\vertex (v1) at (0,-0.4);
\vertex (v2) at (0,0.4);
\vertex (v3) at (0,0);
\vertex (v4) at (0.9*0.3,0.4+0.5*0.3);
\fer{red}{i1}{v2};
\fer{red}{v2}{v3};
\fer{cyan}{v3}{v1};
\fer{red}{v1}{i2};
\draw[brown,thick sca] (v1)--(i3);
\draw[brown,thick bos] (v2)--(v4);
\draw[red,thick bos] (v4)--(i4);
\draw[very thick,rotate=25] plot[mark=x,mark size=2.5] coordinates {(v4)};
\draw[very thick] plot[mark=x,mark size=2.7] coordinates {(v3)};
\end{feynhand} \end{tikzpicture}\quad
&\Rightarrow\quad
\sqrt2(X_1^{W^-} )^{u}_{\bar d}{y'}^{d}_{\bar d}\frac{\tilde m_3}{\mathbf m_3}\mathbf m_{13}\langle13\rangle[\eta_3 2],\\
\begin{tikzpicture}[baseline=-0.1cm] \begin{feynhand}
\setlength{\feynhandarrowsize}{4pt}
\vertex [particle] (i1) at (-0.9,0.9) {$1^-$};
\vertex [particle] (i2) at (-0.9,-0.9) {$2^+$};
\vertex [particle] (i3) at (0.9,-0.9) {$4^0$};
\vertex [particle] (i4) at (0.9,0.9) {$3^-$};
\vertex (v1) at (0,-0.4);
\vertex (v2) at (0,0.4);
\vertex (v3) at (-0.9*0.3,-0.4-0.5*0.3);
\vertex (v4) at (0.9*0.3,0.4+0.5*0.3);
\fer{red}{i1}{v2};
\fer{red}{v2}{v1};
\fer{cyan}{v1}{v3};
\fer{red}{v3}{i2};
\draw[brown,thick sca] (v1)--(i3);
\draw[brown,thick bos] (v2)--(v4);
\draw[red,thick bos] (v4)--(i4);
\draw[very thick,rotate=25] plot[mark=x,mark size=2.5] coordinates {(v4)};
\draw[very thick,rotate=25] plot[mark=x,mark size=2.7] coordinates {(v3)};
\end{feynhand} \end{tikzpicture}\quad
&\Rightarrow\quad
\sqrt2(X_1^{W^-} )^{u}_{\bar d}{y}^{d}_{\bar d}\frac{\tilde m_3}{\mathbf m_3^2}\frac{\tilde m_2}{\mathbf m_2}\langle1\eta_3\rangle[3|P_{13}|\eta_2\rangle,\\
&\;\downarrow\; J_{13}^+ J_{13}^- \nonumber \\
\begin{tikzpicture}[baseline=-0.1cm] \begin{feynhand}
\setlength{\feynhandarrowsize}{4pt}
\vertex [particle] (i1) at (-0.9,0.9) {$1^-$};
\vertex [particle] (i2) at (-0.9,-0.9) {$2^+$};
\vertex [particle] (i3) at (0.9,-0.9) {$4^0$};
\vertex [particle] (i4) at (0.9,0.9) {$3^-$};
\vertex (v1) at (0,-0.4);
\vertex (v2) at (0,0.4);
\vertex (v3) at (-0.9*0.3,-0.4-0.5*0.3);
\vertex (v4) at (0.9*0.3,0.4+0.5*0.3);
\fer{red}{i1}{v2};
\fer{red}{v2}{v1};
\fer{cyan}{v1}{v3};
\fer{red}{v3}{i2};
\draw[brown,thick sca] (v1)--(i3);
\draw[brown,thick bos] (v2)--(v4);
\draw[red,thick bos] (v4)--(i4);
\draw[very thick,rotate=25] plot[mark=x,mark size=2.5] coordinates {(v4)};
\draw[very thick,rotate=25] plot[mark=x,mark size=2.7] coordinates {(v3)};
\draw[very thick,rotate=0] plot[mark=x,mark size=2.7] coordinates {(0,0.2)};
\draw[very thick,rotate=0] plot[mark=x,mark size=2.7] coordinates {(0,-0.2)};
\end{feynhand} \end{tikzpicture}\quad
&\Rightarrow\quad
\sqrt2(X_1^{W^-} )^{u}_{\bar d}{y}^{d}_{\bar d}\frac{\tilde m_3}{\mathbf m_3^2}\frac{\tilde m_2}{\mathbf m_2}\langle1\eta_3\rangle[3|\eta_{13}|\eta_2\rangle.
\end{align}
The results in last two lines can be related by the operator $J_{13}^+ J_{13}^-$. Here the first diagram corresponds to a chirality flip on the internal line, while the second and third diagrams involve no chirality flip.  These diagrams are structurally similar to those in the $(12)$ channel, but with particles 3 and 4 exchanged.

Summing the three contributions and including the propagator denominator, the $(13)$-channel amplitude is given by
\begin{equation} \begin{aligned}
\mathcal{M}_{(13)}
=& \sqrt2(X_1^{W^-} )^{u}_{\bar d}{y'}^{d}_{\bar d}\frac{\tilde m_3}{\mathbf m_3^2}\frac{\mathbf m_{13}\langle13\rangle[\eta_3 2]}{s_{13}}
+\sqrt2(X_1^{W^-} )^{u}_{\bar d}{y}^{d}_{\bar d}\frac{\tilde m_3}{\mathbf m_3^2}\frac{\tilde m_2}{\mathbf m_2}\frac{\langle1\eta_3\rangle[3|P_{13}+\eta_{13}|\eta_2\rangle}{s_{13}}. 
\end{aligned} \end{equation}

\paragraph{$(12)$-channel}

The final contribution comes from the $(12)$ channel, where the internal particle is a $W$ boson. Here, we glue two particle lines with zero, one, or two chirality flips to form the internal vector boson in several different orders:
\begin{align}
&\quad\;\begin{tikzpicture}[baseline=0.7cm] \begin{feynhand}
\setlength{\feynhandarrowsize}{4pt}
\vertex [dot] (v1) at (0,0.8) {};
\vertex (i1) at (1.4,0.8) {$P^0$};
\draw[brown,thick bos] (v1)--(i1);
\end{feynhand} \end{tikzpicture}
\begin{tikzpicture}[baseline=0.7cm] \begin{feynhand}
\setlength{\feynhandarrowsize}{4pt}
\vertex [dot] (v1) at (1.4,0.8) {};
\vertex (i1) at (0,0.8) {$P^0$};
\draw[brown,thick bos] (v1)--(i1);
\end{feynhand} \end{tikzpicture}
= \frac{1}{2} \tilde{\lambda}_{\dot{\alpha}} \lambda_{\alpha}\times\tilde{\lambda}_{\dot{\beta}} \lambda_{\dot{\beta}}, 
 \\
&\left\{\begin{aligned}
\begin{tikzpicture}[baseline=0.7cm] \begin{feynhand}
\setlength{\feynhandarrowsize}{4pt}
\vertex [dot] (v1) at (0,0.8) {};
\vertex (v2) at (0.6,0.8);
\vertex (i1) at (1.4,0.8) {$P^+$};
\draw[brown,thick bos] (v2)--(v1);
\draw[cyan,thick bos] (i1)--(v2);
\draw[very thick] plot[mark=x,mark size=2.5] coordinates {(v2)};
\end{feynhand} \end{tikzpicture}
\begin{tikzpicture}[baseline=0.7cm] \begin{feynhand}
\setlength{\feynhandarrowsize}{4pt}
\vertex [dot] (v1) at (1.4,0.8) {};
\vertex (v2) at (0.9,0.8);
\vertex (i1) at (0,0.8) {$P^-$};
\draw[brown,thick bos] (v1)--(v2);
\draw[cyan,thick bos] (v2)--(i1);
\draw[very thick] plot[mark=x,mark size=2.5] coordinates {(v2)};
\end{feynhand} \end{tikzpicture}
&=-m\tilde{\eta}_{\dot{\alpha}} \lambda_{\alpha} \times\tilde{\lambda}_{\dot{\beta}} \tilde m\eta_{\beta}, \\
\begin{tikzpicture}[baseline=0.7cm] \begin{feynhand}
\setlength{\feynhandarrowsize}{4pt}
\vertex [dot] (v1) at (0,0.8) {};
\vertex (v2) at (0.6,0.8);
\vertex (i1) at (1.4,0.8) {$P^-$};
\draw[brown,thick bos] (v2)--(v1);
\draw[red,thick bos] (i1)--(v2);
\draw[very thick] plot[mark=x,mark size=2.5] coordinates {(v2)};
\end{feynhand} \end{tikzpicture}
\begin{tikzpicture}[baseline=0.7cm] \begin{feynhand}
\setlength{\feynhandarrowsize}{4pt}
\vertex [dot] (v1) at (1.4,0.8) {};
\vertex (v2) at (0.9,0.8);
\vertex (i1) at (0,0.8) {$P^0$};
\draw[brown,thick bos] (v1)--(v2);
\draw[red,thick bos] (v2)--(i1);
\draw[very thick] plot[mark=x,mark size=2.5] coordinates {(v2)};
\end{feynhand} \end{tikzpicture}&=-\tilde{\lambda}_{\dot{\alpha}} \tilde m\eta_{\alpha}\times m \tilde{\eta}_{\dot{\beta}} \lambda_{\beta}, \\ 
\begin{tikzpicture}[baseline=0.7cm] \begin{feynhand}
\setlength{\feynhandarrowsize}{4pt}
\vertex [dot] (v1) at (0,0.8) {};
\vertex (v2) at (0.4,0.8);
\vertex (v3) at (0.7,0.8);
\vertex (i1) at (1.4,0.8) {$P^0$};
\draw[brown,thick bos] (v2)--(v1);
\draw[brown,thick bos] (v3)--(v2);
\draw[brown,thick bos] (i1)--(v3);
\draw[very thick] plot[mark=x,mark size=2.5] coordinates {(v2)};
\draw[very thick] plot[mark=x,mark size=2.5] coordinates {(v3)};
\end{feynhand} \end{tikzpicture}
\begin{tikzpicture}[baseline=0.7cm] \begin{feynhand}
\setlength{\feynhandarrowsize}{4pt}
\vertex [dot] (v1) at (1.4,0.8) {};
\vertex (i1) at (0,0.8) {$P^0$};
\draw[brown,thick bos] (v1)--(i1);
\end{feynhand} \end{tikzpicture}
&=+\frac{1}{2} m\tilde{\eta}_{\dot{\alpha}} \tilde m\eta_{\alpha}\times \tilde{\lambda}_{\dot{\beta}} \lambda_{\beta}, \\
\begin{tikzpicture}[baseline=0.7cm] \begin{feynhand}
\setlength{\feynhandarrowsize}{4pt}
\vertex [dot] (v1) at (0,0.8) {};
\vertex (i1) at (1.4,0.8) {$P^0$};
\draw[brown,thick bos] (i1)--(v1);
\end{feynhand} \end{tikzpicture}
\begin{tikzpicture}[baseline=0.7cm] \begin{feynhand}
\setlength{\feynhandarrowsize}{4pt}
\vertex [dot] (v1) at (1.4,0.8) {};
\vertex (v2) at (1.0,0.8);
\vertex (v3) at (0.7,0.8);
\vertex (i1) at (0,0.8) {$P^0$};
\draw[brown,thick bos] (v1)--(v2);
\draw[brown,thick bos] (v2)--(v3);
\draw[brown,thick bos] (v3)--(i1);
\draw[very thick] plot[mark=x,mark size=2.5] coordinates {(v2)};
\draw[very thick] plot[mark=x,mark size=2.5] coordinates {(v3)};
\end{feynhand} \end{tikzpicture}
&=+\frac{1}{2} \tilde{\lambda}_{\dot{\alpha}} \lambda_{\alpha} \times m\tilde{\eta}_{\dot{\beta}} \tilde m\eta_{\beta} ,
\end{aligned}\right.
 \\
&\quad\;\begin{tikzpicture}[baseline=0.7cm] \begin{feynhand}
\setlength{\feynhandarrowsize}{4pt}
\vertex [dot] (v1) at (0,0.8) {};
\vertex (v2) at (0.4,0.8);
\vertex (v3) at (0.7,0.8);
\vertex (i1) at (1.4,0.8) {$P^0$};
\draw[brown,thick bos] (v2)--(v1);
\draw[brown,thick bos] (v3)--(v2);
\draw[brown,thick bos] (i1)--(v3);
\draw[very thick] plot[mark=x,mark size=2.5] coordinates {(v2)};
\draw[very thick] plot[mark=x,mark size=2.5] coordinates {(v3)};
\end{feynhand} \end{tikzpicture}
\begin{tikzpicture}[baseline=0.7cm] \begin{feynhand}
\setlength{\feynhandarrowsize}{4pt}
\vertex [dot] (v1) at (1.4,0.8) {};
\vertex (v2) at (1.0,0.8);
\vertex (v3) at (0.7,0.8);
\vertex (i1) at (0,0.8) {$P^0$};
\draw[brown,thick bos] (v1)--(v2);
\draw[brown,thick bos] (v2)--(v3);
\draw[brown,thick bos] (v3)--(i1);
\draw[very thick] plot[mark=x,mark size=2.5] coordinates {(v2)};
\draw[very thick] plot[mark=x,mark size=2.5] coordinates {(v3)};
\end{feynhand} \end{tikzpicture}=
\frac{1}{2} m\tilde{\eta}_{\dot{\alpha}} \tilde m\eta_{\alpha} \times m\tilde{\eta}_{\dot{\beta}} \tilde m\eta_{\beta} ,
\label{eq:InternalVector4}
\end{align}
When we glue two external diagrams with helicity $\pm$, the resulting internal diagram has multiple representaions. since $m\tilde{m} = \tilde{m}m$, their effects are quite similar we will use the same color to represent different diagrams, like the following
\begin{equation}
\begin{tikzpicture}[baseline=0.7cm] \begin{feynhand}
\vertex [dot] (v1) at (0,0.8) {};
\vertex [dot] (v2) at (1.6,0.8) {};
\vertex (k3) at (0.5,0.8);
\vertex (k4) at (1.1,0.8);
\draw[brown,thick bos] (v2)--(k4);
\draw[red,thick bos] (k4)--(k3);
\draw[brown,thick bos] (k3)--(v1);
\draw[very thick] plot[mark=x,mark size=2.7] coordinates {(k3)};
\draw[very thick] plot[mark=x,mark size=2.7] coordinates {(k4)};
\end{feynhand} \end{tikzpicture}=
\begin{tikzpicture}[baseline=0.7cm] \begin{feynhand}
\vertex [dot] (v1) at (0,0.8) {};
\vertex [dot] (v2) at (1.6,0.8) {};
\vertex (k3) at (0.5,0.8);
\vertex (k4) at (1.1,0.8);
\draw[brown,thick bos] (v2)--(k4);
\draw[cyan,thick bos] (k3)--(k4);
\draw[brown,thick bos] (k3)--(v1);
\draw[very thick] plot[mark=x,mark size=2.7] coordinates {(k3)};
\draw[very thick] plot[mark=x,mark size=2.7] coordinates {(k4)};
\end{feynhand} \end{tikzpicture}.
\end{equation}
Using this notation, we obtain the following three kinds of propagators
\begin{eqnarray}
\includegraphics[scale=1,valign=c]{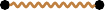}
&=& \frac{1}{2} p_{\alpha \dot{\alpha}} p_{\beta\dot{\beta}},  \\
\includegraphics[scale=1,valign=c]{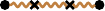}
&=&\mathbf{m}^4 \epsilon_{\alpha\beta} \epsilon_{\dot{\alpha}\dot{\beta}}-\mathbf{m}^2\frac{1}{2} p_{\alpha\dot{\alpha}} \eta_{\beta\dot{\beta}}-\mathbf{m}^2\frac{1}{2} \eta_{\alpha\dot{\alpha}} p_{\beta\dot{\beta}}, \\
\includegraphics[scale=1,valign=c]{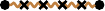}
&=&\frac{1}{2} \mathbf m^4\eta_{\alpha \dot{\alpha}} \eta_{\beta\dot{\beta}}.
\end{eqnarray}

Using these rules, we glue the $W^+ W^- h$ and $u\bar{d}W^-$ sub-amplitudes to obtain the three contributions
\begin{align}
\begin{tikzpicture}[baseline=-0.1cm] \begin{feynhand}
\setlength{\feynhandarrowsize}{4pt}
\vertex [particle] (i1) at (-0.9,0.9) {$1^-$};
\vertex [particle] (i2) at (-0.9,-0.9) {$2^+$};
\vertex [particle] (i3) at (0.9,-0.9) {$3^-$};
\vertex [particle] (i4) at (0.9,0.9) {$4^0$};
\vertex (v1) at (-0.3,0);
\vertex (v2) at (0.3,0);
\vertex (v3) at (0.3+0.6*0.35,-0.9*0.35);
\fer{red}{i1}{v1};
\fer{red}{v1}{i2};
\draw[brown,thick bos] (v1)--(v2);
\draw[brown,thick sca] (v2)--(i4);
\draw[brown,thick bos] (v2)--(v3);
\draw[red,thick bos] (v3)--(i3);
\draw[very thick,rotate=25] plot[mark=x,mark size=2.7] coordinates {(v3)};
\end{feynhand} \end{tikzpicture}
&\Rightarrow
\langle 1|P_{12}|2] [\eta_3|P_{12}|3\rangle,& \\
&\;\downarrow\; J_{12}^+ J_{12}^- \text{ \& add $\mathbf m^2$ contribution} \nonumber \\
\begin{tikzpicture}[baseline=-0.1cm] \begin{feynhand}
\setlength{\feynhandarrowsize}{4pt}
\vertex [particle] (i1) at (-0.9,0.9) {$1^-$};
\vertex [particle] (i2) at (-0.9,-0.9) {$2^+$};
\vertex [particle] (i3) at (0.9,-0.9) {$3^-$};
\vertex [particle] (i4) at (0.9,0.9) {$4^0$};
\vertex (v1) at (-0.3,0);
\vertex (v2) at (0.3,0);
\vertex (v3) at (0.3+0.6*0.35,-0.9*0.35);
\fer{red}{i1}{v1};
\fer{red}{v1}{i2};
\draw[brown,thick bos] (v1)--(v2);
\draw[brown,thick sca] (v2)--(i4);
\draw[brown,thick bos] (v2)--(v3);
\draw[red,thick bos] (v3)--(i3);
\draw[very thick,rotate=25] plot[mark=x,mark size=2.7] coordinates {(v3)};
\draw[very thick,rotate=0] plot[mark=x,mark size=2.7] coordinates {(-0.15,0)};
\draw[very thick,rotate=0] plot[mark=x,mark size=2.7] coordinates {(0.15,0)};
\end{feynhand} \end{tikzpicture}
&\Rightarrow
-2\mathbf{m}^2_{12}\langle13\rangle[2\eta_3]+\langle 1|P_{12}|2] [\eta_3|\eta_{12}|3\rangle+\langle 1|\eta_{12}|2] [\eta_3|P_{12}|3\rangle,& \\
&\;\downarrow\; J_{12}^+ J_{12}^- \nonumber \\
\begin{tikzpicture}[baseline=-0.1cm] \begin{feynhand}
\setlength{\feynhandarrowsize}{4pt}
\vertex [particle] (i1) at (-1,0.9) {$1^-$};
\vertex [particle] (i2) at (-1,-0.9) {$2^+$};
\vertex [particle] (i3) at (1,-0.9) {$3^-$};
\vertex [particle] (i4) at (1,0.9) {$4^0$};
\vertex (v1) at (-0.5,0);
\vertex (v2) at (0.5,0);
\vertex (v3) at (0.5+0.6*0.35,-0.9*0.35);
\fer{red}{i1}{v1};
\fer{red}{v1}{i2};
\draw[brown,thick bos] (v1)--(v2);
\draw[brown,thick sca] (v2)--(i4);
\draw[brown,thick bos] (v2)--(v3);
\draw[red,thick bos] (v3)--(i3);
\draw[very thick,rotate=25] plot[mark=x,mark size=2.7] coordinates {(v3)};
\draw[very thick,rotate=0] plot[mark=x,mark size=2.7] coordinates {(-0.1,0)};
\draw[very thick,rotate=0] plot[mark=x,mark size=2.7] coordinates {(0.1,0)};
\draw[very thick,rotate=0] plot[mark=x,mark size=2.7] coordinates {(-0.3,0)};
\draw[very thick,rotate=0] plot[mark=x,mark size=2.7] coordinates {(0.3,0)};
\end{feynhand} \end{tikzpicture}
&\Rightarrow
\langle 1|\eta_{12}|2] [\eta_3|\eta_{12}|3\rangle,&
\end{align}
where $\mathbf{m}_{12}$ is the physical mass of the internal $W$ boson. These three contributions can be related by the operator $J_{12}^+ J_{12}^-$. This extra term is proportional to the square of the internal particle mass, $\mathbf m^2$. All these contributions share the same coefficient $(X_1^{W^-} )^{u}_{\bar d}\mathbf{g}^{W^+ W^-}$ from the two 3-pt sub-amplitudes.

After restoring the denominator $s_{12}$ and summing over three contributions in this channel, we obtain
\begin{equation} \begin{aligned} \label{eq:FFVS1}
\mathcal{M}_{(12)}
=\sqrt2(X_1^{W^-} )^{u}_{\bar d}\mathbf{g}^{W^+ W^-}\frac{-2\mathbf{m}^2_{12}\langle13\rangle[2\eta_3]+\langle 1|P_{12}+\eta_{12}|2] [\eta_3|P_{12}+\eta_{12}|3\rangle}{s_{12}}.
\end{aligned} \end{equation}

\paragraph{Ladder operator applied to internal particle}

In above calculation, we note that while the external particles are in the same order, while the propagator belongs to different order. Therefore, once the leading-order propagator is known, all the sub-leading order propagators should be obtained by the ladder operator.  

We note that the MHC internal particles are obtained by gluing two external particle states with opposite helicity. Therefore, they always have total helicity $0$. In this case, we must apply the ladder operator twice, defined as
\begin{equation}
\frac{m}{\mathbf m} J^- \frac{\tilde m}{\mathbf m} J^+
=J^- J^+,
\end{equation}
which does not change the helicity.

For an internal fermion, acting with $J^- J^+$ on the primary internal particle state yields both the primary and descendant states,
\begin{equation}
\includegraphics[scale=1,valign=c]{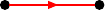}=p_{\alpha\dot\beta}
\quad\overset{J^- J^+}{\to}\quad
\left\{\begin{aligned}
\includegraphics[scale=1,valign=c]{image/internal_F_c--_0.pdf}&=p_{\alpha\dot\beta},\\
\includegraphics[scale=1,valign=c]{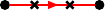}&=\eta_{\alpha\dot\beta}.
\end{aligned}\right.
\end{equation}
Conversely, the primary internal state can also be obtained from the descendant state by acting with $J^- J^+$.

For a vector state, acting with $J^- J^+$ generates all particle states. Starting from a given primary internal vector state, one and two applications yield:
\begin{equation} \begin{aligned}
\includegraphics[scale=1,valign=c]{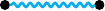}= p_{(\dot{\alpha}_1}^{(\beta_1} p_{\dot{\alpha}_2)}^{\beta_2)}
\quad&\overset{J^- J^+}{\to}\quad 
\left\{\begin{aligned}
\includegraphics[scale=1,valign=c]{image/internal_V_c++_0.pdf}&=p_{(\dot{\alpha}_1}^{(\beta_1} p_{\dot{\alpha}_2)}^{\beta_2)},\\
\includegraphics[scale=1,valign=c]{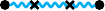}&=p_{(\dot{\alpha}_1}^{(\beta_1} \eta_{\dot{\alpha}_2)}^{\beta_2)},
\end{aligned}\right.\\
\includegraphics[scale=1,valign=c]{image/internal_V_c++_0.pdf}= p_{(\dot{\alpha}_1}^{(\beta_1} p_{\dot{\alpha}_2)}^{\beta_2)}
\quad&\overset{(J^- J^+)^2}{\to}\quad 
\left\{\begin{aligned}
\includegraphics[scale=1,valign=c]{image/internal_V_c++_2.pdf}&= p_{(\dot{\alpha}_1}^{(\beta_1} \eta_{\dot{\alpha}_2)}^{\beta_2)},\\
\includegraphics[scale=1,valign=c]{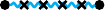}&=\eta_{(\dot{\alpha}_1}^{(\beta_1}\eta_{\dot{\alpha}_2)}^{\beta_2)}.
\end{aligned}\right.
\end{aligned} \end{equation}
Starting from a 1st or 2nd descendant state, one or two applications of $J^- J^+$ similarly produce all states.

Other internal states can be obtained analogously by acting with $J^- J^+$, with one exception: the internal vector state represented by a brown line. In this case, a term $\mathbf m^2\epsilon_{\alpha\beta} \epsilon_{\dot{\alpha}\dot{\beta}}$ appearing in the 1st descendant state cannot be generated by $J^- J^+$ and must be inserted by hand:
\begin{equation}
\includegraphics[scale=1,valign=c]{image/internal_V_c00_0.pdf}= p_{\alpha \dot{\alpha}} p_{\beta\dot{\beta}}
\quad\underset{\text{add }\mathbf{m}^2\epsilon^2}{\overset{J^- J^+}{\to}}\quad 
\left\{\begin{aligned}
\includegraphics[scale=1,valign=c]{image/internal_V_c00_0.pdf}&=p_{\alpha \dot{\alpha}} p_{\beta\dot{\beta}},\\
\includegraphics[scale=1,valign=c]{image/internal_V_c00_2.pdf}&=\mathbf{m}^2 \epsilon_{\alpha\beta} \epsilon_{\dot{\alpha}\dot{\beta}}-p_{\alpha\dot{\alpha}} \eta_{\beta\dot{\beta}}-\eta_{\alpha\dot{\alpha}} p_{\beta\dot{\beta}},
\end{aligned}\right.
\end{equation}

\begin{equation}
\includegraphics[scale=1,valign=c]{image/internal_V_c00_0.pdf}= p_{\alpha \dot{\alpha}} p_{\beta\dot{\beta}}
\quad\underset{\text{add }\mathbf{m}^2\epsilon^2}{\overset{J^- J^+}{\to}}\quad 
\left\{\begin{aligned}
\includegraphics[scale=1,valign=c]{image/internal_V_c00_2.pdf}&=\mathbf{m}^4 \epsilon_{\alpha\beta} \epsilon_{\dot{\alpha}\dot{\beta}}-\mathbf{m}^2 p_{\alpha\dot{\alpha}} \eta_{\beta\dot{\beta}}-\mathbf{m}^2\eta_{\alpha\dot{\alpha}} p_{\beta\dot{\beta}},\\
\includegraphics[scale=1,valign=c]{image/internal_V_c00_4.pdf}&=\mathbf m^4\eta_{\alpha \dot{\alpha}} \eta_{\beta\dot{\beta}}.
\end{aligned}\right.
\end{equation}

\paragraph{total amplitude}

From the above ladder operator action, we can only calculate the leading MHC amplitude by gluing the leading 3-point amplitudes, and then apply the ladder operator to obtain the full MHC results. 
Summing over the contributions in all three channels, the leading MHC amplitudes read
\begin{eqnarray}
&& \begin{tikzpicture}[baseline=-0.1cm] \begin{feynhand}
\setlength{\feynhandarrowsize}{4pt}
\vertex [particle] (i1) at (-0.9,0.9) {$1^-$};
\vertex [particle] (i2) at (-0.9,-0.9) {$2^+$};
\vertex [particle] (i3) at (0.9,-0.9) {$3^-$};
\vertex [particle] (i4) at (0.9,0.9) {$4^0$};
\vertex (v1) at (0,-0.4);
\vertex (v2) at (0,0.4);
\vertex (v3) at (-0.9*0.3,0.4+0.5*0.3);
\vertex (v4) at (0.9*0.3,-0.4-0.5*0.3);
\fer{red}{i1}{v3};
\fer{cyan}{v3}{v2};
\fer{red}{v2}{v1};
\fer{red}{v1}{i2};
\draw[brown,thick sca] (v2)--(i4);
\draw[brown,thick bos] (v1)--(v4);
\draw[red,thick bos] (v4)--(i3);
\draw[very thick,rotate=-25] plot[mark=x,mark size=2.5] coordinates {(v4)};
\draw[very thick,rotate=-25] plot[mark=x,mark size=2.7] coordinates {(v3)};
\end{feynhand} \end{tikzpicture} 
+
\begin{tikzpicture}[baseline=-0.1cm] \begin{feynhand}
\setlength{\feynhandarrowsize}{4pt}
\vertex [particle] (i1) at (-0.9,0.9) {$1^-$};
\vertex [particle] (i2) at (-0.9,-0.9) {$2^+$};
\vertex [particle] (i3) at (0.9,-0.9) {$4^0$};
\vertex [particle] (i4) at (0.9,0.9) {$3^-$};
\vertex (v1) at (0,-0.4);
\vertex (v2) at (0,0.4);
\vertex (v3) at (0,0);
\vertex (v4) at (0.9*0.3,0.4+0.5*0.3);
\fer{red}{i1}{v2};
\fer{red}{v2}{v3};
\fer{cyan}{v3}{v1};
\fer{red}{v1}{i2};
\draw[brown,thick sca] (v1)--(i3);
\draw[brown,thick bos] (v2)--(v4);
\draw[red,thick bos] (v4)--(i4);
\draw[very thick,rotate=25] plot[mark=x,mark size=2.5] coordinates {(v4)};
\draw[very thick] plot[mark=x,mark size=2.7] coordinates {(v3)};
\end{feynhand} \end{tikzpicture}
+
\begin{tikzpicture}[baseline=-0.1cm] \begin{feynhand}
\setlength{\feynhandarrowsize}{4pt}
\vertex [particle] (i1) at (-0.9,0.9) {$1^-$};
\vertex [particle] (i2) at (-0.9,-0.9) {$2^+$};
\vertex [particle] (i3) at (0.9,-0.9) {$4^0$};
\vertex [particle] (i4) at (0.9,0.9) {$3^-$};
\vertex (v1) at (0,-0.4);
\vertex (v2) at (0,0.4);
\vertex (v3) at (-0.9*0.3,-0.4-0.5*0.3);
\vertex (v4) at (0.9*0.3,0.4+0.5*0.3);
\fer{red}{i1}{v2};
\fer{red}{v2}{v1};
\fer{cyan}{v1}{v3};
\fer{red}{v3}{i2};
\draw[brown,thick sca] (v1)--(i3);
\draw[brown,thick bos] (v2)--(v4);
\draw[red,thick bos] (v4)--(i4);
\draw[very thick,rotate=25] plot[mark=x,mark size=2.5] coordinates {(v4)};
\draw[very thick,rotate=25] plot[mark=x,mark size=2.7] coordinates {(v3)};
\end{feynhand} \end{tikzpicture}
+
\begin{tikzpicture}[baseline=-0.1cm] \begin{feynhand}
\setlength{\feynhandarrowsize}{4pt}
\vertex [particle] (i1) at (-0.9,0.9) {$1^-$};
\vertex [particle] (i2) at (-0.9,-0.9) {$2^+$};
\vertex [particle] (i3) at (0.9,-0.9) {$3^-$};
\vertex [particle] (i4) at (0.9,0.9) {$4^0$};
\vertex (v1) at (-0.3,0);
\vertex (v2) at (0.3,0);
\vertex (v3) at (0.3+0.6*0.35,-0.9*0.35);
\fer{red}{i1}{v1};
\fer{red}{v1}{i2};
\draw[brown,thick bos] (v1)--(v2);
\draw[brown,thick sca] (v2)--(i4);
\draw[brown,thick bos] (v2)--(v3);
\draw[red,thick bos] (v3)--(i3);
\draw[very thick,rotate=25] plot[mark=x,mark size=2.7] coordinates {(v3)};
\end{feynhand} \end{tikzpicture} \nonumber
\\
&=&
\sqrt2{y'}^{u}_{\bar u}(X_1^{W^-} )^{u}_{\bar d}\frac{\tilde m_3}{\mathbf m_3^2}\frac{m_1}{\mathbf m_1}[\eta_1|P_{14}|3\rangle[\eta_32] 
+ \sqrt2(X_1^{W^-} )^{u}_{\bar d}{y'}^{d}_{\bar d}\frac{\tilde m_3}{\mathbf m_3}\mathbf m_{13}\langle13\rangle[\eta_3 2] \nonumber \\
&&+
\sqrt2(X_1^{W^-} )^{u}_{\bar d}{y}^{d}_{\bar d}\frac{\tilde m_3}{\mathbf m_3^2}\frac{\tilde m_2}{\mathbf m_2}\langle1\eta_3\rangle[3|P_{13}|\eta_2\rangle + 
\langle 1|P_{12}|2] [\eta_3|P_{12}|3\rangle.
\end{eqnarray}
Add the sub-leading propagator contributions using the ladder operators and add single pole struction, we obtain the total amplitude:
\begin{equation} \begin{aligned}
\mathcal{M}(1^{+\frac{1}{2}},2^{-\frac{1}{2}},3^{+1},4^0)
=&\sqrt2{y}^{u}_{\bar u}(X_1^{W^-} )^{u}_{\bar d}\frac{\tilde m_3}{\mathbf m_3^2}\frac{\mathbf m_{14}\langle13\rangle[\eta_3 2]}{s_{14}}+\sqrt2{y'}^{u}_{\bar u}(X_1^{W^-} )^{u}_{\bar d}\frac{\tilde m_3}{\mathbf m_3^2}\frac{m_1}{\mathbf m_1}\frac{[\eta_1|P_{14}+\eta_{14}|3\rangle[\eta_32]}{s_{14}}\\
&+\sqrt2(X_1^{W^-} )^{u}_{\bar d}{y'}^{d}_{\bar d}\frac{\tilde m_3}{\mathbf m_3^2}\frac{\mathbf m_{13}\langle13\rangle[\eta_3 2]}{s_{13}}+\sqrt2(X_1^{W^-} )^{u}_{\bar d}{y}^{d}_{\bar d}\frac{\tilde m_3}{\mathbf m_3^2}\frac{\tilde m_2}{\mathbf m_2}\frac{\langle1\eta_3\rangle[3|P_{13}+\eta_{13}|\eta_2\rangle}{s_{13}}\\
&+\sqrt2(X_1^{W^-} )^{u}_{\bar d}\mathbf{g}^{W^+ W^-}\frac{-2\mathbf{m}^2_{12}\langle13\rangle[2\eta_3]+\langle 1|P_{12}+\eta_{12}|2] [\eta_3|P_{12}+\eta_{12}|3\rangle}{s_{12}}.
\end{aligned} \end{equation}

The same procedure can be applied to the MHC amplitudes involving in pure gauge sector, such as the $WW \to hh$ amplitude. Similar to the $u\bar{d}\to Wh$ amplitudes, the $WW \to hh$ amplitude has the same topologies
\begin{equation} \begin{aligned}
\mathcal{M}=\underbrace{\mathcal{M}_{(12)}+\mathcal{M}_{(13)}+\mathcal{M}_{(14)}}_{\mathcal{M}_{\text{f}}}.
\end{aligned} \end{equation}
Using the gluing technique, each channel amplitudes can be constructed. For each channel, the corresponding internal particle and sub-amplitudes are
\begin{equation}
\begin{tabular}{c|c|c}
\hline
channel & sub-amplitudes & internal particle \\
\hline
(12)-channel & $WWh$ $\times$ $hhh$ & $h$ \\
\hline
\makecell{(13)-channel\\(14)-channel} & $WWh$ $\times$ $WWh$ & $W$ \\
\hline
\end{tabular}
\end{equation}
This shows that we can use two types of 3-point amplitudes, $WWh$ and $hhh$, to construct the $WWhh$ amplitude. However, the result does not match the correct amplitudes, and we will show a 4-point contact MHC amplitude is needed for $WW\to hh$, while the amplitude $u\bar{d}\to Wh$ does not need any 4-pt contact amplitude.

\subsection{Massless and MHC amplitudes in light-cone gauge}

From the 3-point massless-MHC amplitude matching, we learned that a MHC amplitude should be deformed from massless amplitudes. For the 3-point MHC amplitudes involving in gauge boson $A$, it should be matched from massless amplitudes in the light-cone gauge. We expect a 4-point MHC amplitude should be equivalent to the massless amplitudes in the light-cone gauge. Let us first investigate the on-shell massless amplitudes in two different ways: gauge-independent description, and gauge-dependent (light-cone gauge) description.

In a renormalizable pure gauge theory, all physical information is encoded in the three-point vertices. Four-point vertices are required only to ensure the gauge invariance of the complete scattering amplitude. There exist three distinct conceptual descriptions of scattering amplitudes, distinguished by their treatment of on-shell/off-shell states and gauge (in)dependence:
\begin{itemize}

\item \textit{Off-shell Feynman rule description}: The standard Feynman rules yield off-shell vertices that are not individually gauge invariant. For example, in the Yang-Mills theory, although the cubic term $A^2 \partial A$ captures the information needed for the amplitudes, the quartic term $A^4$ is needed for the requirements of the off-shell gauge invariance of the Lagrangian.

\item \textit{On-shell Gauge invariant description}: In the on-shell description, the scattering amplitude are written entirely in terms of massless spinors $|p]$ and $|p\rangle$. Different from the Feynman rule description, the 3-point on-shell amplitude is gauge invariant, for example, the 3-point $VVV$ amplitudes are
\begin{eqnarray}
    A(+1,+1,-1) = \frac{[12]^3}{[23][31]} ,\qquad  A(-1,-1,+1) = \frac{\langle12\rangle^3}{\langle23\rangle\langle31\rangle},
\end{eqnarray}
and thus the 4-point $VVVV$ amplitude is not needed since the 4-point $VVVV$ contains no new on-shell information. 
Therefore, two 3-point $VVV$ amplitudes are bootstrapped to obtain the four-gluon scattering amplitude, which can be written as 
\begin{equation}
\mathcal{A}[1^-,2^-,3^+,4^+]=\frac{\langle12\rangle^2[34]^2}{s_{12}s_{14}}.
\end{equation}
The form of this 4-pt amplitude is quite compact and the pole structure is obscure, since each $s$, $t$, $u$ channel is not separable.

\item \textit{On-shell Gauge dependence description}: For a massless amplitude involving in gauge boson $A$, the gauge boson state is described using the Lorentz $(\frac12,\frac12)$ representation
\begin{eqnarray}
A_{\alpha\dot{\alpha}}^+: \frac{\xi_{\alpha}\tilde{\lambda}_{\dot{\alpha}}}{\langle \xi \lambda \rangle }, \quad 
A_{\alpha\dot{\alpha}}^-: -\frac{\lambda_{\alpha}\tilde{\xi}_{\dot{\alpha}}}{[ \tilde{\lambda} \tilde{\xi} ] }.
\end{eqnarray}
Here $\xi$ and $\tilde\xi$ are reference spinors introduced to construct the $(\frac12,\frac12)$ representation. These spinors correspond to the reference vector $n$ in the light-cone gauge,
\begin{equation}
n\cdot A^{\pm}\equiv\xi^{\alpha} \tilde\xi^{\dot\alpha} A_{\alpha\dot{\alpha}}^\pm=0. 
\end{equation}
In this description, although the amplitude is on-shell, the 3-point amplitude, made on-shell by complex momenta, is gauge dependent due to the reference spinor $\xi$, and thus the 4-point contact amplitude is needed to ensure the gauge dependence. This is quite similar to the Feynman rule description. However, there are also differences: the propagator is put to on-shell by factorization, while the propagator is off-shell in the Feynman rule description.

\end{itemize}

Since the on-shell gauge-independent description has already been discussed in literature, we now focus on the gauge-dependent description for comparison. 
As an example, we consider scalar QED in this language. In this theory, there exists a two-particle current $J$, which also transforms as a $(\tfrac12,\tfrac12)$ representation:
\begin{equation} \begin{aligned}
J^{{\alpha\dot\alpha}}
=p_1^{\alpha\dot\alpha}-p_2^{\alpha\dot\alpha}.
\end{aligned} \end{equation}
This corresponds to the Noether current of scalar field,  $\phi^* D_\mu\phi-D_\mu\phi^*\phi$.
Thus, the 3-pt massless amplitude can be written as the coupling of this current to the gauge boson,
\begin{equation}
\mathcal{A}(1^0,2^0,3^{\pm})=J\cdot A^\pm,
\end{equation}
where the superscripts denote the helicity. 
Thus in scalar QED, the two 3-pt amplitudes with diferrent helicity are
\begin{equation} \begin{aligned} \label{eq:sQED_3pt}
\begin{tikzpicture}[baseline=0.7cm] \begin{feynhand}
\setlength{\feynhandarrowsize}{3.5pt}
\vertex [particle] (i1) at (0,0.8) {$1^0$}; 
\vertex [particle] (i2) at (1.6,1.6) {$2^0$}; 
\vertex [particle] (i3) at (1.6,0) {$3^+$};  
\vertex (v1) at (0.9,0.8); 
\graph{(i1)--[sca](v1)--[sca](i2)};
\graph{(i3)--[bos] (v1)};  
\end{feynhand} \end{tikzpicture}
=\mathcal{A}(1^0,2^0,3^-)=\frac{e}{\sqrt 2}\frac{\langle\xi_3|1-2|3]}{\langle\xi_3 3\rangle},\\
\begin{tikzpicture}[baseline=0.7cm] \begin{feynhand}
\setlength{\feynhandarrowsize}{3.5pt}
\vertex [particle] (i1) at (0,0.8) {$1^0$}; 
\vertex [particle] (i2) at (1.6,1.6) {$2^0$}; 
\vertex [particle] (i3) at (1.6,0) {$3^-$};  
\vertex (v1) at (0.9,0.8); 
\graph{(i1)--[sca](v1)--[sca](i2)};
\graph{(i3)--[bos] (v1)};  
\end{feynhand} \end{tikzpicture}
=\mathcal{A}(1^0,2^0,3^+)=\frac{e}{\sqrt 2}\frac{\langle3|1-2|\xi_3]}{[3 \xi_3]},    
\end{aligned} \end{equation}
where $e$ is the coupling constant in scalar QED.

Then we consider the 4-pt amplitude in scalar QED. We now examine a $\gamma\gamma\phi\phi^*$ amplitude, which is similar to the massless $WWHH^{\dagger}$ amplitude. The massless $\gamma\gamma\phi\phi^*$ amplitude in helicity $(-1,+1,0,0)$ receives contributions from the $(13)$-channel, the $(14)$-channel, and a contact term. For the $(14)$-channel, the amplitude can be obtained by gluing two 3-pt amplitudes,
\begin{equation} \begin{aligned} 
\begin{tikzpicture}[baseline=0.8cm] \begin{feynhand}
\vertex [particle] (i1) at (0,0) {$2^+$};
\vertex [particle] (i2) at (0,1.6) {$1^-$};
\vertex [particle] (i3) at (1.6,1.6) {$4^0$};
\vertex [particle] (i4) at (1.6,0) {$3^0$};
\vertex (v1) at (0.8,0.4);
\vertex (v2) at (0.8,1.2);
\graph{(i1) --[bos] (v1)--[sca] (i4)};
\graph{(i2) --[bos] (v2)--[sca] (i3)};
\graph{(v1) --[sca] (v2)};
\end{feynhand} \end{tikzpicture}
=\mathcal{A}(1^-,4^0,-P_{14}^0)\times\frac{1}{s_{14}}\times\mathcal{A}(P_{14}^0,2^+,3^0).
\end{aligned} \end{equation}
Substituting Eq.~\eqref{eq:sQED_3pt} with appropriate relabeling gives
\begin{equation} \begin{aligned} 
\frac{e}{\sqrt 2}\frac{\langle\xi_1|4+P_{14}|1]}{\langle\xi_1 1\rangle}\times\frac{1}{s_{14}}\times \frac{e}{\sqrt 2}\frac{\langle2|3+P_{13}|\xi_2]}{[3 \xi_3]}=2 e^2\frac{\langle\xi_1|4|1]\langle2|3|\xi_2]}{s_{14}\langle\xi_1 1\rangle[2 \xi_2]}.
\end{aligned} \end{equation}
Similarly, the (13)-channel gives
\begin{equation} \begin{aligned} 
\begin{tikzpicture}[baseline=0.8cm] \begin{feynhand}
\setlength{\feynhandtopsep}{5pt}
\vertex [particle] (i1) at (0,0) {$2^+$};
\vertex [particle] (i2) at (0,1.6) {$1^-$};
\vertex [particle] (i3) at (1.6,1.6) {$4^0$};
\vertex [particle] (i4) at (1.6,0) {$3^0$};
\vertex (v1) at (0.8,0.4);
\vertex (v2) at (0.8,1.2);
\graph{(i2) --[bos] (v1)--[sca] (i4)};
\graph{(i1) --[bos,top] (v2)--[sca] (i3)};
\graph{(v1) --[sca] (v2)};
\end{feynhand} \end{tikzpicture}
=2e^2\frac{-\langle\xi_1|3|1]\langle2|4|\xi_2]}{s_{13}\langle\xi_1 1\rangle[2 \xi_2]}.
\end{aligned} \end{equation}

These two contributions depend on the gauge parameters $|\xi_1\rangle$ and $|\xi_2]$. To obtain a gauge-invariant amplitude, we must include a contact term,
\begin{equation} \begin{aligned}
\begin{tikzpicture}[baseline=0.8cm] \begin{feynhand}
\vertex [particle] (i1) at (0,0) {$2^+$};
\vertex [particle] (i2) at (0,1.6) {$1^-$};
\vertex [particle] (i3) at (1.6,1.6) {$4^0$};
\vertex [particle] (i4) at (1.6,0) {$3^0$};
\vertex (v1) at (0.8,0.8);
\graph{(i1) --[bos] (v1)--[bos] (i2)};
\graph{(i4) --[sca] (v1)--[sca] (i3)};
\end{feynhand} \end{tikzpicture}=-2 e^2\frac{\langle\xi_12\rangle[1\xi_2]}{\langle\xi_1 1\rangle[2 \xi_2]}.
\end{aligned} \end{equation}

Combining all three contributions, we obtain the total amplitude for $\phi^*\phi\to \gamma\gamma$,
\begin{equation}
\mathcal{A}(1^-,2^+,3,4)=\frac{2 e^2}{\langle\xi_1 1\rangle[2 \xi_2]}\left(\frac{\langle\eta_1|4|1]\langle2|3|\eta_2]}{s_{14}}+\frac{\langle\eta_1|3|1]\langle2|4|\eta_2]}{s_{13}}-\langle\eta_12\rangle[1\eta_2]\right).
\end{equation}
One can check that this amplitude is independent of the gauge choice. After some algebra, it simplifies to
\begin{equation}
\mathcal{A}(1^-,2^+,3^0,4^0)=\frac{e^2}{2}\frac{\langle1|3-4|2]^2}{s_{13}s_{14}}.
\end{equation}
The spurious poles $\langle\xi_1 1\rangle$ and $[2 \xi_2]$ cancel in the final expression. On the other hand, the physical pole structure is not so transparent as the gauge-dependent description.

Similar to the 3-point amplitude matching, the massless contact $VVSS$ amplitude \begin{equation} \begin{aligned}
\begin{tikzpicture}[baseline=0.8cm] \begin{feynhand}
\vertex [particle] (i1) at (0,0) {$2^+$};
\vertex [particle] (i2) at (0,1.6) {$1^-$};
\vertex [particle] (i3) at (1.6,1.6) {$4^0$};
\vertex [particle] (i4) at (1.6,0) {$3^0$};
\vertex (v1) at (0.8,0.8);
\graph{(i1) --[bos] (v1)--[bos] (i2)};
\graph{(i4) --[sca] (v1)--[sca] (i3)};
\end{feynhand} \end{tikzpicture}=
\frac{\langle\xi_12\rangle[1\xi_2]}{\langle\xi_1 1\rangle[2 \xi_2]}
\end{aligned} \end{equation}
should match to the contact $VVSS$ MHC amplitudes, where we ignore the overall coefficient for now. 
Taking the gauge $\xi_i=\eta_i$, we have the MHC amplitude
\begin{equation}
\begin{tikzpicture}[baseline=0.8cm] \begin{feynhand}
\vertex [particle] (i1) at (0,1.8) {$+$};
\vertex [particle] (i2) at (0,0) {$-$};
\vertex [particle] (i3) at (1.8,0) {$0$};
\vertex [particle] (i4) at (1.8,1.8) {$0$};
\vertex (v3) at (0.9,0.9);
\vertex (v4) at (0.9-0.9*0.43,0.9+0.9*0.43);
\vertex (v5) at (0.9-0.9*0.43,0.9-0.9*0.43);
\draw[cyan,thick bos] (i1)--(v4);
\draw[brown,thick bos] (v4)--(v3);
\draw[brown,thick bos] (v5)--(v3);
\draw[red,thick bos] (i2)--(v5);
\draw[brown,thick sca] (i3)--(v3);
\draw[brown,thick sca] (i4)--(v3);
\draw[very thick] plot[mark=x,mark size=3.5,mark options={rotate=45}] coordinates {(v4)};
\draw[very thick] plot[mark=x,mark size=3.5,mark options={rotate=45}] coordinates {(v5)};
\end{feynhand} \end{tikzpicture}=
\frac{\langle\eta_12\rangle[1\eta_2]}{m_1\tilde m_2},
\end{equation}
which contains the gauge dependence. 
Similarly we should obtain other MHC $VVSS$ amplitudes in other helicity categories, by the ladder operators $\frac{m}{\mathbf m} J^-$ and $\frac{\tilde m}{\mathbf m} J^+$. Inversely using the ladder operators, the primary $VVSS$ amplitude is obtained to be
\begin{equation} \begin{aligned}
\begin{tikzpicture}[baseline=0.8cm] \begin{feynhand}
\vertex [particle] (i1) at (0,1.8) {$1^0$};
\vertex [particle] (i2) at (0,0) {$2^0$};
\vertex [particle] (i3) at (1.8,0) {$3^0$};
\vertex [particle] (i4) at (1.8,1.8) {$4^0$};
\vertex (v3) at (0.9,0.9);
\draw[brown,thick bos] (i1)--(v3);
\draw[brown,thick bos] (i2)--(v3);
\draw[brown,thick sca] (i4)--(v3)--(i3);
\end{feynhand} \end{tikzpicture}
=\frac{\langle12\rangle[21]}{\mathbf m_1 \mathbf m_2}.
\end{aligned} \end{equation}
Unlike the 3-point primary MHC amplitude involving the gauge boson, this primary MHC amplitude is non-vanishing. 
Thus there exists $VVSS$ contact amplitude in the light cone gauge. To make sure the $WW \to hh$ amplitude gauge invariant, the contact amplitude is needed.

\subsection{Contact 4-point MHC amplitudes}
\label{sec:contact_4pt}

Similar to the 3-point MHC primary and descendant amplitudes, let us classify all the contact 4-point MHC primary amplitudes and obtain the descendant 4-point contact amplitudes in the SM.

From massless-massive correspondence, the 4-point primary MHC amplitudes should be matched from the 4-point massless amplitudes. If the massless amplitudes with certain helicity belong to the EFT category, we would classify them to be the EFT MHC amplitudes. Otherwise, it would be the 4-point contact MHC amplitudes in the SM. The classification is as follows
\begin{equation}
\begin{tabular}{c|c|c|c}
\hline
Primary MHC & helicity class & helicity category & amplitudes \\
\hline
SSSS & $n_T=0$ & $(0,0,0,0)$ & 1 \\
\hline
 \makecell{FFSS, FFVS\\FFVV, FFFF} & $n_T\ge 0$ & EFT & -\\
\hline
VVSS & \makecell{$n_T=0$\\$n_T> 0$} & \makecell{$(0, 0,0,0)$\\ EFT } & \makecell{$\langle12\rangle[21]$\\  - } \\
 \hline
VVVS & \makecell{$n_T \ge 0$} &  EFT & - \\
\hline
VVVV & \makecell{$n_T=0$  \\$n_T> 0$} & \makecell{$(0, 0,0,0)$ \\ EFT } & \makecell{$(\langle12\rangle[21]\langle34\rangle[34]$+cycle\\ -  }\\
\hline
higher-point &  & All belong to EFT & - \\
\hline

\end{tabular}
\end{equation}
Here all the 4-point contact amplitudes involving in fermions belong to the EFT categories. Only the $(0,0,0,0)$ helicity categories of the $SSSS$, $VVSS$, and $VVVV$ are the SM amplitudes, in which all the vector bosons in the amplitudes should be the Goldstone boson. All others should be the EFT primary amplitudes, or the descendant amplitudes in the SM or EFT. For example, $(+1, -1, 0, 0)$ of the $VVSS$ amplitude is either the descendant amplitude obtained from $(0,0,0,0)$ by ladder operators, or the EFT one. 

Analogous to the construction of 3-point MHC amplitudes, after obtaining the 4-point primary amplitudes, the descendant amplitudes are derived by repeatedly applying the ladder operators $mJ^-$ or $\tilde{m}J^+$ to the primary amplitudes. Applying one such operator to a specific particle flips its helicity and yields the corresponding first descendant current; successive applications generate higher descendants:
\begin{eqnarray}
\begin{aligned}
[\mathcal{M}]_1 &= mJ^- \circ [\mathcal{M}]_0 \;+\; \tilde{m}J^+ \circ [\mathcal{M}]_0,\\
[\mathcal{M}]_2 &= (mJ^-)^2 \circ [\mathcal{M}]_0 \;+\; (\tilde{m}J^+ mJ^-) \circ [\mathcal{M}]_0 \;+\; (\tilde{m}J^+)^2 \circ [\mathcal{M}]_0,\\
&\; \vdots
\end{aligned}
\end{eqnarray}

There are two equivalent ways to obtain the 4-point MHC contact amplitudes:
\begin{itemize}
    \item Starting from one MHC amplitude, we can apply the ladder operators $\frac{\tilde m}{\mathbf m}J^+$ and $\frac{m}{\mathbf m}J^+$ to obtain any other MHC amplitude. First obtain the primary amplitudes with all gauge bosons identified as the Goldstone boson, then use ladder operator to obtain all the descendant ones. In this way, the Lorentz structure is systematically obtained, and the gauge structure is matched from the massless ones.  

    \item First obtain the descendant ones with all the transverse gauge boson, then use ladder operator to obtain others including the primary one. In this way, it is easier to implement the gauge structure from the massless UV SM.
\end{itemize}
Note again the primary 4-point contact amplitudes are non-vanishing, unlike the 3-point amplitudes due to the momentum conservation. 
In the following we will use both ways to obtain the matched gauge structures.

\paragraph{$VVSS$ Contact Amplitudes}

Consider the $WWhh$ MHC contact amplitude. In the SM, the possible helicity categories for the corresponding primary and descendant amplitudes are as follows:
\begin{equation}
\begin{tabular}{c|c|l}
\hline
\multicolumn{2}{c|}{amplitude} & helicity \\
\hline
\multicolumn{2}{c|}{primary} & $(0000)$ \\
\hline
\multirow{4}{*}{descendant} & 1st & $(\pm000),(0\pm00)$ \\
\cline{2-3}
& 2nd & $(0000),(\pm\mp00),(\pm\pm00)$ \\
\cline{2-3}
& 3rd & $(\pm000),(0\pm00)$ \\
\cline{2-3}
& 4th & $(0000)$  \\
\hline
\end{tabular}
\end{equation}

\underline{Primary $\leftrightarrow$ Descendant}

We first take the primary amplitude as the starting point. By repeatedly acting with ladder operators, we can obtain all descendant amplitudes as follows:
\begin{equation}
\text{primary}\quad\overset{\frac{\tilde m}{\mathbf m}J^+,\frac{m}{\mathbf m}J^-}{\to}\quad
\text{1st descendant}\quad\overset{\frac{\tilde m}{\mathbf m}J^+,\frac{m}{\mathbf m}J^-}{\to}\quad
\text{2nd descendant}\quad\to\quad
\cdots
\end{equation}
This shows that by acting with $k$ ladder operators, we obtain the $k$-th descendant amplitude. For the $WWhh$ amplitude, the primary MHC amplitude has helicity $(0000)$. Applying a ladder operator flips the helicity and yields the 1st descendant amplitude with $(+000)$:
\begin{equation}
\includegraphics[scale=1,valign=c]{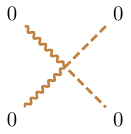}=\frac{\langle12\rangle[12]}{\mathbf m_1\mathbf m_2}
\quad\overset{\frac{\tilde m_1}{\mathbf m_1}J_1^+}{\to}\quad
\includegraphics[scale=1,valign=c]{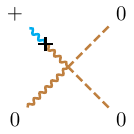}=\frac{\langle\eta_12\rangle[12]}{m_1\mathbf m_2}.
\end{equation}
Other 1st descendant amplitudes can be obtained by acting on the primary amplitude with $\frac{\tilde m_2}{\mathbf m_2}J_2^+$, $\frac{m_1}{\mathbf m_1}J_1^-$, or $\frac{m_2}{\mathbf m_2}J_2^-$. Continuing to apply ladder operators to the 1st descendant amplitude with $(+000)$, we obtain the 2nd descendant amplitudes with $(+-00)$ and $(0000)$,
\begin{equation}
\includegraphics[scale=1,valign=c]{image/VVSS_+000_ct_1st.pdf}
\quad \to
\left\{\begin{aligned}
&\overset{ \frac{m_2}{\mathbf m_2}J_2^-}{\to}\quad
\includegraphics[scale=1,valign=c]{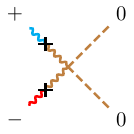}=\frac{\langle\eta_12\rangle[1\eta_2]}{m_1\tilde m_2},\\
&\overset{ \frac{m_1}{\mathbf m_1}J_1^-}{\to}\quad
\includegraphics[scale=1,valign=c]{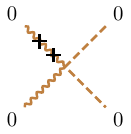}=\frac{\langle\eta_12\rangle[\eta_12]}{\mathbf m_1\mathbf m_2}.
\end{aligned} \right.
\end{equation}
Because there are vector bosons, the helicity of a 2nd descendant amplitude can be the same as that of the primary amplitude, as seen in the second line.

Conversely, we can start from a higher-order descendant amplitude, apply ladder operators to obtain lower-order descendant amplitudes, and finally reach the primary amplitude. Compared with the previous case, we now use the opposite ladder operators:
\begin{equation}
\includegraphics[scale=1,valign=c]{image/VVSS_+-00_ct_2nd.pdf}
\quad\overset{\frac{\tilde m_1}{\mathbf m_1}J_1^+}{\to}\quad
\includegraphics[scale=1,valign=c]{image/VVSS_+000_ct_1st.pdf}
\quad\overset{ \frac{m_2}{\mathbf m_2}J_2^+}{\to}\quad
\includegraphics[scale=1,valign=c]{image/VVSS_0000_ct_pri.pdf}.
\end{equation}

\underline{Descendant $\leftrightarrow$ Descendant}

We now consider conversion between descendant amplitudes. First, we examine conversion within the same order. In most cases, descendant amplitudes of the same order belong to different helicity categories, so we can use ladder operators to flip helicities and obtain different descendant amplitudes. For the 1st descendant of the $WWhh$ amplitude, starting from helicity $(+000)$, we can obtain the other three 1st descendant amplitudes:
\begin{equation}
\includegraphics[scale=1,valign=c]{image/VVSS_+000_ct_1st.pdf}=\frac{\langle\eta_12\rangle[12]}{m_1\mathbf m_2}
\quad \to
\left\{\begin{aligned}
&\overset{\frac{m_1}{\mathbf m_1}J_1^- \frac{\tilde m_1}{\mathbf m_1}J_1^+}{\to}\quad
\includegraphics[scale=1,valign=c]{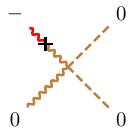}=\frac{\langle12\rangle[\eta_12]}{\tilde m_1\mathbf m_2}
,\\
&\overset{\frac{m_2}{\mathbf m_2}J_2^- \frac{\tilde m_1}{\mathbf m_1}J_1^+}{\to}\quad
\includegraphics[scale=1,valign=c]{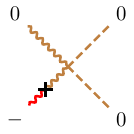}=\frac{\langle12\rangle[1\eta_2]}{\mathbf m_1\tilde m_2},\\
&\overset{\frac{\tilde m_2}{\mathbf m_2}J_2^+ \frac{m_1}{\mathbf m_1}J_1^+}{\to}\quad
\includegraphics[scale=1,valign=c]{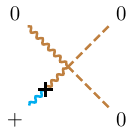}=\frac{\langle\eta_12\rangle[12]}{\mathbf m_1 m_2}.
\end{aligned} \right.
\end{equation}
For the 2nd descendant amplitudes, we can also use ladder operators to flip helicity. Starting from $(+-00)$, we obtain the 2nd descendant amplitudes with helicity $(-+00)$ and $(0000)$:
\begin{equation}
\includegraphics[scale=1,valign=c]{image/VVSS_+-00_ct_2nd.pdf}=\frac{\langle\eta_12\rangle[1\eta_2]}{m_1\tilde m_2}\to
\left\{\begin{aligned}
&\overset{\frac{m_1}{\mathbf m_1}J_1^- \frac{\tilde m_2}{\mathbf m_2}J_2^+}{\to}\quad
\includegraphics[scale=1,valign=c]{image/VVSS_0000_ct_2nd1.pdf}+
\includegraphics[scale=1,valign=c]{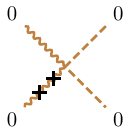}=\frac{\langle\eta_12\rangle[\eta_12]}{\mathbf m_1\mathbf m_2}+\frac{\langle1\eta_2\rangle[1\eta_2]}{\mathbf m_1\mathbf m_2},
\\ 
&\overset{\frac{m_2}{\mathbf m_2}J_2^- \frac{\tilde m_1}{\mathbf m_1}J_1^+}{\to}\quad
\includegraphics[scale=1,valign=c]{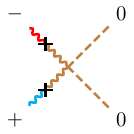}=\frac{\langle1\eta_2\rangle[\eta_12]}{\tilde m_1m_2}.
\end{aligned} \right.
\end{equation}
The helicity $(\pm\pm00)$ can be obtained similarly. Note that the helicity $(0000)$ corresponds to two distinct descendant amplitudes. Starting from one $(0000)$ amplitude, we need four ladder operators to convert it into the other:
\begin{equation}
\includegraphics[scale=1,valign=c]{image/VVSS_0000_ct_2nd1.pdf}=\frac{\langle\eta_12\rangle[\eta_12]}{\mathbf m_1\mathbf m_2}
\quad\overset{\frac{\tilde m_1}{\mathbf m_1}J_1^+ \frac{m_1}{\mathbf m_1} J_1^- \frac{\tilde m_2}{\mathbf m_2} J_2^+ \frac{m_2}{\mathbf m_2} J_2^-}{\to}\quad
\includegraphics[scale=1,valign=c]{image/VVSS_0000_ct_2nd2.pdf}=\frac{\langle1\eta_2\rangle[1\eta_2]}{\mathbf m_1\mathbf m_2}.
\end{equation}
This situation only occurs in amplitudes with two or more vector bosons.

To convert between descendant amplitudes of different orders, we can first use ladder operators to increase or decrease the order, and then convert between amplitudes of the same order to reach the target amplitude.

\paragraph{$VVSS$ Contact Amplitudes by Massless-Massive Matching} 

Although the above ladder operator is quite convenient, but it can not give the exact coefficient inherited from the UV massless contact amplitudes. Now we take the second way to obtain the $VVSS$ contact amplitudes.

The $VVSS$ massless contact amplitude in the SM has the similar structure as in scalar QED, which we have derived in the light-cone gauge. Replace the coefficient by the gauge structure of $SU(2)_W\times U(1)_Y$, we have
\begin{equation} \begin{aligned}
\begin{tikzpicture}[baseline=0.8cm] \begin{feynhand}
\vertex [particle] (i1) at (0,0) {$2^-$};
\vertex [particle] (i2) at (0,1.6) {$1^+$};
\vertex [particle] (i3) at (1.6,1.6) {$4^0$};
\vertex [particle] (i4) at (1.6,0) {$3^0$};
\vertex (v1) at (0.8,0.8);
\graph{(i1) --[bos] (v1)--[bos] (i2)};
\graph{(i4) --[sca] (v1)--[sca] (i3)};
\end{feynhand} \end{tikzpicture}=
\begin{pmatrix}
(T^{I_2}_s T^{I_1}_s)^{i_3}_{i_4}+(T^{I_1}_s T^{I_2}_s)^{i_3}_{i_4}\\
(T^{I_2}_s T^{I_1}_s)^{i_4}_{i_3}+(T^{I_1}_s T^{I_2}_s)^{i_4}_{i_3}
\end{pmatrix}
\frac{\langle\xi_12\rangle[1\xi_2]}{2\langle\xi_1 1\rangle[2 \xi_2]},
\end{aligned} \end{equation}
where $(T^{I_2}_s T^{I_1}_s)^{i_3}_{i_4}\equiv(T^{I_1}_s)^{i_3}_{j}(T^{I_2}_s)^{j}_{i_4}$ is the contraction of two 3-pt coefficients. The coefficients in two rows corresponds to the particle type $WWHH^{\dagger}$ and $WWH^{\dagger} H$. 
Multiplying with the transformation matrices, the coefficient becomes
\begin{equation} \begin{aligned}
\begin{pmatrix}
(T^{I_2}_s T^{I_1}_s)^{i_3}_{i_4}+(T^{I_1}_s T^{I_2}_s)^{i_3}_{i_4} \\
(T^{I_2}_s T^{I_1}_s)^{i_4}_{i_3}+(T^{I_1}_s T^{I_2}_s)^{i_4}_{i_3}
\end{pmatrix}^T
\begin{pmatrix}
O^{I_1 W^+}O^{I_2 W^-}\mathcal{U}^{h}_{i_3}\mathcal{U}^{i_4 h}\\
O^{I_1 W^+}O^{I_2 W^-}\mathcal{U}^{h i_3}\mathcal{U}^{i_4}_{h}
\end{pmatrix}=\mathbf g^{\mathbf I_1\mathbf J}\mathbf g^{\mathbf J \mathbf I_2}.
\end{aligned} \end{equation}
Taking the gauge $\xi_i=\eta_i$, we obtain the MHC contact amplitude
\begin{equation}
\begin{tikzpicture}[baseline=0.8cm] \begin{feynhand}
\vertex [particle] (i1) at (0,1.8) {$+$};
\vertex [particle] (i2) at (0,0) {$-$};
\vertex [particle] (i3) at (1.8,0) {$0$};
\vertex [particle] (i4) at (1.8,1.8) {$0$};
\vertex (v3) at (0.9,0.9);
\vertex (v4) at (0.9-0.9*0.43,0.9+0.9*0.43);
\vertex (v5) at (0.9-0.9*0.43,0.9-0.9*0.43);
\draw[cyan,thick bos] (i1)--(v4);
\draw[brown,thick bos] (v4)--(v3);
\draw[brown,thick bos] (v5)--(v3);
\draw[red,thick bos] (i2)--(v5);
\draw[brown,thick sca] (i3)--(v3);
\draw[brown,thick sca] (i4)--(v3);
\draw[very thick] plot[mark=x,mark size=3.5,mark options={rotate=45}] coordinates {(v4)};
\draw[very thick] plot[mark=x,mark size=3.5,mark options={rotate=45}] coordinates {(v5)};
\end{feynhand} \end{tikzpicture}=
\mathbf g^{\mathbf I_1\mathbf J}\mathbf g^{\mathbf J \mathbf I_2}\frac{\langle\eta_12\rangle[1\eta_2]}{2 m_1\tilde m_2}.
\end{equation}
This is a descendant amplitude in helicity $(+-00)$.
Note that, in the same helicity category, there is no following primary amplitude in the SM
\begin{equation} 
\begin{tikzpicture}[baseline=0.8cm] \begin{feynhand}
\vertex [particle] (i1) at (0,1.8) {$1^+$};
\vertex [particle] (i2) at (0,0) {$2^-$};
\vertex [particle] (i3) at (1.8,0) {$3^0$};
\vertex [particle] (i4) at (1.8,1.8) {$4^0$};
\vertex (v3) at (0.9,0.9);
\draw[cyan,thick bos] (i1)--(v3);
\draw[red,thick bos] (i2)--(v3);
\draw[brown,thick sca] (i4)--(v3)--(i3);
\end{feynhand} \end{tikzpicture}.
\end{equation}

Then we can use ladder operators to obtain the MHC contact amplitude in other helicity. We first consider the contact amplitude with both longitudinal and transverse vectors, which can be obtained by acting one ladder operator $\frac{m}{\mathbf m} J^-$ and $\frac{\tilde m}{\mathbf m} J^+$. For helicity $(+000)$, we obtain the descendant amplitude as below 
\begin{equation}
\mathbf g^{\mathbf I_1\mathbf J}\mathbf g^{\mathbf J \mathbf I_2}\frac{\langle\eta_12\rangle[1\eta_2]}{2 m_1\tilde m_2}
\quad\overset{\frac{\tilde m_2}{\mathbf m_2} J_2^+}{\to}\quad
\left\{ \begin{aligned}
\begin{tikzpicture}[baseline=0.8cm] \begin{feynhand}
\vertex [particle] (i1) at (0,1.8) {$1^+$};
\vertex [particle] (i2) at (0,0) {$2^0$};
\vertex [particle] (i3) at (1.8,0) {$3^0$};
\vertex [particle] (i4) at (1.8,1.8) {$4^0$};
\vertex (v3) at (0.9,0.9);
\vertex (v4) at (0.9-0.9*0.43,0.9+0.9*0.43);
\draw[cyan,thick bos] (i1)--(v4);
\draw[brown,thick bos] (v4)--(v3);
\draw[brown,thick bos] (i2)--(v3);
\draw[brown,thick sca] (i3)--(v3);
\draw[brown,thick sca] (i4)--(v3);
\draw[very thick] plot[mark=x,mark size=3.5,mark options={rotate=45}] coordinates {(v4)};
\end{feynhand} \end{tikzpicture}
&=\mathbf g^{\mathbf I_1\mathbf J}\mathbf g^{\mathbf J \mathbf I_2}\frac{\langle\eta_1 2\rangle[21]}{2m_1\mathbf m_2},\\
\begin{tikzpicture}[baseline=0.8cm] \begin{feynhand}
\vertex [particle] (i1) at (0,1.8) {$1^+$};
\vertex [particle] (i2) at (0,0) {$2^0$};
\vertex [particle] (i3) at (1.8,0) {$3^0$};
\vertex [particle] (i4) at (1.8,1.8) {$4^0$};
\vertex (v3) at (0.9,0.9);
\vertex (v4) at (0.9-0.9*0.43,0.9+0.9*0.43);
\draw[cyan,thick bos] (i1)--(v4);
\draw[brown,thick bos] (v4)--(v3);
\draw[brown,thick bos] (i2)--(v3);
\draw[brown,thick sca] (i3)--(v3);
\draw[brown,thick sca] (i4)--(v3);
\draw[very thick] plot[mark=x,mark size=3.5,mark options={rotate=45}] coordinates {(v4)};
\draw[very thick] plot[mark=x,mark size=3.5,mark options={rotate=45}] coordinates {(0.9-0.9*0.48,0.9-0.9*0.48)};
\draw[very thick] plot[mark=x,mark size=3.5,mark options={rotate=45}] coordinates {(0.9-0.9*0.22,0.9-0.9*0.22)};
\end{feynhand} \end{tikzpicture}
&=\mathbf g^{\mathbf I_1\mathbf J}\mathbf g^{\mathbf J \mathbf I_2}\frac{\langle\eta_1 \eta_2\rangle[\eta_21]}{m_1\mathbf m_2}.
\end{aligned} \right.
\end{equation}
Then we consider the longitudinal contact amplitude, which should be obtained by apply the two ladder operators $m_1 J_1^-$ and $\tilde m_2 J_2^+$, 
\begin{equation}
\begin{tikzpicture}[baseline=0.8cm] \begin{feynhand}
\vertex [particle] (i1) at (0,1.8) {$+$};
\vertex [particle] (i2) at (0,0) {$-$};
\vertex [particle] (i3) at (1.8,0) {$0$};
\vertex [particle] (i4) at (1.8,1.8) {$0$};
\vertex (v3) at (0.9,0.9);
\vertex (v4) at (0.9-0.9*0.43,0.9+0.9*0.43);
\vertex (v5) at (0.9-0.9*0.43,0.9-0.9*0.43);
\draw[cyan,thick bos] (i1)--(v4);
\draw[brown,thick bos] (v4)--(v3);
\draw[brown,thick bos] (v5)--(v3);
\draw[red,thick bos] (i2)--(v5);
\draw[brown,thick sca] (i3)--(v3);
\draw[brown,thick sca] (i4)--(v3);
\draw[very thick] plot[mark=x,mark size=3.5,mark options={rotate=45}] coordinates {(v4)};
\draw[very thick] plot[mark=x,mark size=3.5,mark options={rotate=45}] coordinates {(v5)};
\end{feynhand} \end{tikzpicture}
\overset{\frac{m_1}{\mathbf m_1} J^-_1 \frac{\tilde m_2}{\mathbf m_2} J^+_2}{\to} 
\begin{tikzpicture}[baseline=0.8cm] \begin{feynhand}
\vertex [particle] (i1) at (0,1.8) {$1^0$};
\vertex [particle] (i2) at (0,0) {$2^0$};
\vertex [particle] (i3) at (1.8,0) {$3^0$};
\vertex [particle] (i4) at (1.8,1.8) {$4^0$};
\vertex (v3) at (0.9,0.9);
\draw[brown,thick bos] (i1)--(v3);
\draw[brown,thick bos] (i2)--(v3);
\draw[brown,thick sca] (i4)--(v3)--(i3);
\end{feynhand} \end{tikzpicture}+
\begin{tikzpicture}[baseline=0.8cm] \begin{feynhand}
\vertex [particle] (i1) at (0,1.8) {$1^0$};
\vertex [particle] (i2) at (0,0) {$2^0$};
\vertex [particle] (i3) at (1.8,0) {$3^0$};
\vertex [particle] (i4) at (1.8,1.8) {$4^0$};
\vertex (v3) at (0.9,0.9);
\draw[brown,thick bos] (i1)--(v3);
\draw[brown,thick bos] (i2)--(v3);
\draw[brown,thick sca] (i4)--(v3)--(i3);
\draw[very thick] plot[mark=x,mark size=3.5,mark options={rotate=45}] coordinates {(0.9-0.9*0.48,0.9-0.9*0.48)};
\draw[very thick] plot[mark=x,mark size=3.5,mark options={rotate=45}] coordinates {(0.9-0.9*0.22,0.9-0.9*0.22)}; 
\end{feynhand} \end{tikzpicture}+
\begin{tikzpicture}[baseline=0.8cm] \begin{feynhand}
\vertex [particle] (i1) at (0,1.8) {$1^0$};
\vertex [particle] (i2) at (0,0) {$2^0$};
\vertex [particle] (i3) at (1.8,0) {$3^0$};
\vertex [particle] (i4) at (1.8,1.8) {$4^0$};
\vertex (v3) at (0.9,0.9);
\draw[brown,thick bos] (i1)--(v3);
\draw[brown,thick bos] (i2)--(v3);
\draw[brown,thick sca] (i4)--(v3)--(i3);
\draw[very thick] plot[mark=x,mark size=3.5,mark options={rotate=45}] coordinates {(0.9-0.9*0.48,0.9+0.9*0.48)};
\draw[very thick] plot[mark=x,mark size=3.5,mark options={rotate=45}] coordinates {(0.9-0.9*0.22,0.9+0.9*0.22)};
\end{feynhand} \end{tikzpicture}+
\begin{tikzpicture}[baseline=0.8cm] \begin{feynhand}
\vertex [particle] (i1) at (0,1.8) {$1^0$};
\vertex [particle] (i2) at (0,0) {$2^0$};
\vertex [particle] (i3) at (1.8,0) {$3^0$};
\vertex [particle] (i4) at (1.8,1.8) {$4^0$};
\vertex (v3) at (0.9,0.9);
\draw[brown,thick bos] (i1)--(v3);
\draw[brown,thick bos] (i2)--(v3);
\draw[brown,thick sca] (i4)--(v3)--(i3);
\draw[very thick] plot[mark=x,mark size=3.5,mark options={rotate=45}] coordinates {(0.9-0.9*0.48,0.9-0.9*0.48)};
\draw[very thick] plot[mark=x,mark size=3.5,mark options={rotate=45}] coordinates {(0.9-0.9*0.22,0.9-0.9*0.22)}; 
\draw[very thick] plot[mark=x,mark size=3.5,mark options={rotate=45}] coordinates {(0.9-0.9*0.48,0.9+0.9*0.48)};
\draw[very thick] plot[mark=x,mark size=3.5,mark options={rotate=45}] coordinates {(0.9-0.9*0.22,0.9+0.9*0.22)};
\end{feynhand} \end{tikzpicture}.
\end{equation}
The primary and descendant amplitudes in helicity $(0000)$ are given together. The corresponding amplitude expression is 
\begin{equation}
\mathbf g^{\mathbf I_1\mathbf J}\mathbf g^{\mathbf J \mathbf I_2}\frac{\langle\eta_12\rangle[1\eta_2]}{2m_1\tilde m_2}\to
\mathbf g^{\mathbf I_1\mathbf J}\mathbf g^{\mathbf J \mathbf I_2}\left(\frac{\langle12\rangle[12]}{2\mathbf m_1\mathbf m_2}
-\frac{m_1\tilde m_1 \langle\eta_12\rangle[\eta_12]}{2\mathbf m_1^3\mathbf m_2}
-\frac{m_2\tilde m_2 \langle1\eta_2\rangle[1\eta_2]}{2\mathbf m_1\mathbf m_2^3}
+\frac{m_1\tilde m_1 m_2\tilde m_2\langle\eta_1\eta_2\rangle[\eta_1\eta_2]}{2\mathbf m_1^3\mathbf m_2^3}\right).
\end{equation}

The first term corresponds to the primary $VVSS$ amplitude. Unlike in the 3-pt case, this 4-pt primary contact amplitude is non-vanishing and must therefore have an UV orgin. This origin can be traced to the electroweak chiral Lagrangian $\text{tr}[(D_\mu U)(D^\mu U)^\dagger]$. In this framework, the scalar field $U$ is expressed in an exponential parameterization involving both the Goldstone bosons $\pi$ and the Higgs boson $h$:
\begin{equation}
U=\exp
\left[i\frac{\sqrt2}{v}\begin{pmatrix}
\pi^0 & \pi^- \\
\pi^+ & \pi^0 \\
\end{pmatrix}\right]
\begin{pmatrix}
0\\
\frac{v+h}{\sqrt2}
\end{pmatrix}.
\end{equation}
Here $\pi$ differs from the field $\phi$ used in linear parameterizations. Expanding the chiral Lagrangian yields
\begin{equation}
\text{tr}[(D_\mu U)(D^\mu U)^\dagger]=\frac{1}{v^2}(\partial_\mu\pi^-)(\partial^\mu\pi^+)h^2+\cdots,
\end{equation}
which contains a coupling proportional to $1/v^2$ that is absent in the linear realization. We can therefore establish a correspondence between this non-linearly realized UV theory and the primary $VVSS$ amplitude:
\begin{equation}
\begin{tikzpicture}[baseline=0.8cm] \begin{feynhand}
\vertex [particle] (i1) at (0,0) {$2^0$};
\vertex [particle] (i2) at (0,1.6) {$1^0$};
\vertex [particle] (i3) at (1.6,1.6) {$4^0$};
\vertex [particle] (i4) at (1.6,0) {$3^0$};
\vertex (v1) at (0.8,0.8);
\graph{(i1) --[sca] (v1)--[sca] (i2)};
\graph{(i4) --[sca] (v1)--[sca] (i3)};
\end{feynhand} \end{tikzpicture}=
\frac{1}{v^2}\langle12\rangle[12]
\quad\to\quad
\begin{tikzpicture}[baseline=0.8cm] \begin{feynhand}
\vertex [particle] (i1) at (0,1.8) {$1^0$};
\vertex [particle] (i2) at (0,0) {$2^0$};
\vertex [particle] (i3) at (1.8,0) {$3^0$};
\vertex [particle] (i4) at (1.8,1.8) {$4^0$};
\vertex (v3) at (0.9,0.9);
\draw[brown,thick bos] (i1)--(v3);
\draw[brown,thick bos] (i2)--(v3);
\draw[brown,thick sca] (i4)--(v3)--(i3);
\end{feynhand} \end{tikzpicture}
=\mathbf g^{\mathbf I_1\mathbf J}\mathbf g^{\mathbf J \mathbf I_2}\frac{\langle12\rangle[12]}{2\mathbf m_1\mathbf m_2}\,.
\end{equation}
Note that this UV does not exist in the unbroken linear SM, and thus must vanish in the massless-massive matching due to a cancellation between contact and factorized $(0,0,0,0)$ amplitudes~\footnote{In addition to the contact primary amplitude in the $(0000)$ helicity category, there exists a factorized primary amplitude for the same helicity, which also originates from this chiral Lagrangian. In the (13)-channel, we have
\begin{equation}
\begin{tikzpicture}[baseline=-0.1cm] \begin{feynhand}
\vertex [particle] (i1) at (-0.6,0.6) {$1^0$};
\vertex [particle] (i2) at (-0.6,-0.6) {$2^0$};
\vertex [particle] (i3) at (0.6,-0.6) {$3^0$};
\vertex [particle] (i4) at (0.6,0.6) {$4^0$};
\vertex (v1) at (0,-0.2);
\vertex (v2) at (0,0.2);
\draw[brown,thick bos] (i1)--(v2);
\draw[brown,thick bos] (v2)--(v1);
\draw[brown,thick bos] (i2)--(v1);
\draw[brown,thick sca] (i3)--(v1);
\draw[brown,thick sca] (i4)--(v2);
\end{feynhand} \end{tikzpicture}=\frac{1}{v^2}\frac{\langle13\rangle[13]\langle24\rangle[24]}{s_{13}}.
\end{equation}
The full factorized amplitude includes this term and the corresponding (14)-channel contribution, which is obtained by exchanging particles 3 and 4. Both the factorized and contact primary amplitudes scale as $\frac{E^2}{v^2}$, so their presence can violate the tree-level unitarity. To avoid such unitarity violation, a cancellation between them must occur:
\begin{equation}
\begin{tikzpicture}[baseline=-0.1cm] \begin{feynhand}
\vertex [particle] (i1) at (-0.6,0.6) {$1^0$};
\vertex [particle] (i2) at (-0.6,-0.6) {$2^0$};
\vertex [particle] (i3) at (0.6,-0.6) {$3^0$};
\vertex [particle] (i4) at (0.6,0.6) {$4^0$};
\vertex (v1) at (0,-0.2);
\vertex (v2) at (0,0.2);
\draw[brown,thick bos] (i1)--(v2);
\draw[brown,thick bos] (v2)--(v1);
\draw[brown,thick bos] (i2)--(v1);
\draw[brown,thick sca] (i3)--(v1);
\draw[brown,thick sca] (i4)--(v2);
\end{feynhand} \end{tikzpicture}+
\begin{tikzpicture}[baseline=-0.1cm] \begin{feynhand}
\vertex [particle] (i1) at (-0.6,0.6) {$1^0$};
\vertex [particle] (i2) at (-0.6,-0.6) {$2^0$};
\vertex [particle] (i3) at (0.6,-0.6) {$3^0$};
\vertex [particle] (i4) at (0.6,0.6) {$4^0$};
\vertex (v1) at (0,-0.2);
\vertex (v2) at (0,0.2);
\draw[brown,thick bos] (i1)--(v2);
\draw[brown,thick bos] (i2)--(v1);
\draw[brown,thick sca] (i3)--(v2);
\draw[brown,thick sca] (i4)--(v1);
\draw[brown,thick bos] (v2)--(v1);
\end{feynhand} \end{tikzpicture}+
\begin{tikzpicture}[baseline=-0.1cm] \begin{feynhand}
\vertex [particle] (i1) at (-0.6,0.6) {$1^0$};
\vertex [particle] (i2) at (-0.6,-0.6) {$2^0$};
\vertex [particle] (i3) at (0.6,-0.6) {$3^0$};
\vertex [particle] (i4) at (0.6,0.6) {$4^0$};
\vertex (v3) at (0,0);
\draw[brown,thick bos] (i1)--(v3);
\draw[brown,thick bos] (i2)--(v3);
\draw[brown,thick sca] (i4)--(v3)--(i3);
\end{feynhand} \end{tikzpicture}
=\frac{1}{v^2}\left(\frac{\langle13\rangle[13]\langle24\rangle[24]}{s_{13}}+\frac{\langle14\rangle[14]\langle23\rangle[23]}{s_{14}}+\langle12\rangle[12]\right)=0.
\end{equation}
Note that only the $(0000)$-helicity primary amplitude has mass dimension 4. Primary MHC amplitudes in other helicity categories than the $(0000)$ one, such as
\begin{equation}
    \begin{tikzpicture}[baseline=-0.1cm] \begin{feynhand}
\vertex [particle] (i1) at (-0.6,0.6) {$1^+$};
\vertex [particle] (i2) at (-0.6,-0.6) {$2^-$};
\vertex [particle] (i3) at (0.6,-0.6) {$3^0$};
\vertex [particle] (i4) at (0.6,0.6) {$4^0$};
\vertex (v3) at (0,0);
\draw[cyan,thick bos] (i1)--(v3);
\draw[red,thick bos] (i2)--(v3);
\draw[brown,thick sca] (i4)--(v3)--(i3);
\end{feynhand} \end{tikzpicture},
\end{equation}
belong to the massive EFT.  }. This vanishing can also be explained by the conserved current, which provides an alternative method for obtaining the MHC contact amplitude. This derivation is given in Appendix~\ref{sec:VVS_current}.

Among these MHC terms in helicity $(0000)$, two descendant amplitudes whose diagram has two cross can be matched from the 4-pt massless amplitude. Let us consider the matching from massless four-scalar amplitude to the $W^+W^-hh$ amplitude. In this case, the massless contact amplitude cannot match to the MHC contact amplitude only, we must include the contribution from the massless factorized amplitude. Therefore, we should consider the full massless amplitude, which has four contributions:
\begin{align}
&\begin{tikzpicture}[baseline=0.8cm] \begin{feynhand}
\vertex [particle] (i1) at (0,0) {$2^0$};
\vertex [particle] (i2) at (0,1.6) {$1^0$};
\vertex [particle] (i3) at (1.6,1.6) {$4^0$};
\vertex [particle] (i4) at (1.6,0) {$3^0$};
\vertex (v1) at (0.4,0.8);
\vertex (v2) at (1.2,0.8);
\graph{(i1) --[sca] (v1)--[sca] (i2)};
\graph{(i4) --[sca] (v2)--[sca] (i3)};
\graph{(v1) --[bos] (v2)};
\end{feynhand} \end{tikzpicture}:\quad
\begin{pmatrix}
-(T^{J}_s)^{i_1}_{i_2} (T^{J}_s)^{i_3}_{i_4}\\
(T^{J}_s)^{i_1}_{i_2} (T^{J}_s)^{i_4}_{i_3}
\end{pmatrix}
\frac{s_{13}-s_{14}}{2s_{12}},\\
&\begin{tikzpicture}[baseline=0.8cm] \begin{feynhand}
\vertex [particle] (i1) at (0,0) {$2^0$};
\vertex [particle] (i2) at (0,1.6) {$1^0$};
\vertex [particle] (i3) at (1.6,1.6) {$4^0$};
\vertex [particle] (i4) at (1.6,0) {$3^0$};
\vertex (v1) at (0.8,0.4);
\vertex (v2) at (0.8,1.2);
\graph{(i1) --[sca] (v1)--[sca] (i4)};
\graph{(i2) --[sca] (v2)--[sca] (i3)};
\graph{(v1) --[bos] (v2)};
\end{feynhand} \end{tikzpicture}:\quad
\begin{pmatrix}
-(T^{J}_s)^{i_1}_{i_4} (T^{J}_s)^{i_3}_{i_2}\\
0
\end{pmatrix}
\frac{s_{13}-s_{12}}{2s_{14}},\\
&\begin{tikzpicture}[baseline=0.8cm] \begin{feynhand}
\setlength{\feynhandtopsep}{5pt}
\vertex [particle] (i1) at (0,0) {$2^0$};
\vertex [particle] (i2) at (0,1.6) {$1^0$};
\vertex [particle] (i3) at (1.6,1.6) {$4^0$};
\vertex [particle] (i4) at (1.6,0) {$3^0$};
\vertex (v1) at (0.8,0.4);
\vertex (v2) at (0.8,1.2);
\graph{(i2)--[sca] (v2)--[bos] (v1)-- [sca](i1)};
\graph{(i4)--[sca] (v2)};
\graph{(i3)--[sca,top] (v1)};
\end{feynhand} \end{tikzpicture}:\quad
\begin{pmatrix}
0\\
-(T^{J}_s)^{i_1}_{i_3} (T^{J}_s)^{i_4}_{i_2}
\end{pmatrix}
\frac{s_{14}-s_{12}}{2s_{13}},\\
&\begin{tikzpicture}[baseline=0.8cm] \begin{feynhand}
\vertex [particle] (i1) at (0,0) {$2^0$};
\vertex [particle] (i2) at (0,1.6) {$1^0$};
\vertex [particle] (i3) at (1.6,1.6) {$4^0$};
\vertex [particle] (i4) at (1.6,0) {$3^0$};
\vertex (v1) at (0.8,0.8);
\graph{(i1) --[sca] (v1)--[sca] (i2)};
\graph{(i4) --[sca] (v1)--[sca] (i3)};
\end{feynhand} \end{tikzpicture}:\quad
-\begin{pmatrix}
\delta^{(i_1}_{i_2} \delta^{i_3)}_{i_4}\\
\delta^{(i_1}_{i_2} \delta^{i_4)}_{i_3}
\end{pmatrix}
4\lambda,
\end{align}
where the two rows of coefficients correspond to the particle types $HH^\dagger HH^\dagger$ and $HH^\dagger H^\dagger H$. After multiplying with the transformation matrix, the massless coefficients of the four contributions become
\begin{equation}
\begin{pmatrix}
-(T^{J}_s)^{i_1}_{i_2} (T^{J}_s)^{i_3}_{i_4} &
-(T^{J}_s)^{i_1}_{i_4} (T^{J}_s)^{i_3}_{i_2} & 0 &
\delta^{(i_1}_{i_2} \delta^{i_3)}_{i_4} \\
(T^{J}_s)^{i_1}_{i_2} (T^{J}_s)^{i_4}_{i_3} & 0 &
-(T^{J}_s)^{i_1}_{i_3} (T^{J}_s)^{i_4}_{i_2} &
\delta^{(i_1}_{i_2} \delta^{i_4)}_{i_3}
\end{pmatrix}^T
\begin{pmatrix}
\mathcal{U}^{W^+}_{i_1}\mathcal{U}^{i_2 W^-}\mathcal{U}^{h}_{i_3}\mathcal{U}^{i_4 h}\\
\mathcal{U}^{W^+}_{i_1}\mathcal{U}^{i_2 W^-}\mathcal{U}^{h i_3}\mathcal{U}^{i_4}_{h}
\end{pmatrix}
=\begin{pmatrix}
0\\
g^2\\
g^2\\
-\lambda
\end{pmatrix}.
\end{equation}
The amplitude now becomes
\begin{equation} \begin{aligned}
g^2\frac{s_{13}-s_{12}}{2s_{14}}
+g^2\frac{s_{14}-s_{12}}{2s_{13}}
+4\lambda.
\end{aligned} \end{equation}

\begin{figure}[htbp]
\centering
\includegraphics[width=\linewidth,valign=c]{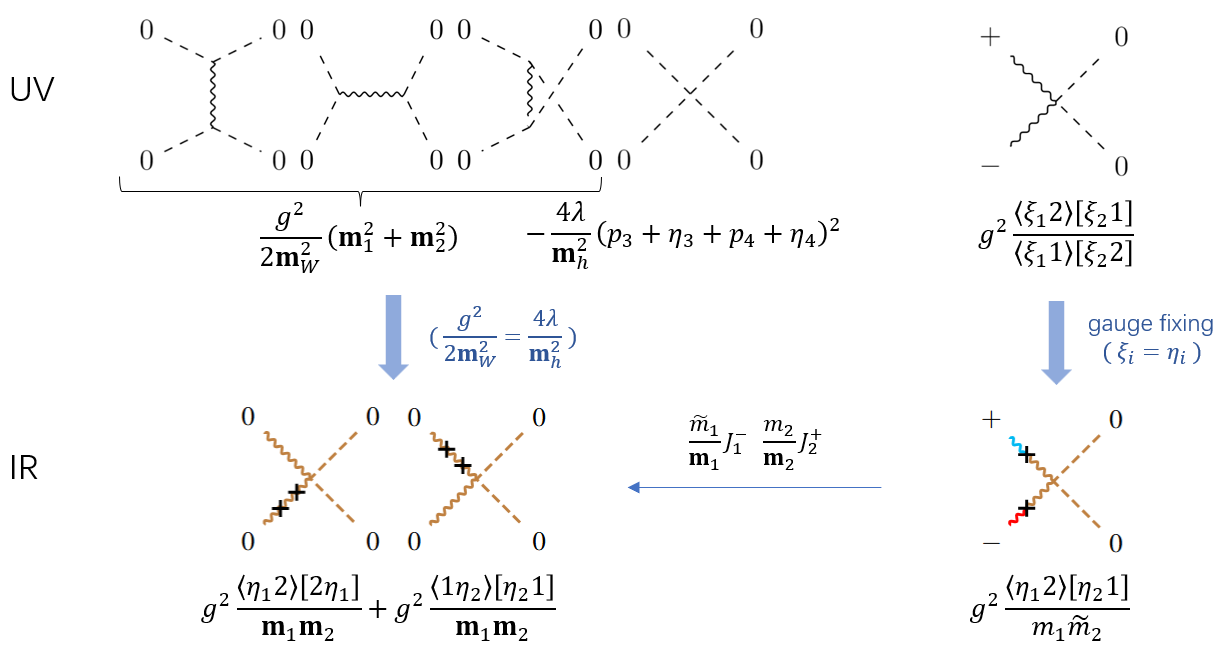}
\caption{The diagrammatic matching for the contact $W^+W^-hh$ amplitude in two helicities, related by the ladder operator.}
\label{fig:contact_match}
\end{figure}

The last term corresponds to the contact amplitude from the $\lambda\phi^4$ interaction. To match the MHC contact amplitude, we must combine contributions from gauge interactions and the $\lambda\phi^4$ interaction. This mixing can be achieved through an amplitude deformation where the factorized terms satisfy the Goldstone behavior for particles 1 and 2:
\begin{equation} \label{eq:4pt_deform}
-g^2\frac{s_{12}}{s_{14}}-g^2\frac{s_{12}}{s_{13}}+g^2+4\lambda,
\end{equation}
The last two terms have no pole structure and can therefore mix. To produce the MHC contact terms $\langle\eta_12\rangle[\eta_12]$ and $\langle1\eta_2\rangle[1\eta_2]$, we convert the coupling constants into mass squares
\begin{equation} \begin{aligned}
g^2 &=\frac{g^2}{2\mathbf m^2_W}(\mathbf m_1^2+\mathbf m_2^2) =\frac{g^2}{2\mathbf m^2_W}[(p_1+\eta_1+p_2+\eta_2)^2\overbrace{-2p_1\cdot\eta_2-2\eta_1\cdot p_2}^{\text{MHC contact term}}],\\
\lambda&=\frac{\lambda}{2\mathbf m^2_h}(\mathbf m_3^2+\mathbf m_4^2)=\frac{\lambda}{2\mathbf m^2_h}[(p_3+\eta_4+p_3+\eta_4)^2-2p_3\cdot\eta_4-2\eta_3\cdot p_4].
\end{aligned} \end{equation}
It shows that the $g^2$ alone does not exactly reproduce the MHC contact term, but a cancellation involving $\lambda$ makes it possible. This requires momentum conservation $p_1+\eta_1+p_2+\eta_2=-p_3-\eta_3-p_4-\eta_4$ and the following relation between UV couplings and IR masses:
\begin{equation}
\frac{g^2}{2\mathbf m_W^2}=\frac{4\lambda}{\mathbf m_h^2},
\end{equation}
Under this condition, the amplitude in Eq.~\eqref{eq:4pt_deform} can be further decomposed as
\begin{equation} \begin{aligned}
&\text{gauge interaction}:& &-g^2\frac{s_{12}}{s_{14}}
-g^2\frac{s_{12}}{s_{13}}+\frac{g^2}{2\mathbf m_W^2}(\mathbf m_1^2+\mathbf m_2^2),\\
&\text{$\lambda\phi^4$ interaction}:& &12\lambda+\frac{8\lambda}{\mathbf m_h^2}(p_3\cdot\eta_4+\eta_3\cdot p_3)-\frac{4\lambda}{\mathbf m_h^2}(p_3+\eta_3+p_4+\eta_4)^2.
\end{aligned} \end{equation}
The last terms in both lines contribute to the MHC contact term, as illustrated in figure~\ref{fig:contact_match}.

Restore the $SU(2)$ little group covariance, we can derive the AHH contact amplitude,
\begin{equation} 
\includegraphics[scale=1,valign=c]{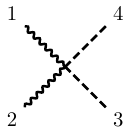}=\frac{\mathbf{g}^{W^+ W^-}\mathbf{g}^{W^+ W^-}}{\mathbf m_W^2}\langle\mathbf{12}\rangle[\mathbf{21}].
\end{equation}

\paragraph{$VVVV$ Contact Amplitudes}

In analogy to the $VVSS$ case, we can write down the massless contact $VVVV$ amplitude in the light-cone gauge,
\begin{equation} \begin{aligned}
\begin{tikzpicture}[baseline=0.8cm] \begin{feynhand}
\vertex [particle] (i1) at (0,0) {$2^+$};
\vertex [particle] (i2) at (0,1.6) {$1^-$};
\vertex [particle] (i3) at (1.6,1.6) {$4^+$};
\vertex [particle] (i4) at (1.6,0) {$3^-$};
\vertex (v1) at (0.8,0.8);
\graph{(i1) --[bos] (v1)--[bos] (i2)};
\graph{(i4) --[bos] (v1)--[bos] (i3)};
\end{feynhand} \end{tikzpicture}=
\frac{1}{[\xi_1 1]\langle 2 \xi_2\rangle[\xi_3 3]\langle4 \xi_4\rangle}
\begin{pmatrix}
f^{I_1 I_2 J}f^{J I_3 I_4}(\langle13\rangle[\xi_3\xi_1]\langle\xi_2\xi_4\rangle[42]-\langle1\xi_4\rangle[4\xi_1]\langle\xi_23\rangle[\xi_32])\\
f^{I_1 I_3 J}f^{J I_4 I_2}(\langle1\xi_4\rangle[4\xi_1]\langle\xi_23\rangle[\xi_32]-\langle1\xi_2\rangle[2\xi_1]\langle3\xi_4\rangle[4\xi_3])\\
f^{I_1 I_4 J}f^{J I_2 I_3}(\langle1\xi_2\rangle[2\xi_1]\langle3\xi_4\rangle[4\xi_3]-\langle13\rangle[\xi_3\xi_1]\langle\xi_2\xi_4\rangle[42])\\
\end{pmatrix}.
\end{aligned} \end{equation}
Multiplying with the transformation matices, the coefficient becomes
\begin{equation}
\begin{pmatrix}
f^{I_1 I_2 J}f^{J I_3 I_4}\\
f^{I_1 I_3 J}f^{J I_4 I_2}\\
f^{I_1 I_4 J}f^{J I_2 I_3}\\
\end{pmatrix}
O^{I_1 \mathbf{I}_1} O^{I_2 \mathbf{I}_2} O^{I_3 \mathbf{I}_3} O^{I_4 \mathbf{I}_4}=
\begin{pmatrix}
\mathbf f^{\mathbf I_1 \mathbf I_2 \mathbf J}\mathbf f^{\mathbf J \mathbf I_3 \mathbf I_4}\\
\mathbf f^{\mathbf I_1 \mathbf I_3 \mathbf J}\mathbf f^{\mathbf J \mathbf I_4 \mathbf I_2}\\
\mathbf f^{\mathbf I_1 \mathbf I_4 \mathbf J}\mathbf f^{\mathbf J \mathbf I_2 \mathbf I_3}\\
\end{pmatrix}.
\end{equation}
Taking the gauge $\xi_i=\eta_i$, we have
\begin{equation}
\Ampfour{-}{+}{-}{+}{\bosflip{1.37}{135}{brown}{cyan}}{\bosflip{1.37}{-135}{brown}{red}}{\bosflip{1.37}{-45}{brown}{cyan}}{\bosflip{1.37}{45}{brown}{red}}
=\frac{1}{\tilde m_1 m_2 \tilde m_3 m_4}
\begin{pmatrix}
\mathbf f^{\mathbf I_1 \mathbf I_2 \mathbf J}\mathbf f^{\mathbf J \mathbf I_3 \mathbf I_4}(\langle13\rangle[\eta_3\eta_1]\langle\eta_2\eta_4\rangle[42]-\langle1\eta_4\rangle[4\eta_1]\langle\eta_23\rangle[\eta_32])\\
\mathbf f^{\mathbf I_1 \mathbf I_3 \mathbf J}\mathbf f^{\mathbf J \mathbf I_4 \mathbf I_2}(\langle1\eta_4\rangle[4\eta_1]\langle\eta_23\rangle[\eta_32]-\langle1\eta_2\rangle[2\eta_1]\langle3\eta_4\rangle[4\eta_3])\\
\mathbf f^{\mathbf I_1 \mathbf I_4 \mathbf J}\mathbf f^{\mathbf J \mathbf I_2 \mathbf I_3}(\langle1\eta_2\rangle[2\eta_1]\langle3\eta_4\rangle[4\eta_3]-\langle13\rangle[\eta_3\eta_1]\langle\eta_2\eta_4\rangle[42])\\
\end{pmatrix}.
\end{equation}
In the SM, the massive particle type can be $W^{\pm}$ and $Z$ boson. For example, we can choose $\mathbf I_1=\mathbf I_3=W^+$ and $\mathbf I_2=\mathbf I_4=W^-$, the amplitude will reduce to
\begin{equation}
\mathbf{f}^{W^+ W^- Z}\mathbf{f}^{ZW^+ W^- }\frac{\langle1\eta_2\rangle[2\eta_1]\langle3\eta_4\rangle[\eta_34]+\langle1\eta_4\rangle[4\eta_1]\langle\eta_23\rangle[\eta_32]-2\langle13\rangle[\eta_3\eta_1]\langle\eta_2\eta_4\rangle[42]}{\tilde m_1 m_2 \tilde m_3 m_4}.
\end{equation}

Reversely using the ladder operator, we recover the primary amplitude
\begin{equation} \begin{aligned}
&\Ampfour{0}{0}{0}{0}{\bos{i1}{brown}}{\bos{i2}{brown}}{\bos{i3}{brown}}{\bos{i4}{brown}}
\begin{aligned}
&=(\mathbf{f}^{W^+ W^- Z})^2\frac{\langle12\rangle[21]\langle34\rangle[34]+\langle14\rangle[41]\langle23\rangle[32]-2\langle13\rangle[31]\langle24\rangle[42]}{\mathbf m_1 \mathbf m_2 \mathbf m_3 \mathbf m_4}.
\end{aligned} 
\end{aligned} \end{equation}
The subleading order contribution in this helicity category is also determined.
\begin{equation} \begin{aligned}
&\Ampfour{0}{0}{0}{0}{\bos{i1}{brown}}{\bos{i2}{brown}}{\bos{i3}{brown}}{\bos{i4}{brown}}+
\Ampfour{0}{0}{0}{0}{\bosflipbrown{1.37}{135}}{\bos{i2}{brown}}{\bos{i3}{brown}}{\bos{i4}{brown}}+
\Ampfour{0}{0}{0}{0}{\bos{i1}{brown}}{\bosflipbrown{1.37}{-135}}{\bos{i3}{brown}}{\bos{i4}{brown}}+
\Ampfour{0}{0}{0}{0}{\bos{i1}{brown}}{\bos{i2}{brown}}{\bosflipbrown{1.37}{-45}}{\bos{i4}{brown}}\\
+&\Ampfour{0}{0}{0}{0}{\bos{i1}{brown}}{\bos{i2}{brown}}{\bos{i3}{brown}}{\bosflipbrown{1.37}{45}}+
\Ampfour{0}{0}{0}{0}{\bosflipbrown{1.37}{135}}{\bosflipbrown{1.37}{-135}}{\bos{i3}{brown}}{\bos{i4}{brown}}+
\Ampfour{0}{0}{0}{0}{\bosflipbrown{1.37}{135}}{\bos{i2}{brown}}{\bosflipbrown{1.37}{-45}}{\bos{i4}{brown}}+
\Ampfour{0}{0}{0}{0}{\bosflipbrown{1.37}{135}}{\bos{i2}{brown}}{\bos{i3}{brown}}{\bosflipbrown{1.37}{45}}\\
+&\Ampfour{0}{0}{0}{0}{\bos{i1}{brown}}{\bosflipbrown{1.37}{-135}}{\bosflipbrown{1.37}{-45}}{\bos{i4}{brown}}+
\Ampfour{0}{0}{0}{0}{\bos{i1}{brown}}{\bosflipbrown{1.37}{-135}}{\bos{i3}{brown}}{\bosflipbrown{1.37}{45}}+
\Ampfour{0}{0}{0}{0}{\bos{i1}{brown}}{\bos{i2}{brown}}{\bosflipbrown{1.37}{-45}}{\bosflipbrown{1.37}{45}}+
\Ampfour{0}{0}{0}{0}{\bosflipbrown{1.37}{135}}{\bosflipbrown{1.37}{-135}}{\bosflipbrown{1.37}{-45}}{\bos{i4}{brown}}\\
+&\Ampfour{0}{0}{0}{0}{\bosflipbrown{1.37}{135}}{\bosflipbrown{1.37}{-135}}{\bos{i3}{brown}}{\bosflipbrown{1.37}{45}}+
\Ampfour{0}{0}{0}{0}{\bosflipbrown{1.37}{135}}{\bos{i2}{brown}}{\bosflipbrown{1.37}{-45}}{\bosflipbrown{1.37}{45}}+
\Ampfour{0}{0}{0}{0}{\bos{i1}{brown}}{\bosflipbrown{1.37}{-135}}{\bosflipbrown{1.37}{-45}}{\bosflipbrown{1.37}{45}}+
\Ampfour{0}{0}{0}{0}{\bosflipbrown{1.37}{135}}{\bosflipbrown{1.37}{-135}}{\bosflipbrown{1.37}{-45}}{\bosflipbrown{1.37}{45}}.
\end{aligned} \end{equation}
The corresponding amplitude is
\begin{equation} \begin{aligned}
\frac{(\mathbf{f}^{W^+ W^- Z})^2}{\mathbf m_1 \mathbf m_2 \mathbf m_3 \mathbf m_4}\times&
\Big((\langle12\rangle[21]-\langle\eta_12\rangle[2\eta_1]-\langle1\eta_2\rangle[\eta_21]+\langle\eta_1\eta_2\rangle[\eta_2\eta_1])\\
&\times(\langle34\rangle[43]-\langle\eta_34\rangle[4\eta_3]-\langle3\eta_4\rangle[\eta_43]+\langle\eta_3\eta_4\rangle[\eta_4\eta_3])\\
&+(\langle14\rangle[41]-\langle\eta_14\rangle[4\eta_1]-\langle1\eta_4\rangle[\eta_41]+\langle\eta_1\eta_4\rangle[\eta_4\eta_1])\\
&\times(\langle23\rangle[32]-\langle\eta_23\rangle[3\eta_2]-\langle2\eta_3\rangle[\eta_32]+\langle\eta_2\eta_3\rangle[\eta_3\eta_2])\\
&+(\langle13\rangle[31]-\langle\eta_13\rangle[3\eta_1]-\langle1\eta_3\rangle[\eta_31]+\langle\eta_1\eta_3\rangle[\eta_3\eta_1])\\
&\times(\langle24\rangle[42]-\langle\eta_24\rangle[4\eta_2]-\langle2\eta_4\rangle[\eta_42]+\langle\eta_2\eta_4\rangle[\eta_4\eta_2])\Big).\\
\end{aligned} \end{equation}

Using the extended little group covariance, we can determine the contact term in all helicity categories. Since it has many diagrams, we only list the leading term in some typical helicity categories,
\begin{equation}
\begin{tabular}{c|c|c}
\hline
helicity category & diagrams & amplitude \\
\hline
$(0,0,0,0)$ & 
\Ampfour{0}{0}{0}{0}{\bos{i1}{brown}}{\bos{i2}{brown}}{\bos{i3}{brown}}{\bos{i4}{brown}}+15 diagrams & 
\makecell{
$\frac{(\mathbf{f}^{W^+ W^- Z})^2}{\mathbf m_1 \mathbf m_2 \mathbf m_3 \mathbf m_4}\times\big(\langle12\rangle[21]\langle34\rangle[34]$\\
$+\langle14\rangle[41]\langle23\rangle[32]-2\langle13\rangle[31]\langle24\rangle[42]\big)$\\
+15 terms} \\
\hline
$(+,0,0,0)$ & 
\Ampfour{+}{0}{0}{0}{\bosflip{1.37}{135}{brown}{cyan}}{\bos{i2}{brown}}{\bos{i3}{brown}}{\bos{i4}{brown}}+7 diagrams & 
\makecell{
$\frac{(\mathbf{f}^{W^+ W^- Z})^2}{m_1 \mathbf m_2 \mathbf m_3 \mathbf m_4}\times\big(\langle\eta_12\rangle[21]\langle34\rangle[34]$\\
$+\langle\eta_14\rangle[41]\langle23\rangle[32]-2\langle\eta_13\rangle[31]\langle24\rangle[42]\big)$\\
+7 terms} \\
\hline
$(+,-,0,0)$ & 
\Ampfour{+}{-}{0}{0}{\bosflip{1.37}{135}{brown}{cyan}}{\bosflip{1.37}{-135}{brown}{red}}{\bos{i3}{brown}}{\bos{i4}{brown}}+3 diagrams & 
\makecell{
$\frac{(\mathbf{f}^{W^+ W^- Z})^2}{m_1 \tilde m_2 \mathbf m_3 \mathbf m_4}\times\big(\langle\eta_12\rangle[\eta_21]\langle34\rangle[34]$\\
$+\langle\eta_14\rangle[41]\langle23\rangle[3\eta_2]-2\langle\eta_13\rangle[31]\langle24\rangle[4\eta_2]\big)$\\
+3 terms} \\
\hline
$(+,-,+,0)$ & 
\Ampfour{+}{-}{+}{0}{\bosflip{1.37}{135}{brown}{cyan}}{\bosflip{1.37}{-135}{brown}{red}}{\bosflip{1.37}{-45}{brown}{cyan}}{\bos{i4}{brown}}+1 diagram & 
\makecell{
$\frac{(\mathbf{f}^{W^+ W^- Z})^2}{m_1 \tilde m_2 m_3 \mathbf m_4}\times\big(\langle\eta_12\rangle[\eta_21]\langle\eta_34\rangle[34]$\\
$+\langle\eta_14\rangle[41]\langle2\eta_3\rangle[3\eta_2]-2\langle\eta_1\eta_3\rangle[31]\langle24\rangle[4\eta_2]\big)$\\
+1 term} \\
\hline
$(+,-,+,-)$ & 
\Ampfour{+}{-}{+}{-}{\bosflip{1.37}{135}{brown}{cyan}}{\bosflip{1.37}{-135}{brown}{red}}{\bosflip{1.37}{-45}{brown}{cyan}}{\bosflip{1.37}{45}{brown}{red}} & 
\makecell{
$\frac{(\mathbf{f}^{W^+ W^- Z})^2}{ m_1 \tilde m_2 m_3 \tilde m_4}\times\big(\langle\eta_12\rangle[\eta_21]\langle\eta_34\rangle[3\eta_4]$\\
$+\langle\eta_14\rangle[\eta_41]\langle2\eta_3\rangle[3\eta_2]-2\langle\eta_1\eta_3\rangle[31]\langle24\rangle[\eta_4\eta_2]\big)$} \\
\hline
\end{tabular}
\end{equation}
Note for each helicity category, there exist several diagrams with increasing double chirality flip, until reaching eight chirality flips for each diagram. Furthermore, the primary and up-to-third descendant diagrams do not contribute to the SM due to unitarity cancellation, and only starting from the 4th descendant contact amplitude, there is a massless-massive matching in the SM, such as
\begin{eqnarray}
    \begin{tikzpicture}[baseline=0.8cm] \begin{feynhand}
\vertex [particle] (i1) at (0,0) {$2^+$};
\vertex [particle] (i2) at (0,1.6) {$1^-$};
\vertex [particle] (i3) at (1.6,1.6) {$4^+$};
\vertex [particle] (i4) at (1.6,0) {$3^-$};
\vertex (v1) at (0.8,0.8);
\graph{(i1) --[bos] (v1)--[bos] (i2)};
\graph{(i4) --[bos] (v1)--[bos] (i3)};
\end{feynhand} \end{tikzpicture} \Rightarrow \Ampfour{+}{-}{+}{-}{\bosflip{1.37}{135}{brown}{cyan}}{\bosflip{1.37}{-135}{brown}{red}}{\bosflip{1.37}{-45}{brown}{cyan}}{\bosflip{1.37}{45}{brown}{red}}. 
\end{eqnarray}
Again, all other primary amplitude than the $(0,0,0,0)$ helicity category, such as
\begin{eqnarray}
    \Ampfour{+}{-}{+}{-}{\bos{i1}{cyan}}{\bos{i2}{red}}{\bos{i3}{cyan}}{\bos{i4}{red}}.
\end{eqnarray}
belong to EFT.

\paragraph{$SSSS$ contact amplitude}
The massless amplitude is
\begin{equation} \begin{aligned}
\begin{tikzpicture}[baseline=0.8cm] \begin{feynhand}
\vertex [particle] (i1) at (0,0) {$2^0$};
\vertex [particle] (i2) at (0,1.6) {$1^0$};
\vertex [particle] (i3) at (1.6,1.6) {$4^0$};
\vertex [particle] (i4) at (1.6,0) {$3^0$};
\vertex (v1) at (0.8,0.8);
\graph{(i1) --[sca] (v1)--[sca] (i2)};
\graph{(i4) --[sca] (v1)--[sca] (i3)};
\end{feynhand} \end{tikzpicture}=-4\lambda 
\begin{pmatrix}
\delta^{(i_1}_{i_3} \delta^{i_2)}_{i_4}\\
\delta^{(i_1}_{i_2} \delta^{i_3)}_{i_4}\\
\delta^{(i_1}_{i_2} \delta^{i_4)}_{i_3}\\
\delta^{(i_2}_{i_1} \delta^{i_3)}_{i_4}\\
\delta^{(i_2}_{i_1} \delta^{i_4)}_{i_3}\\
\delta^{(i_3}_{i_1} \delta^{i_4)}_{i_2}\\
\end{pmatrix}.
\end{aligned} \end{equation}
Multiplying with the transformation matices, the coefficient becomes
\begin{equation}
-4\lambda 
\begin{pmatrix}
\delta^{(i_1}_{i_3} \delta^{i_2)}_{i_4}\\
\delta^{(i_1}_{i_2} \delta^{i_3)}_{i_4}\\
\delta^{(i_1}_{i_2} \delta^{i_4)}_{i_3}\\
\delta^{(i_2}_{i_1} \delta^{i_3)}_{i_4}\\
\delta^{(i_2}_{i_1} \delta^{i_4)}_{i_3}\\
\delta^{(i_3}_{i_1} \delta^{i_4)}_{i_2}\\
\end{pmatrix}^T
\begin{pmatrix}
\mathcal{U}_{i_1}^h \mathcal{U}_{i_2}^h \mathcal{U}^{i_3 h} \mathcal{U}^{i_4 h}\\
\mathcal{U}_{i_1}^h \mathcal{U}^{i_2 h} \mathcal{U}_{i_3}^h \mathcal{U}^{i_4 h}\\
\mathcal{U}_{i_1}^h \mathcal{U}^{i_2 h} \mathcal{U}^{i_3 h} \mathcal{U}_{i_4}^h\\
\mathcal{U}^{i_1 h} \mathcal{U}_{i_2}^h \mathcal{U}_{i_3}^h  \mathcal{U}^{i_4 h}\\
\mathcal{U}^{i_1 h} \mathcal{U}_{i_2}^h \mathcal{U}^{i_3 h} \mathcal{U}_{i_4}^h  \\
\mathcal{U}^{i_1 h} \mathcal{U}^{i_2 h} \mathcal{U}_{i_3}^h \mathcal{U}_{i_4}^h  \\
\end{pmatrix}=-6\lambda .
\end{equation}
Therefore, the massive contact amplitude is
\begin{equation}
\Ampfour{0}{0}{0}{0}{\sca{i1}}{\sca{i2}}{\sca{i3}}{\sca{i4}}
=-6\lambda .
\end{equation}

\section{Constructive Amplitudes in 3 Ways: $WW \to hh$ as Example}
\label{sec:constructive}

In this section, we will perform systematic construction of higher-point MHC amplitudes from the contact MHC amplitudes in the SM. Two kinds of on-shell constructive methods, the bootstrap, and the recursive relation, are proposed for the massless amplitude construction. We will show that both methods can be generalized to the MHC amplitudes and the AHH massive amplitudes.

The building blocks for such construction are the contact MHC amplitudes in the light-cone gauge, containing both the 3-point MHC amplitudes and 4-point contact MHC amplitudes~\footnote{In the SM, there is no need to introduce 5-point contact MHC amplitudes and higher, all belonging to EFT, to obtain higher-point factorized MHC amplitudes. }. Therefore in the bootstrap method, the 4-point amplitudes, would have four contributions
\begin{equation} \begin{aligned}
\mathcal{M}=\mathcal{M}_{(12)}+\mathcal{M}_{(13)}+\mathcal{M}_{(14)}+\mathcal{M}_{\text{ct}}.
\end{aligned} \end{equation}
The first three terms can be obtained by gluing 3-pt amplitudes, while the last contact term is given in previous section. Note that in this section, only leading MHC amplitudes are needed for bootstrap and BCFW construction.

In this section, we will show how to construct higher-point massive amplitudes in three different ways using the same process $WW \to hh$ in the SM:
\begin{itemize}
    \item MHC bootstrap: list all relevant 3-point leading MHC amplitudes and 4-point contact MHC amplitudes, then glue them to obtain 4-point factorized amplitudes. Different from the previous section, we can use the leading MHC contact amplitudes to determine the leading higher point MHC amplitudes, and all sub-leading one can be neglected. 
    \item AHH bootstrap: list all relevant 3-point AHH amplitudes and 4-point contact AHH amplitudes, then glue them to obtain 4-point factorized AHH amplitudes. 
    \item MHC recursion: list only the 3-point leading MHC amplitudes in the light-cone gauge, then use BCFW recursive relation to obtain 4-point leading factorized amplitudes. Note that compare to the massless BCFW, now we can avoid the $\lambda \phi^4$ interaction in the BCFW construction for the MHC amplitudes due to the little group covariance. 
\end{itemize}

\subsection{Bootstrapping MHC amplitudes}

For the MHC amplitudes $WW \to hh$, because of little group covariance, we can focus on one helicity category, such as $(+,-,0,0)$. Once the result for such helicity is obtained, other helicity categories can be obtained by extended little group covariance.

First let us list the relevant 3-pt amplitudes at the leading order. The 3-pt leading MHC amplitudes are
\begin{align}
\begin{tikzpicture}[baseline=-0.1cm] \begin{feynhand}
\setlength{\feynhandarrowsize}{4pt}
\vertex [particle] (i1) at (-0.827,0.579) {$1^+$}; 
\vertex [particle] (i2) at (0.827,0.579) {$4^0$}; 
\vertex [particle] (i3) at (0,-1.01) {$P^0$}; 
\vertex (v1) at (0,0);
\vertex (v2) at (-0.827*0.35,0.579*0.35);
\draw[brown,thick sca] (i2)--(v1);
\draw[brown,thick bos] (i3)--(v1);
\draw[cyan,thick bos] (i1)--(v2);
\draw[brown,thick bos] (v2)--(v1);
\draw[very thick] plot[mark=x,mark size=3.5,mark options={rotate=45}] coordinates {(v2)};
\end{feynhand} \end{tikzpicture} &=\mathbf{g}^{W^+ W^-}\frac{\langle\eta_1 P_{14}\rangle[P_{14} 1]}{m_1}, \label{eq:WWh_sub1}\\
\begin{tikzpicture}[baseline=-0.1cm] \begin{feynhand}
\setlength{\feynhandarrowsize}{4pt}
\vertex [particle] (i1) at (-0.827,-0.579) {$2^-$}; 
\vertex [particle] (i2) at (0.827,-0.579) {$3^0$}; 
\vertex [particle] (i3) at (0,1.01) {$P^0$}; 
\vertex (v1) at (0,0);
\vertex (v2) at (-0.827*0.35,-0.579*0.35);
\draw[brown,thick sca] (i2)--(v1);
\draw[brown,thick bos] (i3)--(v1);
\draw[red,thick bos] (i1)--(v2);
\draw[brown,thick bos] (v2)--(v1);
\draw[very thick] plot[mark=x,mark size=3.5,mark options={rotate=45}] coordinates {(v2)};
\end{feynhand} \end{tikzpicture}
&=\mathbf{g}^{W^+ W^-}\frac{\langle2 P_{14}\rangle[P_{14} \eta_2]}{\tilde m_2}.&  \label{eq:WWh_sub2}
\end{align}
These building blocks can be glued to 4-pt amplitudes in three channels. For the (14)-channel, the internal particle is a W boson. Gluing these two $WWh$ amplitude we can get
\begin{equation} \begin{aligned} 
\begin{tikzpicture}[baseline=1.1cm] \begin{feynhand}
\vertex [particle] (i1) at (0,2.2) {$1^+$};
\vertex [particle] (i2) at (0,0) {$2^-$};
\vertex [particle] (i3) at (1.8,0) {$3^0$};
\vertex [particle] (i4) at (1.8,2.2) {$4^0$};
\vertex (v1) at (0.9,0.6);
\vertex (v2) at (0.9,1.6);
\vertex (v4) at (0.9-0.9*0.4,1.6+0.6*0.4);
\vertex (v5) at (0.9-0.9*0.4,0.6-0.6*0.4);
\draw[cyan,thick bos] (i1)--(v4);
\draw[brown,thick bos] (v4)--(v2);
\draw[brown,thick bos] (v2)--(v1);
\draw[brown,thick bos] (v5)--(v1);
\draw[red,thick bos] (i2)--(v5);
\draw[brown,thick sca] (i3)--(v1);
\draw[brown,thick sca] (i4)--(v2);
\draw[very thick] plot[mark=x,mark size=3.5,mark options={rotate=45}] coordinates {(v4)};
\draw[very thick] plot[mark=x,mark size=3.5,mark options={rotate=45}] coordinates {(v5)};
\end{feynhand} \end{tikzpicture}=(\mathbf{g}^{W^+ W^-})^2\frac{\langle\eta_1|P_{14}|1]\langle2|P_{14}|\eta_2]}{s_{14}m_1 \tilde m_2}.
\end{aligned} \end{equation}

Similarly, we can obtain the contribution of (13)-channel,
\begin{equation} \begin{aligned} 
\begin{tikzpicture}[baseline=1.1cm] \begin{feynhand}
\vertex [particle] (i1) at (0,2.2) {$1^+$};
\vertex [particle] (i2) at (0,0) {$2^-$};
\vertex [particle] (i3) at (1.8,0) {$3^0$};
\vertex [particle] (i4) at (1.8,2.2) {$4^0$};
\vertex (v1) at (0.9,0.6);
\vertex (v2) at (0.9,1.6);
\vertex (v4) at (0.9-0.9*0.4,1.6+0.6*0.4);
\vertex (v5) at (0.9-0.9*0.4,0.6-0.6*0.4);
\draw[cyan,thick bos] (i1)--(v4);
\draw[brown,thick bos] (v4)--(v2);
\draw[brown,thick bos] (v5)--(v1);
\draw[red,thick bos] (i2)--(v5);
\draw[brown,thick sca] (i3)--(v2);
\draw[brown,thick sca, top] (i4)--(v1);
\draw[brown,thick bos] (v2)--(v1);
\draw[very thick] plot[mark=x,mark size=3.5,mark options={rotate=45}] coordinates {(v4)};
\draw[very thick] plot[mark=x,mark size=3.5,mark options={rotate=45}] coordinates {(v5)};
\end{feynhand} \end{tikzpicture}=&(\mathbf{g}^{W^+ W^-})^2
\frac{\langle\eta_1|P_{13}|1]\langle2|P_{13}|\eta_2]}{s_{13}m_1 \tilde m_2}. 
\end{aligned} \end{equation}
It is equivalent to exchange particles 3 and 4 to the result of (14)-channel.
For (12)-channel, the internal particle is a Higgs boson. We glue $WWh$ and $hhh$ amplitude and obtain 
\begin{equation} \begin{aligned}
\begin{tikzpicture}[baseline=0.8cm] \begin{feynhand}
\vertex [particle] (i1) at (0,1.8) {$1^+$};
\vertex [particle] (i2) at (0,0) {$2^-$};
\vertex [particle] (i3) at (1.8,0) {$3^0$};
\vertex [particle] (i4) at (1.8,1.8) {$4^0$};
\vertex (v1) at (0.6,0.9);
\vertex (v2) at (1.2,0.9);
\vertex (v4) at (0.6-0.6*0.4,0.9+0.9*0.4);
\vertex (v5) at (0.6-0.6*0.4,0.9-0.9*0.4);
\draw[cyan,thick bos] (i1)--(v4);
\draw[brown,thick bos] (v4)--(v1);
\draw[red,thick bos] (i2)--(v5);
\draw[brown,thick bos] (v5)--(v1);
\draw[brown,thick sca] (v1)--(v2);
\draw[brown,thick sca] (i4)--(v2)--(i3);
\draw[very thick] plot[mark=x,mark size=3.5,mark options={rotate=45}] coordinates {(v4)};
\draw[very thick] plot[mark=x,mark size=3.5,mark options={rotate=45}] coordinates {(v5)};
\end{feynhand} \end{tikzpicture}=
\frac{\mathbf{g}^{W^+ W^-} \lambda_3}{\mathbf{m}_W^2}\frac{[12]\langle21\rangle}{s_{12} m_1 \tilde m_2}.
\end{aligned} \end{equation}

The contact term is already derived as
\begin{equation}
\begin{tikzpicture}[baseline=0.8cm] \begin{feynhand}
\vertex [particle] (i1) at (0,1.8) {$1^+$};
\vertex [particle] (i2) at (0,0) {$2^-$};
\vertex [particle] (i3) at (1.8,0) {$3^0$};
\vertex [particle] (i4) at (1.8,1.8) {$4^0$};
\vertex (v3) at (0.9,0.9);
\vertex (v4) at (0.9-0.9*0.43,0.9+0.9*0.43);
\vertex (v5) at (0.9-0.9*0.43,0.9-0.9*0.43);
\draw[cyan,thick bos] (i1)--(v4);
\draw[brown,thick bos] (v4)--(v3);
\draw[brown,thick bos] (v5)--(v3);
\draw[red,thick bos] (i2)--(v5);
\draw[brown,thick sca] (i3)--(v3);
\draw[brown,thick sca] (i4)--(v3);
\draw[very thick] plot[mark=x,mark size=3.5,mark options={rotate=45}] coordinates {(v4)};
\draw[very thick] plot[mark=x,mark size=3.5,mark options={rotate=45}] coordinates {(v5)};
\end{feynhand} \end{tikzpicture}
=(\mathbf{g}^{W^+ W^-})^2\frac{\langle\eta_1 2\rangle[\eta_21]}{m_1 \tilde m_2}.
\end{equation}

Combining all these contributions, we obtain 
\begin{equation} \begin{aligned}
&(\mathbf{g}^{W^+ W^-})^2
\frac{\langle\eta_1|P_{14}|1]\langle2|P_{14}|\eta_2]}{s_{14}m_1 \tilde m_2} -(\mathbf{g}^{W^+ W^-})^2
\frac{\langle\eta_1|P_{13}|1]\langle2|P_{13}|\eta_2]}{s_{13}m_1 \tilde m_2} \\
&+\frac{\mathbf{g}^{W^+ W^-} \lambda_3}{\mathbf{m}_W^2}\frac{\langle\eta_1 2\rangle[\eta_21]}{s_{12} m_1 \tilde m_2}
+(\mathbf{g}^{W^+ W^-})^2\frac{\langle\eta_1 2\rangle[\eta_21]}{m_1 \tilde m_2}.
\end{aligned} \end{equation}
There are other contributions from the subleading diagrams with more cross on the internal line. They can be obtained by acting $J^+ J^-$ on the particle line, which we introduced in subsection~\ref{sec:glue_MHC}. For (13)- and (14)-channel, the result is equalvalent to the replacement $p_{\dot{\alpha}_2}^{\beta_1}p_{\alpha_1}^{\dot{\beta}_2}\to -2\mathbf{m}^2\delta_{\alpha_1}^{\beta_1}\delta_{\dot{\alpha}_2}^{\dot{\beta}_2}+(p+\eta)_{\dot{\alpha}_2}^{\beta_1}(p+\eta)_{\alpha_1}^{\dot{\beta}_2}$.

Finally, we derive the total $WWhh$ amplitude,
\begin{equation} \begin{aligned}
\mathcal{M}(1^{+1},2^{-1},3^0,4^0)=
&(\mathbf{g}^{W^+ W^-})^2
\frac{-2\mathbf m^2\langle\eta_1 2\rangle[\eta_21]+\langle\eta_1|P_{14}+\eta_{14}|1]\langle2|P_{14}+\eta_{14}|\eta_2]}{s_{14}m_1 \tilde m_2} \\
&(\mathbf{g}^{W^+ W^-})^2
\frac{-2\mathbf m^2\langle\eta_1 2\rangle[\eta_21]+\langle\eta_1|P_{13}+\eta_{13}|1]\langle2|P_{13}+\eta_{13}|\eta_2]}{s_{13}m_1 \tilde m_2} \\
&+\frac{\mathbf{g}^{W^+ W^-} \lambda_3}{\mathbf{m}_W^2}\frac{\langle\eta_1 2\rangle[\eta_21]}{s_{12} m_1 \tilde m_2}
+(\mathbf{g}^{W^+ W^-})^2\frac{\langle\eta_1 2\rangle[\eta_21]}{m_1 \tilde m_2}.
\end{aligned} \end{equation}

The other 4-pt MHC amplitude can be obtained in a similar way. For instance, the four-vector scattering amplitude is provided in Appendix~\ref{app:4V}. In the SM, the MHC amplitude with more than four external particles do not have new contact term, so we can always glue the lower-point amplitude to construct the  any-point MHC amplitude.

\subsection{Bootstrapping AHH amplitudes}

Similar bootstrap can be applied to the AHH amplitudes. In this case, the building blocks are the 3-pt and 4-pt contact AHH amplitudes, which are obtained by bolding the 3-pt and 4-pt contact MHC amplitudes. The difference from literature is that the 4-pt contact AHH amplitudes should be added in the bootstrap.

We again take the $WWhh$ amplitude as an example to illustrate the bootstrap procedure. The relevant 3-pt contact amplitudes are
\begin{equation} \begin{aligned} 
\begin{tikzpicture}[baseline=0.7cm] \begin{feynhand}
\setlength{\feynhandarrowsize}{3.5pt}
\vertex [particle] (i1) at (0,0.8) {$1$}; 
\vertex [particle] (i2) at (1.6,1.6) {$2$}; 
\vertex [particle] (i3) at (1.6,0) {$3$};  
\vertex (v1) at (0.9,0.8); 
\draw[thick bos] (i1)--(v1);
\draw[thick bos] (i2)--(v1);
\draw[thick sca] (i3)--(v1);
\end{feynhand} \end{tikzpicture}&=
\mathbf M(\mathbf{1}^1,\mathbf{2}^1,\mathbf{3}^1)=\frac{\mathbf{g}^{W^+ W^-}}{\mathbf m_W}\langle\mathbf{12}\rangle[\mathbf{21}],\\
\begin{tikzpicture}[baseline=0.7cm] \begin{feynhand}
\setlength{\feynhandarrowsize}{3.5pt}
\vertex [particle] (i1) at (0,0.8) {$1$}; 
\vertex [particle] (i2) at (1.6,1.6) {$2$}; 
\vertex [particle] (i3) at (1.6,0) {$3$};  
\vertex (v1) at (0.9,0.8); 
\draw[thick sca] (i1)--(v1);
\draw[thick sca] (i2)--(v1);
\draw[thick sca] (i3)--(v1);
\end{feynhand} \end{tikzpicture}&=
\mathbf M(\mathbf{1}^0,\mathbf{2}^0,\mathbf{3}^0)
=\lambda_3,
\end{aligned} \end{equation}
where the superscript represents the particle spin.

The factorized contributions, which correspond to the (12)-, (13)- and (14)-channels, are obtained by gluing two 3-pt AHH amplitudes. We first consider the (14)-channel, in which the internal particle is a $W$ boson. In this case, we glue two $WWh$ amplitudes,
\begin{equation} \begin{aligned} 
\begin{tikzpicture}[baseline=1.1cm] \begin{feynhand}
\vertex [particle] (i1) at (0,2.2) {$1$};
\vertex [particle] (i2) at (0,0) {$2$};
\vertex [particle] (i3) at (1.8,0) {$3$};
\vertex [particle] (i4) at (1.8,2.2) {$4$};
\vertex (v1) at (0.9,0.6);
\vertex (v2) at (0.9,1.6);
\draw[thick bos] (i1)--(v2);
\draw[thick bos] (v2)--(v1);
\draw[thick bos] (i2)--(v1);
\draw[thick sca] (i3)--(v1);
\draw[thick sca] (i4)--(v2);
\end{feynhand} \end{tikzpicture}=
\frac{\mathbf M(\mathbf{1}^1,\mathbf{4}^0,-\mathbf{P}_{14}^1)\otimes
\mathbf M(\mathbf{P}_{14}^1,\mathbf{2}^1,\mathbf{3}^0)}{\mathbf{P}_{14}^2-\mathbf m^2_W},
\end{aligned} \end{equation}
where $\otimes$ denotes the contraction of little group indices of two external particles $\mathbf P_{14}$ and $-\mathbf P_{14}$. Writting the little group indices explictly, this contraction gives
\begin{equation} \begin{aligned}
|\mathbf P]^{\dot\alpha}|\mathbf P\rangle_{\alpha}\otimes[\mathbf P|_{\dot\beta}\langle\mathbf P|^{\beta}
=|\mathbf P]^{\dot\alpha(I_1}|\mathbf P\rangle^{I_2)}_{\alpha}\times [\mathbf P|_{\dot\beta I_1}\langle\mathbf P|_{I_2}^{\beta}
=\mathbf{m}^2\delta^{\dot\alpha}_{\dot\beta}\delta_{\alpha}^{\beta}-\frac12\mathbf{P}^{\dot\alpha\beta}\mathbf{P}_{\alpha\dot\beta}.
\end{aligned} \end{equation}
Thus, the (14)-channel amplitude should be
\begin{equation} \begin{aligned}
&\frac{\mathbf{g}^{W^+ W^-}\mathbf{g}^{W^+ W^-}}{\mathbf m_W^2}\langle\mathbf{1}\mathbf{P}_{14}\rangle[\mathbf{P}_{14} 1]\otimes\langle\mathbf{2}\mathbf{P}_{14}\rangle[\mathbf{P}_{14} 2]\\
=&\frac{\mathbf{g}^{W^+ W^-}\mathbf{g}^{W^+ W^-}}{2\mathbf m_W^2}
\frac{2\mathbf m^2_W\langle\mathbf{12}\rangle[\mathbf{21}]-\langle\mathbf{1}|\mathbf{P}_{14}|\mathbf{1}]\langle\mathbf{2}|\mathbf{P}_{14}|\mathbf{2}]}{\mathbf{P}_{14}^2-\mathbf m^2_W}.
\end{aligned} \end{equation}

Similarly, the (13)-channel amplitude is
\begin{equation} \begin{aligned} 
&\begin{tikzpicture}[baseline=1.1cm] \begin{feynhand}
\vertex [particle] (i1) at (0,2.2) {$1$};
\vertex [particle] (i2) at (0,0) {$2$};
\vertex [particle] (i3) at (1.8,0) {$3$};
\vertex [particle] (i4) at (1.8,2.2) {$4$};
\vertex (v1) at (0.9,0.6);
\vertex (v2) at (0.9,1.6);
\draw[thick bos] (i1)--(v2);
\draw[thick bos] (i2)--(v1);
\draw[thick sca] (i3)--(v2);
\draw[thick sca, top] (i4)--(v1);
\draw[thick bos] (v2)--(v1);
\end{feynhand} \end{tikzpicture}
=\mathbf{g}^{W^+ W^-}\mathbf{g}^{W^+ W^-}
\frac{\mathbf 2m^2_W\langle\mathbf{12}\rangle[\mathbf{21}]-\langle\mathbf{1}|\mathbf{P}_{13}|\mathbf{1}]\langle\mathbf{2}|\mathbf{P}_{13}|\mathbf{2}]}{\mathbf{P}_{13}^2-\mathbf m^2_W}.
\end{aligned} \end{equation}
In the (12)-channel we glue a $WWh$ amplitude with an $hhh$ amplitude. Here the internal particle is a massive scalar, which carries no little group spinor indices. Thus we have
\begin{equation} \begin{aligned}
\begin{tikzpicture}[baseline=0.8cm] \begin{feynhand}
\vertex [particle] (i1) at (0,1.8) {$1$};
\vertex [particle] (i2) at (0,0) {$2$};
\vertex [particle] (i3) at (1.8,0) {$3$};
\vertex [particle] (i4) at (1.8,1.8) {$4$};
\vertex (v1) at (0.6,0.9);
\vertex (v2) at (1.2,0.9);
\draw[thick bos] (i1)--(v1);
\draw[thick bos] (i2)--(v1);
\draw[thick sca] (v1)--(v2);
\draw[thick sca] (i4)--(v2)--(i3);
\end{feynhand} \end{tikzpicture}=
\frac{\mathbf{g}^{W^+ W^-} \lambda_3}{\mathbf{m}_W^2}\frac{[\mathbf{12}]\langle\mathbf{21}\rangle}{\mathbf{P}_{12}^2-\mathbf m^2_h}.
\end{aligned} \end{equation}
The final ingredient is the 4-pt contact amplitude,
\begin{equation}
\includegraphics[scale=1,valign=c]{image/AHH_VVSS_contact.pdf}
=\frac{\mathbf{g}^{W^+ W^-}\mathbf{g}^{W^+ W^-}}{2\mathbf m_W^2}\langle\mathbf{12}\rangle[\mathbf{21}].
\end{equation}

Putting everything together, we obtain the total AHH amplitude,
\begin{equation} \begin{aligned}
\mathbf{M}(1,2,3,4)=
&-\frac{\mathbf{g}^{W^+ W^-}\mathbf{g}^{W^+ W^-}}{2\mathbf m_W^2}
\frac{2\mathbf m^2_W\langle\mathbf{12}\rangle[\mathbf{21}]-\langle\mathbf{1}|\mathbf{P}_{14}|\mathbf{1}]\langle\mathbf{2}|\mathbf{P}_{14}|\mathbf{2}]}{\mathbf{P}_{14}^2-\mathbf m^2_W} \\
&-\frac{\mathbf{g}^{W^+ W^-}\mathbf{g}^{W^+ W^-}}{2\mathbf m_W^2}
\frac{2\mathbf m^2_W\langle\mathbf{12}\rangle[\mathbf{21}]-\langle\mathbf{1}|\mathbf{P}_{13}|\mathbf{1}]\langle\mathbf{2}|\mathbf{P}_{13}|\mathbf{2}]}{\mathbf{P}_{13}^2-\mathbf m^2_W} \\
&-\frac{\mathbf{g}^{W^+ W^-} \lambda_3}{\mathbf{m}_W^2}\frac{\langle\mathbf{12}\rangle[\mathbf{21}]}{\mathbf{P}_{12}^2-\mathbf m^2_h}
-\frac{\mathbf{g}^{W^+ W^-}\mathbf{g}^{W^+ W^-}}{2\mathbf m_W^2}\langle\mathbf{12}\rangle[\mathbf{21}].
\end{aligned} \end{equation}

\subsection{MHC recursive construction}

In addition to the bootstrap method, MHC amplitudes can also be constructed using recursion relations. In Ref.~\cite{Ema:2024vww,Ema:2024rss}, the authors introduced the All-Line Transverse (ALT) shift, which defines different momentum shifts for transverse and longitudinal mode of the spin-1 particles, and thus allows the recursive construction of AHH massive amplitudes. In the AHH amplitudes, different momentum shift for different mode seems strange. On the other hand, it becomes quite natural to consider different shift in the MHC amplitudes for different transversality. Here, we extend this approach to a variant form suitable for the MHC formalism. In the following, we will use the variant ALT shift to construct the $WWhh$ amplitude.

In the variant ALT shift, all external particle spinors are shifted by a complex parameter $z$. The specific shift for each particle depends on its helicity $h$. For a particle $i$ with helicity $h \neq 0$,
\begin{equation}
(h>0):\quad |\hat i\rangle=|i\rangle+z c_i|\eta_i\rangle,\qquad
(h<0):\quad |\hat i]=|i]+z c_i|\eta_i],
\end{equation}
where $c_i$ is a coefficient ensuring momentum conservation.  These coefficients will not appear in the final expression of a correctly constructed amplitude. These shifts leave the fermion and transverse vector states unchanged, consistent with the original ALT shift. The difference lies in the shift for the particle with helicity $h = 0$,
\begin{equation}
(h=0):\quad 
|\hat i\rangle =|i\rangle+z c_i|\eta_i]\\
\quad\text{ or }\quad
|\hat i] =|i]+z c_i|\eta_i].
\end{equation}
Unlike the original ALT shift, we do not require the longitudinal vector states to remain unchanged. Therefore, the scope of application for the variant ALT shift  differs: when an MHC amplitude involves a vector boson, we will only consider shifts for the helicity categories where the vector boson is transverse.

As an example, for the $WWhh$ amplitude, we consider helicity categories where both $W$ bosons are transversely polarized, such as $(+-00)$. A possible variant ALT shift is:
\begin{equation} \begin{aligned}
|\hat 1\rangle &=|1\rangle+z c_1|\eta_1\rangle,\\
|\hat 2] &=|2]+z c_2|\eta_2], \\
|\hat 3\rangle &=|3\rangle+z \frac{c_3}{2}|\eta_3\rangle,&\\
|\hat 4\rangle &=|4\rangle+z \frac{c_4}{2}|\eta_4\rangle.&
\end{aligned} \end{equation}
Under this shift, the momentum of each external particle becomes
\begin{equation}
\hat p_i=p_i+z q_i,\qquad i=1,2,3,4,
\end{equation}
where
\begin{equation} \begin{aligned} \label{eq:shift_momentum}
q_1&=c_1 |1]\langle \eta_1|,&
q_2&=c_2 |\eta_2]\langle 2|,&\\
q_3&=c_3 |3]\langle \eta_3|,&
q_4&=c_4 |4]\langle \eta_4|,&\\
\end{aligned} \end{equation}
and the $q_i$ satisfy momentum conservation
\begin{equation} \begin{aligned}
q_1+q_2+q_3+q_4=0.
\end{aligned} \end{equation}

In the MHC formalism, the momentum of internal particles is also decomposed into the large and small components. For an internal particle in the 4-pt amplitude, the large component is $P_{ij} \equiv (p_i + p_j)$ and the small component is $\eta_{ij}\equiv (\eta_i+\eta_j)$. Under the variant ALT shift, only the large component momentum is shifted,
\begin{align}
\hat P_{ij}&=\hat P_{ij}+z \hat Q_{ij},\\
\hat \eta_{ij}&=\eta_{ij},
\end{align}
where $Q_{1i}\equiv q_1+q_i$. The pole structure of the MHC amplitude consists of  $P_{1i}^2$ for $i=2,3,4$, which do not include $\eta_{1i}$. When a given pole goes on-shell, $\hat P_{1i}^2 = 0$, the shift parameter $z$ takes two solutions:
\begin{equation} \label{eq:z_solution}
\hat P^2_{1i}=0
\quad\to\quad
z^\pm_{1i}=\frac{1}{Q^2_{1i}}(-P_{1i}\cdot Q_{1i}\pm\sqrt{(P_{1i}\cdot Q_{1i})^2-P^2_{1i} Q^2_{1i}}).
\end{equation}
Using these solutions, the shifted pole structure can be written as  
\begin{equation}\label{eq:shift_pole}
\frac{1}{\hat P_{1i}^2}=\frac{1}{P_{1i}^2}\frac{z^+_{1i} z^-_{1i}}{(z-z^+_{1i})(z-z^-_{1i})}.
\end{equation}

Therefore, the MHC amplitude can be calculated from the residue of the shifted amplitude $\hat{\mathcal{M}}(z)$. By Cauchy's residue theorem, the physical amplitude $\hat{\mathcal{M}}(0)$ is determined by the other residues in the complex $z$-plane,
\begin{equation}
\hat{\mathcal{M}}(0)=\frac{1}{2\pi i}\oint \frac{\hat{\mathcal{M}}(z)}{z}=\sum_{i=2,3,4}\left(\text{Res}_{z=z^-_{1i}}\frac{\hat{\mathcal{M}}(z)}{z}+\text{Res}_{z=z^+_{1i}}\frac{\hat{\mathcal{M}}(z)}{z}\right)+B_\infty,
\end{equation}
where $\text{Res}_{z=z_{1i}^{\pm}}$ denotes the residue at finite $z$, and $B_\infty$ denotes the residue at infinity.

The existence of $B_\infty$ depends on the large-$z$ behavior of the amplitude. In the large-$z$ limit, a general amplitude $\hat{\mathcal{M}}(z)$ scales as $z^{\gamma}$. If $\gamma<0$, then $B_\infty=0$. In a renormalizable gauge theory, $B_\infty = 0$ can be ensured by using the Ward identity to analyze the large-$z$ behavior of the scattering amplitude. As an example, consider the $WWhh$ amplitude, which must have the form
\begin{equation}
\mathcal{M}(WWhh)\sim \frac{\varepsilon^2 (P^2+P\eta+\eta^2)}{P^2}+\varepsilon^2,
\end{equation}
where $\varepsilon$ is the polarization vector. The first term corresponds to a factorized term with pole structure $\frac{1}{P^2}$, while the second term corresponds  to a contact term. Here $P^2$, $P\eta$ and $\eta^2$ in the numerator should not be interpreted as inner products of momenta, but rather as direct products. In the large-$z$ limit, $\hat P$ is shifted while $\hat\eta$ and the polarization vector $\hat\epsilon$ remain unchanged. Therefore, the naive large-$z$ behavior of the $WWhh$ amplitude is 
\begin{equation}
\lim_{z\to\infty}\hat{\mathcal{M}}(z)\sim\left(\frac{\hat\epsilon^2 \hat P^2}{\hat P^2}+\hat\epsilon^2\right)\sim z^0.
\end{equation}
This behavior can be further suppressed by consider hidden gauge symmetry. Note that an amplitude involving vector boson can be decomposed as a polarization vector $\epsilon$ contracted with a structure $\Gamma$. As shown in Eq.~\eqref{eq:shift_momentum}, the polarization vector is proportional to the shift momentum $q$, so we have
\begin{equation}
\hat{\mathcal{M}}=\hat\epsilon\cdot \hat\Gamma
\sim q \cdot \hat\Gamma.
\end{equation}
In the large-$z$ limit, the shifted amplitude can be viewed as the amplitude in the high-energy limit, where gauge symmetry is restored, allowing us to use the Ward identity. It states that replacing the polarization vector with the shifted momentum $\hat p=p+zq$ makes the amplitude vanish:
\begin{equation}
(p+zq) \cdot \hat\Gamma=0
\quad\to\quad
q \cdot \Gamma=-\frac{1}{z} p\cdot \Gamma.
\end{equation}
This implies we substitute $q\to-\frac{1}{z} p$ to imporve the large-$z$ behavior. After applying the Ward identity, the large-$z$ bahavior of the $WWhh$ amplitude become $z^{-1}$. Thus, the residue at infinity vanishes, $B_\infty=0$.

The finite-$z$ residues factorize into products of subamplitudes $\mathcal{M}_L$ and $\mathcal{M}_R$. Using Eq.~\eqref{eq:shift_pole}, this factorization can be expressed as
\begin{equation} \begin{aligned} \label{eq:recursive_factorization}
\text{Res}_{z^+}\frac{\hat{\mathcal{M}}(z)}{z}&=-\sum_h\frac{1}{P^2}\frac{z^-}{z^+-z^-}\left.\left(\hat{\mathcal{M}}_L(\cdots,\hat P^h)\times \hat{\mathcal{M}}_R(P^{-h},\cdots)\right)\right|_{z=z^+},\\
\text{Res}_{z^-}\frac{\hat{\mathcal{M}}(z)}{z}&=\sum_h\frac{1}{\hat P^2}\frac{z^+}{z^+-z^-}\left.\left(\hat{\mathcal{M}}_L(\cdots,\hat P^h)\times \hat{\mathcal{M}}_R(P^{-h},\cdots)\right)\right|_{z=z^+},\\
\end{aligned} \end{equation}
where $\hat P$ represents the on-shell internal particle and $h$ denotes its helicity. For convenience, the subscript $1i$ is omitted. It shows that higher-point MHC amplitudes can be calculated from lower-point ones, providing a recursive construction.

Now we return to the $WWhh$ amplitude with helicity $(+-00)$ and use the recursive construction to compute it. The product of two subamplitudes $\mathcal{M}_L$ and $\mathcal{M}_L$ can be treated similarly to the gluing technique, yielding the shifted factorized amplitude without pole structure. For (13) and (14)-channel, we can obtain the numerator of shifted leading amplitude, 
\begin{equation} \begin{aligned} \label{eq:shift_14}
\begin{tikzpicture}[baseline=1.1cm] \begin{feynhand}
\vertex [particle] (i1) at (0,2.2) {$\hat 1^+$};
\vertex [particle] (i2) at (0,0) {$\hat 2^-$};
\vertex [particle] (i3) at (1.8,0) {$\hat 3^0$};
\vertex [particle] (i4) at (1.8,2.2) {$\hat 4^0$};
\vertex (v1) at (0.9,0.6);
\vertex (v2) at (0.9,1.6);
\vertex (v4) at (0.9-0.9*0.4,1.6+0.6*0.4);
\vertex (v5) at (0.9-0.9*0.4,0.6-0.6*0.4);
\draw[cyan,thick bos] (i1)--(v4);
\draw[brown,thick bos] (v4)--(v2);
\draw[brown,thick bos] (v2)--(v1);
\draw[brown,thick bos] (v5)--(v1);
\draw[red,thick bos] (i2)--(v5);
\draw[brown,thick sca] (i3)--(v1);
\draw[brown,thick sca] (i4)--(v2);
\draw[very thick] plot[mark=x,mark size=3.5,mark options={rotate=45}] coordinates {(v4)};
\draw[very thick] plot[mark=x,mark size=3.5,mark options={rotate=45}] coordinates {(v5)};
\end{feynhand} \end{tikzpicture}
&=\frac{\mathbf{g}^{W^+ W^-}\mathbf{g}^{W^+ W^-}}{2}\frac{\langle\eta_1|\hat P_{14}|1]\langle2|\hat P_{14}|\eta_2]}{\tilde{m}_1 m_2},\\
\begin{tikzpicture}[baseline=1.1cm] \begin{feynhand}
\vertex [particle] (i1) at (0,2.2) {$1^+$};
\vertex [particle] (i2) at (0,0) {$2^-$};
\vertex [particle] (i3) at (1.8,0) {$3^0$};
\vertex [particle] (i4) at (1.8,2.2) {$4^0$};
\vertex (v1) at (0.9,0.6);
\vertex (v2) at (0.9,1.6);
\vertex (v4) at (0.9-0.9*0.4,1.6+0.6*0.4);
\vertex (v5) at (0.9-0.9*0.4,0.6-0.6*0.4);
\draw[cyan,thick bos] (i1)--(v4);
\draw[brown,thick bos] (v4)--(v2);
\draw[brown,thick bos] (v5)--(v1);
\draw[red,thick bos] (i2)--(v5);
\draw[brown,thick sca] (i3)--(v2);
\draw[brown,thick sca, top] (i4)--(v1);
\draw[brown,thick bos] (v2)--(v1);
\draw[very thick] plot[mark=x,mark size=3.5,mark options={rotate=45}] coordinates {(v4)};
\draw[very thick] plot[mark=x,mark size=3.5,mark options={rotate=45}] coordinates {(v5)};
\end{feynhand} \end{tikzpicture}
&=\frac{\mathbf{g}^{W^+ W^-}\mathbf{g}^{W^+ W^-}}{2}\frac{\langle\eta_1|\hat P_{13}|1]\langle2|\hat P_{13}|\eta_2]}{\tilde{m}_1 m_2}.\\
\end{aligned} \end{equation}
Since the variant ALT shift does not change the external particle states, we only need to consider the shifted internal particle state for these two diagrams. In general, we write the internal state as $\hat P_{\alpha \dot{\alpha}} \hat P_{\beta\dot{\beta}}$. Using Eq.~\eqref{eq:recursive_factorization} but neglecting the pole structure, the sum of two contributions gives
\begin{equation} \begin{aligned}
\includegraphics[scale=1,valign=c]{image/internal_V_c00_0.pdf}
&\Rightarrow -\frac{z^-}{z^+-z^-}\left.\hat P_{\alpha \dot{\alpha}} \hat P_{\beta\dot{\beta}}\right|_{z=z^+}+\frac{z^+}{z^+-z^-}\left.\hat P_{\alpha \dot{\alpha}} \hat P_{\beta\dot{\beta}}\right|_{z=z^-}\\
&=-\frac{z^-}{z^+-z^-}(P+z^+ Q)_{\alpha \dot{\alpha}} (P+z^+ Q)_{\beta\dot{\beta}}+\frac{z^+}{z^+-z^-}(P+z^- Q)_{\alpha \dot{\alpha}} (P+z^- Q)_{\beta\dot{\beta}}\\
&=P_{\alpha \dot{\alpha}} P_{\beta\dot{\beta}}+z^+ z^- Q_{\alpha \dot{\alpha}} Q_{\beta\dot{\beta}}=P_{\alpha \dot{\alpha}} P_{\beta\dot{\beta}}+\frac{P^2}{Q^2} Q_{\alpha \dot{\alpha}} Q_{\beta\dot{\beta}}.\\
\end{aligned} \end{equation}
The last term $Q_{\alpha \dot{\alpha}} Q_{\beta\dot{\beta}}$ contributes to the contact term, and the value of $z^\pm$ is given in eq.~\eqref{eq:z_solution}. Only the quadratic term in $z^\pm$ survive, as the linear terms cancel when combining the residues at $z^+$ and $z^-$. 

Combining the leading contributions from the (13)- and (14)-channels and restoring the pole structure, we have
\begin{equation} \begin{aligned}
&\mathbf{g}^{W^+ W^-}\mathbf{g}^{W^+ W^-}\left(\frac{\langle\eta_1|P_{14}|1]\langle2|P_{14}|\eta_2]}{P_{14}^2 \tilde{m}_1 m_2}+\frac{\langle\eta_1|Q_{14}|1]\langle2|Q_{14}|\eta_2]}{Q_{14}^2 \tilde{m}_1 m_2}\right)+(3\leftrightarrow 4)\\
=&\mathbf{g}^{W^+ W^-}\mathbf{g}^{W^+ W^-}\left(\frac{\langle\eta_1|P_{14}|1]\langle2|P_{14}|\eta_2]}{P_{14}^2 \tilde{m}_1 m_2}+\frac{q_2\cdot q_4}{c_1 c_2}\right)+(3\leftrightarrow 4)\\
=&\mathbf{g}^{W^+ W^-}\mathbf{g}^{W^+ W^-}\left(\frac{\langle\eta_1|P_{14}|1]\langle2|P_{14}|\eta_2]}{P_{14}^2 \tilde{m}_1 m_2}+\frac{\langle\eta_1|P_{14}|1]\langle2|P_{14}|\eta_2]}{P_{14}^2 \tilde{m}_1 m_2}+\frac{\langle\eta_1 2\rangle[\eta_21]}{\tilde m_1 m_2}\right).\\
\end{aligned} \end{equation}
The last term is the contact term for the $WWhh$ amplitude. Here we use the fact that the shift momenta $q_1$ and $q_2$ are proportional to the transverse polarization vectors, as shown in eq.~\eqref{eq:shift_momentum}. 

For contributions involving subleading internal strucutres, their shift does not contribute to the contact amplitude. This can be verified by examining the following two shifted internal states:
\begin{equation} \begin{aligned}
\includegraphics[scale=1,valign=c]{image/internal_V_c00_2.pdf}
\Rightarrow&-\frac{z^-}{z^+-z^-}\left.(2\mathbf{m}^4 \epsilon_{\alpha\beta} \epsilon_{\dot{\alpha}\dot{\beta}}-\mathbf{m}^2 \hat P_{\alpha\dot{\alpha}} \hat \eta_{\beta\dot{\beta}}-\mathbf{m}^2\hat \eta_{\alpha\dot{\alpha}} \hat P_{\beta\dot{\beta}})\right|_{z=z^+}\\
&+\frac{z^+}{z^+-z^-}\left.(2\mathbf{m}^4 \epsilon_{\alpha\beta} \epsilon_{\dot{\alpha}\dot{\beta}}-\mathbf{m}^2 \hat P_{\alpha\dot{\alpha}} \hat \eta_{\beta\dot{\beta}}-\mathbf{m}^2\hat \eta_{\alpha\dot{\alpha}} \hat P_{\beta\dot{\beta}})\right|_{z=z^-}\\
=&2\mathbf{m}^4 \epsilon_{\alpha\beta} \epsilon_{\dot{\alpha}\dot{\beta}}-\mathbf{m}^2 P_{\alpha\dot{\alpha}} \eta_{\beta\dot{\beta}}-\mathbf{m}^2\eta_{\alpha\dot{\alpha}} P_{\beta\dot{\beta}},\\
\includegraphics[scale=1,valign=c]{image/internal_V_c00_4.pdf}
\Rightarrow& -\frac{z^-}{z^+-z^-}\left.\mathbf{m}^4\hat \eta_{\alpha \dot{\alpha}} \hat \eta_{\beta\dot{\beta}}\right|_{z=z^+}+\frac{z^+}{z^+-z^-}\left.\mathbf{m}^4 \hat \eta_{\alpha \dot{\alpha}} \hat \eta_{\beta\dot{\beta}}\right|_{z=z^-}=\mathbf{m}^4\eta_{\alpha \dot{\alpha}} \eta_{\beta\dot{\beta}}.
\end{aligned} \end{equation}
Both results coincide with the corresponding unshifted internal states, because they contain no quadratic terms in $z^\pm$.

For the (12)-channel, the internal particle is a Higgs boson. Its contribution also equals the unshifted gluing result:
\begin{equation} \begin{aligned}
\begin{tikzpicture}[baseline=0.8cm] \begin{feynhand}
\vertex [particle] (i1) at (0,1.8) {$\hat 1^+$};
\vertex [particle] (i2) at (0,0) {$\hat 2^-$};
\vertex [particle] (i3) at (1.8,0) {$\hat 3^0$};
\vertex [particle] (i4) at (1.8,1.8) {$\hat 4^0$};
\vertex (v1) at (0.6,0.9);
\vertex (v2) at (1.2,0.9);
\vertex (v4) at (0.6-0.6*0.4,0.9+0.9*0.4);
\vertex (v5) at (0.6-0.6*0.4,0.9-0.9*0.4);
\draw[cyan,thick bos] (i1)--(v4);
\draw[brown,thick bos] (v4)--(v1);
\draw[red,thick bos] (i2)--(v5);
\draw[brown,thick bos] (v5)--(v1);
\draw[brown,thick sca] (v1)--(v2);
\draw[brown,thick sca] (i4)--(v2)--(i3);
\draw[very thick] plot[mark=x,mark size=3.5,mark options={rotate=45}] coordinates {(v4)};
\draw[very thick] plot[mark=x,mark size=3.5,mark options={rotate=45}] coordinates {(v5)};
\end{feynhand} \end{tikzpicture}\Rightarrow
\frac{\mathbf{g}^{W^+ W^-} \lambda_3}{\mathbf{m}_W^2}\frac{[12]\langle21\rangle}{s_{12}}.
\end{aligned} \end{equation}

Combining contributions from all channels, we can use the ladder operators to generate the subleading amplitudes. The final result of the total amplitude is
\begin{equation} \begin{aligned}
\mathcal{M}=
&-\frac{\mathbf{g}^{W^+ W^-}\mathbf{g}^{W^+ W^-}}{2}
\frac{2\mathbf m^2\langle\eta_1 2\rangle[\eta_21]-\langle\eta_1|P_{14}+\eta_{14}|1]\langle2|P_{14}+\eta_{14}|\eta_2]}{s_{14}\tilde{m}_1 m_2} \\
&-\frac{\mathbf{g}^{W^+ W^-}\mathbf{g}^{W^+ W^-}}{2}
\frac{2\mathbf m^2\langle\eta_1 2\rangle[\eta_21]-\langle\eta_1|P_{13}+\eta_{13}|1]\langle2|P_{13}+\eta_{13}|\eta_2]}{s_{13}\tilde{m}_1 m_2} \\
&-\frac{\mathbf{g}^{W^+ W^-} \lambda_3}{\mathbf{m}_W}\frac{\langle\eta_1 2\rangle[\eta_21]}{s_{12}\tilde m_1 m_2}
-\frac{\mathbf{g}^{W^+ W^-}\mathbf{g}^{W^+ W^-}}{2}\frac{\langle\eta_1 2\rangle[\eta_21]}{\tilde m_1 m_2}. 
\end{aligned} \end{equation}

\section{Massless-Massive Matching at Leading and Sub-leading Orders}
\label{sec:constructmassiveamp}

In previous sections, since the contact 3-pt and 4-pt MHC amplitudes, are in the light-cone gauge, all the construction of higher-point amplitudes are in the light-cone gauge. In this section, we present an alternative way of constructing higher point amplitudes in the gauge-invariant way.

In this case, the higher-point massless amplitudes are first constructed, then match to the same point massive amplitudes. Let us still use $WW \to hh$ as example. Different from the previous sections, the 4-point massless $WWhh$ amplitudes is constructed by the massless bootstrap method in a gauge-invariant way, and then deform the gauge-invariant massless amplitudes to the MHC amplitudes, which is in the light-cone gauge with pole separation. This is essentially the matching procedure between massless and massive amplitudes for 4-point amplitudes.

For an $n$-point massive amplitude with spin category $\mathcal{S}=(s_1,\cdots,s_n)$, it can be decomposed into a sum of MHC amplitudes 
\begin{eqnarray}
\mathbf{M}(\mathbf{1}^{s_1}, \mathbf{2}^{s_2}, \cdots, \mathbf{n}^{s_n})
&\xrightarrow[]{\text{MHC}} &
\mathcal{M}\equiv\sum_{\mathcal{T} = \mathcal{H}} 
\mathcal{M}^{\mathcal{H},\mathcal{T}}.
\end{eqnarray}
In the high‑energy regime $E\gg v$, these MHC amplitudes can be expanded as
\begin{equation}
\mathcal{M}=\sum_{l} [\mathcal{M}]_l\sim E^{4-n}\left(\frac{v}{E}\right)^{l}.
\end{equation}

There is another expansion, which decomposes the amplitude into a primary part and descendant parts:
\begin{equation} \begin{aligned}
\mathcal{M}=
\mathcal{M}^{\text{pri}}
+\sum_{k}\mathcal{M}^{k\text{-th}},\\
\end{aligned} \end{equation}
where $\mathcal{M}^{\text{pri}}$ denotes the primary amplitude and $\mathcal{M}^{k\text{-th}}$ the $k$-th descendant.

The relation between these two expansions differs for 3-point and higher-point amplitudes:
\begin{itemize}
\item For a 3-point amplitude with total spin $s$, the two expansions are equivalent: all primary amplitudes have the same power counting. We can establish a one‑to‑one correspondence:
\begin{equation} \begin{aligned}
{[\mathcal{M}]}_{-s+1}&=\mathcal{M}^{\text{pri}},\\
[\mathcal{M}]_{-s+2}&=\mathcal{M}^{\text{1st}},\\
&\;\,\vdots\\
[\mathcal{M}]_{s+1}&=\mathcal{M}^{k_{\text{max}}\text{-th}}.\\
\end{aligned} \end{equation}

\item For higher-point amplitudes, internal particles also possess spin, which must be considered. Consequently, primary amplitudes with different internal particles can exhibit different power counting. This means a one‑to‑one correspondence cannot be established for the full amplitude. Instead, we must decompose the amplitude into distinct channels,
\begin{equation} \begin{aligned}
\mathcal{M}=\sum_{\mathcal{I}} \mathcal{M}_{\mathcal{S}_{\mathcal{I}}},\\
\end{aligned} \end{equation}
where $\mathcal{I}$ labels the channel and $\mathcal{S}_\mathcal{I}$ denotes the spin category of the internal particles. For each channel $\mathcal{I}$, the primary and descendant amplitudes scale as
\begin{equation} \begin{aligned}
\mathcal{M}_{\mathcal{S}_{\mathcal{I}}}^{\text{pri}}&\sim v^{4-n} \left(\frac{v}{E}\right)^{-s-2s_{\mathcal{I}}+2n_{\mathcal{I}}}, \\
\mathcal{M}_{\mathcal{S}_{\mathcal{I}}}^{k\text{-th}}&\sim v^{4-n} \left(\frac{v}{E}\right)^{k-s-2s_{\mathcal{I}}+2n_{\mathcal{I}}},
\end{aligned} \end{equation}
where $s$ is the total spin of external particles, while $n_{\mathcal{I}}$ and $s_{\mathcal{I}}$ are number and total spin of internal particles. Pole‑expansion contributions are not considered here.

\end{itemize}

To illustrate the latter case, consider the 4‑point $W^+W^-hh$ amplitude, which has total external spin $s=2$. It receives three types of contributions with different $s_{\mathcal{I}}$ and $n_{\mathcal{I}}$:
\begin{equation}
\begin{tabular}{c|c|c|c|c|c|c}
\hline
\multirow{2}{*}{channel} & internal & \multirow{2}{*}{$s_{\mathcal{I}}$} & \multirow{2}{*}{$n_{\mathcal{I}}$} & \multirow{2}{*}{$k_{\text{max}}$} & \multicolumn{2}{c}{power counting} \\
\cline{6-7}
 & particle &  &  &  & primary & highest order \\
\hline
(13)- and (14)-channel & $W$ boson & $1$ & $1$ & $8$ & $\left(\frac{E}{v}\right)^2$ & $\left(\frac{v}{E}\right)^{-6}$ \\
\hline
(12)-channel & Higgs boson & $0$ & $1$ & $4$ & $1$ & $\left(\frac{E}{v}\right)^{-4}$ \\
\hline
contact term & \mbox{-} & $0$ & $0$ & $4$ & $\left(\frac{E}{v}\right)^2$ & $\left(\frac{E}{v}\right)^{-2}$ \\
\hline
\end{tabular}
\end{equation}
Here $k_{\max}$ is the highest order of descendant amplitudes in each channel. The table shows that primary amplitudes from different channels do not share the same power counting; they belong to $[\mathcal{M}]_{-2}$ and $[\mathcal{M}]_{0}$. Thus we can relate $[\mathcal{M}]_l$ to  the priamry/descendnat ampltiude of each channel:
\begin{equation} \begin{aligned}
{[\mathcal{M}]}_{-2}&=\mathcal{M}^{\text{pri}}_{W}+\mathcal{M}^{\text{pri}}_{\text{ct}},\\
[\mathcal{M}]_{-1}&=\mathcal{M}^{\text{1st}}_{W}+\mathcal{M}^{\text{1st}}_{\text{ct}},\\
[\mathcal{M}]_{0}&=\mathcal{M}^{\text{2nd}}_{W}+\mathcal{M}^{\text{2nd}}_{\text{ct}}+\mathcal{M}^{\text{pri}}_{h},\\
&\;\,\vdots\\
[\mathcal{M}]_{6}&=\mathcal{M}^{\text{8th}}_{W},\\
\end{aligned} \end{equation}
where subscripts $W$, $h$ and $\text{ct}$ refer to channels with an internal $W$ boson, an internal Higgs boson, and the contact term.

An $(n+l)$-point massless amplitude $\mathcal{A}_{n+l}$ scales as $E^{4-n-l}$, so $[\mathcal{M}]_l$ should correspond to such a massless amplitude. This establishes a correspondence between an MHC amplitude with helicity $\mathcal{H}=(h_1,\dots,h_n)$ and a corresponding massless amplitude:
\begin{equation} 
\begin{aligned}
{[\mathcal{M}]_l}\Rightarrow \mathcal{A}^{\mathcal{H}^{l}}_{n+l},\quad l\ge 0,
\end{aligned} \label{eq:MMC} 
\end{equation}
where $\mathcal{H}^{l}=(h_1,\cdots,h_n;\overbrace{0,\cdots,0}^l)$ is the helicity category for the massless amplitude, in containing $l$ extra scalar bosons among the external particles. Now it is the time to reconsider the helicity information on both sides of the matching.

First consider the MHC side. The general procedure is to identify the helicity of the primary amplitude, denoted $\mathcal{H}^{\text{pri}}={h^{\text{pri}}i}$.
For any other helicity category, we define the helicity difference $h_d$ relative to the primary amplitude,
\begin{equation} \begin{aligned}
h_{d}=\sum_i|h^{\text{pri}}_i-h^{k\text{-th}}_i|.
\end{aligned} \end{equation} 
The $k$-th descendant amplitudes then only correspond to helicity categories with $h{d}\le k$, and $h_d$ must have the same parity as $k$. Explicitly,
\begin{equation} \begin{aligned} \label{eq:abc}
\mathcal{M}^{\text{pri}}:&\quad h_d=0,\\
\mathcal{M}^{\text{1st}}:&\quad h_d=1,\\
\mathcal{M}^{\text{2nd}}:&\quad h_d=2,0,\\
\mathcal{M}^{\text{3rd}}:&\quad h_d=3,1,\\
\mathcal{M}^{\text{4th}}:&\quad h_d=4,2,0,\\
&\quad \vdots 
\end{aligned} \end{equation}
Thus we find the helicity difference for MHC amplitude at each order. To see which helicity categories contribute for a given $h_d$, we can return to the $WWhh$ example. Here the primary amplitude for all structures has helicity $\mathcal{H}^{\text{pri}}=(0,0,0,0)$, so we find
\begin{equation} \begin{aligned} \label{eq:hd_helicity}
h_d=0:&\quad (0,0,0,0),\\
h_d=1:&\quad (\pm1,0,0,0),(0,\pm1,0,0),\\
h_d=2:&\quad (\pm1,\pm1,0,0),(\pm1,\mp1,0,0).\\
\end{aligned} \end{equation}
Since the primary amplitude with an internal Higgs boson has a different order from the others, we can list the MHC amplitudes for different channels into a table to distinguish them: 
\begin{equation} \label{eq:MHC_table}
\begin{tabular}{|c|c|c|c|c|c|c|}
\hline
\diagbox{$h_d$}{$l$} & $-2$ & $-1$ & $0$ & $1$ & $2$ & $\cdots$ \\
\hline
$0$ & \makecell{\color{gray} $\mathcal{M}^{\text{pri}}_{W},\mathcal{M}^{\text{pri}}_{\text{ct}}$\\ } & \mbox{-} & \makecell{$\mathcal{M}^{\text{2nd}}_{W},\mathcal{M}^{\text{2nd}}_{\text{ct}}$\\ $\mathcal{M}^{\text{pri}}_{h}$} & & \makecell{$\mathcal{M}^{\text{4th}}_{W},\mathcal{M}^{\text{4th}}_{\text{ct}}$\\ $\mathcal{M}^{\text{2nd}}_{h}$} & \\
\hline
$1$ & \mbox{-} & \makecell{\color{gray} $\mathcal{M}^{\text{1st}}_{W},\mathcal{M}^{\text{1st}}_{\text{ct}}$\\ } & \mbox{-} & \makecell{$\mathcal{M}^{\text{3rd}}_{W},\mathcal{M}^{\text{3rd}}_{\text{ct}}$\\ $\mathcal{M}^{\text{1st}}_{h}$} & & \\
\hline
$2$ & \mbox{-} & \mbox{-} & \makecell{$\mathcal{M}^{\text{2nd}}_{W},\mathcal{M}^{\text{2nd}}_{\text{ct}}$\\ }  & & \makecell{$\mathcal{M}^{\text{4th}}_{W},\mathcal{M}^{\text{4th}}_{\text{ct}}$\\ $\mathcal{M}^{\text{2nd}}_{h}$} & \\
\hline
\end{tabular}
\end{equation}
Entries in gray ($l<0$) do not correspond to massless amplitudes with a linear realization of the scalar field in the SM.

Now consider the helicity category of the $(4+l)$-point massless amplitude. To match the $WWhh$ amplitude, the helicity of first four massless particles are given in Eq.~\eqref{eq:hd_helicity}. Since the number of extra scalars is equal to the order $l$ in Eq.~\eqref{eq:MHC_table}, filling the massless amplitudes into an analogous table gives:
\begin{equation}
\begin{tabular}{|c|c|c|c|c|}
\hline
\diagbox{$h_d$}{$l$} & $0$ & $1$ & $2$ & $\cdots$ \\
\hline
$0$ & $\mathcal{A}^{(0,0,0,0)}$ & & $\mathcal{A}^{(0,0,0,0,0,0)}$ & \\ 
\hline
$1$ & & \makecell{$v\mathcal{A}^{(\pm1,0,0,0,0)}$\\$v\mathcal{A}^{(0,\pm1,0,0,0)}$} &  & \\
\hline
$2$ & $\mathcal{A}^{(\pm1,\mp1,0,0)}$ & & \makecell{$v^2\mathcal{A}^{(\pm1,\mp1,0,0,0,0)}$\\$v^2\mathcal{A}^{(\pm1,\pm1,0,0,0,0)}$}  & \\
\hline
\end{tabular}
\end{equation}

The helicity matching between the MHC amplitude and the UV massless amplitude setup a framework for massless-massive correspondence. This correspondence allows us to reorganize the MHC amplitudes from a top-down matching perspective. We will start with the primary matching, and then discuss matching at sub-leading order.

\subsection{Massless 4-point amplitudes: the gauge-invariant way}

In this subsection, we will construct the 4-point massless amplitudes from the gauge-invariant 3-point massless amplitudes. The building blocks are 3-pt massless amplitudes in the SM, which are given in section~\ref{sec:SMcontact}. Let us briefly review how the recursive and BCFW methods are used for the SM massless amplitudes. The two methods show their own advantages in the SM amplitude construction.

The BCFW recursion relations~\cite{Britto:2004ap,Britto:2005fq} is the most popular method to calculate the higher-point massless amplitudes in gauge theories, which is based on the analytic properties of scattering amplitudes. Consider a holomorphic function $\mathcal{A}(z)$, which is a complexification of the physical amplitude $\mathcal{A}(0)$ with a complex parameter $z$. Although the momenta become complex-valued, the momentum conservation and the on-shell condition are still valid. Therefore, the Cauchy's residue theorem show that
\begin{equation}
\mathcal{A}(0)=\frac{1}{2\pi i}\oint \frac{\mathcal{A}(z)}{z},
\end{equation}
where the poles in the complex plane correspond to the on-shell propagators except for the pole at origin, so $\mathcal{A}(0)$ can be derived from the sum of the residues. In the BCFW recursion relations, two momenta are shifted by to be shifted the complex parameter $z$ (2-line shift).

This method is quite successful to applied to gauge theory, gravity and supersymmetric theory, but fail in the electroweak massless amplitudes.
To show the problem, let us consider a theory includes scalar-QED and $\lambda\phi^4$ interaction, which is contained in the electroweak theory. The 4-scalar amplitude derived by the BCFW recursion relations and Feynman rules \cite{Elvang:2013cua} are given seperately by
\begin{equation} \begin{aligned}
\mathcal{A}^{\text{BCFW}}(\varphi \varphi^* \varphi \varphi^*)
&=\tilde{e}^2\frac{\langle13\rangle^2\langle24\rangle^2}{\langle12\rangle\langle23\rangle\langle34\rangle\langle41\rangle},\\
\mathcal{A}^{\text{Feynman}}(\varphi \varphi^* \varphi \varphi^*)
&=-\lambda+\tilde{e}^2\left(1+\frac{\langle13\rangle^2\langle24\rangle^2}{\langle12\rangle\langle23\rangle\langle34\rangle\langle41\rangle}\right).
\end{aligned} \end{equation}
Since the contact term $-\lambda+\tilde{e}^2$ don't have pole, it cannot be constructed by the simple recursion relations. Therefore, the 4-scalar amplitude should be an input to calculate the massless amplitudes with more than four legs. To calculate the general amplitudes, the generalizations~\cite{Risager:2005vk,Cohen:2010mi,Cheung:2015ota,Franken:2019wqr,Ballav:2020ese,Wu:2021nmq} of BCFW recursion relations were studied, which have more than two shifted momenta. For the SM, people found that the massless amplitudes are 3-line constructible~\cite{Cheung:2015ota}. However, when $n\ge3$, the $n$-line shifts is not convenient to construct the amplitudes. 

Therefore, another approach, known as the bootstrap method, is applied to construct higher-point amplitudes in the SM. The amplitude decomposes into a product of sub-amplitudes corresponding to two distinct processes with the simple pole structure in the massless amplitude
\begin{equation} 
\lim_{P^2\rightarrow 0} P^2 \mathcal{M} = M^L\times M^R,
\end{equation}
where $P^2$ is the on-shell propagators of the immediate massless particle, and the $M^L$ and $M^R$ are sub-amplitudes.
In this method, the final form of the factorized amplitudes should have the following form. The ansatz consists of the denominator $D$, the numerator $N$ and the group tensor $G$
\begin{equation} \label{eq:MasslessAnsatz}
\mathcal{A}=\sum_i G_i\frac{N_i}{D_i},
\end{equation}
where $G$, $D$ are known while $N$ need to be determined. A reasonable ansatz should satisfy the following requirements:
\begin{itemize}
    \item The group tensor $G$ usually form a group structure basis, and thus every group tensor appeared in the amplitude should be independent term.
    \item The denominator $D$ represents the pole structure of the massless amplitudes and must has the form 
\begin{equation}
D\equiv\prod_{i j\cdots} \left(p_i+p_j+\cdots\right)^2.
\end{equation}
\end{itemize}
All the terms in the ansatz should correspond to diagrams. 
Depending on whether the group tensors are independent, the terms in the ansatz may or may not correspond one-to-one with these diagrams. In the following, we will discuss these two cases separately and show how the ansatz can be used to obtain the massless amplitudes. We will construct the primary massless amplitudes in the $(+,-,0,0)$ and $(0,0,0,0)$ helicity categories, in the massless scalar QED and Yang-Mills theory.

\paragraph{$(-1,+1,0,0)$ helicity with non-independent group structure}

In the $WW \to hh$ amplitude, we choose the helicity category to be $(-1,+1,0,0)$. From the bootstrap, this 4-pt amplitude consists of the following three diagrams
\begin{equation} \label{eq:unbroken_diagram_2}
\begin{tikzpicture}[baseline=0.8cm] \begin{feynhand}
\setlength{\feynhandtopsep}{5pt}
\vertex [particle] (i1) at (0,0) {$2^+$};
\vertex [particle] (i2) at (0,1.6) {$1^-$};
\vertex [particle] (i3) at (1.6,1.6) {$4^0$};
\vertex [particle] (i4) at (1.6,0) {$3^0$};
\vertex (v1) at (0.8,0.4);
\vertex (v2) at (0.8,1.2);
\graph{(i2) --[bos] (v1)--[sca] (i4)};
\graph{(i1) --[bos,top] (v2)--[sca] (i3)};
\graph{(v1) --[sca] (v2)};
\end{feynhand} \end{tikzpicture}+
\begin{tikzpicture}[baseline=0.8cm] \begin{feynhand}
\vertex [particle] (i1) at (0,0) {$2^+$};
\vertex [particle] (i2) at (0,1.6) {$1^-$};
\vertex [particle] (i3) at (1.6,1.6) {$4^0$};
\vertex [particle] (i4) at (1.6,0) {$3^0$};
\vertex (v1) at (0.8,0.4);
\vertex (v2) at (0.8,1.2);
\graph{(i1) --[bos] (v1)--[sca] (i4)};
\graph{(i2) --[bos] (v2)--[sca] (i3)};
\graph{(v1) --[sca] (v2)};
\end{feynhand} \end{tikzpicture}+
\begin{tikzpicture}[baseline=0.8cm] \begin{feynhand}
\vertex [particle] (i1) at (0,0) {$2^+$};
\vertex [particle] (i2) at (0,1.6) {$1^-$};
\vertex [particle] (i3) at (1.6,1.6) {$4^0$};
\vertex [particle] (i4) at (1.6,0) {$3^0$};
\vertex (v1) at (0.4,0.8);
\vertex (v2) at (1.2,0.8);
\graph{(i1) --[bos] (v1)--[bos] (i2)};
\graph{(i4) --[sca] (v2)--[sca] (i3)};
\graph{(v1) --[bos] (v2)};
\end{feynhand} \end{tikzpicture}.
\end{equation}
In the three diagrams, we read the corresponding group tensors and denominators
\begin{equation} \begin{aligned}
G_1&=(T^{I_2}_s T^{I_1}_s)^{i_3}_{i_4},&
G_2&=(T^{I_1}_s T^{I_2}_s)^{i_3}_{i_4},&
G_3&=f^{J I_1 I_2 }(T^{J}_s)^{i_3}_{i_4},& \\
D_1&=s_{13},&
D_2&=s_{14},&
D_3&=s_{12}.& \\
\end{aligned} \end{equation} 
One can verify the these three group tensors are not independent. They satisfy the linear following relation
\begin{equation}
G_1-G_2-G_3=0.
\end{equation}

Let us choose two of them to get the ansatz and calculate the massless amplitude. Here we choose ${G_1, G_2}$ as an independent basis, in which case $G_3$ becomes a redundant group structure. The ansatz should therefore only contain the first two gauge tensors. Correspondingly, we define the new denominators $D^\prime$ to incorporate the pole structure $s_{12}$ associated with the redundant gauge tensor $G_3$
\begin{equation}
D_1,D_2,D_3\to
\begin{cases}
D'_1=D_1 D_3= s_{13}s_{12}, \\
D'_2=D_2 D_3= s_{14}s_{12}.
\end{cases}
\end{equation}
Therefore, the full ansatz for the $WWHH^{\dagger}$ amplitude can be written as
\begin{equation} \begin{aligned}
\mathcal{A}(1^{I_1},2^{I_2},3^{i_3},4_{i_4})
&=(T^{I_2}_s T^{I_1}_s)^{i_3}_{i_4}\frac{N_1}{s_{12}s_{13}}
+(T^{I_1}_s T^{I_2}_s)^{i_3}_{i_4}\frac{N_2}{s_{12}s_{14}}.
\end{aligned} \end{equation}
In the factorized limit, we have
\begin{equation} \begin{aligned} \label{eq:s13channel}
\lim_{P_{13}^2\rightarrow 0} P_{13}^2 \mathcal{A}
&=A^L\left(1^{-1,I_1},3^{0}_{i_3},(P_{13})^{0,j}\right)\times A^R\left(2^{+1,I_2},(-P_{13})^{0}_{j},4^{0,i_4}\right)\\
&=(T^{I_1}_s)^{i_3}_{j}\frac{\langle13\rangle\langle P_{13}1\rangle}{\langle 3 P_{13}\rangle}\times (T^{I_2}_s)^{j}_{i_4}\frac{[2(-P_{13})][42]}{[(-P_{13})4]}\\
&=(T^{I_2}_s T^{I_1}_s)^{i_3}_{i_4}\langle13\rangle[32] \frac{[24]}{[14]}=(T^{I_2}_s T^{I_1}_s)^{i_3}_{i_4}\langle13\rangle[32] \frac{[24]\langle41\rangle}{[14]\langle41\rangle}\\
&=(T^{I_2}_s T^{I_1}_s)^{i_3}_{i_4} \frac{\langle1|3|2]^2}{s_{12}},\\
\end{aligned} \end{equation}
where we use $\langle\cdot P_{13}\rangle[(-P_{13})\cdot]=-\langle\cdot |p_1+p_3|\cdot]$ and momentum conservation. For $N_1$, since $\langle1|4|2]=-\langle1|3|2]$, there is only one kinematic structure $\langle1|3|2]^2$ and the coefficient is determined by Eq.~\eqref{eq:s13channel}. Another numerator $N_2$ can be derived by the limit $P_{14}^2\rightarrow 0$. 

Thus, we find the result of the massless $WWHH^{\dagger}$ amplitude in the gauge-invariant form
\begin{equation} \label{eq:WWHH}
\mathcal{A}(1^{-1,I_1},2^{+1,I_2},3^{0,i_3},4^{0}_{i_4})
=(T^{I_2}_s T^{I_1}_s)^{i_3}_{i_4}\frac{\langle1|3|2]^2}{s_{12}s_{13}}
+(T^{I_1}_s T^{I_2}_s)^{i_3}_{i_4}\frac{\langle1|3|2]^2}{s_{12}s_{14}}.
\end{equation}

\paragraph{$(0,0,0,0)$ helicity with independent group structure}

Let us consider the four-scalar amplitude with helicity category $(0,0,0,0)$, in which the external particles do not include gauge bosons.  
Using the bootstrap, the 4-point amplitude involves three diagrams,
\begin{equation}
\begin{tikzpicture}[baseline=0.8cm] \begin{feynhand}
\vertex [particle] (i1) at (0,0) {$2^0$};
\vertex [particle] (i2) at (0,1.6) {$1^0$};
\vertex [particle] (i3) at (1.6,1.6) {$4^0$};
\vertex [particle] (i4) at (1.6,0) {$3^0$};
\vertex (v1) at (0.4,0.8);
\vertex (v2) at (1.2,0.8);
\graph{(i1) --[sca] (v1)--[sca] (i2)};
\graph{(i4) --[sca] (v2)--[sca] (i3)};
\graph{(v1) --[bos] (v2)};
\end{feynhand} \end{tikzpicture}+
\begin{tikzpicture}[baseline=0.8cm] \begin{feynhand}
\vertex [particle] (i1) at (0,0) {$2^0$};
\vertex [particle] (i2) at (0,1.6) {$1^0$};
\vertex [particle] (i3) at (1.6,1.6) {$4^0$};
\vertex [particle] (i4) at (1.6,0) {$3^0$};
\vertex (v1) at (0.8,0.4);
\vertex (v2) at (0.8,1.2);
\graph{(i1) --[sca] (v1)--[sca] (i4)};
\graph{(i2) --[sca] (v2)--[sca] (i3)};
\graph{(v1) --[bos] (v2)};
\end{feynhand} \end{tikzpicture}+
\begin{tikzpicture}[baseline=0.8cm] \begin{feynhand}
\vertex [particle] (i1) at (0,0) {$2^0$};
\vertex [particle] (i2) at (0,1.6) {$1^0$};
\vertex [particle] (i3) at (1.6,1.6) {$4^0$};
\vertex [particle] (i4) at (1.6,0) {$3^0$};
\vertex (v1) at (0.8,0.8);
\graph{(i1) --[sca] (v1)--[sca] (i2)};
\graph{(i4) --[sca] (v1)--[sca] (i3)};
\end{feynhand} \end{tikzpicture},
\end{equation}
where the script $0$ represent the helicity of the external particle. The corresponding group tensors and pole structures are given by:
\begin{equation} \begin{aligned}
G_1&=(T^{J}_s)^{i_1}_{i_2} (T^{J}_s)^{i_3}_{i_4},&
G_2&=(T^{J}_s)^{i_1}_{i_4} (T^{J}_s)^{i_3}_{i_2},&
G_3&=\lambda\delta^{(i_1}_{i_2} \delta^{i_3)}_{i_4}.\\
D_1&=s_{12},&
D_2&=s_{13},&
D_3&=s_{14}.
\end{aligned} \end{equation} 
The first two group tensors, $G_1$ and $G_2$, are simply products of the group tensor from the $VSS$ amplitude, denoted as $T_s$, while the last group tensor $G_3$ originates from the $\lambda\phi^4$ interaction. One can verify that $G_1$, $G_2$, and $G_3$ are independent. Therefore, the ansatz for the four-scalar amplitude can be written as
\begin{equation}
\mathcal{A}(1^{i_1},2_{i_2},3^{i_3},4_{i_4})
=(T^{J}_s)^{i_1}_{i_2} (T^{J}_s)^{i_3}_{i_4}\frac{N_1}{s_{12}}
+(T^{J}_s)^{i_1}_{i_4} (T^{J}_s)^{i_3}_{i_2}\frac{N_2}{s_{14}}+\lambda\delta^{(i_1}_{i_2} \delta^{i_3)}_{i_4}N_3,
\end{equation}
where $s_{ij}=(p_i+p_j)^2$ is the denominator.

To determine the expression for numerator $N_i$, we can consider the factorized limit, in which the corresponding propagator goes on-shell. Therefore, the unknown numerator can factorized into two lower-point amplitudes, 
\begin{equation} \label{eq:FactorizeMassless}
G_i N_i=\lim_{P^2\rightarrow 0} P^2 \mathcal{A}
=\sum_{h_P} A^L\times A^R,
\end{equation}
where $P$ and $h_P$ are the momentum and the helicity of the on-shell propagator, $A^L$ and $A^R$ are massless sub-amplitudes. The building blocks are the 3-pt and 4-scalar massless amplitudes, which are derived in section~\ref{sec:SMcontact}.  In Ref.~\cite{AccettulliHuber:2021uoa}, the authors listed all possible kinematic structures for $N$. Then the coefficients can be determined by Eq.~\eqref{eq:FactorizeMassless}. For simplicity, we will not list all kinematic structure in this subsection. Instead, we will use other argument to find the correct result. 

Return to the four-scalar amplitude. The unknown numerator $N_1$ can be obtained by the we use the the factorized limit $s_{12}\rightarrow0$. We have
\begin{equation} \begin{aligned}
\lim_{P_{12}^2\rightarrow 0} P_{12}^2 \mathcal{A}
&=A^L\left((P_{12})^{-1,J},1^{0,i_1},2^{0}_{i_2}\right)\times A^R\left((-P_{12})^{+1,J},(-P_{12})^{0}_{i_3},4^{0,i_4}\right)\\
&=(T^{J}_s)^{i_1}_{i_2}\frac{\langle1P_{12}\rangle\langle P_{12}2\rangle}{\langle 12\rangle}\times (T^{J}_s)^{i_3}_{i_4}\frac{[3(-P_{12})][(-P_{12})4]}{[34]}\\
&=-(T^{J}_s)^{i_1}_{i_2} (T^{J}_s)^{i_3}_{i_4}\frac{\langle1|P_{12}|3]\langle2|P_{34}|4]+\langle1|P_{12}|4]\langle2|P_{34}|3]}{2\langle 12\rangle[34]}\\
&=(T^{J}_s)^{i_1}_{i_2} (T^{J}_s)^{i_3}_{i_4}\frac{-[23]\langle23\rangle+[24]\langle24\rangle}{2},\\
\end{aligned} \end{equation}
where $P_{12}=p_1+p_2$. In the second line, we should convert $\langle\cdot P\rangle$ and $[P \cdot]$ to $\langle\cdot|P|\cdot]$. There are two ways to convert them, but we don't find any constraint between the two ways. Therefore, we average the two ways in the third line to give the numerator $N_1=\frac12(-[23]\langle23\rangle+[24]\langle24\rangle)$. The other numerator $N_2$ can be computed similarly in the limit $P_{14}^2\rightarrow0$. The last $N_3=-4$ are determined by Eq.~\eqref{eq:4H}.

Finally, we obtain the gauge-invariant four-scalar amplitude in the scalar QED 
\begin{equation} \begin{aligned} \label{eq:A4H}
\mathcal{A}(1^{i_1},2_{i_2},3^{i_3},4_{i_4})
=-(T^{J}_s)^{i_1}_{i_2} (T^{J}_s)^{i_3}_{i_4}\frac{s_{13}-s_{14}}{2s_{12}}
-(T^{J}_s)^{i_1}_{i_4} (T^{J}_s)^{i_3}_{i_2}\frac{s_{13}-s_{12}}{2s_{14}}-4\lambda\delta^{(i_1}_{i_2} \delta^{i_3)}_{i_4}.
\end{aligned} \end{equation}

\subsection{Scaling structure for MHC amplitudes}

Having obtained the gauge-invariant massless amplitudes, we would perform amplitude deformation to obtain massless amplitudes in the light-cone gauge, in which the pole separation should also be done. To guide the amplitude deformation, the scaling behavior of the MHC amplitudes should be known. Let us examine the scaling structure of the massless and MHC amplitudes.

For a general massless amplitude, each term can be characterized by a spinor scaling structure of the form
\begin{equation}
\mathbb A=\mathbb E(\lambda,\tilde\lambda)\times\mathbb I(p)\times \mathbb P,
\end{equation}
where $\mathbb E$ represents the external scaling, which depends on all external particles; $\mathbb I$ is the internal scaling, containing only massless momentum; and $\mathbb P$ denotes the pole structure. This schematic form omits coefficients, particle labels, and spinor indices, thus providing a broad outline of the amplitude's scaling behavior.

To illustrate this, consider the $HHHH$ amplitude obtained in the previous subsection. One of its terms has the spinor behavior
\begin{equation}
\frac{\langle1|3|2]^2}{s_{12}s_{13}}
\to 
\left\{\begin{aligned}
&\text{external scaling:}& E(\lambda^2\tilde\lambda^2)&\sim \lambda_1\tilde\lambda_2^2, \\
&\text{internal scaling:}& I(p^2)&\sim p_3^2, \\
&\text{pole scaling:}& \mathbb P(p^{-4})&\sim \frac{1}{s_{12}s_{13}}.
\end{aligned}\right.
\end{equation}
Similarly, another term in the $HHHH$ amplitude exhibits the same scaling structure
\begin{equation}
\frac{\langle1|3|2]^2}{s_{12}s_{14}}
\sim \mathbb E(\lambda^2\tilde\lambda^2)\times\mathbb I(p^2)\times \mathbb P(p^{-4}).
\end{equation}
In previous subsection, we also obtained the $WWHH$ amplitude. Its three terms display two distinct scaling behaviors
\begin{equation} \begin{aligned}
\frac{s_{13}-s_{14}}{2s_{12}},\frac{s_{13}-s_{12}}{2s_{14}}
&\sim \mathbb E(1)\times\mathbb I(p^2)\times \mathbb P(p^{-2}),\\
1 &\sim \mathbb E(1)\times\mathbb I(1)\times \mathbb P(1).\\
\end{aligned} \end{equation}

Similarly, for general MHC amplitudes, we can also define a scaling structure for each diagram,
\begin{equation}
\mathbb M=\mathbf m^a\times\underbrace{\mathbb E\times\mathbb I}_{\text{MHC diagram}}\times \mathbb P,
\end{equation}
which consists of four parts
\begin{itemize}
\item The external scaling $\mathbb E(\tilde\lambda,\lambda,m\tilde\eta,m\eta)$ captures the spinor scaling behavior of all external particles. Here we list the external particle that will appear in the WWhh amplitude
\begin{equation}
\renewcommand{\arraystretch}{1.2}
\begin{tabular}{c|c|c|c}
\hline
boson & \multicolumn{3}{c}{scaling of external state} \\
\hline
scalar & 
\Ampone{1.5}{0}{\sca{i1}}$\sim 1$ & \mbox{-} & \mbox{-} \\
\hline
vector 
& \Ampone{1.5}{0}{\bos{i1}{brown}}$\sim\tilde\lambda\lambda$ & 
\makecell{
\Ampone{1.5}{-}{\bosflip{1.5}{0}{brown}{red}}$\sim\lambda m\tilde\eta$ \\
\Ampone{1.5}{+}{\bosflip{1.5}{0}{brown}{cyan}}$\sim\tilde\lambda\tilde m\eta$ }
& \Ampone{1.5}{0}{\bosflipbrown{1.5}{0}}$\sim\tilde m\eta m\tilde\eta$  \\
\hline
\end{tabular}
\renewcommand{\arraystretch}{1}
\end{equation}

\item The internal scaling $\mathbb I(p,m\tilde m\eta)$ encodes the scaling information of internal particles. Here $\eta$ is the momentum $\eta_{\dot\alpha\alpha}$, which should not be mixing with the spinor $\eta_{\alpha}$. For WWhh amplitude, the internal particle can be only scalar or vector boson:
\begin{equation}
\renewcommand{\arraystretch}{1.2}
\begin{tabular}{c|c|c|c}
\hline
boson & \multicolumn{3}{c}{scaling of internal state} \\
\hline
scalar & 
\begin{tikzpicture}[baseline=0.7cm] \begin{feynhand}
\setlength{\feynhandarrowsize}{4pt}
\vertex [dot] (i1) at (0,0.8) {};
\vertex [dot] (i2) at (1.6,0.8) {};
\draw[brown,thick sca] (i1)--(i2);
\end{feynhand} \end{tikzpicture}\;$\sim 1$ & \mbox{-} & \mbox{-} \\
\hline
vector & 
\begin{tikzpicture}[baseline=0.7cm] \begin{feynhand}
\setlength{\feynhandarrowsize}{4pt}
\vertex [dot] (i1) at (0,0.8) {};
\vertex [dot] (i2) at (1.6,0.8) {};
\draw[brown,thick bos] (i1)--(i2);
\end{feynhand} \end{tikzpicture}\;$\sim p^2$  & 
\begin{tikzpicture}[baseline=0.7cm] \begin{feynhand}
\setlength{\feynhandarrowsize}{4pt}
\vertex [dot] (i1) at (0,0.8) {};
\vertex [dot] (i2) at (1.6,0.8) {};
\draw[brown,thick bos] (i1)--(i2);
\draw[very thick] plot[mark=x,mark size=2.5] coordinates {(0.5,0.8)};
\draw[very thick] plot[mark=x,mark size=2.5] coordinates {(1.1,0.8)};
\end{feynhand} \end{tikzpicture}\;$\sim m\tilde mp\eta$ & 
\begin{tikzpicture}[baseline=0.7cm] \begin{feynhand}
\setlength{\feynhandarrowsize}{4pt}
\vertex [dot] (i1) at (0,0.8) {};
\vertex [dot] (i2) at (1.6,0.8) {};
\draw[brown,thick bos] (i1)--(i2);
\draw[very thick] plot[mark=x,mark size=2.5] coordinates {(0.4,0.8)};
\draw[very thick] plot[mark=x,mark size=2.5] coordinates {(0.65,0.8)};
\draw[very thick] plot[mark=x,mark size=2.5] coordinates {(0.95,0.8)};
\draw[very thick] plot[mark=x,mark size=2.5] coordinates {(1.2,0.8)};
\end{feynhand} \end{tikzpicture}\;$\sim m^2\tilde m^2\eta^2$ \\
\hline
\end{tabular}
\renewcommand{\arraystretch}{1}
\end{equation}

\item The pole scaling $\mathbb P$ describes a universal scaling behavior of the pole structures, which does not depend strongly on the particle type. For the single-pole term, the pole scaling gives an infinite expansion 
\begin{equation}
\mathbb P(p^{-2})+\mathbb P(\eta p^{-3})+\mathbb P(\eta^2 p^{-4})+\cdots
\end{equation} 
Combined with the internal scaling, it can be represented diagrammatically as
\begin{equation}
\renewcommand{\arraystretch}{1.2}
\begin{tabular}{c|c|c|c}
\hline
boson & \multicolumn{3}{c}{scaling of internal state and pole structure} \\
\hline
scalar & 
\begin{tikzpicture}[baseline=0.7cm] \begin{feynhand}
\setlength{\feynhandarrowsize}{4pt}
\vertex [dot] (i1) at (0,0.8) {};
\vertex [dot] (i2) at (1.6,0.8) {};
\draw[brown,thick sca] (i1)--(i2);
\end{feynhand} \end{tikzpicture}\;$\sim \mathbb I(1)\times \mathbb P(p^{-2})$ & 
\begin{tikzpicture}[baseline=0.7cm] \begin{feynhand}
\setlength{\feynhandarrowsize}{4pt}
\vertex [dot] (i1) at (0,0.8) {};
\vertex [dot] (i2) at (1.6,0.8) {};
\draw[brown,thick sca] (i1)--(i2);
\draw[thick] plot[mark=otimes,mark size=2.5] coordinates {(0.5,0.8)};
\draw[thick] plot[mark=otimes,mark size=2.5] coordinates {(1.1,0.8)};
\end{feynhand} \end{tikzpicture}\;$\sim \mathbb I(1)\times \mathbb P(\eta p^{-3})$ 
& $\cdots$ \\
\hline
vector & 
\begin{tikzpicture}[baseline=0.7cm] \begin{feynhand}
\setlength{\feynhandarrowsize}{4pt}
\vertex [dot] (i1) at (0,0.8) {};
\vertex [dot] (i2) at (1.6,0.8) {};
\draw[brown,thick bos] (i1)--(i2);
\end{feynhand} \end{tikzpicture}\;$\sim \mathbb I(p^2)\times \mathbb P(p^{-2})$  & 
\begin{tikzpicture}[baseline=0.7cm] \begin{feynhand}
\setlength{\feynhandarrowsize}{4pt}
\vertex [dot] (i1) at (0,0.8) {};
\vertex [dot] (i2) at (1.6,0.8) {};
\draw[brown,thick bos] (i1)--(i2);
\draw[thick] plot[mark=otimes,mark size=2.5] coordinates {(0.5,0.8)};
\draw[thick] plot[mark=otimes,mark size=2.5] coordinates {(1.1,0.8)};
\end{feynhand} \end{tikzpicture}\;$\sim \mathbb I(p^2)\times \mathbb P(\eta p^{-3})$ & 
$\cdots$ \\
\hline
\end{tabular}
\renewcommand{\arraystretch}{1}
\end{equation}
where each $\otimes$ on a particle line denotes a subleading correction in the pole expansion. For convenience, we take one $\otimes$ to represent a factor of order $\mathbf m^{2}/E^{2}$ in the expansion. Thus, it is of the same order as a cross term and always appears in pairs.
For contact term, the pole scaling is trivial
\begin{equation}
\mathbb P(1).
\end{equation}

\item The last thing is a power of physical mass $\mathbf m$, which is used to make sure that the $n$-point amplitude satisfy the mass dimension $4-n$.

\end{itemize}

For 3pt amplitudes, there are no internal particle and pole structures, so the scaling structure is simple 
\begin{equation} \begin{aligned}
&\text{primary}:& \mathbb M_3&= \mathbf m^{1-s_1-s_2-s_3}\mathbb E,\\
&\text{k-th descendant}:& \mathbb M_3&= \mathbf m^{1-s_1-s_2-s_3-k}\mathbb E.
\end{aligned} \end{equation}
For example, consider the VSS amplitude, we can list the scaling behavior for each order:
\begin{equation} \begin{aligned}
&\text{primary}:&
\Ampthree{0}{0}{0}{\bos{i1}{brown}}{\bos{i2}{brown}}{\sca{i3}}&\sim \frac{1}{\mathbf m}\mathbb E(\tilde\lambda^2\lambda^2),\\
&\text{1st descendant}:&
\Ampthree{-}{0}{0}{\bosflip{1}{180}{brown}{red}}{\bos{i2}{brown}}{\sca{i3}}&\sim \frac{1}{\mathbf m^2}\mathbb E(\tilde\lambda\lambda^2 m\tilde\eta),\quad
\Ampthree{+}{0}{0}{\bosflip{1}{180}{brown}{cyan}}{\bos{i2}{brown}}{\sca{i3}}\sim\frac{1}{\mathbf m^2}\mathbb E(\tilde\lambda^2\lambda\tilde m\eta), \\
&\text{2nd descendant}:&
\Ampthree{+}{-}{0}{\bosflip{1}{180}{brown}{cyan}}{\bosflip{1}{55}{brown}{red}}{\sca{i3}}&,
\Ampthree{0}{0}{0}{\bosflipflip{1}{-180}{brown}{brown}{brown}}{\bos{i2}{brown}}{\sca{i3}},
\Ampthree{0}{0}{0}{\bos{i1}{brown}}{\bosflipflip{1}{55}{brown}{brown}{brown}}{\sca{i3}}
\sim\frac{1}{\mathbf m^3}\mathbb E(\tilde\lambda\lambda\tilde m\eta m\tilde\eta).
\end{aligned} \end{equation}
For the 2nd descendant, we list only amplitude with total helicity $h_1+h_2+h_3=0$, because it is enough for the $WWhh$ matching. 

For SSS amplitude, the scaling is given by
\begin{equation} \begin{aligned}
&\text{primary}:&
\Ampthree{0}{0}{0}{\sca{i1}}{\sca{i2}}{\sca{i3}}&\sim \mathbf m\mathbb E(1).
\end{aligned} \end{equation}

Then we can consider the scaling of the 4-pt amplitudes. For helicity category $(+1,-1,0,0)$, the external and internal scaling can be obtained by multiplying the scaling of two 3-pt amplitudes,
\begin{equation} \begin{aligned} 
\begin{tikzpicture}[baseline=1.1cm] \begin{feynhand}
\vertex [particle] (i1) at (0,2.2) {$+$};
\vertex [particle] (i2) at (0,0) {$-$};
\vertex [particle] (i3) at (1.8,0) {$0$};
\vertex [particle] (i4) at (1.8,2.2) {$0$};
\vertex (v1) at (0.9,0.6);
\vertex (v2) at (0.9,1.6);
\vertex (v4) at (0.9-0.9*0.4,1.6+0.6*0.4);
\vertex (v5) at (0.9-0.9*0.4,0.6-0.6*0.4);
\draw[cyan,thick bos] (i1)--(v4);
\draw[brown,thick bos] (v4)--(v2);
\draw[brown,thick bos] (v2)--(v1);
\draw[brown,thick bos] (v5)--(v1);
\draw[red,thick bos] (i2)--(v5);
\draw[brown,thick sca] (i3)--(v1);
\draw[brown,thick sca] (i4)--(v2);
\draw[very thick] plot[mark=x,mark size=3.5,mark options={rotate=45}] coordinates {(v4)};
\draw[very thick] plot[mark=x,mark size=3.5,mark options={rotate=45}] coordinates {(v5)};
\end{feynhand} \end{tikzpicture}:
\frac{\mathbb E(\tilde\lambda\lambda^2 m\tilde\eta)}{\mathbf m^2}\times 
\frac{\mathbb E(\tilde\lambda^2\lambda\tilde m\eta)}{\mathbf m^2}
=\frac{\mathbb E(\tilde\lambda\lambda m\tilde\eta\tilde m\eta)}{\mathbf m^4}\times \mathbb I(p^2).
\end{aligned} \end{equation}
Combining with the pole structure, it gives
\begin{equation} \begin{aligned} \label{eq:scaling_example}
\begin{tikzpicture}[baseline=1.1cm] \begin{feynhand}
\vertex [particle] (i1) at (0,2.2) {$+$};
\vertex [particle] (i2) at (0,0) {$-$};
\vertex [particle] (i3) at (1.8,0) {$0$};
\vertex [particle] (i4) at (1.8,2.2) {$0$};
\vertex (v1) at (0.9,0.6);
\vertex (v2) at (0.9,1.6);
\vertex (v4) at (0.9-0.9*0.4,1.6+0.6*0.4);
\vertex (v5) at (0.9-0.9*0.4,0.6-0.6*0.4);
\draw[cyan,thick bos] (i1)--(v4);
\draw[brown,thick bos] (v4)--(v2);
\draw[brown,thick bos] (v2)--(v1);
\draw[brown,thick bos] (v5)--(v1);
\draw[red,thick bos] (i2)--(v5);
\draw[brown,thick sca] (i3)--(v1);
\draw[brown,thick sca] (i4)--(v2);
\draw[very thick] plot[mark=x,mark size=3.5,mark options={rotate=45}] coordinates {(v4)};
\draw[very thick] plot[mark=x,mark size=3.5,mark options={rotate=45}] coordinates {(v5)};
\end{feynhand} \end{tikzpicture}\sim
\frac{1}{\mathbf m^4}\mathbb E(\tilde\lambda\lambda m\tilde\eta\tilde m\eta)\times \mathbb I(p^2)\times\left(\mathbb P(p^{-2})+\mathbb P(\eta p^{-3})+\cdots\right).
\end{aligned} \end{equation}

The massless amplitude does not match all these MHC diagrams. To determine which MHC diagrams can be matched, we can carry out a power counting analysis. The massive amplitude involves two scales: the energy scale $E$ and mass scale $\mathbf m$. In the high-energy regime $E\gg \mathbf m$, only terms scaling as $E^0\mathbf m^0$ can be matched from 4-pt massless amplitude. The building blocks appearing in the scaling structure have the following power counting
\begin{equation} \begin{aligned}
\lambda,\tilde{\lambda} &\sim \sqrt{E},&  \tilde m\eta,m\tilde{\eta} &\sim \frac{\mathbf m^2}{\sqrt{E}},& \\
p &\sim E,& \eta &\sim \frac{\mathbf m^2}{E}.& 
\end{aligned} \end{equation}
Using these, we can determine the power counting of eq.~\eqref{eq:scaling_example},
\begin{equation} \begin{aligned}
\frac{1}{\mathbf m^4}\mathbb E(\tilde\lambda\lambda m\tilde\eta\tilde m\eta)&\quad\sim\quad 1,\\
\mathbb I(p^2)&\quad\sim\quad E^2,\\
\mathbb P(p^{-2})+\mathbb P(\eta p^{-3})+\cdots &\quad\sim\quad
\frac{1}{E^2}+\frac{\mathbf m^2}{E^4}+\cdots.
\end{aligned} \end{equation}
Therefore, only the leading term of the pole structure contributes at $E^0\mathbf m^0$, which corresponds to
\begin{equation} \begin{aligned} \label{eq:scaling_example}
\frac{1}{\mathbf m^4}\mathbb E(\tilde\lambda\lambda m\tilde\eta\tilde m\eta)\times \mathbb I(p^2)\times\mathbb P(p^{-2}).
\end{aligned} \end{equation}

Similarly, we can derive the leading scaling of all MHC diagrams in helicity category $(+,-,0,0)$ and the corresponding power counting.
\begin{equation}
\begin{tabular}{c|c|c}
\hline
diagram & leading scaling & leading power counting \\
\hline
\begin{tikzpicture}[baseline=1.1cm] \begin{feynhand}
\vertex [particle] (i1) at (0,2.2) {$+$};
\vertex [particle] (i2) at (0,0) {$-$};
\vertex [particle] (i3) at (1.8,0) {$0$};
\vertex [particle] (i4) at (1.8,2.2) {$0$};
\vertex (v1) at (0.9,0.6);
\vertex (v2) at (0.9,1.6);
\vertex (v4) at (0.9-0.9*0.4,1.6+0.6*0.4);
\vertex (v5) at (0.9-0.9*0.4,0.6-0.6*0.4);
\draw[cyan,thick bos] (i1)--(v4);
\draw[brown,thick bos] (v4)--(v2);
\draw[brown,thick bos] (v2)--(v1);
\draw[brown,thick bos] (v5)--(v1);
\draw[red,thick bos] (i2)--(v5);
\draw[brown,thick sca] (i3)--(v1);
\draw[brown,thick sca] (i4)--(v2);
\draw[very thick] plot[mark=x,mark size=3.5,mark options={rotate=45}] coordinates {(v4)};
\draw[very thick] plot[mark=x,mark size=3.5,mark options={rotate=45}] coordinates {(v5)};
\end{feynhand} \end{tikzpicture} & 
$\frac{1}{\mathbf m^4}\mathbb E(\tilde\lambda\lambda m\tilde\eta\tilde m\eta)\times \mathbb I(p^2)\times\mathbb P(p^{-2})$ & 
\multirow{2}{*}{$1$} \\
\begin{tikzpicture}[baseline=0.8cm] \begin{feynhand}
\vertex [particle] (i1) at (0,1.8) {$+$};
\vertex [particle] (i2) at (0,0) {$-$};
\vertex [particle] (i3) at (1.8,0) {$0$};
\vertex [particle] (i4) at (1.8,1.8) {$0$};
\vertex (v3) at (0.9,0.9);
\vertex (v4) at (0.9-0.9*0.43,0.9+0.9*0.43);
\vertex (v5) at (0.9-0.9*0.43,0.9-0.9*0.43);
\draw[cyan,thick bos] (i1)--(v4);
\draw[brown,thick bos] (v4)--(v3);
\draw[brown,thick bos] (v5)--(v3);
\draw[red,thick bos] (i2)--(v5);
\draw[brown,thick sca] (i3)--(v3);
\draw[brown,thick sca] (i4)--(v3);
\draw[very thick] plot[mark=x,mark size=3.5,mark options={rotate=45}] coordinates {(v4)};
\draw[very thick] plot[mark=x,mark size=3.5,mark options={rotate=45}] coordinates {(v5)};
\end{feynhand} \end{tikzpicture} & 
$\frac{1}{\mathbf m^4}\mathbb E(\tilde\lambda\lambda m\tilde\eta\tilde m\eta)\times \mathbb I(1)\times\mathbb P(1)$ & \\
\hline
\begin{tikzpicture}[baseline=1.1cm] \begin{feynhand}
\vertex [particle] (i1) at (0,2.2) {$+$};
\vertex [particle] (i2) at (0,0) {$-$};
\vertex [particle] (i3) at (1.8,0) {$0$};
\vertex [particle] (i4) at (1.8,2.2) {$0$};
\vertex (v1) at (0.9,0.6);
\vertex (v2) at (0.9,1.6);
\vertex (v4) at (0.9-0.9*0.4,1.6+0.6*0.4);
\vertex (v5) at (0.9-0.9*0.4,0.6-0.6*0.4);
\draw[cyan,thick bos] (i1)--(v4);
\draw[brown,thick bos] (v4)--(v2);
\draw[brown,thick bos] (v2)--(v1);
\draw[brown,thick bos] (v5)--(v1);
\draw[red,thick bos] (i2)--(v5);
\draw[brown,thick sca] (i3)--(v1);
\draw[brown,thick sca] (i4)--(v2);
\draw[very thick] plot[mark=x,mark size=3.5,mark options={rotate=45}] coordinates {(v4)};
\draw[very thick] plot[mark=x,mark size=3.5,mark options={rotate=45}] coordinates {(v5)};
\draw[very thick] plot[mark=x,mark size=3.5,mark options={rotate=0}] coordinates {(0.9,0.9)};
\draw[very thick] plot[mark=x,mark size=3.5,mark options={rotate=0}] coordinates {(0.9,1.3)};
\end{feynhand} \end{tikzpicture} &
$\frac{1}{\mathbf m^4}\mathbb E(\tilde\lambda\lambda m\tilde\eta\tilde m\eta)\times \mathbb I(p\eta)\times\mathbb P(p^{-2})$ & 
\multirow{2}{*}{$\displaystyle\frac{\mathbf m^2}{E^2}$} \\
\begin{tikzpicture}[baseline=0.8cm] \begin{feynhand}
\vertex [particle] (i1) at (0,1.8) {$+$};
\vertex [particle] (i2) at (0,0) {$-$};
\vertex [particle] (i3) at (1.8,0) {$0$};
\vertex [particle] (i4) at (1.8,1.8) {$0$};
\vertex (v1) at (0.6,0.9);
\vertex (v2) at (1.2,0.9);
\vertex (v3) at (0.9,0.9);
\vertex (v4) at (0.6-0.6*0.43,0.9+0.9*0.43);
\vertex (v5) at (0.6-0.6*0.43,0.9-0.9*0.43);
\draw[cyan,thick bos] (i1)--(v4);
\draw[brown,thick bos] (v4)--(v1);
\draw[brown,thick bos] (v5)--(v1);
\draw[red,thick bos] (i2)--(v5);
\draw[brown,thick sca] (v1)--(v2);
\draw[brown,thick sca] (i3)--(v2);
\draw[brown,thick sca] (i4)--(v2);
\draw[very thick] plot[mark=x,mark size=3.5,mark options={rotate=45}] coordinates {(v4)};
\draw[very thick] plot[mark=x,mark size=3.5,mark options={rotate=45}] coordinates {(v5)};
\end{feynhand} \end{tikzpicture}
& $\frac{1}{\mathbf m^2}\mathbb E(\tilde\lambda\lambda m\tilde\eta\tilde m\eta)\times \mathbb I(1)\times\mathbb P(p^{-2})$ & \\
\hline
\begin{tikzpicture}[baseline=1.1cm] \begin{feynhand}
\vertex [particle] (i1) at (0,2.2) {$+$};
\vertex [particle] (i2) at (0,0) {$-$};
\vertex [particle] (i3) at (1.8,0) {$0$};
\vertex [particle] (i4) at (1.8,2.2) {$0$};
\vertex (v1) at (0.9,0.6);
\vertex (v2) at (0.9,1.6);
\vertex (v4) at (0.9-0.9*0.4,1.6+0.6*0.4);
\vertex (v5) at (0.9-0.9*0.4,0.6-0.6*0.4);
\draw[cyan,thick bos] (i1)--(v4);
\draw[brown,thick bos] (v4)--(v2);
\draw[brown,thick bos] (v2)--(v1);
\draw[brown,thick bos] (v5)--(v1);
\draw[red,thick bos] (i2)--(v5);
\draw[brown,thick sca] (i3)--(v1);
\draw[brown,thick sca] (i4)--(v2);
\draw[very thick] plot[mark=x,mark size=3.5,mark options={rotate=45}] coordinates {(v4)};
\draw[very thick] plot[mark=x,mark size=3.5,mark options={rotate=45}] coordinates {(v5)};
\draw[very thick] plot[mark=x,mark size=3.5,mark options={rotate=0}] coordinates {(0.9,0.76)};
\draw[very thick] plot[mark=x,mark size=3.5,mark options={rotate=0}] coordinates {(0.9,0.98)};
\draw[very thick] plot[mark=x,mark size=3.5,mark options={rotate=0}] coordinates {(0.9,1.22)};
\draw[very thick] plot[mark=x,mark size=3.5,mark options={rotate=0}] coordinates {(0.9,1.44)};
\end{feynhand} \end{tikzpicture} &
$\frac{1}{\mathbf m^4}\mathbb E(\tilde\lambda\lambda m\tilde\eta\tilde m\eta)\times \mathbb I(\eta^2)\times\mathbb P(p^{-2})$ & $\displaystyle\frac{\mathbf m^4}{E^4}$ \\
\hline
\end{tabular}
\end{equation}
Since the 4-pt massless amplitude has power counting $E^0$, so it can match to the first two MHC diagrams with leading power counting $1$. The massless matching 
\begin{equation}
\mathcal{A}\to
\begin{tikzpicture}[baseline=1.1cm] \begin{feynhand}
\vertex [particle] (i1) at (0,2.2) {$+$};
\vertex [particle] (i2) at (0,0) {$-$};
\vertex [particle] (i3) at (1.8,0) {$0$};
\vertex [particle] (i4) at (1.8,2.2) {$0$};
\vertex (v1) at (0.9,0.6);
\vertex (v2) at (0.9,1.6);
\vertex (v4) at (0.9-0.9*0.4,1.6+0.6*0.4);
\vertex (v5) at (0.9-0.9*0.4,0.6-0.6*0.4);
\draw[cyan,thick bos] (i1)--(v4);
\draw[brown,thick bos] (v4)--(v2);
\draw[brown,thick bos] (v2)--(v1);
\draw[brown,thick bos] (v5)--(v1);
\draw[red,thick bos] (i2)--(v5);
\draw[brown,thick sca] (i3)--(v1);
\draw[brown,thick sca] (i4)--(v2);
\draw[very thick] plot[mark=x,mark size=3.5,mark options={rotate=45}] coordinates {(v4)};
\draw[very thick] plot[mark=x,mark size=3.5,mark options={rotate=45}] coordinates {(v5)};
\end{feynhand} \end{tikzpicture} +
\begin{tikzpicture}[baseline=0.8cm] \begin{feynhand}
\vertex [particle] (i1) at (0,1.8) {$+$};
\vertex [particle] (i2) at (0,0) {$-$};
\vertex [particle] (i3) at (1.8,0) {$0$};
\vertex [particle] (i4) at (1.8,1.8) {$0$};
\vertex (v3) at (0.9,0.9);
\vertex (v4) at (0.9-0.9*0.43,0.9+0.9*0.43);
\vertex (v5) at (0.9-0.9*0.43,0.9-0.9*0.43);
\draw[cyan,thick bos] (i1)--(v4);
\draw[brown,thick bos] (v4)--(v3);
\draw[brown,thick bos] (v5)--(v3);
\draw[red,thick bos] (i2)--(v5);
\draw[brown,thick sca] (i3)--(v3);
\draw[brown,thick sca] (i4)--(v3);
\draw[very thick] plot[mark=x,mark size=3.5,mark options={rotate=45}] coordinates {(v4)};
\draw[very thick] plot[mark=x,mark size=3.5,mark options={rotate=45}] coordinates {(v5)};
\end{feynhand} \end{tikzpicture}.
\end{equation}

Similarly, we can obtain the MHC scaling for $(0,0,0,0)$ helicity category, and select the ones that have same power counting as 4-pt masselss amplitude.

For $s_{13}$ and $s_{14}$ channel, we have 
\begin{equation} \begin{aligned}
\begin{tikzpicture}[baseline=0.8cm] \begin{feynhand}
\vertex [particle] (i1) at (0,1.8) {$0$};
\vertex [particle] (i2) at (0,0) {$0$};
\vertex [particle] (i3) at (1.8,0) {$0$};
\vertex [particle] (i4) at (1.8,1.8) {$0$};
\vertex (v1) at (0.9,0.6);
\vertex (v2) at (0.9,1.2);
\draw[brown,thick bos] (i1)--(v2);
\draw[brown,thick bos] (v2)--(v1);
\draw[brown,thick bos] (i2)--(v1);
\draw[brown,thick sca] (i3)--(v1);
\draw[brown,thick sca] (i4)--(v2);
\draw[thick] plot[mark=otimes,mark size=3,mark options={rotate=0}] coordinates {(0.9,0.75)};
\draw[thick] plot[mark=otimes,mark size=3,mark options={rotate=0}] coordinates {(0.9,1.05)};
\end{feynhand} \end{tikzpicture}&\sim
\frac{1}{\mathbf m^2}\mathbb E(\tilde\lambda^2\lambda^2 )\times \mathbb I(p^2)\times\mathbb P(\eta p^{-3}), \\
\begin{tikzpicture}[baseline=0.8cm] \begin{feynhand}
\vertex [particle] (i1) at (0,1.8) {$0$};
\vertex [particle] (i2) at (0,0) {$0$};
\vertex [particle] (i3) at (1.8,0) {$0$};
\vertex [particle] (i4) at (1.8,1.8) {$0$};
\vertex (v1) at (0.9,0.6);
\vertex (v2) at (0.9,1.2);
\draw[brown,thick bos] (i1)--(v2);
\draw[brown,thick bos] (v2)--(v1);
\draw[brown,thick bos] (i2)--(v1);
\draw[brown,thick sca] (i3)--(v1);
\draw[brown,thick sca] (i4)--(v2);
\draw[very thick] plot[mark=x,mark size=3.5,mark options={rotate=0}] coordinates {(0.9,0.75)};
\draw[very thick] plot[mark=x,mark size=3.5,mark options={rotate=0}] coordinates {(0.9,1.05)};
\end{feynhand} \end{tikzpicture}&\sim
\frac{1}{\mathbf m^2}\mathbb E(\tilde\lambda^2\lambda^2 )\times \mathbb I(p\eta)\times\mathbb P(p^{-2}), \\
\begin{tikzpicture}[baseline=0.8cm] \begin{feynhand}
\vertex [particle] (i1) at (0,1.8) {$0$};
\vertex [particle] (i2) at (0,0) {$0$};
\vertex [particle] (i3) at (1.8,0) {$0$};
\vertex [particle] (i4) at (1.8,1.8) {$0$};
\vertex (v1) at (0.9,0.6);
\vertex (v2) at (0.9,1.2);
\draw[brown,thick bos] (i1)--(v2);
\draw[brown,thick bos] (v2)--(v1);
\draw[brown,thick bos] (i2)--(v1);
\draw[brown,thick sca] (i3)--(v1);
\draw[brown,thick sca] (i4)--(v2);
\draw[very thick] plot[mark=x,mark size=3.5,mark options={rotate=56.3}] coordinates {(0.9-0.9*0.48,1.2+0.6*0.48)};
\draw[very thick] plot[mark=x,mark size=3.5,mark options={rotate=56.3}] coordinates {(0.9-0.9*0.22,1.2+0.6*0.22)};
\end{feynhand} \end{tikzpicture}+
\begin{tikzpicture}[baseline=0.8cm] \begin{feynhand}
\vertex [particle] (i1) at (0,1.8) {$0$};
\vertex [particle] (i2) at (0,0) {$0$};
\vertex [particle] (i3) at (1.8,0) {$0$};
\vertex [particle] (i4) at (1.8,1.8) {$0$};
\vertex (v1) at (0.9,0.6);
\vertex (v2) at (0.9,1.2);
\draw[brown,thick bos] (i1)--(v2);
\draw[brown,thick bos] (v2)--(v1);
\draw[brown,thick bos] (i2)--(v1);
\draw[brown,thick sca] (i3)--(v1);
\draw[brown,thick sca] (i4)--(v2);
\draw[very thick] plot[mark=x,mark size=3.5,mark options={rotate=33.7}] coordinates {(0.9-0.9*0.48,0.6-0.6*0.48)};
\draw[very thick] plot[mark=x,mark size=3.5,mark options={rotate=33.7}] coordinates {(0.9-0.9*0.22,0.6-0.6*0.22)}; 
\end{feynhand} \end{tikzpicture} &\sim
\frac{1}{\mathbf m^4}\mathbb E(\tilde\lambda\lambda m\tilde\eta\tilde m\eta )\times \mathbb I(p^2)\times\mathbb P(p^{-2}). 
\end{aligned} \end{equation}
For $s_{12}$ channel and contact term, we have
\begin{equation} \begin{aligned}
\begin{tikzpicture}[baseline=0.8cm] \begin{feynhand}
\vertex [particle] (i1) at (0,1.8) {$0$};
\vertex [particle] (i2) at (0,0) {$0$};
\vertex [particle] (i3) at (1.8,0) {$0$};
\vertex [particle] (i4) at (1.8,1.8) {$0$};
\vertex (v3) at (0.9,0.9);
\draw[brown,thick bos] (i1)--(v3);
\draw[brown,thick bos] (i2)--(v3);
\draw[brown,thick sca] (i4)--(v3)--(i3);
\draw[very thick] plot[mark=x,mark size=3.5,mark options={rotate=45}] coordinates {(0.9-0.9*0.48,0.9-0.9*0.48)};
\draw[very thick] plot[mark=x,mark size=3.5,mark options={rotate=45}] coordinates {(0.9-0.9*0.22,0.9-0.9*0.22)}; 
\end{feynhand} \end{tikzpicture}+
\begin{tikzpicture}[baseline=0.8cm] \begin{feynhand}
\vertex [particle] (i1) at (0,1.8) {$0$};
\vertex [particle] (i2) at (0,0) {$0$};
\vertex [particle] (i3) at (1.8,0) {$0$};
\vertex [particle] (i4) at (1.8,1.8) {$0$};
\vertex (v3) at (0.9,0.9);
\draw[brown,thick bos] (i1)--(v3);
\draw[brown,thick bos] (i2)--(v3);
\draw[brown,thick sca] (i4)--(v3)--(i3);
\draw[very thick] plot[mark=x,mark size=3.5,mark options={rotate=45}] coordinates {(0.9-0.9*0.48,0.9+0.9*0.48)};
\draw[very thick] plot[mark=x,mark size=3.5,mark options={rotate=45}] coordinates {(0.9-0.9*0.22,0.9+0.9*0.22)}; 
\end{feynhand} \end{tikzpicture}&\sim 
\frac{1}{\mathbf m^4}\mathbb E(\tilde\lambda\lambda m\tilde\eta\tilde m\eta)\times \mathbb I(1)\times\mathbb P(1),\\
\begin{tikzpicture}[baseline=0.8cm] \begin{feynhand}
\vertex [particle] (i1) at (0,1.8) {$0$};
\vertex [particle] (i2) at (0,0) {$0$};
\vertex [particle] (i3) at (1.8,0) {$0$};
\vertex [particle] (i4) at (1.8,1.8) {$0$};
\vertex (v1) at (0.6,0.9);
\vertex (v2) at (1.2,0.9);
\draw[brown,thick bos] (i1)--(v1);
\draw[brown,thick bos] (i2)--(v1);
\draw[brown,thick sca] (v1)--(v2);
\draw[brown,thick sca] (i4)--(v2)--(i3);
\end{feynhand} \end{tikzpicture}&\sim 
\mathbb E(\tilde\lambda^2\lambda^2)\times \mathbb I(1)\times\mathbb P(p^{-2}).
\end{aligned} \end{equation}

The scaling change will guide us to perform the amplitude matching in the following disucssion.

\subsection{Amplitude deformation and pole separation}

In this subsection, we match the massless amplitude to the MHC amplitude by amplitude deformation. This deformation has two steps. First match the coefficient, denoted as group tensor, from unbroken phase to the broken phase. Then we deform the massless amplitude in the gauge invariant form, to the massless amplitude in the light cone gauge, so the massless pole structure can match the MHC ones, which is guided by the analysis of scaling structure between massless and MHC amplitude. This pole matching will involve the cases of the pole separation and extraction, which will be discussed in detail with the example of $W^+W^-hh$ matching.

\paragraph{helicity $(+,-,0,0)$}

We fist consider the helicity category $(+,-,0,0)$. The corresponding massless amplitude has been derived. Since both Higgs doublets $H$ and $H^\dagger$ can match to the massive Higgs boson $h$, we need the following two amplitude expressions:
\begin{equation} \begin{aligned} \label{eq:massless_1}
\mathcal{A}(1^{-1,I_1},2^{+1,I_2},3^{0,i_3},4^{0}_{i_4})
&=(T^{I_2}_s T^{I_1}_s)^{i_3}_{i_4}\frac{[1|3|2\rangle[1|4|2\rangle}{s_{12}s_{13}}
+(T^{I_1}_s T^{I_2}_s)^{i_3}_{i_4}\frac{[1|3|2\rangle[1|4|2\rangle}{s_{12}s_{14}},\\
\mathcal{A}(1^{-1,I_1},2^{+1,I_2},3^{0}_{i_3},4^{0,i_4})
&=(T^{I_2}_s T^{I_1}_s)^{i_4}_{i_3}\frac{[1|3|2\rangle[1|4|2\rangle}{s_{12}s_{14}}
+(T^{I_1}_s T^{I_2}_s)^{i_4}_{i_3}\frac{[1|3|2\rangle[1|4|2\rangle}{s_{12}s_{13}}.
\end{aligned} \end{equation}
These correspond to the $WWHH^\dagger$ and $WWH^\dagger H$ amplitudes. 

The massless amplitude contains double poles, such as $s_{12}s_{13}$, which do not appear in the massive case. Therefore, we need to separate this multipole structure. This can be analyzed through scaling behavior. All terms in eq.~\eqref{eq:massless_1} share the same scaling $\mathbb E(\lambda^2\tilde\lambda^2)\times\mathbb I(p^2)\times \mathbb P(p^{-4})$, where the double pole structure is contained in $\mathbb P(p^{-4})$. To separate the poles, we must increase the power of the pole scaling from $\mathbb P(p^{-4})$ to something like $\mathbb P(p^{-2})$. However, this scaling cannot be changed directly as
\begin{equation} \begin{aligned}
\mathbb E(\lambda^2\tilde\lambda^2)\times\mathbb I(p^2)\times \mathbb P(p^{-4})\not\to \mathbb E(\lambda^2\tilde\lambda^2)\times\mathbb I(1)\times \mathbb P(p^{-2}),
\end{aligned} \end{equation}
because the only spinor structure satisfying $\mathbb E(\lambda^2\tilde\lambda^2)\times\mathbb I(1)$ is $\langle11\rangle[22]=0$. A better way to achieve a sensible pole separation is to convert the massless scaling to the MHC scaling.  Multiplying by $\frac{\langle\lambda\eta\rangle}{m}\frac{[\lambda\eta]}{\tilde m}$ transforms the massless scaling as follows,
\begin{equation} \begin{aligned}
&\mathbb E(\lambda^2\tilde\lambda^2)\times\mathbb I(p^2)\times \mathbb P(p^{-4})\\
&\xrightarrow{\times\frac{\langle\lambda\eta\rangle}{m}\frac{[\lambda\eta]}{\tilde m}} \frac{1}{\mathbf m^4}\mathbb E(\tilde\lambda^3\lambda^3 m\tilde\eta\tilde m\eta)\times \mathbb I(p^2)\times\mathbb P(p^{-4})\\
&\to \frac{1}{\mathbf m^4}\mathbb E(\tilde\lambda\lambda m\tilde\eta\tilde m\eta)\times \mathbb I(p^4)\times\mathbb P(p^{-4})\\
&\xrightarrow{\text{pole isolation}} \frac{1}{\mathbf m^4}\mathbb E(\tilde\lambda\lambda m\tilde\eta\tilde m\eta)\times \mathbb I(p^2)\times\mathbb P(p^{-2})+\frac{1}{\mathbf m^4}\mathbb E(\tilde\lambda\lambda m\tilde\eta\tilde m\eta)\times \mathbb I(1)\times\mathbb P(1).
\end{aligned} \end{equation}
In the third line, part of the spinor scaling is transferred from the external scaling $\mathbb E$ to the internal scaling $\mathbb I$, which acquires an inverse scaling relative to $\mathbb P(p^{-4})$. As a result, the pole separation yields not only a term with a single pole but also a contact term with no poles.

The corresponding amplitude deformation is given by
\begin{equation} \begin{aligned}
\frac{[1|3|2\rangle[1|4|2\rangle}{s_{12}s_{13}}
&\xrightarrow{\times\frac{\langle\lambda\eta\rangle}{m}\frac{[\lambda\eta]}{\tilde m}}
\frac{\langle\eta_1 1\rangle[1|3|2\rangle[1|4|2\rangle[2\eta_2]}{\tilde m_1 m_2 s_{12}s_{13}} \\
&=\frac{\langle\eta_1 |13|2\rangle[1|42|\eta_2]}{\tilde m_1 m_2 s_{12}s_{13}}.\\
\end{aligned} \end{equation}
To isolate the pole structure, we can deform the numerator. By commuting the momenta in the spinor chains, we obtain
\begin{equation} \begin{aligned}
\langle\eta_1 |13|2\rangle &=-\langle\eta_1|3|1]\langle12\rangle+s_{13}\langle\eta_12\rangle, \\
[1|42|\eta_2] &=-[12]\langle2|4|\eta_2]+s_{24}[1\eta_2] .
\end{aligned} \end{equation}
Using this relation, we find
\begin{equation} \begin{aligned}
\frac{[1|3|2\rangle[1|4|2\rangle}{s_{12}s_{13}}
=&\frac{(-\langle\eta_1|3|1]\langle12\rangle+s_{13}\langle\eta_12\rangle)(-[12]\langle2|4|\eta_2]+s_{24}[1\eta_2])}{s_{12}s_{13}}\\
=&\frac{\langle\eta_1|2|1]\langle2|4|\eta_2]-\langle\eta_1|3|1]\langle2|1|\eta_2]+s_{13}\langle\eta_12\rangle[1\eta_2]}{s_{12}}\\
&-\frac{\langle\eta_1|3|1]\langle2|4|\eta_2]}{s_{13}}+\langle\eta_12\rangle[1\eta_2]\\
=&\frac{\langle\eta_1|2|1]\langle2|4-3|\eta_2]-\langle\eta_1|3-4|1]\langle2|1|\eta_2]+(s_{13}-s_{14})\langle\eta_12\rangle[1\eta_2]}{2s_{12}}\\
&-\frac{\langle\eta_1|3|1]\langle2|4|\eta_2]}{s_{13}}+\frac12\langle\eta_12\rangle[1\eta_2].\\
\end{aligned} \end{equation}
In the last step, we require the spinor structure in the term containing the $s_{12}$ pole to be proportional to $p_3-p_4$, which ensures that the total angular momentum of this channel is $1$.

Now the massless amplitudes reduce to a form containing at most one pole, which become the massless amplitudes in the light-cone gauge. Therefore, it can be decomposed into four channel contributions:
\begin{align}
\begin{tikzpicture}[baseline=0.8cm] \begin{feynhand}
\setlength{\feynhandtopsep}{5pt}
\vertex [particle] (i1) at (0,0) {$2^-$};
\vertex [particle] (i2) at (0,1.6) {$1^+$};
\vertex [particle] (i3) at (1.6,1.6) {$4^0$};
\vertex [particle] (i4) at (1.6,0) {$3^0$};
\vertex (v1) at (0.8,0.4);
\vertex (v2) at (0.8,1.2);
\graph{(i2)--[bos] (v2)--[sca] (v1)-- [bos](i1)};
\graph{(i4)--[sca] (v2)};
\graph{(i3)--[sca,top] (v1)};
\end{feynhand} \end{tikzpicture}:\quad 
&\begin{pmatrix}
(T^{I_2}_s T^{I_1}_s)^{i_3}_{i_4}\\
(T^{I_1}_s T^{I_2}_s)^{i_4}_{i_3}
\end{pmatrix}
\frac{-\langle\eta_1|3|1]\langle2|4|\eta_2]}{s_{13}},\\
\begin{tikzpicture}[baseline=0.8cm] \begin{feynhand}
\vertex [particle] (i1) at (0,0) {$2^-$};
\vertex [particle] (i2) at (0,1.6) {$1^+$};
\vertex [particle] (i3) at (1.6,1.6) {$4^0$};
\vertex [particle] (i4) at (1.6,0) {$3^0$};
\vertex (v1) at (0.8,0.4);
\vertex (v2) at (0.8,1.2);
\graph{(i1) --[bos] (v1)--[sca] (i4)};
\graph{(i2) --[bos] (v2)--[sca] (i3)};
\graph{(v1) --[sca] (v2)};
\end{feynhand} \end{tikzpicture}:\quad 
&\begin{pmatrix}
(T^{I_1}_s T^{I_2}_s)^{i_3}_{i_4}\\
(T^{I_2}_s T^{I_1}_s)^{i_4}_{i_3}
\end{pmatrix}\frac{-\langle\eta_1|4|1]\langle2|3|\eta_2]}{s_{14}},\\
\begin{tikzpicture}[baseline=0.8cm] \begin{feynhand}
\vertex [particle] (i1) at (0,0) {$2^-$};
\vertex [particle] (i2) at (0,1.6) {$1^+$};
\vertex [particle] (i3) at (1.6,1.6) {$4^0$};
\vertex [particle] (i4) at (1.6,0) {$3^0$};
\vertex (v1) at (0.4,0.8);
\vertex (v2) at (1.2,0.8);
\graph{(i1) --[bos] (v1)--[bos] (i2)};
\graph{(i4) --[sca] (v2)--[sca] (i3)};
\graph{(v1) --[bos] (v2)};
\end{feynhand} \end{tikzpicture}:\quad 
&\begin{pmatrix}
(T^{I_2}_s T^{I_1}_s)^{i_3}_{i_4}-(T^{I_1}_s T^{I_2}_s)^{i_3}_{i_4}\\
(T^{I_1}_s T^{I_2}_s)^{i_4}_{i_3}-(T^{I_2}_s T^{I_1}_s)^{i_4}_{i_3}
\end{pmatrix}\nonumber \\
&\times\frac{\langle\eta_1|2|1]\langle2|4-3|\eta_2]-\langle\eta_1|3-4|1]\langle2|1|\eta_2]+(s_{13}-s_{14})\langle\eta_12\rangle[1\eta_2]}{2s_{12}},\\
\begin{tikzpicture}[baseline=0.8cm] \begin{feynhand}
\vertex [particle] (i1) at (0,0) {$2^-$};
\vertex [particle] (i2) at (0,1.6) {$1^+$};
\vertex [particle] (i3) at (1.6,1.6) {$4^0$};
\vertex [particle] (i4) at (1.6,0) {$3^0$};
\vertex (v1) at (0.8,0.8);
\graph{(i1) --[bos] (v1)--[bos] (i2)};
\graph{(i4) --[sca] (v1)--[sca] (i3)};
\end{feynhand} \end{tikzpicture}:\quad
&\begin{pmatrix}
(T^{I_2}_s T^{I_1}_s)^{i_3}_{i_4}+(T^{I_1}_s T^{I_2}_s)^{i_3}_{i_4}\\
(T^{I_2}_s T^{I_1}_s)^{i_4}_{i_3}+(T^{I_1}_s T^{I_2}_s)^{i_4}_{i_3}
\end{pmatrix}
\langle\eta_12\rangle[1\eta_2].
\end{align}

Then we can match the massless light-cone gauge amplitude to the MHC amplitude. The conversion from unbroken-phase gauge tensors to broken ones is performed using two transformation matrices, $\mathcal{T}_1$ and $\mathcal{T}_2$:
\begin{equation} \begin{aligned}
\mathcal{T}_1=O^{I_1 W^+}O^{I_2 W^-}\mathcal{U}^{h}_{i_3}\mathcal{U}^{i_4 h},\quad
\mathcal{T}_2=O^{I_1 W^+}O^{I_2 W^-}\mathcal{U}^{h i_3}\mathcal{U}^{i_4}_{h}.
\end{aligned} \end{equation}
These matrices convert the particle types from $WWHH^\dagger$ or $WWH^\dagger H$ to $W^+W^-hh$. The transformation shows that only half of the coefficients survive
\begin{equation} \begin{aligned}
&(T^{I_2}_s T^{I_1}_s)^{i_3}_{i_4}\xrightarrow{\mathcal{T}_1}\frac12 g^2,\qquad
(T^{I_1}_s T^{I_2}_s)^{i_3}_{i_4}\xrightarrow{\mathcal{T}_1}0,\\
&(T^{I_2}_s T^{I_1}_s)^{i_4}_{i_3}\xrightarrow{\mathcal{T}_2}\frac12 g^2,\qquad
(T^{I_1}_s T^{I_2}_s)^{i_4}_{i_3}\xrightarrow{\mathcal{T}_2}0.
\end{aligned} \end{equation}

\begin{figure}[htbp]
\centering
\includegraphics[width=0.6\linewidth,valign=c]{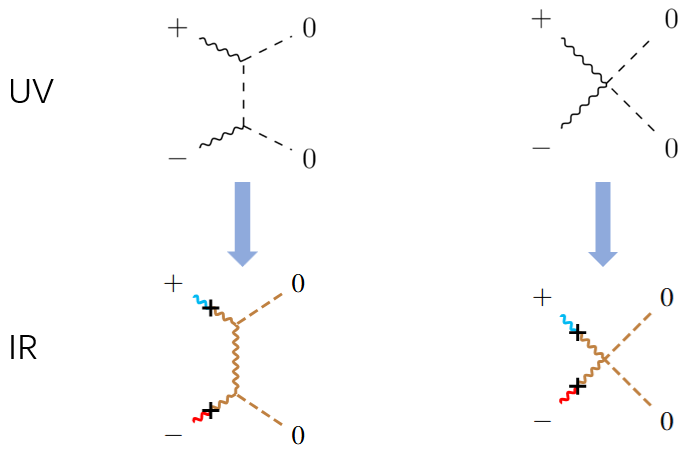}
\caption{The diagrammatic matching for helicity category $(-1,+1,0,0)$ of $W^+W^-hh$ amplitude.}
\label{fig:diagram_match_1}
\end{figure}

Substituting these broken gauge structures into the massless amplitude establishes a matching between massless diagrams and MHC diagrams. As shown in Fig.~\ref{fig:diagram_match_1}, the diagrammatic correspondence is straightforward: the $(13)$-channel, $(14)$-channel and contact term in the massless amplitude match directly to their MHC counterparts.

Finally, we obtain the MHC amplitude for this helicity category
\begin{equation} \begin{aligned}
[\mathcal{M}(1^{+1},2^{-1},3^0,4^0)]_0=
&\frac12 g^2
\frac{\langle\eta_1|P_{14}|1]\langle2|P_{14}|\eta_2]}{s_{14}\tilde{m}_1 m_2}+\frac12 g^2
\frac{\langle\eta_1|P_{13}|1]\langle2|P_{13}|\eta_2]}{s_{13}\tilde{m}_1 m_2} \\
&+\frac12 g^2\frac{\langle\eta_1 2\rangle[\eta_21]}{\tilde m_1 m_2}.
\end{aligned} \end{equation}

\paragraph{helicity (0,0,0,0)}

Then we consider (0,0,0,0) helicity category. Similarly, in this case, we should also consider two amplitude expressions
\begin{equation} \begin{aligned} \label{eq:A4H}
\mathcal{A}(1^{i_1},2_{i_2},3^{i_3},4_{i_4})
&=-(T^{J}_s)^{i_1}_{i_2} (T^{J}_s)^{i_3}_{i_4}\frac{s_{13}-s_{14}}{2s_{12}}
-(T^{J}_s)^{i_1}_{i_4} (T^{J}_s)^{i_3}_{i_2}\frac{s_{13}-s_{12}}{2s_{14}}-4\lambda\delta^{(i_1}_{i_2} \delta^{i_3)}_{i_4}, \\
\mathcal{A}(1^{i_1},2_{i_2},3_{i_3},4^{i_4})
&=-(T^{J}_s)^{i_1}_{i_2} (T^{J}_s)^{i_4}_{i_3}\frac{s_{14}-s_{13}}{2s_{12}}
-(T^{J}_s)^{i_1}_{i_3} (T^{J}_s)^{i_4}_{i_2}\frac{s_{14}-s_{12}}{2s_{13}}-4\lambda\delta^{(i_1}_{i_2} \delta^{i_4)}_{i_3}. \\
\end{aligned} \end{equation}
In this case, we do not need to separate pole structure. These massless amplitudes can be decomposed into four contributions:
\begin{align}
&\begin{tikzpicture}[baseline=0.8cm] \begin{feynhand}
\vertex [particle] (i1) at (0,0) {$2^0$};
\vertex [particle] (i2) at (0,1.6) {$1^0$};
\vertex [particle] (i3) at (1.6,1.6) {$4^0$};
\vertex [particle] (i4) at (1.6,0) {$3^0$};
\vertex (v1) at (0.4,0.8);
\vertex (v2) at (1.2,0.8);
\graph{(i1) --[sca] (v1)--[sca] (i2)};
\graph{(i4) --[sca] (v2)--[sca] (i3)};
\graph{(v1) --[bos] (v2)};
\end{feynhand} \end{tikzpicture}:\quad
\begin{pmatrix}
-(T^{J}_s)^{i_1}_{i_2} (T^{J}_s)^{i_3}_{i_4}\\
(T^{J}_s)^{i_1}_{i_2} (T^{J}_s)^{i_4}_{i_3}
\end{pmatrix}
\frac{s_{13}-s_{14}}{2s_{12}},\\
&\begin{tikzpicture}[baseline=0.8cm] \begin{feynhand}
\vertex [particle] (i1) at (0,0) {$2^0$};
\vertex [particle] (i2) at (0,1.6) {$1^0$};
\vertex [particle] (i3) at (1.6,1.6) {$4^0$};
\vertex [particle] (i4) at (1.6,0) {$3^0$};
\vertex (v1) at (0.8,0.4);
\vertex (v2) at (0.8,1.2);
\graph{(i1) --[sca] (v1)--[sca] (i4)};
\graph{(i2) --[sca] (v2)--[sca] (i3)};
\graph{(v1) --[bos] (v2)};
\end{feynhand} \end{tikzpicture}:\quad
\begin{pmatrix}
-(T^{J}_s)^{i_1}_{i_4} (T^{J}_s)^{i_3}_{i_2}\\
0
\end{pmatrix}
\frac{s_{13}-s_{12}}{2s_{14}},\\
&\begin{tikzpicture}[baseline=0.8cm] \begin{feynhand}
\setlength{\feynhandtopsep}{5pt}
\vertex [particle] (i1) at (0,0) {$2^0$};
\vertex [particle] (i2) at (0,1.6) {$1^0$};
\vertex [particle] (i3) at (1.6,1.6) {$4^0$};
\vertex [particle] (i4) at (1.6,0) {$3^0$};
\vertex (v1) at (0.8,0.4);
\vertex (v2) at (0.8,1.2);
\graph{(i2)--[sca] (v2)--[bos] (v1)-- [sca](i1)};
\graph{(i4)--[sca] (v2)};
\graph{(i3)--[sca,top] (v1)};
\end{feynhand} \end{tikzpicture}:\quad
\begin{pmatrix}
0\\
-(T^{J}_s)^{i_1}_{i_3} (T^{J}_s)^{i_4}_{i_2}
\end{pmatrix}
\frac{s_{14}-s_{12}}{2s_{13}},\\
&\begin{tikzpicture}[baseline=0.8cm] \begin{feynhand}
\vertex [particle] (i1) at (0,0) {$2^0$};
\vertex [particle] (i2) at (0,1.6) {$1^0$};
\vertex [particle] (i3) at (1.6,1.6) {$4^0$};
\vertex [particle] (i4) at (1.6,0) {$3^0$};
\vertex (v1) at (0.8,0.8);
\graph{(i1) --[sca] (v1)--[sca] (i2)};
\graph{(i4) --[sca] (v1)--[sca] (i3)};
\end{feynhand} \end{tikzpicture}:\quad
-\begin{pmatrix}
\delta^{(i_1}_{i_2} \delta^{i_3)}_{i_4}\\
\delta^{(i_1}_{i_2} \delta^{i_4)}_{i_3}
\end{pmatrix}
4\lambda.
\end{align}

Then we multiply the transformation matrix $\mathcal T_1$ and $\mathcal T_2$,
\begin{equation} \begin{aligned}
\mathcal{T}_1=\mathcal{U}^{W^+}_{i_1}\mathcal{U}^{i_2 W^-}\mathcal{U}^{h}_{i_3}\mathcal{U}^{i_4 h},\quad
\mathcal{T}_2=\mathcal{U}^{W^+}_{i_1}\mathcal{U}^{i_2 W^-}\mathcal{U}^{h i_3}\mathcal{U}^{i_4}_{h}.
\end{aligned} \end{equation}
The group tensors become
\begin{equation} \begin{aligned} \label{eq:A4H}
(T^{J}_s)^{i_1}_{i_2} (T^{J}_s)^{i_3}_{i_4}&\xrightarrow{\mathcal{T}_1}\frac{g^2}{2},\qquad
(T^{J}_s)^{i_1}_{i_4} (T^{J}_s)^{i_3}_{i_2}\xrightarrow{\mathcal{T}_1}-g^2,\qquad
\lambda\delta^{(i_1}_{i_2} \delta^{i_3)}_{i_4}\xrightarrow{\mathcal{T}_1}-\frac12\lambda,\\
(T^{J}_s)^{i_1}_{i_2} (T^{J}_s)^{i_4}_{i_3}&\xrightarrow{\mathcal{T}_2}\frac{g^2}{2},\qquad
(T^{J}_s)^{i_1}_{i_3} (T^{J}_s)^{i_4}_{i_2}\xrightarrow{\mathcal{T}_2}-g^2,\qquad
\lambda\delta^{(i_1}_{i_2} \delta^{i_4)}_{i_3}\xrightarrow{\mathcal{T}_2}-\frac12\lambda.
\end{aligned} \end{equation}
The amplitude now becomes
\begin{equation} \begin{aligned}
g^2\frac{s_{13}-s_{12}}{2s_{14}}
+g^2\frac{s_{14}-s_{12}}{2s_{13}}
+4\lambda.
\end{aligned} \end{equation}
It shows the (12)-channel contribution vanishes.

\begin{figure}[htbp]
\centering
\includegraphics[width=\linewidth,valign=c]{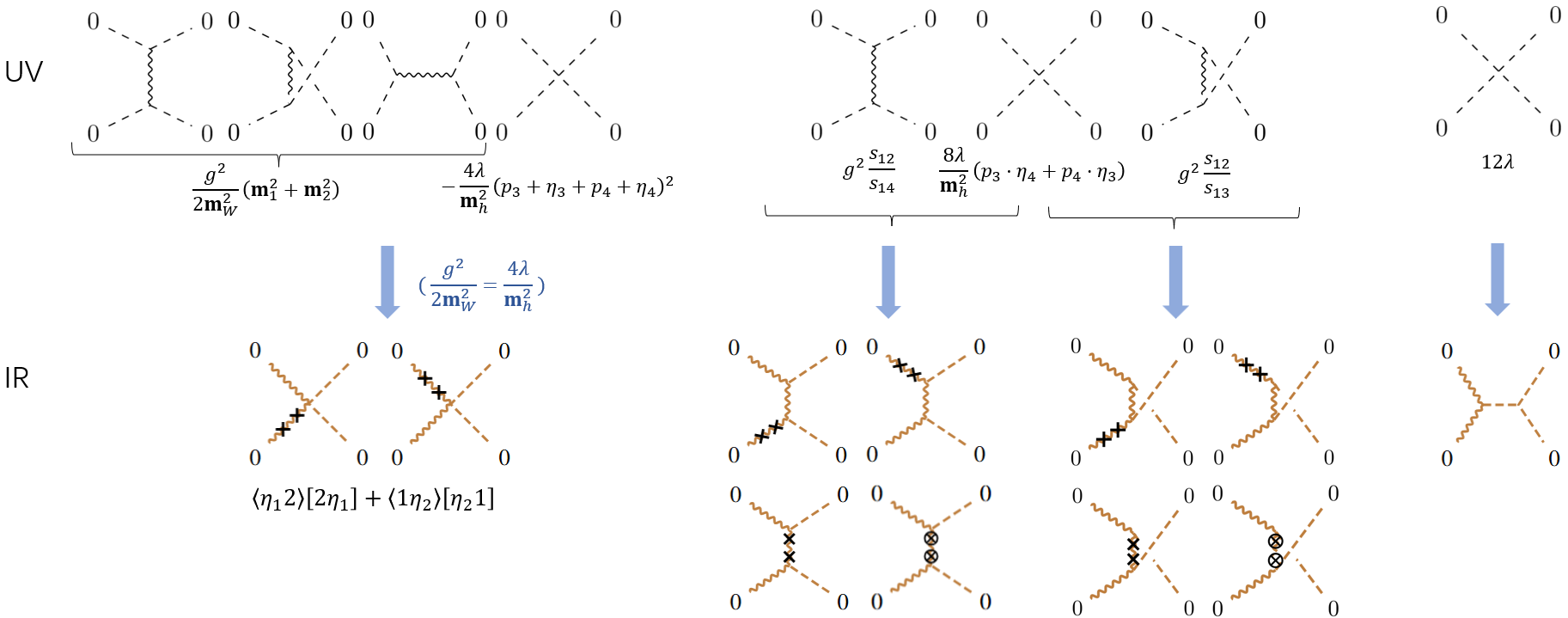}
\caption{The diagrammatic matching for helicity category $(0,0,0,0)$ of $W^+W^-hh$ amplitude.}
\label{fig:diagram_match_2}
\end{figure}

As discussed in section~\ref{sec:contact_4pt}, we can separate this amplitude expression into two contributions, 
\begin{equation} \begin{aligned} \label{eq:two_term}
&\text{matched factorized term}:& &-g^2\frac{s_{12}}{s_{14}}
-g^2\frac{s_{12}}{s_{13}}
+\frac{8\lambda}{\mathbf m_h^2}(p_3\cdot\eta_4+\eta_3\cdot p_3)^2+12\lambda,\\
&\text{matched contact term}:& &\frac{g^2}{2\mathbf m_W^2}(\mathbf m_1^2+\mathbf m_2^2)-\frac{4\lambda}{\mathbf m_h^2}(p_3+\eta_3+p_4+\eta_4)^2.
\end{aligned} \end{equation}
As shown in Fig.~\ref{fig:diagram_match_2}, these contributions belongs to different massless diagrams, and they will match to the following terms of the MHC amplitude
\begin{equation}
[\mathcal{M}]_0=
[\mathcal{M}_{(12)}]_0+
[\mathcal{M}_{(13)}]_0+
[\mathcal{M}_{(14)}]_0+
[\mathcal{M}_{\text{ct}}]_0.
\end{equation}
They correspond to (12), (13), (14)-channel and contact amplitude separately.

\begin{itemize}
\item contact term: Using the condition $\frac{g^2}{2\mathbf m_W^2}=\frac{4\lambda}{\mathbf m_h^2}$, we can convert the amplitude in eq.~\eqref{eq:two_term} to the MHC contact term
\begin{equation} \begin{aligned}
[\mathcal{M}_{\text{ct}}]_0=-\frac{g^2}{2}([1\eta_2]\langle\eta_21\rangle+[\eta_12]\langle2\eta_1\rangle).\\
\end{aligned} \end{equation}

\item (12)-channel: There is no term with the pole $s_{12}$ now, so we need to extract a such pole from it. This extraction can be only applied to the term with no pole, which corresponds to the following scaling change
\begin{equation} \begin{aligned} \label{eq:scaling_extract}
\mathbb E(1)\times\mathbb I(1)\times \mathbb P(1)
\quad\to\quad
\mathbb E(\tilde\lambda^2\lambda^2)\times\mathbb I(1)\times \mathbb P(p^{-2}).\\
\end{aligned} \end{equation}
For the term with $s_{13}$ and $s_{14}$ pole, such extraction will give the double pole contribution, which is not we need. The corresponding amplitude deformation is
\begin{equation} \begin{aligned}
12 \lambda\quad \Rightarrow \quad [\mathcal{M}_{(12)}]_0=12 \lambda\frac{[12]\langle21\rangle}{s_{12}}.
\end{aligned} \end{equation}
The term $(p_3\cdot\eta_4+\eta_3\cdot p_3)^2$ also do not have pole, but it does not satisfy the scaling in eq.~\eqref{eq:scaling_extract}, so it will contribute to not (12)-channel, but (13) and (14)-channels.

\item (13) and (14)-channel:
These two channels inlcude more diagrams than the former contributions. Analyze each scaling change can give the corresponding amplitude deformation is possible, but here we can use a more convenient method. It is to use the amplitude expression in helicity (+1,-1,0,0), and use the extended little group  covariance to convert them into helicity (0,0,0,0). The result of (13)-channel is 
\begin{equation} \begin{aligned} \label{eq:Asub}
[\mathcal{M}_{(13)}]_0=&\frac{g^2}{2}\bigg(\frac{[13]\langle 31\rangle([2\eta_4]\langle \eta_42\rangle - [\eta_24]\langle 4\eta_2\rangle)+([1\eta_3]\langle \eta_31\rangle - [\eta_13]\langle 3\eta_1\rangle)[24]\langle 42\rangle}{\mathbf{m}_1 \mathbf{m}_2 s_{13}}\\
&+\frac{2\mathbf m_W^2 [12]\langle 21\rangle}{\mathbf{m}_1 \mathbf{m}_2 s_{13}}-\frac{[13]\langle 31\rangle [24]\langle 42\rangle}{s_{13}\mathbf{m}_1 \mathbf{m}_2}\frac{2p_1\cdot \eta_3+2p_3\cdot \eta_1+\mathbf{m}_1^2+\mathbf{m}_3^2-\mathbf{m}_W^2}{s_{13}}\bigg).\\
\end{aligned} \end{equation}
The (14)-channel can be obtained by exchange particles 3 and 4.

\end{itemize}

Combining these contritbuion, the total amplitude is
\begin{equation} \begin{aligned}
[\mathcal{M}]_0=
&\frac{g^2}{2}\bigg(\frac{[13]\langle 31\rangle([2\eta_4]\langle \eta_42\rangle - [\eta_24]\langle 4\eta_2\rangle)+([1\eta_3]\langle \eta_31\rangle - [\eta_13]\langle 3\eta_1\rangle)[24]\langle 42\rangle}{\mathbf{m}_1 \mathbf{m}_2 s_{13}}\\
&\qquad+\frac{2\mathbf m_W^2 [12]\langle 21\rangle}{\mathbf{m}_1 \mathbf{m}_2 s_{13}}-\frac{[13]\langle 31\rangle [24]\langle 42\rangle}{s_{13}\mathbf{m}_1 \mathbf{m}_2}\frac{2p_1\cdot \eta_3+2p_3\cdot \eta_1+\mathbf{m}_1^2+\mathbf{m}_3^2-\mathbf{m}_W^2}{s_{13}}\bigg)\\
&+\frac{g^2}{2}\bigg(\frac{[14]\langle 41\rangle([2\eta_3]\langle \eta_32\rangle - [\eta_23]\langle 3\eta_2\rangle)+([1\eta_4]\langle \eta_41\rangle - [\eta_14]\langle 4\eta_1\rangle)[23]\langle 32\rangle}{\mathbf{m}_1 \mathbf{m}_2 s_{14}}\\
&\qquad+\frac{2\mathbf m_W^2 [12]\langle 21\rangle}{\mathbf{m}_1 \mathbf{m}_2 s_{14}}-\frac{[14]\langle 41\rangle [23]\langle 32\rangle}{s_{14}\mathbf{m}_1 \mathbf{m}_2}\frac{2p_1\cdot \eta_4+2p_4\cdot \eta_1+\mathbf{m}_1^2+\mathbf{m}_4^2-\mathbf{m}_W^2}{s_{14}}\bigg)\\
&+12 \lambda\frac{[12]\langle21\rangle}{s_{12}}-\frac{g^2}{2}([1\eta_2]\langle\eta_21\rangle+[\eta_12]\langle2\eta_1\rangle).
\end{aligned} \end{equation}

\paragraph{Matching Results for Unitarized Amplitudes}

Now we have derived the MHC amplitude from massless amplitudes in the two helicity category $(+,-,0,0)$ and $(0,0,0,0)$. Restore the $SU(2)$ little group covariance, these matching result can simply gives the AHH amplitude.

In $(+,-,0,0)$, we can rewrite the gauge coupling to massive coefficient as $g^2\to (\mathbf{g}^{W^+ W^-})^2$, so the MHC amplitude is
\begin{equation} \begin{aligned}
[\mathcal{M}(1^{+1},2^{-1},3^0,4^0)]_0=
&\frac{(\mathbf{g}^{W^+ W^-})^2}{2}
\frac{\langle\eta_1|P_{14}|1]\langle2|P_{14}|\eta_2]}{s_{14}\tilde{m}_1 m_2} +\frac{(\mathbf{g}^{W^+ W^-})^2}{2}
\frac{\langle\eta_1|P_{13}|1]\langle2|P_{13}|\eta_2]}{s_{13}\tilde{m}_1 m_2} \\
&+\frac{(\mathbf{g}^{W^+ W^-})^2}{2}\frac{\langle\eta_1 2\rangle[\eta_21]}{\tilde m_1 m_2}.
\end{aligned} \end{equation}
Bolding it, we obtain
\begin{equation} \begin{aligned} \label{eq:AHH_1}
\mathbf{M}_{1}
=&\frac{(\mathbf{g}^{W^+ W^-})^2}{2\mathbf{m}_W^2}\frac{\langle\mathbf{1}|\mathbf{p}_3|\mathbf{1}]\langle\mathbf{2}|\mathbf{p}_4|\mathbf{2}]}{\mathbf{P}_{13}^2-\mathbf{m}_W^2}
+\frac{(\mathbf{g}^{W^+ W^-})^2}{2\mathbf{m}_W^2}\frac{\langle\mathbf{1}|\mathbf{p}_4|\mathbf{1}]\langle\mathbf{2}|\mathbf{p}_3|\mathbf{2}]}{\mathbf{P}_{14}^2-\mathbf{m}_W^2}
+\frac{(\mathbf{g}^{W^+ W^-})^2}{2\mathbf{m}_W^2}[\mathbf{12}]\langle\mathbf{21}\rangle.
\end{aligned} \end{equation}

In $(0,0,0,0)$, we need to rewrite another coupling to massive coefficient as $12\lambda\to \mathbf{g}^{W^+ W^-}\lambda_3$. The MHC amplitude include the following terms
\begin{equation} \begin{aligned}
[\mathcal{M}(1^{0},2^{0},3^0,4^0)]_0 \supset
\frac{\mathbf{g}^{W^+ W^-} \lambda_3}{\mathbf m_W}\frac{[12]\langle21\rangle}{s_{12}}.
\end{aligned} \end{equation}
Bolding it, we get the following AHH structure 
\begin{equation} \begin{aligned}
\mathbf{M}_{2}
&=\frac{\mathbf{g}^{W^+ W^-} \lambda_3}{\mathbf m_W}\frac{[\mathbf{12}]\langle\mathbf{21}\rangle}{\mathbf{P}_{12}^2-\mathbf{m}_h^2}+\frac{(\mathbf{g}^{W^+ W^-})^2}{2\mathbf{m}_W^2}\frac{2\mathbf{m}_W^2[\mathbf{12}]\langle\mathbf{21}\rangle}{\mathbf{P}_{13}^2-\mathbf{m}_W^2}+\frac{(\mathbf{g}^{W^+ W^-})^2}{2\mathbf{m}_W^2}\frac{2\mathbf{m}_W^2[\mathbf{12}]\langle\mathbf{21}\rangle}{\mathbf{P}_{14}^2-\mathbf{m}_W^2}.
\end{aligned} \end{equation}
The other MHC terms in $(0,0,0,0)$ will match to Eq.~\eqref{eq:AHH_1}.

Sum over these two contributions, we obtain the total AHH amplitude,
\begin{equation}\label{eq:WWhhresult} \begin{aligned}
\mathbf{M}(W^+W^-hh)
=&\mathbf{M}_1+\mathbf{M}_2\\
=&\frac{\mathbf{g}^{W^+ W^-}\mathbf{g}^{W^+ W^-}}{2\mathbf{m}_W^2}\frac{\langle\mathbf{1}|\mathbf{p}_3|\mathbf{1}]\langle\mathbf{2}|\mathbf{p}_4|\mathbf{2}]+2\mathbf{m}_W^2[\mathbf{12}]\langle\mathbf{21}\rangle}{\mathbf{P}_{13}^2-\mathbf{m}_W^2}\\
&+\frac{\mathbf{g}^{W^+ W^-}\mathbf{g}^{W^+ W^-}}{2\mathbf{m}_W^2}\frac{\langle\mathbf{1}|\mathbf{p}_4|\mathbf{1}]\langle\mathbf{2}|\mathbf{p}_3|\mathbf{2}]+2\mathbf{m}_W^2[\mathbf{12}]\langle\mathbf{21}\rangle}{\mathbf{P}_{14}^2-\mathbf{m}_W^2}\\
&+\frac{\mathbf{g}^{W^+ W^-} \lambda_3}{\mathbf m_W}\frac{[\mathbf{12}]\langle\mathbf{21}\rangle}{\mathbf{P}_{12}^2-\mathbf{m}_h^2}
+\frac{\mathbf{g}^{W^+ W^-}\mathbf{g}^{W^+ W^-}}{2\mathbf{m}_W^2}[\mathbf{12}]\langle\mathbf{21}\rangle.
\end{aligned} \end{equation}

\subsection{Sub-leading MHC matching}

After obtained the leading matching, let us consider the sub-leading matching for MHC amplitudes. We continue to use the $WWhh$ amplitude as an example to illustrate the matching procedure.

In the leading matching, the 4-pt massless amplitude is matched to the leading MHC amplitude $[\mathcal{M}]_0$. For the WWhh amplitude, this gives
\begin{equation} \begin{aligned}
\mathcal{A}(1^{+}, 2^{-}, 3^{0}, 4^{0})&\rightarrow [\mathcal{M}(\mathbf{1}^{+}, \mathbf{2}^{-}, \mathbf{3}^{0}, \mathbf{4}^{0})]_{0},\\
\mathcal{A}(1^{0}, 2^{0}, 3^{0}, 4^{0})&\rightarrow [\mathcal{M}(\mathbf{1}^{0}, \mathbf{2}^{0}, \mathbf{3}^{0}, \mathbf{4}^{0})]_{0},
\end{aligned} \end{equation}
where the superscripts denote the helicity.

In the subleading matching, we match $(4+l)$-pt massless amplitudes with $l$ additional Higgs bosons to subleading MHC amplitude $[\mathcal{M}]_{l>0}$. Once the leading matching ($l=0$) is known, the sub-leading matching ($l>0$) can be obtained by attaching additional Higgs bosons to particle lines, following what we call \textit{Higgsing rules}.  Depending on whether the additional Higgs bosons are attached to external or internal particles, there are two types of Higgsing rules. 

Below, we illustrate these two kinds of rules using two typical sub-leading matchings for $l=1$ and $l=2$:
\begin{equation} \begin{aligned}
\mathcal{A}(1^{+}, 2^{0}, 3^{0}, 4^{0};5^0)&\rightarrow [\mathcal{M}(\mathbf{1}^{+}, \mathbf{2}^{0}, \mathbf{3}^{0}, \mathbf{4}^{0})]_{1},\\
\mathcal{A}(1^{+}, 2^{-}, 3^{0}, 4^{0};5^0,6^0)&\rightarrow [\mathcal{M}(\mathbf{1}^{+}, \mathbf{2}^{-}, \mathbf{3}^{0}, \mathbf{4}^{0})]_{2}.
\end{aligned} \end{equation}
where massless particles 5 and 6 are the additional Higgs bosons. Subleading matching for other helicity categories and higher orders can be obtained similarly using these two Higgsing rules.

\paragraph{External Higgsing rule}

We first consider the subleading matching for massless amplitude with helicity $(+0000)$, which contains one additional Higgs boson. This additional Higgs boson can be viewed as having split from one of the external particles of the 4-pt massless amplitude used in the leading matching. The 4-pt amplitude can have helicity $(+-00)$ and $(0000)$. To obtain the complete subleading MHC amplitude, both cases must be considered.

\begin{figure}[htbp]
\centering
\includegraphics[width=0.95\linewidth,valign=c]{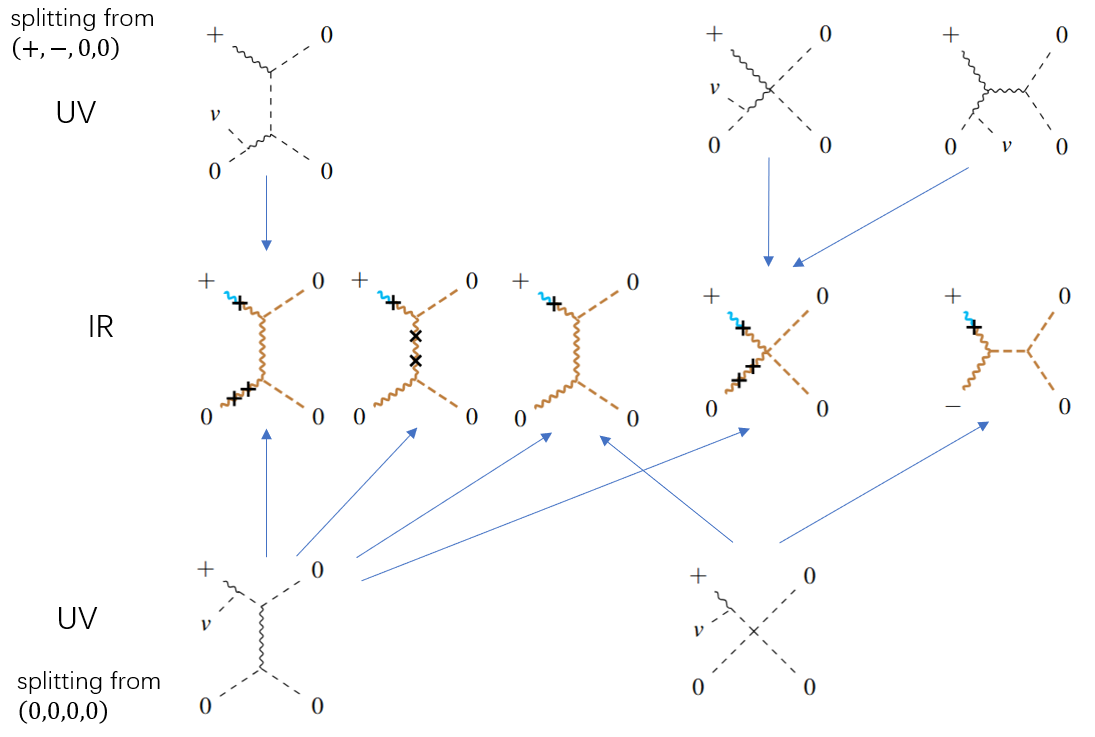}
\caption{The diagrammatic matching for helicity category $(+1,0,0,0)$ of $W^+W^-hh$ amplitude.}
\label{fig:diagram_match_3}
\end{figure}

Let us begin with the $(+-00)$ 4-pt amplitude. Focusing on particle 2, we recall that in the leading matching this gauge boson with negative helicity corresponds to
\begin{equation}
\begin{tikzpicture}[baseline=0.7cm]
\begin{feynhand}
\setlength{\feynhandblobsize}{6mm}
\setlength{\feynhandarrowsize}{3.5pt}
\vertex [dot] (v1) at (-0.4,0.8) {};
\vertex [particle] (i2) at (1.3,0.8) {$-$};
\graph{(i2) --[bos] (v1)};
\end{feynhand}
\end{tikzpicture}
\quad\to\quad 
\begin{tikzpicture}[baseline=0.7cm]
\begin{feynhand}
\setlength{\feynhandblobsize}{6mm}
\vertex [dot] (v1) at (-0.4,0.8) {};
\vertex (v2) at (0.4,0.8);
\vertex [particle] (i2) at (1.3,0.8) {$-$};
\draw[brown,thick bos] (v1)--(v2);
\draw[red,thick bos] (v2)--(i2);
\draw[very thick] plot[mark=x,mark size=3.5,mark options={rotate=0}] coordinates {(v2)};
\end{feynhand}
\end{tikzpicture}=\frac{1}{\tilde m_2}|2\rangle|\eta_2].
\end{equation}
Going from $(+-00)$ to $(+0000)$, the helicity of particle 2 changes from $-1$ to $0$, while particle 5 (the additional Higgs boson) is created. This process can be interpreted as a \textit{Higgs splitting}:  a gauge boson with negative helicity splits into a scalar boson (zero helicity) and an extra Higgs boson, illustrated as
\begin{equation}
\begin{tikzpicture}[baseline=0.7cm]
\begin{feynhand}
\setlength{\feynhandblobsize}{6mm}
\setlength{\feynhandarrowsize}{3.5pt}
\vertex [dot] (v1) at (-0.4,0.8) {};
\vertex [particle] (i2) at (1.3,0.8) {$-$};
\graph{(i2) --[bos] (v1)};
\end{feynhand}
\end{tikzpicture}
\quad\overset{\text{split}}{\to}\quad
\begin{tikzpicture}[baseline=-0.1cm] \begin{feynhand}
\setlength{\feynhandblobsize}{6mm}
\setlength{\feynhandarrowsize}{5pt}
\vertex [particle] (i1) at (2,0) {$0$};
\vertex [particle] (i2) at (1.9,0.7) {$h$};
\vertex (v2) at (1,0);
\vertex [dot] (v1) at (0,0) {};
\graph{(i1)--[sca](v2)--[bos](v1)};
\graph{(i2)--[sca](v2)};
\end{feynhand} \end{tikzpicture}.
\end{equation}
The internal particle need to be slightly off-shell, so its momentum $P_{25}$ can be decomposed as
\begin{equation}
P_{25}=p_\chi+q=|\chi]\langle\chi|+q,
\end{equation}
where $p_\chi$ is the on-shell part and $q$ is the off-shell part. Assuming $p_\chi\gg q$, the line with the Higgs insertion factorizes into the original line multiplied by a 3-pt amplitude,
\begin{equation} \begin{aligned}
\frac{|\eta_{\chi}]_{\dot\alpha}|\chi\rangle_{\alpha}}{[\eta_{\chi}\xi]}\times\frac{[5\chi][\chi 2]}{[52]}\sim \frac{1}{2}\left(\frac{(|5])_{\dot\alpha}(P_{25}|2])_{\alpha}}{[52]}-(2\leftrightarrow 5)\right).
\end{aligned} \end{equation}
We now treat the two massless particles 2 and 5 as a single massive particle. Taking the on-shell limit $p_5\to \eta_2$, the momentum of the extra Higgs boson becomes the subleading part of the massive momentum. Consequently, the splitting structure matches the massive MHC state,
\begin{equation} \begin{aligned}
\begin{tikzpicture}[baseline=-0.1cm] \begin{feynhand}
\setlength{\feynhandblobsize}{6mm}
\setlength{\feynhandarrowsize}{5pt}
\vertex [particle] (i1) at (2,0) {$0$};
\vertex [particle] (i2) at (1.9,0.7) {$h$};
\vertex (v2) at (1,0);
\vertex [dot] (v1) at (0,0) {};
\graph{(i1)--[sca](v2)--[bos](v1)};
\graph{(i2)--[sca](v2)};
\end{feynhand} \end{tikzpicture}
&= \frac{1}{2}\lim_{p_5\rightarrow \eta_2} \left(\frac{(|5])_{\dot\alpha}(P|2])_{\alpha}}{[52]}-(2\leftrightarrow 5)\right) \\
&\sim \frac{1}{2}(|\eta_2]_{\dot\alpha}|\eta_2\rangle_{\alpha}-|2]_{\dot\alpha}|2\rangle_{\alpha})  \\
& \rightarrow 
\begin{tikzpicture}[baseline=-0.1cm] \begin{feynhand}
\vertex [particle] (i1) at (2,0) {$0$};
\vertex [dot] (v1) at (0,0) {};
\vertex (v2) at (0.95,0);
\vertex (v3) at (1.4,0);
\draw[brown,thick bos] (v1)--(i1);
\draw[very thick] plot[mark=x,mark size=3.5] coordinates {(v2)};
\draw[very thick] plot[mark=x,mark size=3.5] coordinates {(v3)};
\end{feynhand} \end{tikzpicture}
+\begin{tikzpicture}[baseline=-0.1cm] \begin{feynhand}
\setlength{\feynhandblobsize}{6mm}
\setlength{\feynhandarrowsize}{5pt}
\vertex [particle] (i1) at (2,0) {$0$};
\vertex [dot] (v1) at (0,0) {};
\draw[brown,thick bos] (i1)--(v1);
\end{feynhand} \end{tikzpicture}.
\end{aligned} \end{equation}
Here both the leading and subleading terms appear. Since we are focusing on the subleading matching, only the term $|\eta]|\eta\rangle$ contributes to the subleading MHC amplitude. Thus we can extract a matching rule from leading to subleading order
\begin{equation}
\begin{array}{ccc}
\begin{tikzpicture}[baseline=0.7cm]
\begin{feynhand}
\setlength{\feynhandblobsize}{6mm}
\setlength{\feynhandarrowsize}{3.5pt}
\vertex [dot] (v1) at (-0.4,0.8) {};
\vertex [particle] (i2) at (1.3,0.8) {$-$};
\graph{(i2) --[bos] (v1)};
\end{feynhand}
\end{tikzpicture}  & 
\to &
\begin{tikzpicture}[baseline=0.7cm]
\begin{feynhand}
\setlength{\feynhandblobsize}{6mm}
\vertex [dot] (v1) at (-0.4,0.8) {};
\vertex (v2) at (0.4,0.8);
\vertex [particle] (i2) at (1.3,0.8) {$-$};
\draw[brown,thick bos] (v1)--(v2);
\draw[red,thick bos] (v2)--(i2);
\draw[very thick] plot[mark=x,mark size=3.5,mark options={rotate=0}] coordinates {(v2)};
\end{feynhand} \end{tikzpicture}=|\eta_2]|2\rangle \\
\hspace{-1.3cm}\text{split}\downarrow &  &  \\
\begin{tikzpicture}[baseline=-0.1cm] \begin{feynhand}
\setlength{\feynhandblobsize}{6mm}
\setlength{\feynhandarrowsize}{5pt}
\vertex [particle] (i1) at (2,0) {$0$};
\vertex [particle] (i2) at (1.9,0.7) {$h$};
\vertex (v2) at (1,0);
\vertex [dot] (v1) at (0,0) {};
\graph{(i1)--[sca](v2)--[bos](v1)};
\graph{(i2)--[sca](v2)};
\end{feynhand} \end{tikzpicture}
 & \to &
\begin{tikzpicture}[baseline=0.7cm]
\begin{feynhand}
\setlength{\feynhandblobsize}{6mm}
\vertex [dot] (v1) at (-0.4,0.8) {};
\vertex [particle] (i2) at (1.3,0.8) {$0$};
\draw[brown,thick bos] (v1)--(i2);
\draw[very thick] plot[mark=x,mark size=3.5,mark options={rotate=0}] coordinates {(0.1,0.8)};
\draw[very thick] plot[mark=x,mark size=3.5,mark options={rotate=0}] coordinates {(0.6,0.8)};
\end{feynhand}
\end{tikzpicture}=|\eta_2]|\eta_2\rangle
\end{array}
\end{equation}
This is called the external Higgsing rule. Using it, we can match the massless amplitude with helicity $(+0000)$ to the subleading MHC amplitude. As an explicit  example, let us consider the following diagrammatic matching: 
\begin{equation}
\begin{array}{ccc}
\begin{tikzpicture}[baseline=0.8cm] \begin{feynhand}
\vertex [particle] (i1) at (0,0) {$-$};
\vertex [particle] (i2) at (0,1.6) {$+$};
\vertex [particle] (i3) at (1.6,1.6) {$0$};
\vertex [particle] (i4) at (1.6,0) {$0$};
\vertex (v1) at (0.8,0.4);
\vertex (v2) at (0.8,1.2);
\graph{(i1) --[bos] (v1)--[sca] (i4)};
\graph{(i2) --[bos] (v2)--[sca] (i3)};
\graph{(v1) --[sca] (v2)};
\end{feynhand} \end{tikzpicture}  & 
\to &
\begin{tikzpicture}[baseline=1.1cm] \begin{feynhand}
\vertex [particle] (i1) at (0,2.2) {$+$};
\vertex [particle] (i2) at (0,0) {$-$};
\vertex [particle] (i3) at (1.8,0) {$0$};
\vertex [particle] (i4) at (1.8,2.2) {$0$};
\vertex (v1) at (0.9,0.6);
\vertex (v2) at (0.9,1.6);
\vertex (v4) at (0.9-0.9*0.4,1.6+0.6*0.4);
\vertex (v5) at (0.9-0.9*0.4,0.6-0.6*0.4);
\draw[cyan,thick bos] (i1)--(v4);
\draw[brown,thick bos] (v4)--(v2);
\draw[brown,thick bos] (v2)--(v1);
\draw[brown,thick bos] (v5)--(v1);
\draw[red,thick bos] (i2)--(v5);
\draw[brown,thick sca] (i3)--(v1);
\draw[brown,thick sca] (i4)--(v2);
\draw[very thick] plot[mark=x,mark size=3.5,mark options={rotate=45}] coordinates {(v4)};
\draw[very thick] plot[mark=x,mark size=3.5,mark options={rotate=45}] coordinates {(v5)};
\end{feynhand} \end{tikzpicture}
=\frac12 g^2 \frac{\langle\eta_1|P_{14}|1]\langle 2|P_{14}|\eta_2]}{s_{14}\tilde{m}_1 m_2} \\
\hspace{-1.3cm}\text{split}\downarrow &  &  \\
\begin{tikzpicture}[baseline=1.1cm] \begin{feynhand}
\vertex [particle] (i1) at (0,2.2) {$+$};
\vertex [particle] (i2) at (0,0) {$0$};
\vertex [particle] (i3) at (1.8,0) {$0$};
\vertex [particle] (i4) at (1.8,2.2) {$0$};
\vertex [particle] (i5) at (0,0.9) {$v$};
\vertex (v1) at (0.9,0.6);
\vertex (v2) at (0.9,1.6);
\vertex (v4) at (0.9-0.9*0.4,1.6+0.6*0.4);
\vertex (v5) at (0.9-0.9*0.4,0.6-0.6*0.4);
\graph{(i1)--[bos] (v2)--[sca] (i4)};
\graph{(v2)--[sca] (v1)};
\graph{(i2)--[sca] (v5)--[bos] (v1)--[sca] (i3)};
\graph{(i5)--[sca] (v5)};
\end{feynhand} \end{tikzpicture}
 & \to &
\begin{tikzpicture}[baseline=1.1cm] \begin{feynhand}
\vertex [particle] (i1) at (0,2.2) {$+$};
\vertex [particle] (i2) at (0,0) {$0$};
\vertex [particle] (i3) at (1.8,0) {$0$};
\vertex [particle] (i4) at (1.8,2.2) {$0$};
\vertex (v1) at (0.9,0.6);
\vertex (v2) at (0.9,1.6);
\vertex (v4) at (0.9-0.9*0.4,1.6+0.6*0.4);
\draw[cyan,thick bos] (i1)--(v4);
\draw[brown,thick bos] (v4)--(v2);
\draw[brown,thick bos] (v2)--(v1);
\draw[brown,thick bos] (i2)--(v1);
\draw[brown,thick sca] (i3)--(v1);
\draw[brown,thick sca] (i4)--(v2);
\draw[very thick] plot[mark=x,mark size=3.5,mark options={rotate=45}] coordinates {(v4)};
\draw[very thick] plot[mark=x,mark size=3.5,mark options={rotate=45}] coordinates {(0.9-0.9*0.25,0.6-0.6*0.25)};
\draw[very thick] plot[mark=x,mark size=3.5,mark options={rotate=45}] coordinates {(0.9-0.9*0.5,0.6-0.6*0.5)};
\end{feynhand} \end{tikzpicture}
=\frac12 g^2
\frac{\langle\eta_1|P_{14}|1]\langle\eta_2|P_{14}|\eta_2]}{s_{14}\tilde{m}_1 \mathbf m_2}
\end{array}
\end{equation}
Other subleading MHC terms can be derived in a similar way, as illustrated in figure~\ref{fig:diagram_match_3}. Note, however, that this external Higgsing rule does not produce all subleading MHC amplitudes. Contributions from other helicity must also be taken into account.

Then we consider the helicity $(0000)$. From $(0000)$ to $(+0000)$, the helicity of particle 1 is changed from $0$ to $+1$. This requires another kind of external Higgsing rules:
\begin{equation}
\begin{array}{ccc}
\begin{tikzpicture}[baseline=0.7cm]
\begin{feynhand}
\setlength{\feynhandblobsize}{6mm}
\setlength{\feynhandarrowsize}{3.5pt}
\vertex [dot] (v1) at (-0.4,0.8) {};
\vertex [particle] (i2) at (1.3,0.8) {$0$};
\graph{(i2) --[sca] (v1)};
\end{feynhand}
\end{tikzpicture}  & 
\to &
\begin{tikzpicture}[baseline=0.7cm]
\begin{feynhand}
\setlength{\feynhandblobsize}{6mm}
\vertex [dot] (v1) at (-0.4,0.8) {};
\vertex [particle] (i2) at (1.3,0.8) {$0$};
\draw[brown,thick bos] (v1)--(i2);
\end{feynhand}
\end{tikzpicture}
=|p\rangle|p] \\
\hspace{-1.3cm}\text{split}\downarrow &  &  \\
\begin{tikzpicture}[baseline=-0.1cm] \begin{feynhand}
\setlength{\feynhandblobsize}{6mm}
\setlength{\feynhandarrowsize}{5pt}
\vertex [particle] (i1) at (2,0) {$+$};
\vertex [particle] (i2) at (1.9,0.7) {$h$};
\vertex (v2) at (1,0);
\vertex [dot] (v1) at (0,0) {};
\graph{(i1)--[bos](v2)--[sca](v1)};
\graph{(i2)--[sca](v2)};
\end{feynhand} \end{tikzpicture}
 & \to &
\begin{tikzpicture}[baseline=0.7cm] \begin{feynhand}
\setlength{\feynhandblobsize}{6mm}
\vertex [dot] (v1) at (-0.4,0.8) {};
\vertex (v2) at (0.4,0.8);
\vertex [particle] (i2) at (1.3,0.8) {$+$};
\draw[brown,thick bos] (v1)--(v2);
\draw[cyan,thick bos] (v2)--(i2);
\draw[very thick] plot[mark=x,mark size=3.5,mark options={rotate=0}] coordinates {(v2)};
\end{feynhand} \end{tikzpicture} 
=|\eta\rangle|p]
\end{array}
\end{equation}
Using this Higgsing rule, we can obtain the other subleading MHC amplitudes that are cannot be given from $(+-00)$, such as 
\begin{equation}
\begin{array}{ccc}
\begin{tikzpicture}[baseline=0.8cm] \begin{feynhand}
\vertex [particle] (i1) at (0,0) {$0$};
\vertex [particle] (i2) at (0,1.6) {$0$};
\vertex [particle] (i3) at (1.6,1.6) {$0$};
\vertex [particle] (i4) at (1.6,0) {$0$};
\vertex (v1) at (0.8,0.8);
\graph{(i1) --[sca] (v1)--[sca] (i2)};
\graph{(i4) --[sca] (v1)--[sca] (i3)};
\end{feynhand} \end{tikzpicture}  & 
\to &
\begin{tikzpicture}[baseline=0.8cm] \begin{feynhand}
\vertex [particle] (i1) at (0,1.8) {$0$};
\vertex [particle] (i2) at (0,0) {$0$};
\vertex [particle] (i3) at (1.8,0) {$0$};
\vertex [particle] (i4) at (1.8,1.8) {$0$};
\vertex (v1) at (0.6,0.9);
\vertex (v2) at (1.2,0.9);
\draw[brown,thick bos] (i1)--(v1);
\draw[brown,thick bos] (i2)--(v1);
\draw[brown,thick sca] (v1)--(v2);
\draw[brown,thick sca] (i4)--(v2)--(i3);
\end{feynhand} \end{tikzpicture}
=-12 \lambda\frac{\langle12\rangle [2 1]}{s_{12}} \\
\hspace{-1.3cm}\text{split}\downarrow &  &  \\
\begin{tikzpicture}[baseline=0.8cm] \begin{feynhand}
\vertex [particle] (i1) at (0,0) {$0$};
\vertex [particle] (i2) at (0,1.8) {$+$};
\vertex [particle] (i3) at (1.8,1.8) {$0$};
\vertex [particle] (i4) at (1.8,0) {$0$};
\vertex [particle] (i5) at (0,0.9) {$v$};
\vertex (v1) at (0.9,0.9);
\vertex (v5) at (0.9-0.9*0.4,0.9+0.9*0.4);
\graph{(i1)--[sca] (v1)--[sca] (v5)--[bos] (i2)};
\graph{(i4)--[sca] (v1)--[sca] (i3)};
\graph{(i5)--[sca] (v5)};
\end{feynhand} \end{tikzpicture}
 & \to &
\begin{tikzpicture}[baseline=0.8cm] \begin{feynhand}
\vertex [particle] (i1) at (0,1.8) {$+$};
\vertex [particle] (i2) at (0,0) {$0$};
\vertex [particle] (i3) at (1.8,0) {$0$};
\vertex [particle] (i4) at (1.8,1.8) {$0$};
\vertex (v1) at (0.6,0.9);
\vertex (v2) at (1.2,0.9);
\vertex (v3) at (0.9,0.9);
\vertex (v4) at (0.6-0.6*0.43,0.9+0.9*0.43);
\draw[cyan,thick bos] (i1)--(v4);
\draw[brown,thick bos] (v4)--(v1);
\draw[brown,thick bos] (i2)--(v1);
\draw[brown,thick sca] (v1)--(v2);
\draw[brown,thick sca] (i3)--(v2);
\draw[brown,thick sca] (i4)--(v2);
\draw[very thick] plot[mark=x,mark size=3.5,mark options={rotate=45}] coordinates {(v4)};
\end{feynhand} \end{tikzpicture}
=-12 \lambda\frac{\mathbf m_1}{m_1}\frac{\langle\eta_12\rangle [2 1]}{s_{12}}
\end{array}
\end{equation}

Using the external Higgsing rules, we can find all MHC subleading terms. The final result is 
\begin{align}
{[\mathcal{M}]_1}=
&\frac{g^2}{2}\bigg(\frac{\langle\eta_1|P_{13}|1](\langle2|\eta_{24}|2]-\langle\eta_2|P_{24}|\eta_2])+\langle\eta_1|\eta_{13}|1]\langle2|P_{24}|2]+2\mathbf m_W^2 \langle\eta_12\rangle [2 1]}{m_1 \mathbf{m}_2 s_{13}}\nonumber \\
&\quad-\frac{\langle\eta_1|P_{13}|1]\langle2|P_{24}|2]}{s_{13}m_1 \mathbf{m}_2}\frac{2p_1\cdot \eta_3+2p_3\cdot \eta_1+\mathbf{m}_1^2+\mathbf{m}_3^2-\mathbf{m}_W^2}{s_{13}}\bigg)\nonumber \\
&+\frac{g^2}{2}\bigg(\frac{\langle\eta_1|P_{14}|1](\langle2|\eta_{23}|2]-\langle\eta_2|P_{23}|\eta_2])+\langle\eta_1|\eta_{14}|1]\langle2|P_{23}|2]+2\mathbf m_W^2 \langle\eta_12\rangle [2 1]}{m_1 \mathbf{m}_2 s_{14}}\nonumber \\
&\quad-\frac{\langle\eta_1|P_{14}|1]\langle2|P_{23}|2]}{s_{14}m_1 \mathbf{m}_2}\frac{2p_1\cdot \eta_4+2p_4\cdot \eta_1+\mathbf{m}_1^2+\mathbf{m}_4^2-\mathbf{m}_W^2}{s_{14}}\bigg)
\nonumber \\
&-12 \lambda\frac{\mathbf m_1}{m_1}\frac{\langle\eta_12\rangle [2 1]}{s_{12}}-\frac{g^2}{2}\frac{\langle\eta_1\eta_2\rangle [\eta_2 1]}{m_1 \mathbf m_2}.
\end{align}

\paragraph{Internal Higgsing rule}

Then we consider the matching for 6-pt massless amplitudes with helicity $(+-0000)$, which contains two additional Higgs boson. In this case, the additional Higgs boson can be split not only from external particles, but also from internal ones. This require us to consider the internal Higgsing rules.

We start from the 4-pt amplitude with helicity $(+-00)$. In the leading matching, the massless internal particle in the (14)-channel match to 
\begin{equation}
\begin{tikzpicture}[baseline=0.7cm]
\begin{feynhand}
\setlength{\feynhandblobsize}{6mm}
\setlength{\feynhandarrowsize}{3.5pt}
\vertex [dot] (v1) at (-0.4,0.8) {};
\vertex [dot] (i2) at (1.3,0.8) {};
\graph{(i2) --[sca] (v1)};
\end{feynhand}
\end{tikzpicture}  
\quad\to\quad 
\begin{tikzpicture}[baseline=0.7cm] \begin{feynhand}
\setlength{\feynhandarrowsize}{4pt}
\vertex [dot] (i1) at (0,0.8) {};
\vertex [dot] (i2) at (1.6,0.8) {};
\draw[brown,thick bos] (i1)--(i2);
\end{feynhand} \end{tikzpicture}=P_{14} P_{14}.
\end{equation}
When going from $(+-00)$ to $(+-0000)$, two additional Higgs bosons are split from this internal massless, shown as
\begin{equation}  
\begin{tikzpicture}[baseline=0.7cm]
\begin{feynhand}
\setlength{\feynhandblobsize}{6mm}
\setlength{\feynhandarrowsize}{3.5pt}
\vertex [dot] (v1) at (-0.4,0.8) {};
\vertex [dot] (i2) at (1.3,0.8) {};
\graph{(i2) --[sca] (v1)};
\end{feynhand}
\end{tikzpicture}  
\quad\to\quad 
\begin{tikzpicture}[baseline=-0.1cm] \begin{feynhand}
\vertex [dot] (d1) at (0,0) {};
\vertex [dot] (d2) at (2,0) {};
\vertex [particle] (i1) at (0.7,0.7) {$5$};
\vertex [particle] (i2) at (1.3,0.7) {$6$};
\vertex (v1) at (0.7,0);
\vertex (v2) at (1.3,0);
\graph{(d2)--[sca](v2)--[bos](v1)--[sca](d1)};
\graph{(i1)--[sca](v1)};
\graph{(i2)--[sca](v2)};
\end{feynhand} \end{tikzpicture}.
\end{equation}
Here, particles 5 and 6 are the two extra Higgs bosons. This massless structure involves three propagators
\begin{equation}
P_{14}^2, P_{145}^2, P_{1456}^2,
\end{equation}
corresponding to the three internal particles in the massless diagrams. To match this to the MHC internal line, we combine the additional Higgs bosons with one of the internal particles into a massive particle. A convenient way to select this internal particle is to take its momentum to be nearly on-shell:
\begin{equation}
P_{ij\cdots}\sim p_\chi\equiv|\chi]\langle\chi|,
\end{equation}
where $p_{\chi}$ represents the on-shell internal momentum. The additional Higgs bosons and this on-shell internal particle are then packaged into a massive state. Since there are three internal particles, this yields three contributions:
\begin{itemize}
\item First we take the momentum $P_{14}\sim p_\chi$ to be nearly on-shell, which corresponds a scalar boson. The internal structure factorizes into two external pieces as
\begin{equation} \begin{aligned}
\begin{tikzpicture}[baseline=-0.1cm] \begin{feynhand}
\vertex [dot] (d1) at (0,0) {};
\vertex [dot] (d2) at (2,0) {};
\vertex [particle] (i1) at (0.7,0.7) {$5$};
\vertex [particle] (i2) at (1.3,0.7) {$6$};
\vertex (v1) at (0.7,0);
\vertex (v2) at (1.3,0);
\graph{(d2)--[sca](v2)--[bos](v1)--[sca,edge label=$p_\chi$](d1)};
\graph{(i1)--[sca](v1)};
\graph{(i2)--[sca](v2)};
\end{feynhand} \end{tikzpicture}
\quad\to\quad
\begin{tikzpicture}[baseline=-0.1cm] \begin{feynhand}
\vertex [dot] (d1) at (-0.9,0) {};
\vertex [dot] (d2) at (2,0) {};
\vertex [particle] (i1) at (0.7,0.7) {$5$};
\vertex [particle] (i2) at (1.3,0.7) {$6$};
\vertex (v1) at (0.7,0);
\vertex (v2) at (1.3,0);
\vertex (v3) at (0.1,0);
\vertex (v4) at (-0.3,0);
\graph{(d2)--[sca](v2)--[bos](v1)--[sca](v3)};
\graph{(v4)--[sca](d1)};
\graph{(i1)--[sca](v1)};
\graph{(i2)--[sca](v2)};
\end{feynhand} \end{tikzpicture}.
\end{aligned} \end{equation}
The first piece (without Higgs splitting) corresponds to the Goldstone structure $|\chi]|\chi\rangle$, while the second gives
\begin{equation} \begin{aligned}
\begin{tikzpicture}[baseline=-0.1cm] \begin{feynhand}
\vertex [dot] (d2) at (2,0) {};
\vertex [particle] (i1) at (0.7,0.7) {$5$};
\vertex [particle] (i2) at (1.3,0.7) {$6$};
\vertex (v1) at (0.7,0);
\vertex (v2) at (1.3,0);
\vertex (v3) at (0.1,0);
\graph{(d2)--[sca](v2)--[bos](v1)--[sca](v3)};
\graph{(i1)--[sca](v1)};
\graph{(i2)--[sca](v2)};
\end{feynhand} \end{tikzpicture}
=\frac{(\langle \chi|[\chi|-\langle 5|[5|)}{(p_\chi+p_5)^2}.
\end{aligned} \end{equation}
Therefore we obtain
\begin{equation} \begin{aligned}
\mathcal{I}_{(14)}=
\begin{tikzpicture}[baseline=-0.1cm] \begin{feynhand}
\vertex [dot] (d1) at (0,0) {};
\vertex [dot] (d2) at (2,0) {};
\vertex [particle] (i1) at (0.7,0.7) {$5$};
\vertex [particle] (i2) at (1.3,0.7) {$6$};
\vertex (v1) at (0.7,0);
\vertex (v2) at (1.3,0);
\graph{(d2)--[sca](v2)--[bos](v1)--[sca,edge label=$p_\chi$](d1)};
\graph{(i1)--[sca](v1)};
\graph{(i2)--[sca](v2)};
\end{feynhand} \end{tikzpicture}
\sim |\chi]|\chi\rangle\times
\frac{(\langle \chi|[\chi|-\langle 5|[5|)}{(p_\chi+p_5)^2}.
\end{aligned} \end{equation}
Here, particle 5 acts as the subleading spinor $\eta_{14}$ of the internal momentum. Taking the appropriate on-shell limit gives
\begin{equation} \begin{aligned}
\lim_{\substack{p_5\to \eta_{14}\\p_6\to 0}}\mathcal{I}_{(14)}
\sim &|\chi]|\chi\rangle\times
\frac{(\langle \chi|[\chi|-\langle \eta_{14}|[\eta_{14}|)}{(p_\chi+p_5)^2}.
\end{aligned} \end{equation}

\item Then we take the momentum $P_{145}\sim p_\chi$ to be nearly on-shell, which corresponds to an internal gauge boson. Since the gauge boson can have helicity $\pm 1$, the internal line factorizes into two terms
\begin{equation} \begin{aligned}
\mathcal{I}_{(145)}=
\begin{tikzpicture}[baseline=-0.1cm] \begin{feynhand}
\vertex [dot] (d1) at (0,0) {};
\vertex [dot] (d2) at (2,0) {};
\vertex [particle] (i1) at (0.7,0.7) {$5$};
\vertex [particle] (i2) at (1.3,0.7) {$6$};
\vertex (v1) at (0.7,0);
\vertex (v2) at (1.3,0);
\graph{(d2)--[sca](v2)--[bos,edge label=$p_\chi$](v1)--[sca](d1)};
\graph{(i1)--[sca](v1)};
\graph{(i2)--[sca](v2)};
\end{feynhand} \end{tikzpicture}
\sim \frac{|\chi]|5\rangle}{\langle\chi5\rangle}\times\frac{[6|\langle \chi|}{[6\chi]}+
\frac{|5]|\chi\rangle}{[5\chi]}\times\frac{[\chi|\langle 6|}{\langle\chi6\rangle}.
\end{aligned} \end{equation}
In this case, both particles 5 and 6 serve as the subleading spinor $\eta_{14}$, requiring a different on-shell limit
\begin{equation} \begin{aligned}
\lim_{\substack{p_5\to -\eta_{14}\\ p_6\to \eta_{14}}}\mathcal{I}_{(14)}
\sim \frac{|\chi]|\eta_{14}\rangle\times[\eta_{14}|\langle \chi|+|\eta_{14}]|\chi\rangle\times[\chi|\langle \eta_{14}|}{[\eta_{14}\chi]\langle\chi\eta_{14}\rangle}.
\end{aligned} \end{equation}

\item The last contribution is to take the momentum $P_{1456}\sim p_\chi$ to be nearly on-shell, which also corresponds to a scalar boson. The internal line factorizes into
\begin{equation} \begin{aligned}
\mathcal{I}_{(1456)}=
\begin{tikzpicture}[baseline=-0.1cm] \begin{feynhand}
\vertex [dot] (d1) at (0,0) {};
\vertex [dot] (d2) at (2,0) {};
\vertex [particle] (i1) at (0.7,0.7) {$5$};
\vertex [particle] (i2) at (1.3,0.7) {$6$};
\vertex (v1) at (0.7,0);
\vertex (v2) at (1.3,0);
\graph{(d2)--[sca,edge label=$p_\chi$](v2)--[bos](v1)--[sca](d1)};
\graph{(i1)--[sca](v1)};
\graph{(i2)--[sca](v2)};
\end{feynhand} \end{tikzpicture}
\sim \frac{|\chi]|\chi\rangle-|6]|6\rangle}{(-p_\chi+p_6)^2}\times
\langle \chi|[\chi|.
\end{aligned} \end{equation}
Particle 6 provides the subleading structure, leading to
\begin{equation} \begin{aligned}
\lim_{\substack{p_5\to 0\\p_6\to -\eta_{14}}}\mathcal{I}_{(1456)}
\sim &\frac{|\chi]|\chi\rangle-|6]|6\rangle}{(-p_\chi+p_6)^2}\times
\langle \chi|[\chi|.
\end{aligned} \end{equation}

\end{itemize}

In the above, the three on-shell momenta $p_\chi$ can be identified with the same internal momentum
\begin{equation}
p_\chi\to P_{14}.
\end{equation}
Combining the three contributions, we obtain
\begin{equation} \begin{aligned}
\begin{tikzpicture}[baseline=-0.1cm] \begin{feynhand}
\vertex [dot] (d1) at (0,0) {};
\vertex [dot] (d2) at (2,0) {};
\vertex [particle] (i1) at (0.7,0.7) {$5$};
\vertex [particle] (i2) at (1.3,0.7) {$6$};
\vertex (v1) at (0.7,0);
\vertex (v2) at (1.3,0);
\graph{(d2)--[sca](v2)--[bos](v1)--[sca](d1)};
\graph{(i1)--[sca](v1)};
\graph{(i2)--[sca](v2)};
\end{feynhand} \end{tikzpicture}
&\to |P]\langle P||\eta]\langle \eta|+|P]|\eta\rangle[\eta|\langle P|+|\eta]|P\rangle[P|\langle \eta|+|\eta]\langle \eta||P]\langle P|\\
&= P\eta+\eta P-\mathbf m^2.
\end{aligned} \end{equation}

This analysis can be summarized as the following internal Higgsing rules
\begin{equation}
\begin{array}{ccc}
\begin{tikzpicture}[baseline=0.7cm]
\begin{feynhand}
\setlength{\feynhandblobsize}{6mm}
\setlength{\feynhandarrowsize}{3.5pt}
\vertex [dot] (v1) at (-0.4,0.8) {};
\vertex [dot] (i2) at (1.3,0.8) {};
\graph{(i2) --[sca] (v1)};
\end{feynhand}
\end{tikzpicture}  & 
\to &
\begin{tikzpicture}[baseline=0.7cm] \begin{feynhand}
\setlength{\feynhandarrowsize}{4pt}
\vertex [dot] (i1) at (0,0.8) {};
\vertex [dot] (i2) at (1.6,0.8) {};
\draw[brown,thick bos] (i1)--(i2);
\end{feynhand} \end{tikzpicture}=PP \\
\hspace{-1.3cm}\text{split}\downarrow &  &  \\
\begin{tikzpicture}[baseline=-0.1cm] \begin{feynhand}
\setlength{\feynhandblobsize}{6mm}
\setlength{\feynhandarrowsize}{5pt}
\vertex [dot] (d1) at (0,0) {};
\vertex [dot] (d2) at (2,0) {};
\vertex [particle] (i1) at (0.1,0.7) {$h$};
\vertex [particle] (i2) at (1.9,0.7) {$h$};
\vertex (v1) at (0.8,0);
\vertex (v2) at (1.2,0);
\graph{(d1)--[sca](v1)--[bos](v2)--[sca](d2)};
\graph{(i1)--[sca](v1)};
\graph{(i2)--[sca](v2)};
\end{feynhand} \end{tikzpicture}
 & \to &
\begin{tikzpicture}[baseline=0.7cm] \begin{feynhand}
\setlength{\feynhandarrowsize}{4pt}
\vertex [dot] (i1) at (0,0.8) {};
\vertex [dot] (i2) at (1.6,0.8) {};
\draw[brown,thick bos] (i1)--(i2);
\draw[very thick] plot[mark=x,mark size=2.5] coordinates {(0.5,0.8)};
\draw[very thick] plot[mark=x,mark size=2.5] coordinates {(1.1,0.8)};
\end{feynhand} \end{tikzpicture}=P\eta+\eta P-\mathbf m^2
\end{array}
\end{equation}
A more detail discussion about the internal Higgsing rules for other states is given in Appendix~\ref{app:internal_Higgsing}.

Using this internal Higgsing rule, we can derive the subleading matching for the $(+-00)$ amplitude. Recall that the leading matching in the (14)-channel gives
\begin{equation}
\begin{tikzpicture}[baseline=1.1cm] \begin{feynhand}
\vertex [particle] (i1) at (0,2.2) {$+$};
\vertex [particle] (i2) at (0,0) {$-$};
\vertex [particle] (i3) at (1.8,0) {$0$};
\vertex [particle] (i4) at (1.8,2.2) {$0$};
\vertex (v1) at (0.9,0.6);
\vertex (v2) at (0.9,1.6);
\vertex (v4) at (0.9-0.9*0.4,1.6+0.6*0.4);
\graph{(i1)--[bos] (v2)--[sca] (i4)};
\graph{(v2)--[sca] (v1)};
\graph{(i2)--[sca] (v1)--[sca] (i3)};
\end{feynhand} \end{tikzpicture}
\quad\to\quad
\begin{tikzpicture}[baseline=1.1cm] \begin{feynhand}
\vertex [particle] (i1) at (0,2.2) {$+$};
\vertex [particle] (i2) at (0,0) {$-$};
\vertex [particle] (i3) at (1.8,0) {$0$};
\vertex [particle] (i4) at (1.8,2.2) {$0$};
\vertex (v1) at (0.9,0.6);
\vertex (v2) at (0.9,1.6);
\vertex (v4) at (0.9-0.9*0.4,1.6+0.6*0.4);
\vertex (v5) at (0.9-0.9*0.4,0.6-0.6*0.4);
\draw[cyan,thick bos] (i1)--(v4);
\draw[brown,thick bos] (v4)--(v2);
\draw[brown,thick bos] (v2)--(v1);
\draw[brown,thick bos] (v5)--(v1);
\draw[red,thick bos] (i2)--(v5);
\draw[brown,thick sca] (i3)--(v1);
\draw[brown,thick sca] (i4)--(v2);
\draw[very thick] plot[mark=x,mark size=3.5,mark options={rotate=45}] coordinates {(v4)};
\draw[very thick] plot[mark=x,mark size=3.5,mark options={rotate=45}] coordinates {(v5)};
\end{feynhand} \end{tikzpicture}=
\frac{\langle\eta_1|P_{14}|1]\langle2|P_{14}|\eta_2]}{s_{14}\tilde{m}_1 \mathbf m_2}.
\end{equation}
When two Higgs bosons are split from the internal line, we obtain the subleading MHC amplitude with the same helicity category,
\begin{equation}
\begin{tikzpicture}[baseline=1.1cm] \begin{feynhand}
\vertex [particle] (i1) at (0,2.2) {$+$};
\vertex [particle] (i2) at (0,0) {$-$};
\vertex [particle] (i3) at (1.8,0) {$0$};
\vertex [particle] (i4) at (1.8,2.2) {$0$};
\vertex [particle] (i5) at (0,1.6) {$v$};
\vertex [particle] (i6) at (0,0.6) {$v$};
\vertex (v1) at (0.9,0.6);
\vertex (v2) at (0.9,1.6);
\vertex (v4) at (0.9-0.9*0.4,1.6+0.6*0.4);
\vertex (v5) at (0.9,1.3);
\vertex (v6) at (0.9,0.9);
\graph{(i1)--[bos] (v2)--[sca] (i4)};
\graph{(v2)--[sca] (v5)--[bos] (v6)--[sca] (v1)};
\graph{(i2)--[sca] (v1)--[sca] (i3)};
\graph{(i5)--[sca] (v5)};
\graph{(i6)--[sca] (v6)};
\end{feynhand} \end{tikzpicture}
\quad\to\quad 
\begin{tikzpicture}[baseline=1.1cm] \begin{feynhand}
\vertex [particle] (i1) at (0,2.2) {$+$};
\vertex [particle] (i2) at (0,0) {$-$};
\vertex [particle] (i3) at (1.8,0) {$0$};
\vertex [particle] (i4) at (1.8,2.2) {$0$};
\vertex (v1) at (0.9,0.6);
\vertex (v2) at (0.9,1.6);
\vertex (v4) at (0.9-0.9*0.4,1.6+0.6*0.4);
\vertex (v5) at (0.9-0.9*0.4,0.6-0.6*0.4);
\draw[cyan,thick bos] (i1)--(v4);
\draw[brown,thick bos] (v4)--(v2);
\draw[brown,thick bos] (v2)--(v1);
\draw[brown,thick bos] (v5)--(v1);
\draw[red,thick bos] (i2)--(v5);
\draw[brown,thick sca] (i3)--(v1);
\draw[brown,thick sca] (i4)--(v2);
\draw[very thick] plot[mark=x,mark size=3.5,mark options={rotate=45}] coordinates {(v4)};
\draw[very thick] plot[mark=x,mark size=3.5,mark options={rotate=45}] coordinates {(v5)};
\draw[very thick] plot[mark=x,mark size=3.5,mark options={rotate=0}] coordinates {(0.9,0.9)};
\draw[very thick] plot[mark=x,mark size=3.5,mark options={rotate=0}] coordinates {(0.9,1.3)};
\end{feynhand} \end{tikzpicture}
=\frac{\langle\eta_1|\eta_{14}|1]\langle2|P_{14}|\eta_2]+\langle\eta_1|P_{14}|1]\langle2|\eta_{14}|\eta_2]+\mathbf m^2\langle\eta_12\rangle[\eta_21]}{s_{14}\tilde{m}_1 \mathbf m_2}.
\end{equation}
Similarly, we can derive the subleading amplitude in the (13)-channel. The subleading amplitude in the (12)-channel can be obtained by applying the external Higgsing rule to the $(0000)$ helicity amplitude.

\section{Summary and Outlook}
\label{sec:sum}

In this work, we have systematically developed the on-shell Minimal Helicity-Chirality (MHC) formalism for higher-point amplitudes to enable the efficient calculation of scattering amplitudes involving all massive particles, with a particular focus on massive gauge bosons in the spontaneous broken standard model. By establishing a constructive massless-massive correspondence, we bridge the gap between the highly streamlined methods of massless amplitude calculations and the phenomenological necessity of treating constructive amplitudes with massive external states.

The MHC framework achieves this by working with amplitudes of definite helicity that are directly analogous to massless amplitudes in the light-cone gauge, thereby inheriting their computational advantages. Through the extended little-group covariance, these MHC amplitudes are then combined to reconstruct the complete physical massive amplitude. This approach resolves key technical challenges intrinsic to massive calculations, such as the need for distinct treatments of longitudinal and transverse polarizations and provides motivation for necessary contact terms in the light-cone gauge, and systematic derivation of all contact 4-point SM amplitudes.


We have provided a complete prescription on the constructive massive amplitudes for the MHC formalism within the SM context and the following results are obtained
\begin{itemize}
    \item The complete standard model MHC contact 3-point and 4-piont amplitudes are derived as the building blocks of the constructive higher-point massive amplitudes. For the same spin amplitudes, different orders of MHC amplitudes should be related by the ladder operators.

    \item Analogy to the light-cone gauge massless amplitude construction,  the light-cone gauge choice of the MHC amplitudes is specified. The AHH amplitudes should also contain such gauge dependence in the spontaneous broken theory, and thus 4-point contact AHH amplitudes is also needed.

    \item Using the contact 3-point and 4-point MHC amplitudes at the leading order as the building blocks, explicit constructive (bootstrap) and recursive (BCFW) methods are utilized to construct the higher-point massive amplitudes, analogous to massless amplitude construction in the light-cone gauge. Once the contact 4-point AHH amplitudes is identified, similar construction can be applied to the AHH amplitudes.

    \item A robust procedure for massless-massive amplitude matching is performed. First bootstrapping 3-point gauge-invariant massless amplitudes to obtain the higher point gauge-invariant ones, and then performing the amplitude deformation and pole separation to match to the massive amplitudes

\end{itemize}
All the construction and matching are at the leading MHC order, and the sub-leading matching is obtained with the generalized Higgsing rules. This procedure can be able to apply to the higher point massive amplitude construction than 4-point ones in the standard model.

In summary, the MHC formalism successfully extends the conceptual clarity and computational power of modern on-shell methods to the massive sector. It transforms the calculation of massive amplitudes, particularly those involving massive gauge boson, into a more transparent, efficient, and systematic endeavor. This work provides a unified and powerful toolkit for precision calculations in the electroweak sector and beyond, with direct applications to collider phenomenology and the study of effective field theories.

\begin{acknowledgments}
This work is supported by the National Science Foundation of China under Grants No. 12347105, No. 12375099 and No. 12447101, and the National Key Research and Development Program of China Grant No. 2020YFC2201501, No. 2021YFA0718304. 
\end{acknowledgments}

\appendix

\section{Higgsing Rules for Internal Particles}
\label{app:internal_Higgsing}

\subsection{MHC internal particles}

In section 4, we showed that gluing two MHC external particle states gives an internal particle states. For completeness, we list all possible internal particle states with spin $s\le 1$.

For a scalar boson, there is only one type of internal particle state:
\begin{equation}
\begin{tikzpicture}[baseline=0.7cm] \begin{feynhand}
\setlength{\feynhandarrowsize}{4pt}
\vertex [dot] (v1) at (0,0.8) {};
\vertex [dot] (i1) at (1.6,0.8) {};
\sca{i1};
\end{feynhand} \end{tikzpicture}=1.
\end{equation}

For a fermion, there are two chirality, denoted by red and cyan lines. Therefore, the internal fermion line can change chirality according to
\begin{align}
\begin{tikzpicture}[baseline=0.7cm] \begin{feynhand}
\setlength{\feynhandarrowsize}{4pt}
\vertex [dot] (i1) at (0,0.8) {};
\vertex (v2) at (0.8,0.8);
\vertex [dot] (i2) at (1.6,0.8) {};
\fer{red}{i1}{v2};
\fer{cyan}{v2}{i2};
\draw[very thick] plot[mark=x,mark size=2.5] coordinates {(v2)};
\end{feynhand} \end{tikzpicture}&=\mathbf m^2\delta_\alpha^\beta,\\
\begin{tikzpicture}[baseline=0.7cm] \begin{feynhand}
\setlength{\feynhandarrowsize}{4pt}
\vertex [dot] (i1) at (0,0.8) {};
\vertex (v2) at (0.8,0.8);
\vertex [dot] (i2) at (1.6,0.8) {};
\fer{cyan}{i1}{v2};
\fer{red}{v2}{i2};
\draw[very thick] plot[mark=x,mark size=2.5] coordinates {(v2)};
\end{feynhand} \end{tikzpicture}
&=\mathbf m^2\delta_{\dot{\alpha}}^{\dot{\beta}},
\end{align}
where the upper and lower indices correspond to the left- and right-hand sides of the diagrams, respectively. When the chirality is not flipped, there are two contributions:
\begin{align}
\begin{tikzpicture}[baseline=0.7cm] \begin{feynhand}
\setlength{\feynhandarrowsize}{4pt}
\vertex [dot] (i1) at (0,0.8) {};
\vertex [dot] (i2) at (1.6,0.8) {};
\fer{cyan}{i1}{i2};
\end{feynhand} \end{tikzpicture}&=p_{\dot{\alpha}}^\beta,&
\begin{tikzpicture}[baseline=0.7cm] \begin{feynhand}
\setlength{\feynhandarrowsize}{4pt}
\vertex [dot] (i1) at (0,0.8) {};
\vertex [dot] (i2) at (1.6,0.8) {};
\fer{cyan}{i1}{i2};
\draw[very thick] plot[mark=x,mark size=2.5] coordinates {(0.5,0.8)};
\draw[very thick] plot[mark=x,mark size=2.5] coordinates {(1.1,0.8)};
\end{feynhand} \end{tikzpicture}&=\mathbf m^2 \eta_{\dot{\alpha}}^\beta,&\\
\begin{tikzpicture}[baseline=0.7cm] \begin{feynhand}
\setlength{\feynhandarrowsize}{4pt}
\vertex [dot] (i1) at (0,0.8) {};
\vertex [dot] (i2) at (1.6,0.8) {};
\fer{red}{i1}{i2};
\end{feynhand} \end{tikzpicture}&=p_{\alpha}^{\dot{\beta}},&
\begin{tikzpicture}[baseline=0.7cm] \begin{feynhand}
\setlength{\feynhandarrowsize}{4pt}
\vertex [dot] (i1) at (0,0.8) {};
\vertex [dot] (i2) at (1.6,0.8) {};
\fer{red}{i1}{i2};
\draw[very thick] plot[mark=x,mark size=2.5] coordinates {(0.5,0.8)};
\draw[very thick] plot[mark=x,mark size=2.5] coordinates {(1.1,0.8)};
\end{feynhand} \end{tikzpicture}&=\mathbf m^2 \eta_{\alpha}^{\dot{\beta}}.&
\end{align}
Including the appropriate coefficients, these two contributions are not independent and combine into a $p+\eta$ structure, which smoothly matches the $SU(2)$ LG covarirant momentum $\mathbf p$.

For a vector boson, there are three chirality, denoted by red, brown and cyan lines. In this case, the chirality of the internal state can be flipped twice as
\begin{align}
\begin{tikzpicture}[baseline=0.7cm] \begin{feynhand}
\vertex [dot] (i1) at (1,0.8) {};
\vertex [dot] (v1) at (0,0.8) {};
\vertex (v2) at (0.333,0.8);
\vertex (v3) at (0.667,0.8);
\draw[red,thick bos] (v1)--(v2);
\draw[brown,thick bos] (v2)--(v3);
\draw[cyan,thick bos] (v3)--(i1);
\draw[very thick] plot[mark=x,mark size=2.7] coordinates {(v2)};
\draw[very thick] plot[mark=x,mark size=2.7] coordinates {(v3)};
\end{feynhand} \end{tikzpicture}&=\mathbf m^4\delta_{(\alpha_1}^{(\beta_1}\delta_{\alpha_2)}^{\beta_2)},\\
\begin{tikzpicture}[baseline=0.7cm] \begin{feynhand}
\vertex [dot] (i1) at (1,0.8) {};
\vertex [dot] (v1) at (0,0.8) {};
\vertex (v2) at (0.333,0.8);
\vertex (v3) at (0.667,0.8);
\draw[cyan,thick bos] (v1)--(v2);
\draw[brown,thick bos] (v2)--(v3);
\draw[red,thick bos] (v3)--(i1);
\draw[very thick] plot[mark=x,mark size=2.7] coordinates {(v2)};
\draw[very thick] plot[mark=x,mark size=2.7] coordinates {(v3)};
\end{feynhand} \end{tikzpicture}&=\mathbf m^4\delta_{(\dot{\alpha}_1}^{(\dot{\beta}_2}\delta_{\dot{\alpha}_2)}^{\dot{\beta}_2)}.
\end{align}
When the chirality is flipped once, the internal particle state receives two contributions,
\begin{align}
\begin{tikzpicture}[baseline=0.7cm] \begin{feynhand}
\vertex [dot] (v1) at (0,0.8) {};
\vertex [dot] (v2) at (1,0.8) {};
\vertex (v3) at (0.5,0.8);
\draw[cyan,thick bos] (v2)--(v3);
\draw[brown,thick bos] (v3)--(v1);
\draw[very thick] plot[mark=x,mark size=2.7] coordinates {(v3)};
\end{feynhand} \end{tikzpicture}&=\mathbf m^2 p_{\dot{\alpha}_2}^{(\beta_2}\delta_{\alpha_1}^{\beta_1)},&
\begin{tikzpicture}[baseline=0.7cm] \begin{feynhand}
\vertex [dot] (v1) at (0,0.8) {};
\vertex [dot] (v2) at (1,0.8) {};
\vertex (v3) at (0.5,0.8);
\draw[cyan,thick bos] (v2)--(v3);
\draw[brown,thick bos] (v3)--(v1);
\draw[very thick] plot[mark=x,mark size=2.7] coordinates {(v3)};
\draw[very thick] plot[mark=x,mark size=2.7] coordinates {(0.25,0.8)};
\draw[very thick] plot[mark=x,mark size=2.7] coordinates {(0.75,0.8)};
\end{feynhand} \end{tikzpicture}&=\mathbf m^4 \eta_{\dot{\alpha}_2}^{(\beta_2}\delta_{\alpha_1}^{\beta_1)},&\\
\begin{tikzpicture}[baseline=0.7cm] \begin{feynhand}
\vertex [dot] (v1) at (0,0.8) {};
\vertex [dot] (v2) at (1,0.8) {};
\vertex (v3) at (0.5,0.8);
\draw[brown,thick bos] (v2)--(v3);
\draw[red,thick bos] (v3)--(v1);
\draw[very thick] plot[mark=x,mark size=2.7] coordinates {(v3)};
\end{feynhand} \end{tikzpicture}&=\mathbf m^2 p_{(\alpha_1}^{\dot{\beta}_1}\delta_{\alpha_2)}^{\beta_2},&
\begin{tikzpicture}[baseline=0.7cm] \begin{feynhand}
\vertex [dot] (v1) at (0,0.8) {};
\vertex [dot] (v2) at (1,0.8) {};
\vertex (v3) at (0.5,0.8);
\draw[brown,thick bos] (v2)--(v3);
\draw[red,thick bos] (v3)--(v1);
\draw[very thick] plot[mark=x,mark size=2.7] coordinates {(v3)};
\draw[very thick] plot[mark=x,mark size=2.7] coordinates {(0.25,0.8)};
\draw[very thick] plot[mark=x,mark size=2.7] coordinates {(0.75,0.8)};
\end{feynhand} \end{tikzpicture}&=\mathbf m^4 \eta_{(\alpha_1}^{\dot{\beta}_1}\delta_{\alpha_2)}^{\beta_2},&\\
\begin{tikzpicture}[baseline=0.7cm] \begin{feynhand}
\vertex [dot] (v1) at (0,0.8) {};
\vertex [dot] (v2) at (1,0.8) {};
\vertex (v3) at (0.5,0.8);
\draw[brown,thick bos] (v2)--(v3);
\draw[cyan,thick bos] (v3)--(v1);
\draw[very thick] plot[mark=x,mark size=2.7] coordinates {(v3)};
\end{feynhand} \end{tikzpicture}&=\mathbf m^2 p_{(\dot{\alpha}_1}^{\beta_1}\delta_{\dot{\alpha}_2)}^{\dot{\beta}_2},& 
\begin{tikzpicture}[baseline=0.7cm] \begin{feynhand}
\vertex [dot] (v1) at (0,0.8) {};
\vertex [dot] (v2) at (1,0.8) {};
\vertex (v3) at (0.5,0.8);
\draw[brown,thick bos] (v2)--(v3);
\draw[cyan,thick bos] (v3)--(v1);
\draw[very thick] plot[mark=x,mark size=2.7] coordinates {(v3)};
\draw[very thick] plot[mark=x,mark size=2.7] coordinates {(0.25,0.8)};
\draw[very thick] plot[mark=x,mark size=2.7] coordinates {(0.75,0.8)};
\end{feynhand} \end{tikzpicture}&=\mathbf m^4 \eta_{(\dot{\alpha}_1}^{\beta_1}\delta_{\dot{\alpha}_2)}^{\dot{\beta}_2},&\\
\begin{tikzpicture}[baseline=0.7cm] \begin{feynhand}
\vertex [dot] (v1) at (0,0.8) {};
\vertex [dot] (v2) at (1,0.8) {};
\vertex (v3) at (0.5,0.8);
\draw[red,thick bos] (v2)--(v3);
\draw[brown,thick bos] (v3)--(v1);
\graph{(v2)--[plain,red,very thick] (v3)--[plain,brown,very thick] (v1)};
\draw[very thick] plot[mark=x,mark size=2.7] coordinates {(v3)};
\end{feynhand} \end{tikzpicture}&=\mathbf m^2 p_{\alpha_1}^{(\dot{\beta}_1}\delta_{\dot{\alpha}_2}^{\dot{\beta}_2)},&
\begin{tikzpicture}[baseline=0.7cm] \begin{feynhand}
\vertex [dot] (v1) at (0,0.8) {};
\vertex [dot] (v2) at (1,0.8) {};
\vertex (v3) at (0.5,0.8);
\draw[red,thick bos] (v2)--(v3);
\draw[brown,thick bos] (v3)--(v1);
\draw[very thick] plot[mark=x,mark size=2.7] coordinates {(v3)};
\draw[very thick] plot[mark=x,mark size=2.7] coordinates {(0.25,0.8)};
\draw[very thick] plot[mark=x,mark size=2.7] coordinates {(0.75,0.8)};
\end{feynhand} \end{tikzpicture}&=\mathbf m^4 \eta_{\alpha_1}^{(\dot{\beta}_1}\delta_{\dot{\alpha}_2}^{\dot{\beta}_2)}.&
\end{align}
If there is no chirality flip, three contributions appear,
\begin{align}
\begin{tikzpicture}[baseline=0.7cm] \begin{feynhand}
\vertex [dot] (v1) at (0,0.8) {};
\vertex [dot] (v2) at (1,0.8) {};
\draw[cyan,thick bos] (v2)--(v1);
\end{feynhand} \end{tikzpicture}&=p_{(\dot{\alpha}_1}^{(\beta_1} p_{\dot{\alpha}_2)}^{\beta_2)},&
\begin{tikzpicture}[baseline=0.7cm] \begin{feynhand}
\vertex [dot] (v1) at (0,0.8) {};
\vertex [dot] (v2) at (1,0.8) {};
\draw[cyan,thick bos] (v2)--(v1);
\draw[very thick] plot[mark=x,mark size=2.7] coordinates {(0.3,0.8)};
\draw[very thick] plot[mark=x,mark size=2.7] coordinates {(0.7,0.8)};
\end{feynhand} \end{tikzpicture}&=\mathbf m^2 p_{(\dot{\alpha}_1}^{(\beta_1} \eta_{\dot{\alpha}_2)}^{\beta_2)}+\mathbf m^2 \eta_{(\dot{\alpha}_1}^{(\beta_1} p_{\dot{\alpha}_2)}^{\beta_2)},&
\begin{tikzpicture}[baseline=0.7cm] \begin{feynhand}
\vertex [dot] (v1) at (0,0.8) {};
\vertex [dot] (v2) at (1,0.8) {};
\draw[cyan,thick bos] (v2)--(v1);
\graph{(v2)--[plain,cyan,very thick] (v3)--[plain,cyan,very thick] (v1)};
\draw[very thick] plot[mark=x,mark size=2.7] coordinates {(0.2,0.8)};
\draw[very thick] plot[mark=x,mark size=2.7] coordinates {(0.4,0.8)};
\draw[very thick] plot[mark=x,mark size=2.7] coordinates {(0.6,0.8)};
\draw[very thick] plot[mark=x,mark size=2.7] coordinates {(0.8,0.8)};
\end{feynhand} \end{tikzpicture}&=\mathbf m^4\eta_{(\dot{\alpha}_1}^{(\beta_1}\eta_{\dot{\alpha}_2)}^{\beta_2)},&\\
\begin{tikzpicture}[baseline=0.7cm] \begin{feynhand}
\vertex [dot] (v1) at (0,0.8) {};
\vertex [dot] (v2) at (1,0.8) {};
\draw[brown,thick bos] (v2)--(v1);
\end{feynhand} \end{tikzpicture}&=-p_{\dot{\alpha}_2}^{\beta_1}p_{\alpha_1}^{\dot{\beta}_2},&
\begin{tikzpicture}[baseline=0.7cm] \begin{feynhand}
\vertex [dot] (v1) at (0,0.8) {};
\vertex [dot] (v2) at (1,0.8) {};
\draw[brown,thick bos] (v2)--(v1);
\draw[very thick] plot[mark=x,mark size=2.7] coordinates {(0.333,0.8)};
\draw[very thick] plot[mark=x,mark size=2.7] coordinates {(0.667,0.8)};
\end{feynhand} \end{tikzpicture}&=2\mathbf{m}^4\delta_{\alpha_1}^{\beta_1}\delta_{\dot{\alpha}_2}^{\dot{\beta}_2}-\mathbf{m}^2 p_{\dot{\alpha}_2}^{\beta_1}\eta_{\alpha_1}^{\dot{\beta}_2}-\mathbf{m}^2 \eta_{\dot{\alpha}_2}^{\beta_1}p_{\alpha_1}^{\dot{\beta}_2},&
\begin{tikzpicture}[baseline=0.7cm] \begin{feynhand}
\vertex [dot] (v1) at (0,0.8) {};
\vertex [dot] (v2) at (1,0.8) {};
\draw[brown,thick bos] (v2)--(v1);
\draw[very thick] plot[mark=x,mark size=2.7] coordinates {(0.2,0.8)};
\draw[very thick] plot[mark=x,mark size=2.7] coordinates {(0.4,0.8)};
\draw[very thick] plot[mark=x,mark size=2.7] coordinates {(0.6,0.8)};
\draw[very thick] plot[mark=x,mark size=2.7] coordinates {(0.8,0.8)};
\end{feynhand} \end{tikzpicture}&=-\mathbf m^4\eta_{\dot{\alpha}_2}^{\beta_1}\eta_{\alpha_1}^{\dot{\beta}_2},&\\
\begin{tikzpicture}[baseline=0.7cm] \begin{feynhand}
\vertex [dot] (v1) at (0,0.8) {};
\vertex [dot] (v2) at (1,0.8) {};
\draw[red,thick bos] (v2)--(v1);
\end{feynhand} \end{tikzpicture}&=p_{(\alpha_1}^{(\dot{\beta}_1}p_{\alpha_2)}^{\dot{\beta}_2)},&
\begin{tikzpicture}[baseline=0.7cm] \begin{feynhand}
\vertex [dot] (v1) at (0,0.8) {};
\vertex [dot] (v2) at (1,0.8) {};
\draw[red,thick bos] (v2)--(v1);
\draw[very thick] plot[mark=x,mark size=2.7] coordinates {(0.333,0.8)};
\draw[very thick] plot[mark=x,mark size=2.7] coordinates {(0.667,0.8)};
\end{feynhand} \end{tikzpicture}&=\mathbf m^2 p_{(\alpha_1}^{(\dot{\beta}_1}\eta_{\alpha_2)}^{\dot{\beta}_2)}+\mathbf m^2\eta_{(\alpha_1}^{(\dot{\beta}_1} p_{\alpha_2)}^{\dot{\beta}_2)},&
\begin{tikzpicture}[baseline=0.7cm] \begin{feynhand}
\vertex [dot] (v1) at (0,0.8) {};
\vertex [dot] (v2) at (1,0.8) {};
\draw[red,thick bos] (v2)--(v1);
\draw[very thick] plot[mark=x,mark size=2.7] coordinates {(0.2,0.8)};
\draw[very thick] plot[mark=x,mark size=2.7] coordinates {(0.4,0.8)};
\draw[very thick] plot[mark=x,mark size=2.7] coordinates {(0.6,0.8)};
\draw[very thick] plot[mark=x,mark size=2.7] coordinates {(0.8,0.8)};
\end{feynhand} \end{tikzpicture}&=\mathbf m^4\eta_{(\alpha_1}^{(\dot{\beta}_1}\eta_{\alpha_2)}^{\dot{\beta}_2)}.&
\end{align}

\subsection{Internal particle Higgsing}

After enumerating all MHC internal particle states, we now  present a systematic set of Higgsing rules for internal particles. This rules map the massless internal structure with additional Higgs bosons splitting from it, to the corresponding MHC internal particle states. Given the large number of MHC states, the rules are best illustrated diagrammatically.

\paragraph{Internal fermion}

We first consider the fermion case. Compared to the leading matching, the Higgs splitting introduces a chirality flip, represented by a cross in MHC diagrams. Therefore, the Higgsing rules will depends on the MHC diagram from the leading matching. Below we list three typical leading matching and their resulting Higgs splittings. Other cases can be derived similarly by changing the colors of the MHC diagrams.
\begin{equation} \label{eq:internal_f_table}
\begin{tabular}{c|c|c|c}
\hline
Leading matching &
\begin{tikzpicture}[baseline=-0.1cm] \begin{feynhand}
\setlength{\feynhandblobsize}{6mm}
\setlength{\feynhandarrowsize}{5pt}
\vertex (v1) at (1,0);
\vertex [dot] (d1) at (0,0) {};
\vertex [dot] (d2) at (2,0) {};
\graph{(d1)--[fer](d2)};
\end{feynhand} \end{tikzpicture}$\to$
\begin{tikzpicture}[baseline=0.7cm] \begin{feynhand}
\setlength{\feynhandarrowsize}{4pt}
\vertex [dot] (i1) at (0,0.8) {};
\vertex [dot] (i2) at (1.6,0.8) {};
\fer{cyan}{i1}{i2};
\end{feynhand} \end{tikzpicture} & 
\begin{tikzpicture}[baseline=-0.1cm] \begin{feynhand}
\setlength{\feynhandblobsize}{6mm}
\setlength{\feynhandarrowsize}{5pt}
\vertex (v1) at (1,0);
\vertex [dot] (d1) at (0,0) {};
\vertex [dot] (d2) at (2,0) {};
\graph{(d1)--[fer](d2)};
\end{feynhand} \end{tikzpicture}$\to$
\begin{tikzpicture}[baseline=0.7cm] \begin{feynhand}
\setlength{\feynhandarrowsize}{4pt}
\vertex [dot] (i1) at (0,0.8) {};
\vertex (v2) at (0.8,0.8);
\vertex [dot] (i2) at (1.6,0.8) {};
\fer{red}{i1}{v2};
\fer{cyan}{v2}{i2};
\draw[very thick] plot[mark=x,mark size=2.5] coordinates {(v2)};
\end{feynhand} \end{tikzpicture} & 
\begin{tikzpicture}[baseline=-0.1cm] \begin{feynhand}
\setlength{\feynhandblobsize}{6mm}
\setlength{\feynhandarrowsize}{5pt}
\vertex (v1) at (1,0);
\vertex [dot] (d1) at (0,0) {};
\vertex [dot] (d2) at (2,0) {};
\graph{(d1)--[fer](d2)};
\end{feynhand} \end{tikzpicture}$\to$
\begin{tikzpicture}[baseline=0.7cm] \begin{feynhand}
\setlength{\feynhandarrowsize}{4pt}
\vertex [dot] (i1) at (0,0.8) {};
\vertex [dot] (i2) at (1.6,0.8) {};
\fer{cyan}{i1}{i2};
\draw[very thick] plot[mark=x,mark size=2.5] coordinates {(0.5,0.8)};
\draw[very thick] plot[mark=x,mark size=2.5] coordinates {(1.1,0.8)};
\end{feynhand} \end{tikzpicture} \\
One Higgs splitting &
\begin{tikzpicture}[baseline=-0.1cm] \begin{feynhand}
\setlength{\feynhandblobsize}{6mm}
\setlength{\feynhandarrowsize}{5pt}
\vertex [particle] (i1) at (1,0.7) {$h$};
\vertex (v1) at (1,0);
\vertex [dot] (d1) at (0,0) {};
\vertex [dot] (d2) at (2,0) {};
\graph{(d1)--[fer](v1)--[fer](d2)};
\graph{(i1)--[sca](v1)};
\end{feynhand} \end{tikzpicture}$\to$
\makecell{\begin{tikzpicture}[baseline=0.7cm] \begin{feynhand}
\setlength{\feynhandarrowsize}{4pt}
\vertex [dot] (i1) at (0,0.8) {};
\vertex (v2) at (0.8,0.8);
\vertex [dot] (i2) at (1.6,0.8) {};
\fer{red}{i1}{v2};
\fer{cyan}{v2}{i2};
\draw[very thick] plot[mark=x,mark size=2.5] coordinates {(v2)};
\end{feynhand} \end{tikzpicture}\\
\begin{tikzpicture}[baseline=0.7cm] \begin{feynhand}
\setlength{\feynhandarrowsize}{4pt}
\vertex [dot] (i1) at (0,0.8) {};
\vertex (v2) at (0.8,0.8);
\vertex [dot] (i2) at (1.6,0.8) {};
\fer{cyan}{i1}{v2};
\fer{red}{v2}{i2};
\draw[very thick] plot[mark=x,mark size=2.5] coordinates {(v2)};
\end{feynhand} \end{tikzpicture}} & 
\begin{tikzpicture}[baseline=-0.1cm] \begin{feynhand}
\setlength{\feynhandblobsize}{6mm}
\setlength{\feynhandarrowsize}{5pt}
\vertex [particle] (i1) at (1,0.7) {$h$};
\vertex (v1) at (1,0);
\vertex [dot] (d1) at (0,0) {};
\vertex [dot] (d2) at (2,0) {};
\graph{(d1)--[fer](v1)--[fer](d2)};
\graph{(i1)--[sca](v1)};
\end{feynhand} \end{tikzpicture}$\to$
\makecell{
\begin{tikzpicture}[baseline=0.7cm] \begin{feynhand}
\setlength{\feynhandarrowsize}{4pt}
\vertex [dot] (i1) at (0,0.8) {};
\vertex [dot] (i2) at (1.6,0.8) {};
\fer{cyan}{i1}{i2};
\draw[very thick] plot[mark=x,mark size=2.5] coordinates {(0.5,0.8)};
\draw[very thick] plot[mark=x,mark size=2.5] coordinates {(1.1,0.8)};
\end{feynhand} \end{tikzpicture}\\
\begin{tikzpicture}[baseline=0.7cm] \begin{feynhand}
\setlength{\feynhandarrowsize}{4pt}
\vertex [dot] (i1) at (0,0.8) {};
\vertex [dot] (i2) at (1.6,0.8) {};
\fer{red}{i1}{i2};
\draw[very thick] plot[mark=x,mark size=2.5] coordinates {(0.5,0.8)};
\draw[very thick] plot[mark=x,mark size=2.5] coordinates {(1.1,0.8)};
\end{feynhand} \end{tikzpicture}} 
& \mbox{-}
\\
Two Higgs splitting &
\begin{tikzpicture}[baseline=-0.1cm] \begin{feynhand}
\setlength{\feynhandblobsize}{6mm}
\setlength{\feynhandarrowsize}{5pt}
\vertex [particle] (i1) at (0.7,0.7) {$h$};
\vertex [particle] (i2) at (1.3,0.7) {$h$};
\vertex (v1) at (0.7,0);
\vertex (v2) at (1.3,0);
\vertex [dot] (d1) at (0,0) {};
\vertex [dot] (d2) at (2,0) {};
\graph{(d1)--[fer](v1)--[fer](v2)--[fer](d2)};
\graph{(i1)--[sca](v1)};
\graph{(i2)--[sca](v2)};
\end{feynhand} \end{tikzpicture}$\to$
\begin{tikzpicture}[baseline=0.7cm] \begin{feynhand}
\setlength{\feynhandarrowsize}{4pt}
\vertex [dot] (i1) at (0,0.8) {};
\vertex [dot] (i2) at (1.6,0.8) {};
\fer{cyan}{i1}{i2};
\draw[very thick] plot[mark=x,mark size=2.5] coordinates {(0.5,0.8)};
\draw[very thick] plot[mark=x,mark size=2.5] coordinates {(1.1,0.8)};
\end{feynhand} \end{tikzpicture} & \mbox{-} & \mbox{-} \\
\hline
\end{tabular}
\end{equation}
Note that a single Higgs splitting changes the helicity of the external particles. Since an external particle can attach to either the left or the right side of the internal line, the corresponding MHC diagrams appear in two variants, as shown in the table above. In contrast, two Higgs splittings do not change the external helicity, so only one MHC diagram corresponds to that case.

\paragraph{Internal vector boson}

Then we consider internal vector bosons. These have more particle states than fermions, so we restrict ourselves to the leading matching without chirality flips. Other cases can be obtained by shifting the MHC diagrams upward within a column, analogous to the fermion case in Eq.~\eqref{eq:internal_f_table}. Below we show the leading matching with two chirality and the corresponding Higgsing rules,
\begin{equation}
\begin{tabular}{c|c|c}
\hline
Leading matching &
\begin{tikzpicture}[baseline=-0.1cm] \begin{feynhand}
\setlength{\feynhandblobsize}{6mm}
\setlength{\feynhandarrowsize}{5pt}
\vertex (v1) at (1,0);
\vertex [dot] (d1) at (0,0) {};
\vertex [dot] (d2) at (2,0) {};
\graph{(d1)--[sca](d2)};
\end{feynhand} \end{tikzpicture}$\to$
\includegraphics[scale=1,valign=c]{image/internal_V_c00_0.pdf}
 & 
\begin{tikzpicture}[baseline=-0.1cm] \begin{feynhand}
\setlength{\feynhandblobsize}{6mm}
\setlength{\feynhandarrowsize}{5pt}
\vertex (v1) at (1,0);
\vertex [dot] (d1) at (0,0) {};
\vertex [dot] (d2) at (2,0) {};
\graph{(d1)--[bos](d2)};
\end{feynhand} \end{tikzpicture}$\to$
\begin{tikzpicture}[baseline=0.7cm] \begin{feynhand}
\vertex [dot] (v1) at (0,0.8) {};
\vertex [dot] (v2) at (1.6,0.8) {};
\draw[cyan,thick bos] (v2)--(v1);
\end{feynhand} \end{tikzpicture}
 \\
One Higgs splitting &
\begin{tikzpicture}[baseline=-0.1cm] \begin{feynhand}
\setlength{\feynhandblobsize}{6mm}
\setlength{\feynhandarrowsize}{5pt}
\vertex [particle] (i1) at (1,0.7) {$h$};
\vertex (v1) at (1,0);
\vertex [dot] (d1) at (0,0) {};
\vertex [dot] (d2) at (2,0) {};
\graph{(d1)--[bos](v1)--[sca](d2)};
\graph{(i1)--[sca](v1)};
\end{feynhand} \end{tikzpicture}$\to$
\makecell{
\begin{tikzpicture}[baseline=0.7cm] \begin{feynhand}
\vertex [dot] (v1) at (0,0.8) {};
\vertex [dot] (v2) at (1.6,0.8) {};
\vertex (v3) at (0.8,0.8);
\draw[brown,thick bos] (v2)--(v3);
\draw[red,thick bos] (v3)--(v1);
\draw[very thick] plot[mark=x,mark size=2.7] coordinates {(v3)};
\end{feynhand} \end{tikzpicture}\\
\begin{tikzpicture}[baseline=0.7cm] \begin{feynhand}
\vertex [dot] (v1) at (0,0.8) {};
\vertex [dot] (v2) at (1.6,0.8) {};
\vertex (v3) at (0.8,0.8);
\draw[brown,thick bos] (v2)--(v3);
\draw[cyan,thick bos] (v3)--(v1);
\draw[very thick] plot[mark=x,mark size=2.7] coordinates {(v3)};
\end{feynhand} \end{tikzpicture}} & 
\begin{tikzpicture}[baseline=-0.1cm] \begin{feynhand}
\setlength{\feynhandblobsize}{6mm}
\setlength{\feynhandarrowsize}{5pt}
\vertex [particle] (i1) at (1,0.7) {$h$};
\vertex (v1) at (1,0);
\vertex [dot] (d1) at (0,0) {};
\vertex [dot] (d2) at (2,0) {};
\graph{(d1)--[bos](v1)--[sca](d2)};
\graph{(i1)--[sca](v1)};
\end{feynhand} \end{tikzpicture}$\to$
\begin{tikzpicture}[baseline=0.7cm] \begin{feynhand}
\vertex [dot] (v1) at (0,0.8) {};
\vertex [dot] (v2) at (1.6,0.8) {};
\vertex (v3) at (0.8,0.8);
\draw[brown,thick bos] (v2)--(v3);
\draw[cyan,thick bos] (v3)--(v1);
\draw[very thick] plot[mark=x,mark size=2.7] coordinates {(v3)};
\end{feynhand} \end{tikzpicture}
\\
Two Higgs splitting &
\begin{tikzpicture}[baseline=-0.1cm] \begin{feynhand}
\setlength{\feynhandblobsize}{6mm}
\setlength{\feynhandarrowsize}{5pt}
\vertex [particle] (i1) at (0.7,0.7) {$h$};
\vertex [particle] (i2) at (1.3,0.7) {$h$};
\vertex (v1) at (0.7,0);
\vertex (v2) at (1.3,0);
\vertex [dot] (d1) at (0,0) {};
\vertex [dot] (d2) at (2,0) {};
\graph{(d1)--[sca](v1)--[bos](v2)--[sca](d2)};
\graph{(i1)--[sca](v1)};
\graph{(i2)--[sca](v2)};
\end{feynhand} \end{tikzpicture}$\to$
\includegraphics[scale=1,valign=c]{image/internal_V_c00_2.pdf} & 
\begin{tikzpicture}[baseline=-0.1cm] \begin{feynhand}
\setlength{\feynhandblobsize}{6mm}
\setlength{\feynhandarrowsize}{5pt}
\vertex [particle] (i1) at (0.7,0.7) {$h$};
\vertex [particle] (i2) at (1.3,0.7) {$h$};
\vertex (v1) at (0.7,0);
\vertex (v2) at (1.3,0);
\vertex [dot] (d1) at (0,0) {};
\vertex [dot] (d2) at (2,0) {};
\graph{(d1)--[bos](v1)--[sca](v2)--[bos](d2)};
\graph{(i1)--[sca](v1)};
\graph{(i2)--[sca](v2)};
\end{feynhand} \end{tikzpicture}$\to$
\begin{tikzpicture}[baseline=0.7cm] \begin{feynhand}
\vertex [dot] (v1) at (0,0.8) {};
\vertex [dot] (v2) at (1.6,0.8) {};
\draw[cyan,thick bos] (v2)--(v1);
\draw[very thick] plot[mark=x,mark size=2.7] coordinates {(0.5,0.8)};
\draw[very thick] plot[mark=x,mark size=2.7] coordinates {(1.1,0.8)};
\end{feynhand} \end{tikzpicture} \\
Three Higgs splitting & 
\begin{tikzpicture}[baseline=-0.1cm] \begin{feynhand}
\setlength{\feynhandblobsize}{6mm}
\setlength{\feynhandarrowsize}{5pt}
\vertex [particle] (i1) at (0.5,0.7) {$h$};
\vertex [particle] (i2) at (1,0.7) {$h$};
\vertex [particle] (i3) at (1.5,0.7) {$h$};
\vertex (v1) at (0.5,0);
\vertex (v2) at (1,0);
\vertex (v3) at (1.5,0);
\vertex [dot] (d1) at (0,0) {};
\vertex [dot] (d2) at (2,0) {};
\graph{(d1)--[bos](v1)--[sca](v2)--[bos](v3)--[sca](d2)};
\graph{(i1)--[sca](v1)};
\graph{(i2)--[sca](v2)};
\graph{(i3)--[sca](v3)};
\end{feynhand} \end{tikzpicture}$\to$
\makecell{
\begin{tikzpicture}[baseline=0.7cm] \begin{feynhand}
\vertex [dot] (v1) at (0,0.8) {};
\vertex [dot] (v2) at (1.6,0.8) {};
\vertex (v3) at (0.8,0.8);
\draw[brown,thick bos] (v2)--(v3);
\draw[red,thick bos] (v3)--(v1);
\draw[very thick] plot[mark=x,mark size=2.7] coordinates {(v3)};
\draw[very thick] plot[mark=x,mark size=2.7] coordinates {(0.4,0.8)};
\draw[very thick] plot[mark=x,mark size=2.7] coordinates {(1.2,0.8)};
\end{feynhand} \end{tikzpicture}\\
\begin{tikzpicture}[baseline=0.7cm] \begin{feynhand}
\vertex [dot] (v1) at (0,0.8) {};
\vertex [dot] (v2) at (1.6,0.8) {};
\vertex (v3) at (0.8,0.8);
\draw[brown,thick bos] (v2)--(v3);
\draw[cyan,thick bos] (v3)--(v1);
\draw[very thick] plot[mark=x,mark size=2.7] coordinates {(v3)};
\draw[very thick] plot[mark=x,mark size=2.7] coordinates {(0.4,0.8)};
\draw[very thick] plot[mark=x,mark size=2.7] coordinates {(1.2,0.8)};
\end{feynhand} \end{tikzpicture}
} & 
\begin{tikzpicture}[baseline=-0.1cm] \begin{feynhand}
\setlength{\feynhandblobsize}{6mm}
\setlength{\feynhandarrowsize}{5pt}
\vertex [particle] (i1) at (0.5,0.7) {$h$};
\vertex [particle] (i2) at (1,0.7) {$h$};
\vertex [particle] (i3) at (1.5,0.7) {$h$};
\vertex (v1) at (0.5,0);
\vertex (v2) at (1,0);
\vertex (v3) at (1.5,0);
\vertex [dot] (d1) at (0,0) {};
\vertex [dot] (d2) at (2,0) {};
\graph{(d1)--[bos](v1)--[sca](v2)--[bos](v3)--[sca](d2)};
\graph{(i1)--[sca](v1)};
\graph{(i2)--[sca](v2)};
\graph{(i3)--[sca](v3)};
\end{feynhand} \end{tikzpicture}$\to$
\begin{tikzpicture}[baseline=0.7cm] \begin{feynhand}
\vertex [dot] (v1) at (0,0.8) {};
\vertex [dot] (v2) at (1.6,0.8) {};
\vertex (v3) at (0.8,0.8);
\draw[brown,thick bos] (v2)--(v3);
\draw[cyan,thick bos] (v3)--(v1);
\draw[very thick] plot[mark=x,mark size=2.7] coordinates {(v3)};
\draw[very thick] plot[mark=x,mark size=2.7] coordinates {(0.4,0.8)};
\draw[very thick] plot[mark=x,mark size=2.7] coordinates {(1.2,0.8)};
\end{feynhand} \end{tikzpicture} \\
Four Higgs splitting &
\begin{tikzpicture}[baseline=-0.1cm] \begin{feynhand}
\setlength{\feynhandblobsize}{6mm}
\setlength{\feynhandarrowsize}{5pt}
\vertex [particle] (i1) at (0.4,0.7) {$h$};
\vertex [particle] (i2) at (0.8,0.7) {$h$};
\vertex [particle] (i3) at (1.2,0.7) {$h$};
\vertex [particle] (i4) at (1.6,0.7) {$h$};
\vertex (v1) at (0.4,0);
\vertex (v2) at (0.8,0);
\vertex (v3) at (1.2,0);
\vertex (v4) at (1.6,0);
\vertex [dot] (d1) at (0,0) {};
\vertex [dot] (d2) at (2,0) {};
\graph{(d1)--[sca](v1)--[bos](v2)--[sca](v3)--[bos](v4)--[sca](d2)};
\graph{(i1)--[sca](v1)};
\graph{(i2)--[sca](v2)};
\graph{(i3)--[sca](v3)};
\graph{(i4)--[sca](v4)};
\end{feynhand} \end{tikzpicture}$\to$
\includegraphics[scale=1,valign=c]{image/internal_V_c00_4.pdf} & 
\begin{tikzpicture}[baseline=-0.1cm] \begin{feynhand}
\setlength{\feynhandblobsize}{6mm}
\setlength{\feynhandarrowsize}{5pt}
\vertex [particle] (i1) at (0.4,0.7) {$h$};
\vertex [particle] (i2) at (0.8,0.7) {$h$};
\vertex [particle] (i3) at (1.2,0.7) {$h$};
\vertex [particle] (i4) at (1.6,0.7) {$h$};
\vertex (v1) at (0.4,0);
\vertex (v2) at (0.8,0);
\vertex (v3) at (1.2,0);
\vertex (v4) at (1.6,0);
\vertex [dot] (d1) at (0,0) {};
\vertex [dot] (d2) at (2,0) {};
\graph{(d1)--[bos](v1)--[sca](v2)--[bos](v3)--[sca](v4)--[bos](d2)};
\graph{(i1)--[sca](v1)};
\graph{(i2)--[sca](v2)};
\graph{(i3)--[sca](v3)};
\graph{(i4)--[sca](v4)};
\end{feynhand} \end{tikzpicture}$\to$
\begin{tikzpicture}[baseline=0.7cm] \begin{feynhand}
\vertex [dot] (v1) at (0,0.8) {};
\vertex [dot] (v2) at (1.6,0.8) {};
\draw[cyan,thick bos] (v2)--(v1);
\draw[very thick] plot[mark=x,mark size=2.7] coordinates {(0.325,0.8)};
\draw[very thick] plot[mark=x,mark size=2.7] coordinates {(0.65,0.8)};
\draw[very thick] plot[mark=x,mark size=2.7] coordinates {(0.975,0.8)};
\draw[very thick] plot[mark=x,mark size=2.7] coordinates {(1.3,0.8)};
\end{feynhand} \end{tikzpicture} \\
\hline
\end{tabular}
\end{equation}
As in the fermion case, splittings involving an odd number of Higgs bosons change the helicity of the external particles, so the corresponding MHC diagrams may have more than one variant.

\section{Contact 4-point Amplitudes from Conserved Current}
\label{sec:VVS_current}

In this section, we verify the contact amplitude using the conserved current structure. The 4-point amplitude can be written in the form of factorized and contact terms, and then it can be rewritten as certain sub-amplitudes with the conserved currents
\begin{equation}
\mathcal{M}=\mathcal{M}_{\text{f}}+\mathcal{M}_{\text{ct}} 
\xrightarrow{\text{deform}} 
\mathcal{M}_{\text{f}}^{\text{conserved}}
\end{equation}
where $\mathcal{M}_{\text{f}}$ denotes the factorized term containing poles, and $\mathcal{M}_{\text{ct}}$ denotes the contact term. 
By gluing these conserved current MHC sub-amplitudes, we can determine the contact term at leading order.

\subsection{Contact $VVSS$ amplitudes}

Let us first calculate the $VVSS$ contact amplitude using the $WWhh$ amplitude. The $WWhh$ amplitude has the following structure
\begin{equation} \begin{aligned}
\mathcal{M}=\underbrace{\mathcal{M}_{(12)}+\mathcal{M}_{(13)}+\mathcal{M}_{(14)}}_{\mathcal{M}_{\text{f}}}+\mathcal{M}_{\text{ct}}.
\end{aligned} \end{equation}
Using the gluing technique, the first three terms can be constructed. For each term, the corresponding internal particle and sub-amplitudes are listed below
\begin{equation}
\begin{tabular}{c|c|c}
\hline
channel & sub-amplitudes & internal particle \\
\hline
(12)-channel & WWh$\times$hhh & h \\
\hline
\makecell{(13)-channel\\(14)-channel} & WWh$\times$WWh & W \\
\hline
\end{tabular}
\end{equation}
This shows that we can use two types of 3-point amplitudes, $WWh$ and $hhh$, to construct the $WWhh$ amplitude. The only MHC amplitude related to a conserved current is the primary $WWh$ amplitude,
\begin{equation} \begin{aligned} \label{eq:WWh}
\Ampthree{1^0}{P^0}{2^0}{\bos{i1}{brown}}{\bos{i2}{brown}}{\sca{i3}}&=\mathbf{g}^{W^+ W^-}\frac{\langle1P\rangle[P1]}{\mathbf m_1}.\\
\end{aligned} \end{equation}
This amplitude vanishes when momentum conservation $p_1 + p_2 + P = 0$ and the on-shell conditions for all particles are imposed. This primary amplitude can be decomposed into a product of an MHC current $[\mathbf J]_0$ and a vector $[\mathbf A]_0$,
\begin{equation}
    \langle1P\rangle[P1]=[\mathbf J]_0\cdot [\mathbf A]_0,
\end{equation}
where the primary current and vector are defined as 
\begin{equation}
[\mathbf J]_0=|1]|1\rangle,\quad
[\mathbf A]_0=|P]|P\rangle,
\end{equation}
Note that $\partial\cdot[\mathbf J]_0\neq 0$, so it does not directly correspond to a conserved current.

To find a truly conserved current, we can examine the massless 2-point structure in the UV theory. For two scalar particles, the Noether current is given by
\begin{equation}
    J=\frac12(|1]|1\rangle-|2]|2\rangle),
\end{equation}
which corresponds to $\phi^* D\phi-\phi D\phi^*$. This current satisfies the conservation condition
\begin{equation}
    \partial\cdot J=\frac12(\langle1|1+2|1]-\langle2|1+2|2])=0.
\end{equation}
We can match this Noether current $J$ and a Goldstone mode $\partial\phi$ to the MHC current and vector 
\begin{equation} \begin{aligned} 
\partial\phi&=|P]|P\rangle& &\to& [\mathbf A]_0&=|P]|P\rangle,\\
J&=\frac12(|1]|1\rangle-|2]|2\rangle)& &\to& [\mathbf J]_0+\frac12 P&=|1]|1\rangle+\frac12|P]|P\rangle.
\end{aligned} \end{equation}
Therefore, the contraction $J \cdot \partial\phi$ corresponds to the MHC primary amplitude
\begin{equation} \begin{aligned} \label{eq:VVS_conserved_form}
J\cdot\partial\phi 
\quad\to\quad
\Ampthree{1^0}{P^0}{2^0}{\bos{i1}{brown}}{\bos{i2}{brown}}{\sca{i3}}=\mathbf{g}^{W^+ W^-}\frac{\langle1P\rangle[P1]+P^2}{\mathbf m_1},\\
\end{aligned} \end{equation}
which contains the off-shell quantity $P^2$. When particle $P$ is on-shell ($P^2=0$), this quantity vanishes and the amplitude returns to eq.~\eqref{eq:WWh}. However, when $P$ is glued as an internal particle, such a term must be retained.

We can now use this knowledge of the conserved current to determine the contact $WWhh$ amplitude. The $WWhh$ amplitude can be categorized into nine helicity categories, classified by the number of transverse vectors $n_T$:
\begin{equation} \begin{aligned}
n_T=0:\quad&(0,0,0,0),\\
n_T=1:\quad&(\pm,0,0,0),(0,\pm,0,0),\\
n_T=2:\quad&(\pm,\mp,0,0),(\pm,\pm,0,0).    
\end{aligned} \end{equation}
We will examine each case to see whether the primary $WWh$ amplitude appears and whether the contact amplitude can be determined.

For $n_T=0$, there is only one helicity category $(0,0,0,0)$. We first consider the $(14)$-channel, where the internal particle is a $W$ boson. In this channel, we glue two $WWh$ amplitudes of the form given in Eq.~\eqref{eq:VVS_conserved_form},
\begin{equation}
\makecell{
\begin{tikzpicture}[baseline=-0.1cm] \begin{feynhand}
\setlength{\feynhandarrowsize}{4pt}
\vertex [particle] (i1) at (-0.827,0.579) {$1^0$}; 
\vertex [particle] (i2) at (0.827,0.579) {$4^0$}; 
\vertex [particle] (i3) at (0,-1.01) {$P^0$}; 
\vertex (v1) at (0,0);
\draw[brown,thick bos] (i1)--(v1);
\draw[brown,thick sca] (i2)--(v1);
\draw[brown,thick bos] (v1)--(i3);
\end{feynhand} \end{tikzpicture}\\
\begin{tikzpicture}[baseline=-0.1cm] \begin{feynhand}
\setlength{\feynhandarrowsize}{4pt}
\vertex [particle] (i1) at (-0.827,-0.579) {$2^0$}; 
\vertex [particle] (i2) at (0.827,-0.579) {$3^0$}; 
\vertex [particle] (i3) at (0,1.01) {$P^0$}; 
\vertex (v1) at (0,0);
\draw[brown,thick bos] (i1)--(v1);
\draw[brown,thick sca] (i2)--(v1);
\draw[brown,thick bos] (v1)--(i3);
\end{feynhand} \end{tikzpicture}}
\begin{aligned}
&\to \mathbf{g}^{W^+ W^-}\mathbf{g}^{W^+ W^-}\frac{(-\langle1P_{14}\rangle[P_{14}1]+P^2_{14})\times (-\langle2P_{23}\rangle[P_{23}2]+P^2_{23})}{s_{14}\mathbf m_1\mathbf m_2}\\
&=\mathbf{g}^{W^+ W^-}\mathbf{g}^{W^+ W^-}\frac{\langle1|P_{14}|1]\langle2|P_{23}|2]}{s_{14}}-\mathbf{g}^{W^+ W^-}\mathbf{g}^{W^+ W^-}\frac{s_{14}}{\mathbf m_1\mathbf m_2}.
\end{aligned} 
\end{equation}
The second term contributes to the contact amplitude. Combining it with the contribution from the $(13)$-channel gives the full contact term
\begin{equation} \begin{aligned}
-\mathbf{g}^{W^+ W^-}\mathbf{g}^{W^+ W^-}\frac{s_{14}}{\mathbf m_1\mathbf m_2}+(3\leftrightarrow 4)
\quad\to\quad
\begin{tikzpicture}[baseline=0.8cm] \begin{feynhand}
\vertex [particle] (i1) at (0,1.8) {$1^0$};
\vertex [particle] (i2) at (0,0) {$2^0$};
\vertex [particle] (i3) at (1.8,0) {$3^0$};
\vertex [particle] (i4) at (1.8,1.8) {$4^0$};
\vertex (v3) at (0.9,0.9);
\draw[brown,thick bos] (i1)--(v3);
\draw[brown,thick bos] (i2)--(v3);
\draw[brown,thick sca] (i4)--(v3)--(i3);
\end{feynhand} \end{tikzpicture}
=\mathbf{g}^{W^+ W^-}\mathbf{g}^{W^+ W^-}\frac{\langle12\rangle[21]}{\mathbf m_1\mathbf m_2}.
\end{aligned} \end{equation}
which exhibits the correct Goldstone behavior $\tilde\lambda\lambda$ for both particles 1 and 2. Once this primary contact amplitude is known, descendant contact amplitudes can also be determined by the extended LG covariance.
So far we have not used the $(12)$-channel. Although this channel also contains the primary $WWh$ sub-amplitude in the factorized limit, the internal particle is a scalar boson. The corresponding diagram is
\begin{equation} \begin{aligned}
\begin{tikzpicture}[baseline=0.8cm] \begin{feynhand}
\vertex [particle] (i1) at (0,1.8) {$1^0$};
\vertex [particle] (i2) at (0,0) {$2^0$};
\vertex [particle] (i3) at (1.8,0) {$3^0$};
\vertex [particle] (i4) at (1.8,1.8) {$4^0$};
\vertex (v1) at (0.6,0.9);
\vertex (v2) at (1.2,0.9);
\draw[brown,thick bos] (i1)--(v1);
\draw[brown,thick bos] (i2)--(v1);
\draw[brown,thick sca] (v1)--(v2);
\draw[brown,thick sca] (i4)--(v2)--(i3);
\end{feynhand} \end{tikzpicture}
\end{aligned} \end{equation}
This differs from the case in Eq.~\eqref{eq:VVS_conserved_form}, where the internal particle is a vector boson. Therefore, the $(12)$-channel does not contribute to the contact term.

Similarly, we can determine the contact term for helicity categories with $n_T=1$. As an example, we consider the $(+1,0,0,0)$ as an example. In the $(14)$-channel, gluing the two sub-amplitudes gives,
\begin{equation}
\makecell{
\begin{tikzpicture}[baseline=-0.1cm] \begin{feynhand}
\setlength{\feynhandarrowsize}{4pt}
\vertex [particle] (i1) at (-0.827,0.579) {$1^+$}; 
\vertex [particle] (i2) at (0.827,0.579) {$4^0$}; 
\vertex [particle] (i3) at (0,-1.01) {$P^0$}; 
\vertex (v2) at (-0.827*0.3,0.579*0.3); 
\vertex (v1) at (0,0);
\draw[cyan,thick bos] (i1)--(v2);
\draw[brown,thick bos] (v2)--(v1);
\draw[brown,thick sca] (i2)--(v1);
\draw[brown,thick bos] (v1)--(i3);
\draw[very thick] plot[mark=x,mark size=3.5,mark options={rotate=45}] coordinates {(v2)};
\end{feynhand} \end{tikzpicture}\\
\begin{tikzpicture}[baseline=-0.1cm] \begin{feynhand}
\setlength{\feynhandarrowsize}{4pt}
\vertex [particle] (i1) at (-0.827,-0.579) {$2^0$}; 
\vertex [particle] (i2) at (0.827,-0.579) {$3^0$}; 
\vertex [particle] (i3) at (0,1.01) {$P^0$}; 
\vertex (v1) at (0,0);
\draw[brown,thick bos] (i1)--(v1);
\draw[brown,thick sca] (i2)--(v1);
\draw[brown,thick bos] (v1)--(i3);
\end{feynhand} \end{tikzpicture}}
\begin{aligned}
&\to \mathbf{g}^{W^+ W^-}\mathbf{g}^{W^+ W^-}\frac{\langle\eta_1P_{14}\rangle[P_{14}1]\times (-\langle2P_{23}\rangle[P_{23}2]+P^2_{23})}{m_1 \mathbf m_2 s_{14}}\\
&=\mathbf{g}^{W^+ W^-}\mathbf{g}^{W^+ W^-}\frac{\langle\eta_1|P_{14}|1]\langle2|P_{23}|2]}{m_1 \mathbf m_2 s_{14}}-\mathbf{g}^{W^+ W^-}\mathbf{g}^{W^+ W^-}\frac{\langle\eta_14\rangle[41]}{m_1 \mathbf m_2}.
\end{aligned} 
\end{equation}
Combining this with the contribution from the $(13)$-channel yields the contact term
\begin{equation} \begin{aligned}
-\mathbf{g}^{W^+ W^-}\mathbf{g}^{W^+ W^-}\frac{\langle\eta_14\rangle[41]}{m_1 \mathbf m_2}+(3\leftrightarrow 4)
\quad\to\quad
\begin{tikzpicture}[baseline=0.8cm] \begin{feynhand}
\vertex [particle] (i1) at (0,1.8) {$1^+$};
\vertex [particle] (i2) at (0,0) {$2^0$};
\vertex [particle] (i3) at (1.8,0) {$3^0$};
\vertex [particle] (i4) at (1.8,1.8) {$4^0$};
\vertex (v3) at (0.9,0.9);
\vertex (v4) at (0.9-0.9*0.43,0.9+0.9*0.43);
\draw[cyan,thick bos] (i1)--(v4);
\draw[brown,thick bos] (v4)--(v3);
\draw[brown,thick bos] (i2)--(v3);
\draw[brown,thick sca] (i3)--(v3);
\draw[brown,thick sca] (i4)--(v3);
\draw[very thick] plot[mark=x,mark size=3.5,mark options={rotate=45}] coordinates {(v4)};
\end{feynhand} \end{tikzpicture}
=\mathbf{g}^{W^+ W^-}\mathbf{g}^{W^+ W^-}\frac{\langle\eta_1 2\rangle[21]}{m_1 \mathbf m_2}.
\end{aligned} \end{equation}
The corresponding descendant contact amplitude is
\begin{equation} \begin{aligned}
\begin{tikzpicture}[baseline=0.8cm] \begin{feynhand}
\vertex [particle] (i1) at (0,1.8) {$1^+$};
\vertex [particle] (i2) at (0,0) {$2^0$};
\vertex [particle] (i3) at (1.8,0) {$3^0$};
\vertex [particle] (i4) at (1.8,1.8) {$4^0$};
\vertex (v3) at (0.9,0.9);
\vertex (v4) at (0.9-0.9*0.43,0.9+0.9*0.43);
\draw[cyan,thick bos] (i1)--(v4);
\draw[brown,thick bos] (v4)--(v3);
\draw[brown,thick bos] (i2)--(v3);
\draw[brown,thick sca] (i3)--(v3);
\draw[brown,thick sca] (i4)--(v3);
\draw[very thick] plot[mark=x,mark size=3.5,mark options={rotate=45}] coordinates {(v4)};
\draw[very thick] plot[mark=x,mark size=3.5,mark options={rotate=45}] coordinates {(0.9-0.9*0.48,0.9-0.9*0.48)};
\draw[very thick] plot[mark=x,mark size=3.5,mark options={rotate=45}] coordinates {(0.9-0.9*0.22,0.9-0.9*0.22)};
\end{feynhand} \end{tikzpicture}=\mathbf{g}^{W^+ W^-}\mathbf{g}^{W^+ W^-}\frac{\langle\eta_1 \eta_2\rangle[\eta_21]}{m_1 \mathbf m_2}.
\end{aligned} \end{equation}
For helicity categories with $n_T=2$, no priamry $WWh$ amplitude exists, so the conserved-current method cannot be used to fix the contact amplitude.

In summary, by rewriting the vanishing 3-point MHC amplitudes in terms of conserved currents from the UV theory, and employing these modified forms in the gluing procedure, we can determine certain contact amplitudes. This approach applies only to helicity categories where such vanishing MHC amplitudes are present.

\subsection{Contact $VVVV$ amplitudes}

Now we will determine $VVVV$ contact massive amplitude in the SM using the conserved current analysis in $WW \to ZZ$ amplitudes. The contact term has two origins: the massless contact term and the conserved current. The first origin only relate to the $\lambda\phi^4$ interaction, so we discuss this later. For now, we consider the second origin, the conserved current. 

As previously discussion, the amplitude can be decomposed to a product of the current and the vector. This current may not equal to the massless Noether, so their difference can create a contact term. In the massless gauge theory with spin$\le$1, there are three kinds of Noether current $J$. Let us examine whether they equal to the MHC current $\mathbf J$ in a specific order,
\begin{itemize}
\item The fermion Noether current $\bar\psi\gamma^\mu\psi$. It equals to priamry $FF$ current
\begin{equation}
    J=|1]|2\rangle\to [\mathbf J]=|1]|2\rangle.
\end{equation}
Therefore, $FFV$ amplitude will not generate the contact term.

\item The scalar Noether current $\phi^*D\phi$. It equals is not equal to priamry $VS$ current
\begin{equation}
    J=\frac{1}{2}(|1]\langle1|-|2]\langle2|)\to [\mathbf J]+\frac12 P=|1]\langle1|+\frac12 P.
\end{equation}
Therefore, $VVS$ amplitude will generate the contact term.

\item The Yang-Mills Noether current $F^{\mu\nu}A_{\mu}$. It equals to the 1st descendant $VV$ current.
\begin{equation} \begin{aligned}
&J(1^{-1},2^{-1})=\frac{\langle12\rangle}{[\xi_2 2]} |\xi_2]^{\dot\alpha}\langle1|^{\alpha}+\frac{\langle12\rangle}{[\xi_1 1]} |\xi_1]^{\dot\alpha}\langle2|^{\alpha}\\
\to&
[\mathbf J(1^{-1},2^{-1})]_1= -\frac{1}{\mathbf m_2^2} m_2\langle12\rangle |\eta_2]^{\dot\alpha}\langle1|^{\alpha}
-\frac{1}{\mathbf m_1^2} m_1\langle12\rangle |\eta_1]^{\dot\alpha}\langle2|^{\alpha}.
\end{aligned}\end{equation}
Therefore, the $VVV$ amplitude will not generate contact term, if the internal particle is in $(\frac12,\frac12)$ representation.

\end{itemize}

However, the above analysis do not consider the case that the internal vector boson can be a chiral particle, which has Lorentz representation $(1,0)$ or $(0,1)$. As an example, let us consider the $W^+ W^- ZZ$ amplitude. It has three channels. THe corresponding internal particles and sub-amplitudes in the factorized limit are
\begin{equation}
\begin{tabular}{c|c|c}
\hline
channel & sub-amplitudes & internal particle \\
\hline
(12)-channel & WWh$\times$ZZh & h \\
\hline
\makecell{(13)-channel\\(14)-channel} & WWZ$\times$WWZ & W \\
\hline
\end{tabular}
\end{equation}
Since we are consider the case with internal vector boson, we can only pay attention to (13) and (14)-channel. The corresponding 3-pt amplitude is $W^+ W^- Z$ amplitude, with one particle will be labeled as $P$. This particle $P$ will become internal vector, and it can be chiral or anti-chiral. We first consider the chiral case, the primary $W^+ W^- Z$ amplitude is 
\begin{equation} \begin{aligned} \label{eq:WWZ}
\Ampthree{1^0}{P^-}{2^0}{\bos{i1}{brown}}{\bos{i2}{red}}{\bos{i3}{brown}}&=\mathbf{f}^{W^+ W^- Z}\frac{\langle P1\rangle[12]\langle 2P\rangle}{\mathbf m_1 \mathbf m_2}.
\end{aligned} \end{equation}
This primary amplitude can be decomposed into a product of an field strength tensor $[\mathbf F^\pm]_0$ and two vector $[\mathbf A]_0$. 
\begin{equation} \begin{aligned}
\langle P1\rangle[12]\langle 2P\rangle&=[\mathbf F^{\alpha_1}_{\alpha_2}]_0 \times[\mathbf A_{\alpha_1 \dot\alpha}]_0 \times[\mathbf A^{\dot\alpha\alpha_2}]_0,\\
\end{aligned} \end{equation}
where the field strength tensor and vectors are defined as
\begin{equation}
[\mathbf F^{\alpha_1}_{\alpha_2}]_0 =\langle P|^{\alpha_1} |P\rangle_{\alpha_2},\quad
[\mathbf A_{\alpha_1 \dot\alpha}]_0 =|1\rangle_{\alpha_1}[1|_{\dot\alpha},\quad
[\mathbf A^{\dot\alpha\alpha_2}]_0 =|2]^{\dot\alpha}\langle2|^{\alpha_2}.
\end{equation}

Now particles 1 and 2 are scalar boson, so they may relate to the scalar Noether current. For each particle, we associate a Noether current,
\begin{equation} \begin{aligned}
J_{\alpha_1\dot\alpha}&=\frac12(|1\rangle_{\alpha_1}[1|_{\dot\alpha}-|2\rangle_{\alpha_1}[2|_{\dot\alpha}),\\
J^{\dot\alpha\alpha_2}&=\frac12(|1]^{\dot\alpha}\langle1|^{\alpha_2}-|2]^{\dot\alpha}\langle2|^{\alpha_2}).
\end{aligned} \end{equation}
Then we can match these two Noether current $J$ and a field strength tensor $\mathcal F$ to two MHC vectors and the MHC field strength tensor,
\begin{equation} \begin{aligned} 
\mathcal F^{\alpha_1}_{\alpha_2}&=\langle P|^{\alpha_1} |P\rangle_{\alpha_2}& &\to& [\mathbf F^{\alpha_1}_{\alpha_2}]_0&=\langle P|^{\alpha_1} |P\rangle_{\alpha_2},\\
J_{\alpha_1\dot\alpha}&=\frac12(|1\rangle_{\alpha_1}[1|_{\dot\alpha}-|2\rangle_{\alpha_1}[2|_{\dot\alpha})& &\to& [\mathbf A_{\alpha_1 \dot\alpha}]_0+\frac12 P_{\alpha_1 \dot\alpha}&=|1\rangle_{\alpha_1}[1|_{\dot\alpha}+\frac12P_{\alpha_1 \dot\alpha},\\
-J^{\dot\alpha\alpha_2}&=\frac12(|2]^{\dot\alpha}\langle2|^{\alpha_2}-|1]^{\dot\alpha}\langle1|^{\alpha_2})& &\to& [\mathbf A^{\dot\alpha\alpha_2}]_0+\frac12 P^{\dot\alpha\alpha_2}&=|2]^{\dot\alpha}\langle2|^{\alpha_2}+\frac12 P^{\dot\alpha\alpha_2},\\
\end{aligned} \end{equation}
Therefore, the contraction $J J \mathcal F$ corresponds to the MHC primary amplitude,
\begin{equation}
\Ampthree{1^0}{P^-}{2^0}{\bos{i1}{brown}}{\bos{i2}{red}}{\bos{i3}{brown}}=
\frac{[\mathbf F^{\alpha_1}_{\alpha_2}]_0}{\mathbf m_1 \mathbf m_2}\times \left(|1\rangle_{\alpha_1}[12]\langle2|^{\alpha_2}+\frac{|1\rangle_{\alpha_1}([1|P)^{\alpha_2}}{2}+\frac{(P|2])_{\alpha_1}\langle2|^{\alpha_2}}{2}+\frac{P^2 \delta^{\alpha_2}_{\alpha_1}}{4}  \right).
\end{equation}
Contract with $[\mathbf F^{\alpha_1}_{\alpha_2}]_0$, the first term in the bracket return to the amplitude in eq.~\eqref{eq:WWZ}, while other three terms will contribute to the contact term.

Similarly, we can give the priamry amplitude with particle $P$ is anti-chiral,
\begin{equation}
\Ampthree{1^0}{P^+}{2^0}{\bos{i1}{brown}}{\bos{i2}{cyan}}{\bos{i3}{brown}}=
\frac{[\mathbf F^{\dot\alpha_2}_{\dot\alpha_1}]_0}{\mathbf m_1 \mathbf m_2}\times \left(|1]^{\dot\alpha_1}\langle12\rangle[2|_{\dot\alpha_2}+\frac{|1]^{\dot\alpha_1}(\langle 1|P)_{\dot\alpha_2}}{2}+\frac{(P|2\rangle)^{\dot\alpha_1}[2|_{\dot\alpha_2}}{2}+\frac{P_{12}^2 \delta^{\dot\alpha_1}_{\dot\alpha_2}}{4} \right).
\end{equation}

Now we can gluing the $W^+W^-Z$ amplitude to determine the contact term. Here we choose the helicity category $(0,0,0,0)$. Gluing the two field strength tensor can give an internal vector boson
\begin{equation}
[\mathbf F^{\alpha_1}_{\alpha_2}]_0\times [\mathbf F^{\dot\beta_1}_{\dot\beta_2}]_0=\frac{1}{2}(P^{\alpha_1\dot\beta_1} P_{\alpha_2\dot\beta_2}+P^{\alpha_1}_{\dot\beta_2} P_{\alpha_2}^{\dot\beta_1}).
\end{equation}
Therefore, contracting two primary amplitudes, we get
\begin{equation} \begin{aligned}
&\begin{tikzpicture}[baseline=-0.1cm] \begin{feynhand}
\setlength{\feynhandarrowsize}{4pt}
\vertex [particle] (i1) at (-0.579,-0.827) {$2^0$}; 
\vertex [particle] (i2) at (-0.579,0.827) {$1^0$}; 
\vertex [particle] (i3) at (1.01,0) {$P^-$}; 
\vertex (v1) at (0,0);
\draw[brown,thick bos] (i1)--(v1);
\draw[brown,thick bos] (i2)--(v1);
\draw[red,thick bos] (v1)--(i3);
\end{feynhand} \end{tikzpicture}
\begin{tikzpicture}[baseline=-0.1cm] \begin{feynhand}
\setlength{\feynhandarrowsize}{4pt}
\vertex [particle] (i1) at (0.579,-0.827) {$3^0$}; 
\vertex [particle] (i2) at (0.579,0.827) {$4^0$}; 
\vertex [particle] (i3) at (-1.01,0) {$P^+$}; 
\vertex (v1) at (0,0);
\draw[brown,thick bos] (i1)--(v1);
\draw[brown,thick bos] (i2)--(v1);
\draw[cyan,thick bos] (v1)--(i3);
\end{feynhand} \end{tikzpicture}\\
=&\frac{\mathbf{f}^{W^+ W^- Z}}{\mathbf m_1 \mathbf m_2}\left(|1\rangle_{\alpha_1}[12]\langle2|^{\alpha_2}-\frac{|1\rangle_{\alpha_1}([1|P_{12})^{\alpha_2}}{2}-\frac{(P_{12}|2])_{\alpha_1}\langle2|^{\alpha_2}}{2}+\frac{P_{12}^2 \delta^{\alpha_2}_{\alpha_1}}{4} \right)\\
&\times \frac{P_{12}^{\alpha_1\dot\beta_1} P_{12,\alpha_2\dot\beta_2}+P^{\alpha_1}_{12\dot\beta_2} P_{12\alpha_2}^{\dot\beta_1}}{2s_{12}}\\
&\times \frac{\mathbf{f}^{W^+ W^- Z}}{\mathbf m_3 \mathbf m_4}\left(|3]^{\dot\beta_1}\langle34\rangle[4|_{\dot\beta_2}+\frac{|3]^{\dot\beta_1}(\langle 3|P_{12})_{\dot\beta_2}}{2}+\frac{(P_{12}|4\rangle)^{\dot\beta_1}[4|_{\dot\beta_2}}{2}+\frac{P_{12}^2 \delta^{\dot\beta_1}_{\dot\beta_2}}{4} \right)\\
=&\frac{\mathbf{f}^{W^+ W^- Z}\mathbf{f}^{W^+ W^- Z}}{\mathbf m_1 \mathbf m_2 \mathbf m_3 \mathbf m_4}\frac{[12](\langle1|P_{12}|3]\langle2|P_{12}|4]+\langle1|P_{12}|4]\langle2|P_{12}|3])\langle34\rangle}{2s_{12}}\\
&+\frac{\mathbf{f}^{W^+ W^- Z}\mathbf{f}^{W^+ W^- Z}}{\mathbf m_1 \mathbf m_2 \mathbf m_3 \mathbf m_4}\frac{s_{12}(s_{14}-s_{13})}{2}.
\end{aligned} \end{equation}
The second term contributes to the contact term. Similarly, the other helicity of internal line gives
\begin{equation} \begin{aligned}
&\begin{tikzpicture}[baseline=-0.1cm] \begin{feynhand}
\setlength{\feynhandarrowsize}{4pt}
\vertex [particle] (i1) at (-0.579,-0.827) {$2^0$}; 
\vertex [particle] (i2) at (-0.579,0.827) {$1^0$}; 
\vertex [particle] (i3) at (1.01,0) {$P^+$}; 
\vertex (v1) at (0,0);
\draw[brown,thick bos] (i1)--(v1);
\draw[brown,thick bos] (i2)--(v1);
\draw[cyan,thick bos] (v1)--(i3);
\end{feynhand} \end{tikzpicture}
\begin{tikzpicture}[baseline=-0.1cm] \begin{feynhand}
\setlength{\feynhandarrowsize}{4pt}
\vertex [particle] (i1) at (0.579,-0.827) {$3^0$}; 
\vertex [particle] (i2) at (0.579,0.827) {$4^0$}; 
\vertex [particle] (i3) at (-1.01,0) {$P^-$}; 
\vertex (v1) at (0,0);
\draw[brown,thick bos] (i1)--(v1);
\draw[brown,thick bos] (i2)--(v1);
\draw[red,thick bos] (v1)--(i3);
\end{feynhand} \end{tikzpicture}.\\
=&\frac{mathbf{f}^{W^+ W^- Z}\mathbf{f}^{W^+ W^- Z}}{\mathbf m_1 \mathbf m_2 \mathbf m_3 \mathbf m_4}\frac{\langle12\rangle([1|P_{12}|3\rangle[2|P_{12}|4\rangle+[1|P_{12}|4\rangle[2|P_{12}|3\rangle)[34]}{2s_{12}}\\
&+\frac{\mathbf{f}^{W^+ W^- Z}\mathbf{f}^{W^+ W^- Z}}{\mathbf m_1 \mathbf m_2 \mathbf m_3 \mathbf m_4}\frac{s_{12}(s_{14}-s_{13})}{2}
\end{aligned} \end{equation}
Combining it with the contribution from (14)-channel gives the full contact term,
\begin{equation} \begin{aligned}
&\frac{\mathbf{f}^{W^+ W^- Z}\mathbf{f}^{W^+ W^- Z}}{\mathbf m_1 \mathbf m_2 \mathbf m_3 \mathbf m_4}s_{12}(s_{14}-s_{13})+(2\leftrightarrow 4)
\quad\to\quad \\
&\Ampfour{0}{0}{0}{0}{\bos{i1}{brown}}{\bos{i2}{brown}}{\bos{i3}{brown}}{\bos{i4}{brown}}
\begin{aligned}
&=\frac{\mathbf{f}^{W^+ W^- Z}\mathbf{f}^{W^+ W^- Z}}{\mathbf m_1 \mathbf m_2 \mathbf m_3 \mathbf m_4}(\langle12\rangle[21]\langle34\rangle[34]+\langle14\rangle[41]\langle23\rangle[32]-2\langle13\rangle[31]\langle24\rangle[42]).
\end{aligned} 
\end{aligned} \end{equation}
The subleading order contribution in this helicity category can be obtained by applying ladder operators.

\section{$WW \to ZZ$ Amplitudes}
\label{app:4V}

In this section, we briefly calculate the $WW \to ZZ$ amplitudes in the MHC and AHH formalism.

\subsection{Bootstrapping MHC amplitudes}

We can use gluing technique to obtain the $WW \to ZZ$ amplitudes. This amplitude has four contributions,
\begin{equation} \begin{aligned}
\mathcal{M}(WWZZ)=\mathcal{M}_{(12)}+\mathcal{M}_{(13)}+\mathcal{M}_{(14)}+\mathcal{M}_{\text{ct}}.
\end{aligned} \end{equation}
We choose helicity category $(+,+,-,-)$ to gluing the amplitude.

For (12)-channel, the internal particle is Higgs boson. We glue $WWh$ and $ZZh$ amplitude and get
\begin{equation}
\begin{tikzpicture}[baseline=0.8cm] \begin{feynhand}
\vertex [particle] (i1) at (0,1.8) {$1^+$};
\vertex [particle] (i2) at (0,0) {$2^+$};
\vertex [particle] (i3) at (1.8,0) {$3^-$};
\vertex [particle] (i4) at (1.8,1.8) {$4^-$};
\vertex (v1) at (0.6,0.9);
\vertex (v2) at (1.2,0.9);
\vertex (v4) at (0.6-0.6*0.4,0.9+0.9*0.4);
\vertex (v5) at (0.6-0.6*0.4,0.9-0.9*0.4);
\vertex (v6) at (1.2+0.6*0.4,0.9+0.9*0.4);
\vertex (v7) at (1.2+0.6*0.4,0.9-0.9*0.4);
\draw[red,thick bos] (i3)--(v7);
\draw[red,thick bos] (i4)--(v6);
\draw[brown,thick bos] (v6)--(v2)--(v7);
\draw[brown,thick sca] (v1)--(v2);
\draw[brown,thick bos] (v4)--(v1)--(v5);
\draw[cyan,thick bos] (i1)--(v4);
\draw[cyan,thick bos] (i2)--(v5);
\draw[very thick] plot[mark=x,mark size=3.5,mark options={rotate=45}] coordinates {(v4)};
\draw[very thick] plot[mark=x,mark size=3.5,mark options={rotate=45}] coordinates {(v5)};
\draw[very thick] plot[mark=x,mark size=3.5,mark options={rotate=45}] coordinates {(v6)};
\draw[very thick] plot[mark=x,mark size=3.5,mark options={rotate=45}] coordinates {(v7)};
\end{feynhand} \end{tikzpicture}=\mathbf g^{W^+ W^-} \mathbf g^{ZZ} \frac{\tilde m_1 \tilde m_2 m_3 m_4 \langle\eta_1\eta_2\rangle[21]\langle34\rangle[\eta_4\eta_3]}{s_{12} \mathbf m_1^2 \mathbf m_2^2 \mathbf m_3^2 \mathbf m_4^2}.
\end{equation}

For the (14)-channel, the internal particle is a W boson. In this case, we glue two $WWZ$ amplitudes. Since the $WWZ$ amplitude has six primary vertices, this results in a total of 36 terms. These terms fall into three categories, which we illustrate with the following examples. They can be grouped by the number of internal chirality flip:

\begin{itemize}
\item Three internal chirality flips:
\begin{equation} \begin{aligned} \label{eq:glue_VVVV_1}
&\begin{tikzpicture}[baseline=0.8cm] \begin{feynhand}
\vertex [particle] (i1) at (0,2.0) {$1^+$};
\vertex [particle] (i2) at (0,0) {$2^+$};
\vertex [particle] (i3) at (1.8,0) {$3^-$};
\vertex [particle] (i4) at (1.8,2.0) {$4^-$};
\vertex (v1) at (0.9,0.6);
\vertex (v2) at (0.9,1.4);
\vertex (v6) at (0.9+0.9*0.4,1.4+0.6*0.4);
\vertex (v8) at (0.9-0.9*0.4,0.6-0.6*0.4);
\draw[red,thick bos] (i3)--(v1);
\draw[red,thick bos] (i4)--(v6);
\draw[brown,thick bos] (v6)--(v2);
\draw[brown,thick bos] (v1)--(v2);
\draw[brown,thick bos] (v8)--(v1);
\draw[cyan,thick bos] (i1)--(v2);
\draw[cyan,thick bos] (i2)--(v8);
\draw[very thick] plot[mark=x,mark size=3.5,mark options={rotate=45}] coordinates {(v6)};
\draw[very thick] plot[mark=x,mark size=3.5,mark options={rotate=45}] coordinates {(v8)};
\end{feynhand} \end{tikzpicture}=\frac{(\mathbf{f}^{W^+ W^- Z})^2}{2\mathbf m_2^2 \mathbf m_4^2 \mathbf{m}^2 s_{14}}  \tilde m_2  m_4 [1\eta_4]\langle\eta_23\rangle\langle 4|P_{14}|1]\langle 3|P_{14}|2].
\end{aligned} \end{equation}

\item Two internal chirality flips:
\begin{equation} \begin{aligned} \label{eq:glue_VVVV_2}
&\begin{tikzpicture}[baseline=0.8cm] \begin{feynhand}
\vertex [particle] (i1) at (0,2.0) {$1^+$};
\vertex [particle] (i2) at (0,0) {$2^+$};
\vertex [particle] (i3) at (1.8,0) {$3^-$};
\vertex [particle] (i4) at (1.8,2.0) {$4^-$};
\vertex (v1) at (0.9,0.6);
\vertex (v2) at (0.9,1.4);
\vertex (v3) at (0.9,1.0);
\vertex (v5) at (0.9-0.9*0.4,1.4+0.6*0.4);
\vertex (v6) at (0.9+0.9*0.4,1.4+0.6*0.4);
\vertex (v7) at (0.9+0.9*0.4,0.6-0.6*0.4);
\vertex (v8) at (0.9-0.9*0.4,0.6-0.6*0.4);
\draw[red,thick bos] (i3)--(v1);
\draw[red,thick bos] (i4)--(v6);
\draw[brown,thick bos] (v6)--(v2)--(v5);
\draw[brown,thick bos] (v1)--(v3);
\draw[cyan,thick bos] (v3)--(v2);
\draw[brown,thick bos] (v8)--(v1);
\draw[cyan,thick bos] (i1)--(v5);
\draw[cyan,thick bos] (i2)--(v8);
\draw[very thick] plot[mark=x,mark size=3.5,mark options={rotate=0}] coordinates {(v3)};
\draw[very thick] plot[mark=x,mark size=3.5,mark options={rotate=45}] coordinates {(v5)};
\draw[very thick] plot[mark=x,mark size=3.5,mark options={rotate=45}] coordinates {(v6)};
\draw[very thick] plot[mark=x,mark size=3.5,mark options={rotate=45}] coordinates {(v8)};
\end{feynhand} \end{tikzpicture}
=\frac{(\mathbf{f}^{W^+ W^- Z})^2}{2\mathbf m_1^2 \mathbf m_2^2 \mathbf m_4^2 s_{14}}  \tilde m_1 \tilde m_2  m_4 \langle \eta_14\rangle \langle\eta_23\rangle([\eta_4|P_{14}|3\rangle[12]+[1|P_{14}|3\rangle[\eta_42]).
\end{aligned} \end{equation}

\item One internal chirality flip:
\begin{equation} \begin{aligned}  \label{eq:glue_VVVV_3}
&\begin{tikzpicture}[baseline=0.8cm] \begin{feynhand}
\vertex [particle] (i1) at (0,2.0) {$1^+$};
\vertex [particle] (i2) at (0,0) {$2^+$};
\vertex [particle] (i3) at (1.8,0) {$3^-$};
\vertex [particle] (i4) at (1.8,2.0) {$4^-$};
\vertex (v1) at (0.9,0.6);
\vertex (v2) at (0.9,1.4);
\vertex (v3) at (0.9,0.85);
\vertex (v4) at (0.9,1.15);
\vertex (v5) at (0.9-0.9*0.4,1.4+0.6*0.4);
\vertex (v6) at (0.9+0.9*0.4,1.4+0.6*0.4);
\vertex (v7) at (0.9+0.9*0.4,0.6-0.6*0.4);
\vertex (v8) at (0.9-0.9*0.4,0.6-0.6*0.4);
\draw[red,thick bos] (i3)--(v7);
\draw[red,thick bos] (i4)--(v6);
\draw[brown,thick bos] (v6)--(v2)--(v5);
\draw[red,thick bos] (v1)--(v3);
\draw[brown,thick bos] (v3)--(v4);
\draw[cyan,thick bos] (v4)--(v2);
\draw[brown,thick bos] (v8)--(v1)--(v7);
\draw[cyan,thick bos] (i1)--(v5);
\draw[cyan,thick bos] (i2)--(v8);
\draw[very thick] plot[mark=x,mark size=3.5,mark options={rotate=0}] coordinates {(v3)};
\draw[very thick] plot[mark=x,mark size=3.5,mark options={rotate=0}] coordinates {(v4)};
\draw[very thick] plot[mark=x,mark size=3.5,mark options={rotate=45}] coordinates {(v5)};
\draw[very thick] plot[mark=x,mark size=3.5,mark options={rotate=45}] coordinates {(v6)};
\draw[very thick] plot[mark=x,mark size=3.5,mark options={rotate=45}] coordinates {(v7)};
\draw[very thick] plot[mark=x,mark size=3.5,mark options={rotate=45}] coordinates {(v8)};
\end{feynhand} \end{tikzpicture}
=\frac{(\mathbf{f}^{W^+ W^- Z})^2}{2\mathbf m_1^2 \mathbf m_2^2 \mathbf m_3^2 \mathbf m_4^2 s_{14}} \tilde m_1 \tilde m_2 m_3 m_4 \mathbf m^2 \langle \eta_14\rangle[2\eta_3]([4\eta_2][\eta_13]+[43][\eta_1\eta_2]).
\end{aligned} \end{equation}
\end{itemize}

The internal line with more cross can be obtained by applying $J^+ J^-$. Similarly, we can derive other terms in the (14)-channel. 
For the (13)-channel, we can apply a symmetry argument to obtain the amplitude directly. This is possible because particles 3 and 4 are both Z bosons and have identical helicities. The relation is given by
\begin{equation}
\mathcal{M}_{(13)}(1,2,3,4)=\mathcal{M}_{(14)}(1,2,4,3).
\end{equation}

The contact term is given by
\begin{equation}
\Ampfour{1^+}{2^+}{3^-}{4^-}{\bosflip{1.37}{135}{brown}{cyan}}{\bosflip{1.37}{-135}{brown}{cyan}}{\bosflip{1.37}{-45}{brown}{red}}{\bosflip{1.37}{45}{brown}{red}}=(
\mathbf{f}^{W^+ W^- Z})^2\times\big(\langle\eta_1\eta_2\rangle[21]\langle34\rangle[\eta_3\eta_4]
+\langle\eta_14\rangle[\eta_41]\langle\eta_23\rangle[\eta_32]-2\langle\eta_13\rangle[\eta_31]\langle\eta_24\rangle[\eta_42]\big).
\end{equation}

The final MHC result is just summing over the above four contributions.

\subsection{Bootstrapping AHH amplitudes}

We can also derive the $WW \to ZZ$ amplitude within the AHH formalism. The building blocks are the 3-pt and 4-pt contact AHH amplitudes. In this formalism, the 3-pt $VVV$ amplitude  can be expressed in two forms.
\begin{itemize}
\item When all three vector bosons are described by the $(\tfrac12,\tfrac12)$ Lorentz representation, we have
\begin{equation}
\mathbf M(\mathbf{1},\mathbf{2},\mathbf{3})=
\frac{\langle\mathbf{12}\rangle[\mathbf{21}]\langle\mathbf3|\mathbf 1|\mathbf3]+\langle\mathbf{23}\rangle[\mathbf{32}]\langle\mathbf1|\mathbf 2|\mathbf1]+\langle\mathbf{31}\rangle[\mathbf{13}]\langle\mathbf2|\mathbf 3|\mathbf2]}{\mathbf m_1 \mathbf m_2 \mathbf m_3}.
\end{equation}
This form is similar to the Feynman amplitude, where each term corresponds to a contraction of three polarization vector $ \varepsilon$ and one massive momentum $\mathbf p$.

\item Alternatively, two vector bosons can be assigned the $(\tfrac12,\tfrac12)$ representation while the third is in the $(1,0)$ or $(0,1)$ representation:
\begin{equation}
\mathbf M(\mathbf{1},\mathbf{2},\mathbf{3})=
\left(\frac{[\mathbf{12}]\langle\mathbf{23}\rangle[\mathbf{31}]+\langle\mathbf{12}\rangle[\mathbf{23}]\langle\mathbf{31}\rangle}{\mathbf m_2 \mathbf m_3}+\frac{[\mathbf{12}][\mathbf{23}]\langle\mathbf{31}\rangle+\langle\mathbf{12}\rangle\langle\mathbf{23}\rangle[\mathbf{31}]}{\mathbf m_1 \mathbf m_3}+\frac{[\mathbf{12}] \langle\mathbf{23}\rangle\langle\mathbf{31}\rangle+\langle\mathbf{12}\rangle[\mathbf{23}][\mathbf{31}]}{\mathbf m_1 \mathbf m_2}\right).
\end{equation}
This form can be obtained by bolding the MHC $VVV$ amplitude.

\end{itemize}

Then, let us discuss the gluing technique for AHH amplitudes, using $WW \to ZZ$ as an example.  In the (12)-channel, two $VVV$ amplitudes can be glued together as
\begin{equation}
\begin{tikzpicture}[baseline=0.8cm] \begin{feynhand}
\vertex [particle] (i1) at (0,2.0) {$1$};
\vertex [particle] (i2) at (0,0) {$2$};
\vertex [particle] (i3) at (1.8,0) {$3$};
\vertex [particle] (i4) at (1.8,2.0) {$4$};
\vertex (v1) at (0.9,0.6);
\vertex (v2) at (0.9,1.4);
\draw[thick bos] (i3)--(v1);
\draw[thick bos] (i4)--(v2);
\draw[thick bos] (v1)--(v2);
\draw[thick bos] (i1)--(v2);
\draw[thick bos] (i2)--(v1);
\end{feynhand} \end{tikzpicture}
=\mathbf M(\mathbf{1},\mathbf{4},-\mathbf{P}_{14})\otimes
\mathbf M(\mathbf{P}_{14},\mathbf{2},\mathbf{3}),
\end{equation}
where $\mathbf{P}_{14}$  denotes the momentum of the on-shell internal line, and $\otimes$ indicates the contraction over the corresponding LG indices from the 3-pt amplitudes.

Writting the LG indices explictly, the on-shell gluing gives
\begin{equation} \begin{aligned}
|\mathbf P]^{\dot\alpha}|\mathbf P\rangle^{\alpha}\otimes[\mathbf P|_{\dot\beta}\langle\mathbf P|_{\beta}
&=|\mathbf P]^{\dot\alpha(I_1}|\mathbf P\rangle^{\alpha I_2)}\times [\mathbf P|_{\dot\beta I_1}\langle\mathbf P|_{\beta I_2}
=\mathbf{m}^2\delta^{\dot\alpha}_{\dot\beta}\delta^{\alpha}_{\beta}-\frac12\mathbf{P}^{\dot\alpha\alpha}\mathbf{P}_{\beta\dot\beta}, \\
|\mathbf P]^{\dot\alpha_1}|\mathbf P]^{\dot\alpha_2}\otimes[\mathbf P|_{\dot\beta}\langle\mathbf P|_{\beta}
&=|\mathbf P]^{\dot{\alpha}_1(I_1}|\mathbf P]^{\dot{\alpha}_2 I_2)} \times [\mathbf P|_{\dot\beta I_1}\langle\mathbf P|_{\beta I_2}
=\mathbf m\delta^{(\dot\alpha_1}_{\dot\beta} \mathbf{P}^{\dot\alpha_2)\beta}, \\
|\mathbf P]^{\dot\alpha_1}|\mathbf P]^{\dot\alpha_2}\otimes[\mathbf P|_{\dot\beta_1}[\mathbf P|_{\dot\beta_2}
&=|\mathbf P]^{\dot{\alpha}_1(I_1}|\mathbf P]^{\dot{\alpha}_2 I_2)}\times \mathbf m[\mathbf P|_{\dot\beta_1 I_1}[\mathbf P|_{\dot\beta_2 I_2}
=\mathbf{m}^2\delta^{(\dot\alpha_1}_{\dot\beta_1}\delta^{\dot\alpha_2)}_{\dot\beta_2}, \\
&\vdots
\end{aligned} \end{equation}

To compare with the gluing results from MHC amplitudes, we select specific terms of the 3-pt massive amplitude and glue them in three cases:
\begin{align}
[\mathbf{14}]\langle\mathbf{4}\mathbf{P}\rangle[\mathbf{P}\mathbf{1}]\otimes\langle\mathbf{23}\rangle\langle\mathbf{3}\mathbf{P}\rangle[\mathbf{P}\mathbf{2}]
&=[\mathbf{14}]\langle\mathbf{23}\rangle(\mathbf m^2\langle\mathbf{43}\rangle[\mathbf{21}]-\frac12\langle\mathbf4|\mathbf{P}_{14}|\mathbf1]\langle\mathbf3|\mathbf{P}_{14}|\mathbf2]),\\
\langle\mathbf{1}\mathbf{4}\rangle[\mathbf{4}\mathbf{P}][\mathbf{P}\mathbf{1}]\otimes\langle\mathbf{23}\rangle\langle\mathbf{3}\mathbf{P}\rangle[\mathbf{P}\mathbf{2}]
&=\frac12\mathbf m \langle\mathbf{14}\rangle \langle\mathbf{23}\rangle([\mathbf{4}|\mathbf{P}_{14}|\mathbf{3}\rangle[\mathbf{12}]+[\mathbf{1}|\mathbf{P}_{14}|\mathbf{3}\rangle[\mathbf{42}]),\\
\langle\mathbf{1}\mathbf{4}\rangle[\mathbf{4}\mathbf{P}][\mathbf{P}\mathbf{1}]\otimes\langle\mathbf{2}\mathbf{3}\rangle[\mathbf{3}\mathbf{P}][\mathbf{P}\mathbf{2}]
&=\frac12\mathbf m^2\langle\mathbf{14}\rangle[\mathbf{23}]([\mathbf{42}][\mathbf{13}]+[\mathbf{43}][\mathbf{12}]).
\end{align}
These results are consistent with Eqs.~(\ref{eq:glue_VVVV_1}--\ref{eq:glue_VVVV_3}), if we bold the massless spinors.

\bibliographystyle{JHEP}
\bibliography{ref}

\end{document}